\documentclass[smallcondensed]{svjour3}     
\usepackage[ruled,linesnumbered]{algorithm2e}
\usepackage{amsmath}
 
\usepackage{tabularx}
\usepackage{booktabs}
\usepackage{threeparttable}
\usepackage{multicol}
\usepackage{multirow}
\usepackage{graphicx}
\usepackage{bm}	
\usepackage{latexsym}
\usepackage{listings}
\usepackage{xcolor}
\usepackage{enumitem}
\usepackage{longtable}
\graphicspath{{figure/}} 
\usepackage{makecell}
\usepackage{caption}
\usepackage[labelfont=bf,justification=raggedright]{caption}
\usepackage{subfigure}
\usepackage{threeparttable}
\definecolor{dkgreen}{rgb}{0,0.6,0}
\definecolor{gray}{rgb}{0.5,0.5,0.5}
\definecolor{mauve}{rgb}{0.58,0,0.82}
\usepackage{comment}
\usepackage{adjustbox}
\usepackage{sverb, longtable}
\usepackage{rotating, multirow}
\usepackage{amsmath}
\usepackage{rotating}
\usepackage{setspace}

\usepackage{ragged2e}
\usepackage{booktabs}

\usepackage{graphicx}

\usepackage{tablefootnote}
\usepackage[numbers,sort&compress]{natbib}

\usepackage[misc]{ifsym}
\usepackage{bbding}

\usepackage{wrapfig}
\usepackage{picinpar}
\usepackage{lipsum}
\usepackage{picins}
\usepackage{url}
\usepackage{amssymb}
\newcommand{\tabincell}[2]{\begin{tabular}{@{}#1@{}}#2\end{tabular}}

\smartqed  
\usepackage[switch]{lineno}
\usepackage{soul}
\soulregister\cite7
\soulregister\ref7 

\usepackage{xcolor}
\usepackage{soul}

\usepackage{graphicx}
\usepackage{soul}
\usepackage{longtable}

\definecolor{bluee}{RGB}{0,161,214}
\usepackage{xcolor}

\usepackage{lscape}
\usepackage{rotating, graphicx}
\usepackage[title]{appendix}

\usepackage{booktabs}
\begin{document}

    \title{Embedding API Dependency Graph for Neural Code Generation}
	
	\author{Chen Lyu$^1$ \and
		Ruyun Wang$^1$         \and
		\\
		Hongyu Zhang$^2$   \and
		Hanwen Zhang$^3$         \and
		\\
		Songlin Hu$^{4,5}$
	}

	\institute{	Chen Lyu (\Letter)\\
		\email{lvchen@sdnu.edu.cn}\at
		\and
		Ruyun Wang \\
		\email{ruyunw@outlook.com}\at
		\and  
		Hongyu Zhang \\
		\email{hongyu.zhang@newcastle.edu.au}\at
		\and   	            
		Hanwen Zhang \\
		\email{zhanghanwen0726@gmail.com}\at
		\and  
		Songlin Hu \\
		\email{husonglin@iie.ac.cn}\at
		\and   
		$1.$School of Information Science and Engineering, Shandong Normal University, Jinan, China.
		\\ 
		$2.$The University of Newcastle, Callaghan, NSW, Australia.
		\\
		$3.$Big Data Center of Shandong Province, Jinan, China.\\
		$4.$Institute of Information Engineering, Chinese Academy of Sciences, Beijing, China.
		\\
		$5.$School of Cyber Security, University of Chinese Academy of Sciences, Beijing, China.
	}

	\date{Received: date / Accepted: date}

	\maketitle
	
	\begin{abstract}
 \justifying The problem of code generation from textual program descriptions has long been viewed as a grand challenge in software engineering. In recent years, many deep learning based approaches have been proposed, which can generate a sequence of code from a sequence of textual program description. However, the existing approaches ignore the global relationships among API methods, which are important for understanding the usage of APIs. In this paper, we propose to model the dependencies among API methods as an API dependency graph (ADG) and incorporate the graph embedding into a  sequence-to-sequence (Seq2Seq) model. In addition to the existing encoder-decoder structure, a new module named ``embedder" is introduced. In this way, the decoder can utilize both global structural dependencies and textual program description to predict the target code. We conduct extensive code generation experiments on three public datasets and in two programming languages (Python and Java). Our proposed approach, called ADG-Seq2Seq,  yields significant improvements over existing state-of-the-art methods and maintains its performance as the length of the target code increases. Extensive ablation tests show that the proposed ADG embedding is effective and outperforms the baselines. 
		
	\end{abstract}
	
	\section{Introduction}
	\label{sec:introduction}
		 An ultimate goal of computer programming is to let computers to generate programs automatically. Automatic code generation from textual program descriptions\footnote{A textual program description can be a natural language description of requirements or a structured specification.} is a very challenging task.  For example, given a textual program description ``\texttt{Binds a Hibernate Session to the current thread}'', it is challenging for a computer to automatically generate the target code. Despite the challenge, researchers are exerting considerable effort towards achieving this goal.

Recently, neural machine translation (NMT) based approaches have been increasingly applied to code generation \cite{LingW, HayatiSA, YinP, SunZ, SunZ2, RabinovichM, MouL, LiG, QuirkC, DongL}. These approaches generally consider code generation as a sequence learning problem. A mature deep learning model called encoder-decoder, also known as the sequence-to-sequence (Seq2Seq) model \cite{SutskeverI, GehringJ, LuongMT, ChoK}, has been used. 
    In a Seq2Seq model, the textual program description and the target code are regarded as two sequences. The model is trained by feeding the description sequence, and then predicts the target code sequence.
	In this way, a textual program description can be translated into the corresponding target code through a deep neural network (i.e., neural code generation).

Due to the differences between textual program descriptions (e.g., in English) and target code (e.g., in Java), the existing approaches focus on improving the traditional Seq2Seq model to better support neural code generation \cite{LuongMT, ChoK, HuangPY, ZhangJ, BahdanauD, Costa-JussaMR}.
For example, Sun et al. \cite{SunZ} proposed a convolutional neural network (CNN) module in their Seq2Seq-based code generation approach. In particular, the authors used the abstract syntax tree (AST) to represent the target code and added three tree-based CNN decoders. As a result, the performance of their approach was greatly improved. To alleviate the long-range dependency problem, Sun et al. \cite{SunZ2} further proposed TreeGen, a tree-based transformer architecture for neural code generation. 
However, the existing approaches mostly use AST to represent the target code and lack in-depth exploration of program structure. They ignore the dependencies among the API methods, which were found useful for deep learning models to understand program semantics~\cite{hu2018summarizing}.

	In this paper, to better incorporate the inherent semantics of APIs in neural code generation, we introduce a model based on API dependency graph (ADG), which embeds the information about the API dependency structure into the Seq2Seq model for code generation. We focus on the dependencies among the APIs, including the following: 1) The invocation constraints for each API (i.e., if and only if all input parameters of the current API are provided can the API be invoked). Embedding the constraints into the program representation can guide the model to learn the correct API usage patterns so that illegal API calls (e.g., those with incomplete input parameters) can be removed.
	 2) The invocation order of the API sequence. Embedding API invocation order  enables the model to generate the target code in the correct order. From a graph perspective, the above API dependency structure of source code usually appears in the form of directed acyclic graph (DAG), which brings new challenges.

	In this paper, we propose an ADG-based Seq2Seq (ADG-Seq2Seq for short) approach to generating target code from textual program descriptions. The ADG-Seq2Seq architecture consists of three components: an encoder, an embedder, and a decoder. The encoder generates a vector representation of a textual program description. The embedder embeds the nodes of the ADG  into vector representations. The decoder then decodes the corresponding vector representations generated by the encoder and the embedder to predict the target code sequence. Unlike the existing research that represents an entire API sequence as a vector, this paper aims to encode the structural information about API dependencies into each API. In particular, we adopt a more fine-grained representation of API node embedding so that the decoder can learn more sophisticated program semantic information.

		To evaluate the effectiveness of our model, we perform extensive experiments on public benchmarks and compare our model quantitatively and qualitatively with the state-of-the-art models. The evaluation results show that the proposed ADG-Seq2Seq model significantly outperforms the state-of-the-art models in  terms of eight metrics (accuracy, BLEU, F1, CIDEr, RougeL, Rouge1, Rouge2 and RIBES) by large margins (ranging approximately from 9\% to 15\%) on two Java datasets. It also achieves  comparable results on the Python dataset. Furthermore, 
		extensive ablation tests show that our ADG embedding algorithm is effective and can significantly improve the performance of code generation. 
	
	This paper makes the following contributions.
	
	\begin{itemize}
		\item We propose ADG-Seq2Seq, a new approach to neural code generation, which builds an end-to-end encoder-embedder-decoder architecture that utilizes both textual program descriptions and API dependency information.
		
		\item We develop a novel ADG embedding algorithm that can convert an ADG into vectors while preserving the structural information implied in code. 
	\end{itemize}

	\noindent \textbf{Paper Organization}  \quad The rest of this paper is organized as follows. Section~\ref{sec:Work} surveys the related work and describes the background of our research. Section~\ref{sec:Moedl} presents an overview of the proposed model. Section~\ref{modeling} describes the details of the ADG-based embedder. Section~\ref{sec:set} provides the experimental settings. Section~\ref{sec:rq} states three  research questions (RQs), followed by experiments in Section~\ref{Results}. Section~\ref{sec:dicc} presents the qualitative analysis. Section~\ref{sec:threats} discusses the strengths and weaknesses of our approach and threats to its validity. Finally, Section~\ref{sec:Conclusion} summarizes this paper with some concluding remarks.

	\section{Background and Related Work}
	
	\label{sec:Work}

	\subsection{Graph Embedding}

    Graph embedding has proven to be very useful in a variety of prediction tasks. Many graph embedding algorithms for learning over graphs have been proposed, such as graph neural network (GNN) \cite{ScarselliF}, DeepWalk \cite{perozzi2014deepwalk}, Node2vec \cite{grover2016node2vec}, graph convolutional network (GCN) \cite{kipf2016semi}, graph sample and aggregate (GraphSAGE) \cite{HamiltonW}, and graph attention network (GAT) \cite{velivckovic2017graph}. These methods embed a node into a vector that contains the graph information and can even infer unseen nodes or graphs by aggregating the information of  neighbourhoods. 
    
	Encouraged by the development of graph embedding algorithms, researchers \cite{zhang-graph, LiY, AllamanisM} proposed to embed source code structure by treating it as a graph. For example, Gu et al. \cite{zhang-graph} used a graph embedding algorithm to select API usage examples. In this study, object usage graphs were explored to model the source code; the graph kernel is used to embed the structural semantics of graphs into a high-dimensional continuous space, and it is demonstrated that such an embedding can preserve many aspects of the original graph and capture structural information such as loops, branches, and third-party method invocations in source code. Li et al. \cite{LiY} proposed a gated-GNNetwork (GGNN) using gated recurrent units to detect null pointers. The researchers represented the memory state of a program as a graph, with a graph node representing the memory address where the pointer is stored and a graph edge representing the value of the pointer. GGNN is then used to predict the program's Hoare logic to determine whether the program is correct or not: e.g., whether the memory is safe. Allamanis et al. \cite{AllamanisM} used GGNN to learn representations of C$\#$ programs. These authors represented the source code as an AST-based graph, where each node in the graph represents a token in the code. The edges of the graph represent the associations between nodes in the AST and the data dependencies between variables. The method uses GGNN to learn the graph representation of the code and obtain a feature vector representation of each token in the code. The framework based on this representation model achieves satisfactory accuracy in the tasks of variable naming and variable misuse correction.

   Some researchers also applied graph embedding algorithms to control/data flow graph of source code. For example, Phan et al. \cite{PhanA} converted C programs to assembly language code and built control flow graphs, which were then analysed using GCN and applied to program defect detection.  Li et al. \cite{LiY2} proposed a deep learning-based graph matching network (GMN) to determine graph similarity. GMN is applied to check the similarity between control flow graphs and shows better detection ability than the GNN-based model. Wang et al. \cite{WangW} proposed a GNN-based framework for detecting code clones. The framework constructs a graph representing the flow-augmented AST by adding edges representing the control and data flows to the AST. Brockschmidt et al. \cite{BrockschmidtM} added edges to an AST to build a program graph that can model the structure and data flow of a partial program. GNNs are then used to learn a program representation on such a graph for code completion.
			
	In the models discussed above, the graph embedding algorithms were mainly applied to object-usage graph, control/data flow graph, and flow-augmented ASTs.
	In contrast to these work, we focus more on two types of API dependencies in source code, namely API invocation constraints and invocation order. The goal is to build a new graph model to describe the unique structural and semantic information implied in API dependencies. Inspired by GNN, GCN, and GraphSAGE, we design an API dependency graph (ADG) based model, which can be used together with a Seq2Seq model for code generation.

	\subsection{Neural Machine Translation Models}
	
	Neural machine translation models in natural language processing (NLP) field provide helpful inspiration for the code generation task. 
	Recurrent neural network (RNN) \cite{williams1989learning} and it variants, especially long short-term memory (LSTM) \cite{KalchbrennerN}, are commonly used as the encoder and decoder in the Seq2Seq model~\cite{SutskeverI}. The Seq2Seq model transforms a source sequence into hidden vectors from which the decoder generates the corresponding target sequence word by word.
	Cho et al. \cite{ChoK} were the first to propose an RNN-based encoder-decoder neural network and used it for machine translation, laying a solid foundation for subsequent studies on Seq2Seq. Sutskever et al. \cite{SutskeverI} presented a general end-to-end approach to sequence learning and used two layers of LSTM as the encoder and decoder. The seq2seq model has achieved good results in machine translation.
	
	However, one disadvantage of the Seq2Seq model is that the last hidden state of the RNN in the encoder, namely the context vector, contains only a limited amount of information. The greater the text length is, the more the information could be lost; thus, the Seq2Seq model does not perform well on long text generation tasks \cite{xu2015show}.  The attention mechanism \cite{Mnih2014} can
	improve neural machine translation by selectively focusing on parts of the source sentence during translation. For example, Luong et al. \cite{LuongM} proposed an attention-based Seq2Seq approach, which applied both global and local attention mechanisms.
	Vaswani et al. \cite{VaswaniA} proposed the Transformer architecture, which is based solely on attention mechanisms. 
	Currently, the Seq2Seq model with attention has been widely adopted in neural machine translation.

	\subsection{Code Generation Models}

	Over the years, many grammar-based code generation models have been proposed. As early as 1989, Alfred V. Aho at AT\&T Bell Labs proposed a code generation approach using grammar-tree matching \cite{AhoAV}. 
	AST \cite{ZhangJ} and other grammar based models \cite{DongL, RabinovichM, YinP, SunZ} have also been developed. Dong and Lapata \cite{DongL} proposed an approach to generating target code that could capture the syntactical structure of a program along with the AST model. Quirk et al. \cite{QuirkC} presented an approach that learned to map a textual program description from simple ``if-then" rules to executable code. 
	All the above-mentioned approaches  have achieved relatively good performance for specific program languages. However, these approaches are primarily based on rules and human-defined features and are restricted to specific applications.
	
	Recently, deep learning based approaches have been proposed for code generation. For example, Mou et al. \cite{MouL} envisioned an end-to-end program generation scenario using RNN. However, the RNN-based model encountered difficulty in handling long-term dependencies. The long dependency problem is a common dilemma, even with LSTM. Therefore, Sun et al. \cite{SunZ} replaced LSTM in the decoder with a CNN to capture features in different regions via a sliding window. Sun et al. \cite{SunZ2} further proposed a tree-based transformer architecture for code generation that alleviated the long dependency problem with the attention mechanism in Transformer. The above methods mainly use AST as the source code representation and learn syntactical features from the ASTs. Source code is then generated based on the learned features. 
	
	Although the current approaches have achieved good results in code generation tasks, they ignore the importance of API dependencies in source code. They do not adequately consider whether the generated program satisfies API dependencies, i.e., a set of structural and semantic constraints such as ``do not use undeclared or unprovided variables as arguments for method invocations'', ``invoke APIs with the correct types of variables/parameters'', and ``follow the legal order in which the program invokes API methods''. Representing and learning these constraints is challenging. Murali et al.~\cite{murali2017neural} discussed this challenge and proposed to generate programs based on code skeletons. Their proposed method first generates a code skeleton from a defined semantic label and then converts it into code. However, their method does not accomplish the task of generating code from textual program descriptions. Furthermore, the dependency between the two APIs in the code skeleton is specified by sibling and descendant relationships, which is different from our definition of API dependencies. In this paper, our proposed ADG-Seq2Seq takes a global perspective and uses ADG to model API dependencies. A new ADG embedding algorithm is proposed to integrate such structural information, which is shown to be effective for learning API usage and code generation.

	\section{Proposed Model}\label{sec:Moedl}    
	
	\begin{figure*}[htbp]
		\centering
		\includegraphics[width=1.2\linewidth]{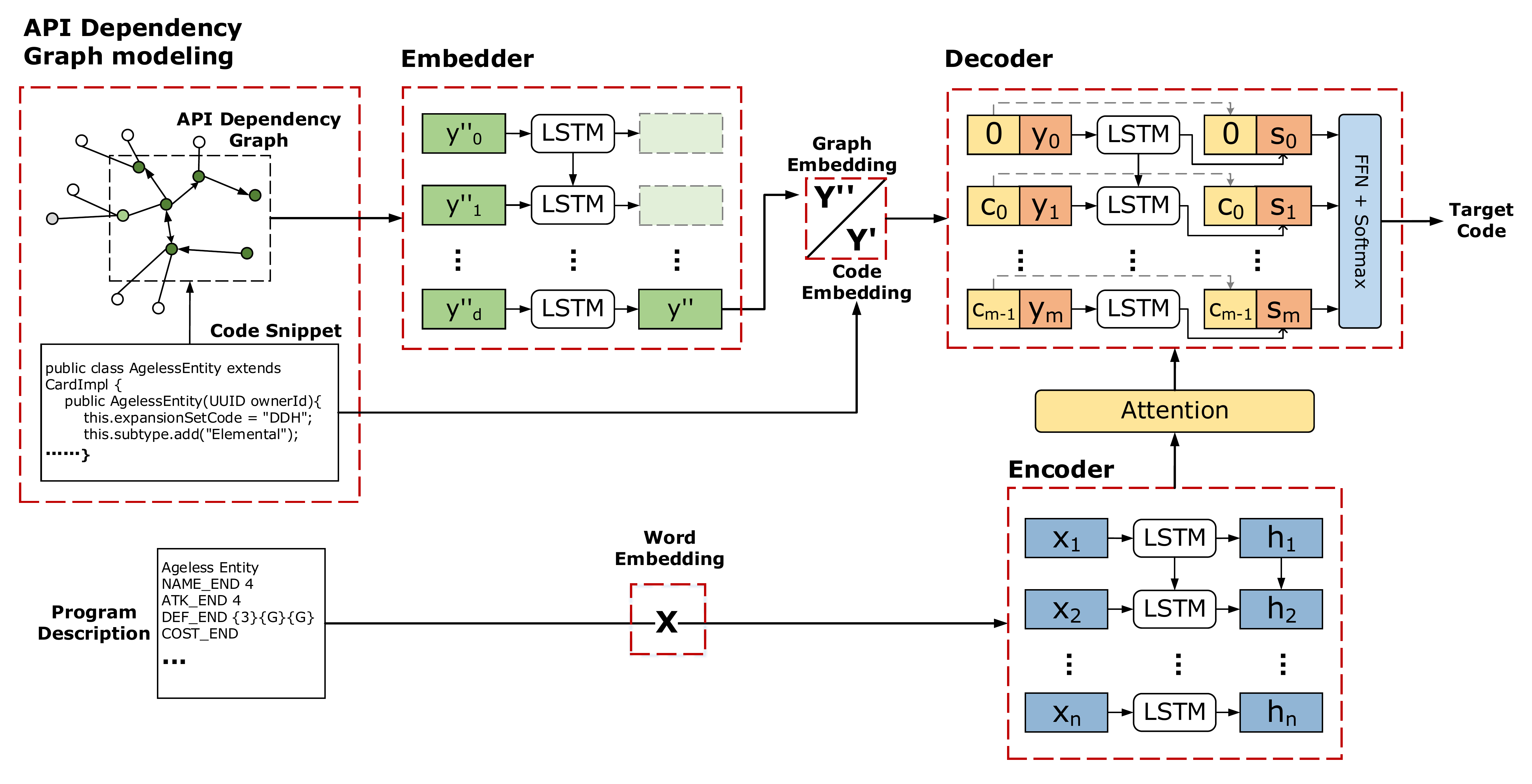}
		\caption{\textbf{Overall framework of our approach.} This ADG-embedding-based Seq2Seq framework can learn the structure of the program-oriented graph to generate target code automatically from its corresponding textual program description.}
		\label{fig:figure1}
	\end{figure*}
	
	In this paper, we propose an ADG-Seq2Seq model for code generation. Our model integrates Seq2Seq with a new module called  $embedder$ to incorporate API dependency information that is captured by an ADG. The overall structure of the proposed approach is shown in Fig.~\ref{fig:figure1}. Specifically, the encoder accepts the embeddings of a textural program description. The decoder fuses the target code embeddings (involving node embeddings produced by the embedder for API methods  and code embeddings for other program tokens) with the context vectors (generated by an attention mechanism) to generate the corresponding code sequence.
    In an ADG, each node represents an API method, and each node embedding is a node vector representation generated by the ADG embedding algorithm. Since node embeddings incorporate neighbouring nodes, associated edges and their dependencies, properties such as ADG graph structure and node dependencies are preserved. Therefore, using node embeddings as inputs to the decoder helps capture accurate information about API dependencies and program structure.
    Such an encoder-embedder-decoder framework enables the decoder to embed more information from the input and, consequently, generate a more semantically inherent method invoking sequences to form the entire program. We describe the proposed model in detail in this section.

	\subsection{Encoder}
	The tokens $ w_1, w_2, \cdots, w_t $  in a textual program description sentence will be embedded into the vectors $ x_1, x_2, \cdots, x_t $ through a lookup table $ LUT \in \mathbb{R}^{V \times d} $, where $ V $ is the vocabulary size. 
	The encoder is defined as 
	\begin{equation}
		\bm{h}_{t}=LSTM_{enc}({x}_{t},\bm{h}_{t-1})
	\end{equation}
	where $ \bm{h}_{t} $ is the hidden state of input vector $ {x}_{t} $, $ t $ is the length of $ {x}_{t} $, and $ LSTM_{enc} $ is the mapping function between the hidden states and the input sequence. The encoder receives each word embedding from the input sequence along with the hidden state at the previous time and then outputs the hidden state at the current time.
	
	Specifically, we embed the input sequence by indexing each word in the dictionary and then obtain the vector sequence $ \bm{x}_{1},\bm{x}_2,...,\bm{x}_{t} $. Moreover, we use a rectified linear unit (ReLU) activation function to connect each neural network layer:
	\begin{equation}
		\bm{x}^{l}_{t}=ReLU(\bm{W}^{l}[\bm{x}^{l-1}_{t-s},...,\bm{x}^{l-1}_{t+s}])
	\end{equation}
	where $ l=1,2,...,L $ indicates the layer in the neural network.

	To reinforce the memory of future and historical information in the input sequence, we feed the extracted features $ \bm{x}_{1},\bm{x}_{2},...,\bm{x}_{t} $ into LSTM. This approach can connect input vectors to the cell state, which can receive information processed by LSTM. The above process of LSTM can be formulated as:
	\begin{equation}
		\begin{pmatrix} \bm{i}_{t} \\ \bm{f}_{t} \\ \bm{o}_{t} \\ \widetilde{{\bm{C}}_{t}} \end{pmatrix} = \begin{pmatrix} \sigma \\ \sigma \\ \sigma \\ \tanh \end{pmatrix} \cdot \begin{pmatrix} [\bm{h}_{t-1},\bm{x}_{t}]\cdot \begin{bmatrix} \bm{W}_{i} \\ \bm{W}_{f} \\ \bm{W}_{o} \\ \bm{W}_{c} \end{bmatrix} + \begin{bmatrix} b_{i} \\ b_{f} \\ b_{o} \\ b_{c} \end{bmatrix} \end{pmatrix}
	\end{equation}
	\begin{equation}
		\bm{C}_{t}=\bm{f}_{t}*\bm{C}_{t-1}+\bm{i}_{t}*\widetilde{\bm{C}_{t}} 
	\end{equation}
	\begin{equation}
		\bm{h}_{t}=\bm{o}_{t}*\tanh (\bm{C}_{t})
	\end{equation}
	where 	$ \bm{i}_{t}, \bm{f}_{t}, \bm{o}_{t}, \widetilde{{\bm{C}}_{t}}, \bm{h}_{t} $ denote the input, forget, output, memory gates and hidden states of each LSTM unit; $ \bm{W}_{i}, \bm{W}_{f}, \bm{W}_{o}, \bm{W}_{c} $ denote the weight matrices  of input, forget, output and memory gates of each LSTM unit; and $ b_{i}, b_{f}, b_{o}, b_{c} $ denote the bias terms of the input, forget, output and memory gates of each LSTM unit, respectively. $ \sigma $ is a sigmoid function used to calculate a number between 0 and 1, where 1 is fully reserved and 0 is completely discarded.

	\subsection{Embedder}
	\label{sec:embedder}
	The embedder aims to generate graph embeddings by converting each node in the ADG into a vector expression; then, the decoder uses the vector expressions to gather program's structural information, i.e., API dependencies. To further exploit the global and sequential information of programs, we design an embedder to represent the API dependencies implied in the target code based on its dependent class libraries and utilize the ADG embedding algorithm to vectorize the nodes in the graph.
	
	Each node $ m \in \mathcal{M} $ in a code snippet is represented through a one-hot encoding $ h_m^0 $. To incorporate the graph information, we present the ADG embedding, which fuses node $ m $ and its neighbours' information; this process is described in detail  in Section \ref{modeling}. The graph embedding of node $ m $ that contains graph information is represented as $ h_m^k $, where $ k $ is the hop size. 

	\begin{equation}
		h_m^k = LSTM_{emb}(\bm{h_{\mathcal{U}}^{k-1}})
	\end{equation}
	Here, $ \bm{h_{\mathcal{U}}^{k-1}} = (h_{u1}^{k-1}, h_{u2}^{k-1}, \cdots, h_{un}^{k-1}) $ represents the graph embeddings of neighbours of node $ m $ in the $ k\text{-}1_{th} $ update; $ u1, u2, \cdots, un \in orderset_m $; and $ h_m^k $ is the graph embedding of node $ m $ in the $ k_{th} $ update. In this study, we set the hop size to $ k=2 $. 
	
	To incorporate the embedder into the decoder, the decoder selectively adopts the node embedding of the embedder's output as its query  (the token to be expanded in the code sequence). Specifically, when the token in the query represents an API method, the query adopts a node embedding vector (generated by the embedder); otherwise, the query adopts a word embedding vector.

	\subsection{Decoder}

	The decoder is designed to generate the target code sequence based on its textual program description. The core network of the decoder is LSTM, which produces the t-th token $y_t$ by predicting its probability as follows:
	\begin{equation}
		p(\bm{y_{t}} \mid \bm{y}_{<t}, \bm{x})\sim Softmax(Linear(f(\bm{y}_{t-1},\bm{h}_{t},\bm{c}_{t},\bm{s}_{t})))
	\end{equation}
	where $y_{<t}$ (i.e., $\{y_1, y_2, ..., y_{t-1}\}$ denotes the previously generated partial code, $ Linear $ is a linear transformation function in the decoder, $ f $ represents the LSTM activation function, $ \bm{y}_{t} $ is a word token feature or a node/method feature generated by the graph embedding algorithm of the ADG (see Section \ref{ADG}), $ \bm{h}_{t} $ is the $ t_{th} $ hidden state computed by the encoder, $ \bm{c}_{t} $ denotes the $ t_{th} $ context vector computed by the attention mechanism, and $\bm{s}_{t} $ is the
	$ t_{th} $ hidden state computed by the decoder. 
	
	The attention mechanism proposed by Mnih et al. \cite{Mnih2014} utilizes a matching function and the tangent change to calculate the weight values corresponding to textual program description features and target code features, thereby emphasizing the part that has the greatest impact on the current target code feature.
	
\begin{figure}[!htbp]
		\centering
		\includegraphics[width=0.75\linewidth]{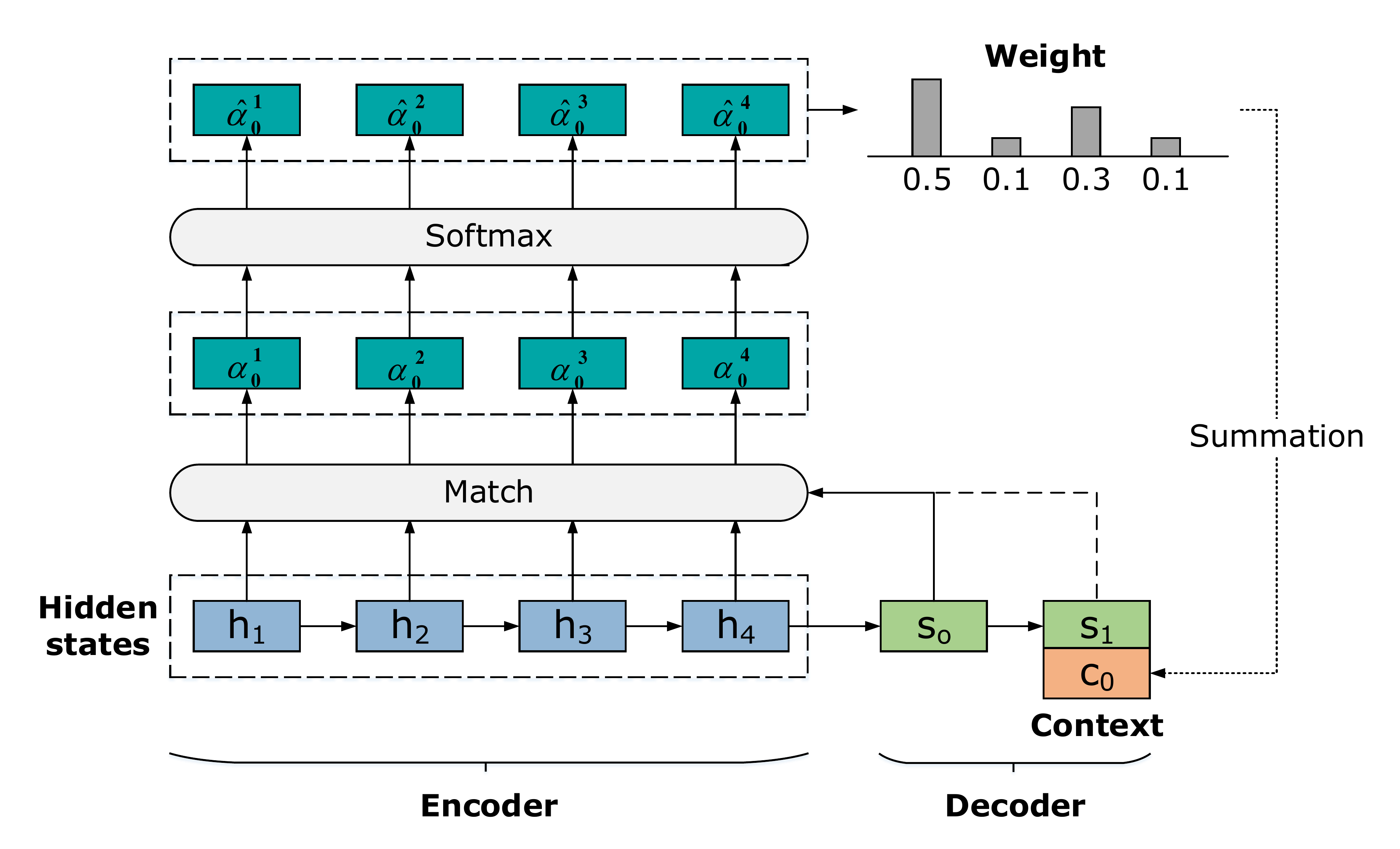}
		\caption{\textbf{Overview of the attention mechanism.} The input of the encoder is distributed with different weights according to each input of the decoder.}
		\label{fig:figure8}
	\end{figure}

	In the traditional Seq2Seq neural network, the encoder of the input sequence is a fixed-length hidden vector. Nevertheless, it might be inadequate to encode the input sequence into a fixed length and assign the same weight to each word when decoding. Hence, we add a global attention mechanism \cite{Mnih2014} that calculates the weight for the importance-based current input to our network. This mechanism is shown in Fig.~\ref{fig:figure8}.
	
	In our approach, the attention mechanism is used to calculate a match for the output of the current decoder and encoder at each moment. Given the hidden states
	$ {\bm{h}_{1},\bm{h}_{2},...,\bm{h}_{t}} $ 
	for each output of the encoder and the current state $ {\bm{s}_{i},\forall i\in \{0,...,t\}} $ for the output of the decoder, where $ \bm{s}_{0} $ is the output vector initialized by the first time step of the decoder, the matching value is computed as:
	\begin{equation}
		\bm{u}^{t}_{i}=\bm{h}^{T}Ws_{i}
	\end{equation}
	where $ \textbf{W} $ represents a weight matrix. We then can normalize the matching degree and calculate the weight of each hidden state by:
	\begin{equation}
		\bm{\alpha}^{t}_{i} =softmax(\bm{u}^{t}_{i})
	\end{equation}
	where $ \bm{\alpha}^{t}_{i} $ is a number between 0 and 1, and is the weight value for each input. Afterwards, we calculate the sum of weight vectors by:
	\begin{equation}
		\bm{c}_{t}=\Sigma _{i}\bm{\alpha} ^{t}_{i}\bm{h}_{i}
	\end{equation}
	The calculated vector sum is concatenated with the input vector in the next time step of the decoder to obtain a new input to the decoder. This completes the loop process of the attention mechanism. 
	
	With the attention mechanism, for each input vector of the decoder, different weights are assigned to the hidden state of the encoder at each moment. By retaining the weighted hidden states of the encoder LSTM at each time step, the attentive encoder can selectively learn the current input textual program description to correlate the output code of the attentive decoder with its most relevant description word. 
	\begin{equation}
		s_t = LSTM_{dec}(y_{t-1}, h_t,c_t)
	\end{equation}
	where $ s_t $ is the output of the decoder at time step  $ t $. 
    If $ y_{t-1} $ indicates to a node $ m \in \mathcal{M} $ in the ADG, the input to the decoder of the next time step is replaced with node $ m $'s graph embedding $ h_m^k $, which has the same dimension as vector $ y_{t} $. 
	
	Finally, a fully connected feedforward network with softmax follows:
	\begin{equation}
		p(y_t \mid y_{t-1}) = Softmax(FFN(s_t,c_t))
	\end{equation}
	Using this network, we can obtain the generated corresponding token by mapping. Specifically, we concatenate the outputs of the decoder with the corresponding context vectors. They are fed to a two-layer perceptron, where a softmax activation function is used in the last layer to predict the probability of the next token, as shown below: 
	\begin{equation}
		p(y_{i} \mid \cdot)=\frac{exp{c^{(MLP)_{i}}}}{\Sigma ^{R}_{j=1}exp{c^{(MLP)_{j}}}}
	\end{equation}
	where $ C^{(MLP)_{i}} $ is the input logic of the softmax function, and $ R $ is the number of candidate tokens. By mapping to the corresponding token, the source code is generated eventually. The evaluation stops if token ``$ \langle $EOS$ \rangle $'' is generated or the output reaches the maximum length. 

	\subsection{Model Training}
	We pre-process the inputs of the encoder as follows: replace special characters with spaces; number and count each word simultaneously. In addition, for the target code input by the decoder, we analyse and extract the API dependencies from the target code.\footnote{Specifically, we resort to Spoon (http://spoon.gforge.inria.fr/) and Javassist (http://www.javassist.org/) for Java-based programs' analysis.} The resulting APIs are classified according to their input and output parameter types. This classification information is used in the subsequent training for graph embedding. 
	
	\noindent\textbf{Loss Function}\quad During training, a unified loss function is used to perform end-to-end joint training of the encoder-decoder network and the ADG embedding network. The loss is backpropagated to the encoder, embedder, and decoder simultaneously. In other words, the parameters in the three networks are adjusted together according to a single cross-entropy loss function defined by:
	\begin{equation}
		Loss=-\frac{1}{|\boldsymbol{y}|}\sum_{t=1}^{|\boldsymbol{y}|} \log p(y_{t} \mid {y}_{<t}, \boldsymbol{x})
	\end{equation}	
	
	The essence of joint training is a gradient descent regression of parameters using uniform losses for the three networks, i.e., encoder, embedder, and decoder. Joint training implies that the graph embedding network is not trained separately. The graph embedding vector is different for the same node in each iteration, and the embedder's weights are updated with the entire model for each batch. Take one training batch as an example: the encoder receives the textual program description as an input, and the decoder receives code as an input query. We iterate through the current sequence of input code tokens. If the current token is an API method, we use the node embedding as input for the current time step of the decoder. Otherwise, we use the code embedding of the token as the input. The cross-entropy loss function is then computed, and the gradient descent is performed uniformly at the end of one batch. The embedding vector for each node is then updated by the embedder.

	\noindent \textbf{Optimization}\quad Our model is optimized by the Adam optimizer \cite{KingmaDP} with its default settings. According to \cite{VaswaniA}, we change the learning rate as follows:
	\begin{equation}
		lrate = d^{-0.5}_{model} \cdot min(step_{num}^{-0.5}, step_{num} \cdot warmup_{steps}^{-1.5})
	\end{equation}
	where $d_{model}$ is the dimension of the output layer (dense), $step_{num}$ is the number of training steps, and $warmip_{steps}$ is set to 4000 following \cite{VaswaniA}. The above settings can increase the learning rate at the beginning of the training process and then reduce the learning rate proportionally. In addition, a beam search of width 5 is utilized to approximate the global inference \cite{SunZ} during testing. 
	
	\noindent\textbf{Initialization} \quad We initialize the weight parameters of the encoder, the decoder, and the embedder randomly. The program description features, which are pre-processed into a two-dimensional vector sequence, constitute the input of the encoder. The node embedding features generated by the initialized embedder are the input of the decoder, and the feature vectors of nodes are the input of the embedder.

	\section{ADG-based Embedder}
	\label{modeling}
	\subsection{ADG Modelling}

    A more comprehensive representation of the program structure can improve the effectiveness of a code generation model. To this end, an ADG model is constructed to represent the dependencies among API methods. 

	\begin{figure}[htbp]
		\centering
		\subfigure[A snippet of the Java program. ]{
			\includegraphics[width=0.75\linewidth]{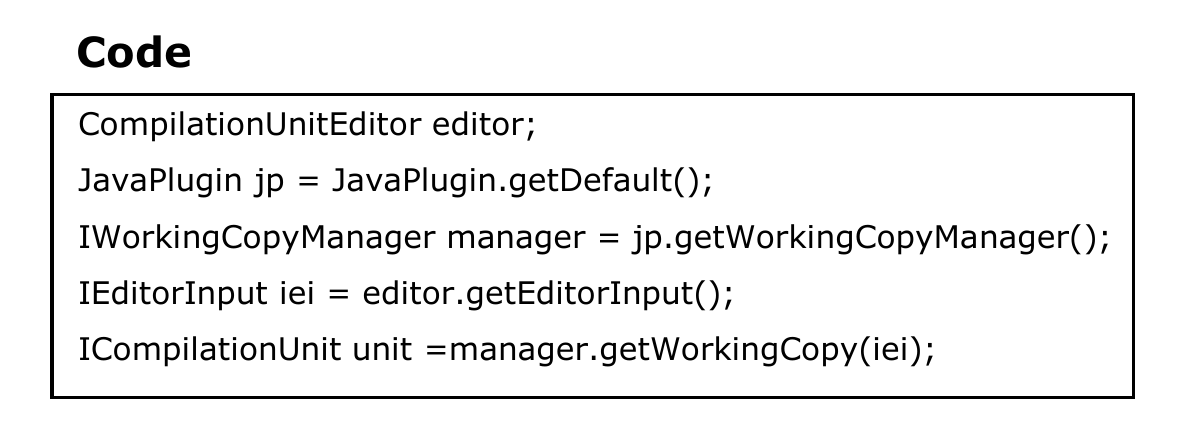}
			\label{fig:Figure2a}
		}
		
		\subfigure[ADG representation of API dependencies involved in (a). The labels of the edges are ``tags" that denote the data types.	]{
			\includegraphics[width=0.8\linewidth]{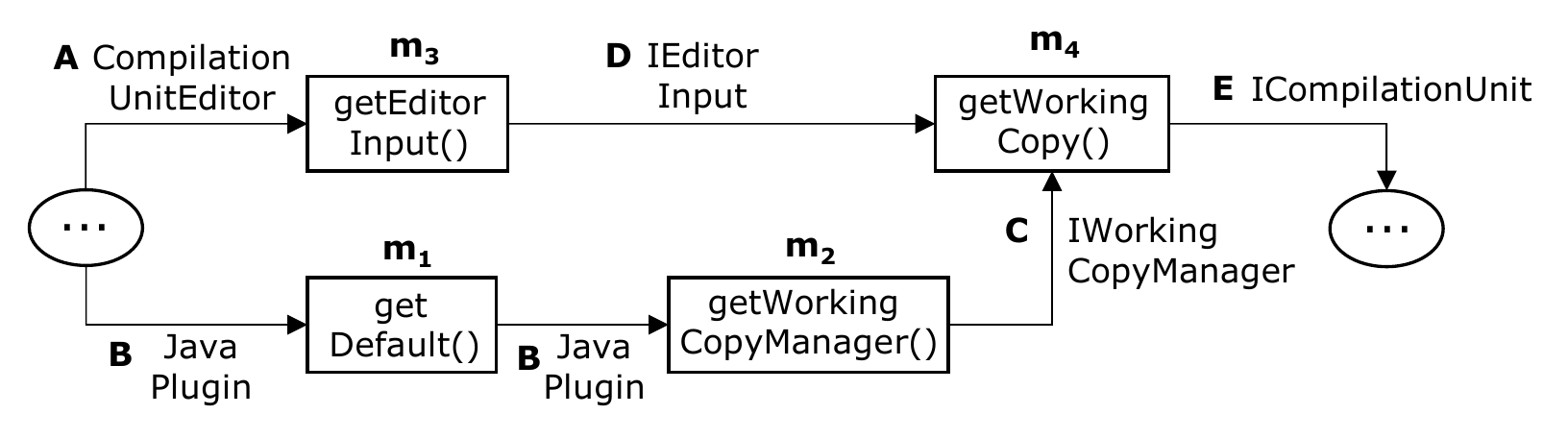}
			\label{fig:Figure2b}
		}
		
		\caption{\textbf{Example of ADG modelling.}}
		\label{fig:figure2}
	\end{figure}
	
    An illustrative example of ADG modelling is shown in Fig.~\ref{fig:figure2}. A sample code snippet from the codebase is shown in Fig. \ref{fig:Figure2a}. Fig. \ref{fig:Figure2b} illustrates a part of the ADG that represents a partial view of the API methods and their dependencies involved in this code snippet.  
    In the ADG shown in Fig.~\ref{fig:Figure2b}, method ``$m_4$" corresponds to node  ``$m_4$" with input types ``$C$, $D$" and output type``$E$". Each node has some dependent parameters as declared by its inputs, e.g., node $m_4$ can be invoked and perform its function after its two inputs become available, and both instances of ``$C$" and ``$D$" are provided by nodes $m_2$ and $m_3$. A dependent parameter or a constraint resembles the real world in that many things can be moved on only after their dependent parameters have been satisfied. Therefore, ``$m_4$" has two dependent parameters,  ``$C$" and ``$D$", which are the input types of ``$m_4$", so ``$IN_{m_4}$" equals \{$C,D$\}. Similarly,  ``$m_4$" has one return type, ``$E$", which is the output type of ``$m_4$", so ``$OUT_{m_4}$" equals \{$E$\}. Node ``$m_2$" can satisfy one of the dependent parameters of ``$m_4$", ``$C$". Thus, there is an edge from ``$m_2$" to ``$m_4$", and the tag is ``$C$".
	
	Compared with other source code representations such as AST and control flow graph (CFG), ADG has three main advantages: (1) it can describe the API  dependencies from a global perspective; (2) it is more capable of capturing long-range structural dependencies among methods than a tree-based model; and (3) the reachability of the ADG allows it to holistically reflect the invocation dependencies between methods, thereby improving performance (as described in the next subsection).

	\subsubsection{ Graph Construction and Reachability Judgement}

	\begin{table}[htbp]
		\centering
		\small
		\caption{General terms. $1\leq i \leq N$, where $N$ is the number of API methods}
		\label{tab:terms}
		\begin{tabular}{|l|l|} \hline
			{\em Terms} & {\em Meaning}  \\ \hline \hline
			
			\tabincell{l}{API~ Method\\ $(m_i)$  }    & \parbox{6cm}{\rule{0pt}{2ex}An API method $m_i$ $(1 \leq i \leq n)$ is considered as a tuple. $m_i =\{{I}_{m_i}, {O}_{m_i}\}$.	        
			${I}_{m_i}$ denotes the input parameters of $m_i$, and ${O}_{m_i}$ denotes the output parameters of $m_i$.  }     \\ \hline
			
			\tabincell{l}{Parameter \\Match}  &\parbox{6cm}{ \rule{0pt}{2ex}Two API method parameters, $p_{a}$ and $p_{b}$, are matched by not only their types but also inheritance relationships. We use the ``exact" match to judge the parameter matching relationship.
				Let $Type(p) $ be the type to which parameter $p$ belongs.
				Here, $p_a$ and $p_b$ can be exactly matched if $Type(p_a) $ is $subClassof$ or the same as $Type(p_b) $, where $subClassof$ is defined as the relationships of objects or classes through inheritance.}
			\\\hline
			
			\tabincell{l}{API ~Method \\ Dependency}   & \parbox{6cm}{ \rule{0pt}{2ex}Given two API methods, $m_{a}$  has dependencies with $m_{b}$ if some outputs of $m_{a}$ can match some inputs of $m_{b}$. This case is labelled as
				${O}_{m_{a}} \bigcap {I}_{m_{b}} \neq  \emptyset.$
				Method $m_{a}$ has full dependencies with $m_{b}$ if ${O}_{m_{a}}\supseteq {I}_{m_{b}}$, and has partial dependencies with  $m_{b}$  if $({O}_{m_{a}} \bigcap {I}_{m_{b}} \neq  \emptyset )\wedge ( {O}_{m_{a}} \nsupseteq  {I}_{m_{b}}).$}\\ \hline
		\end{tabular}
	\end{table}

	\noindent \textbf{Construction}  \quad For a specific source code dataset, we first extract a set of API methods and their dependencies by retrieving all API signatures of the frameworks/libraries. The general terms are listed in Table~\ref{tab:terms}.

	We then can build an ADG, $\mathcal{G}=(\mathcal{M},\mathcal{E})$ for these API methods from a graph perspective.  Recall that the set contains $n$ API methods. Intuitively, we denote the $ith$ API method by $m_i$, $1 \leq i \leq n$. The graph $\mathcal{G} $ corresponding to the set is constructed as follows.
	\begin{itemize}
    \item Regard each API method as a node in $ \mathcal{M}$, $\forall m_i \in \mathcal{M} $, $(1 \leq i \leq n)$, where $i$ is the identifier $(id)$ of $m_i$ and $n$ is the number of nodes in the graph.
    
    \item Connect two nodes, $m_i$ and $m_j$ a directed edge in $\mathcal{E}$ from $m_i$ and $m_j$ if one of $m_i$'s outputs matches one of  $m_j$'s inputs, where each edge $e_{i,tag,j}$ = $(m_i,m_j,tag)$ $(1 \leq i, j \leq n)$. This edge is from node $m_i$ to node $m_j$ with the identifier $tag$ that denotes the output parameter type of $m_i$ and the input parameter type of $m_j$. Nodes $m_i$ and $m_j$ are the head and the tail of the edge, respectively. Node $m_i$ is the direct successor of $m_j$, and $m_j$ is the direct predecessor of $m_i$.
    
    \item Express the graph $\mathcal{G}$ as an inverted index table and a nodes table (see Section~\ref{sec:Data Structures}).
    \end{itemize}
	
    The definitions of the key concepts in the ADG are given below.
    
    $IN_i=\{ tag|e_{j,tag,i} \in E\}$~ $((1 \leq j \leq n) \wedge (i
	\neq j)).$ It is a union set of all tags of edges with
	tails of $m_i$. Variable $indegree_i$ is the number of its incoming edges, and $intagdegree_i$ is the number of different tags of its incoming edges, which is equal to  $|IN_i|$. Note that $intagdegree_i \leq indegree_i$. 
	
	$OUT_i=\{ tag|e_{i,tag,k} \in E\}$~ $((1 \leq k \leq n) \wedge
	(i \neq k)).$ It is a union set of all tags of edges with heads of $m_i$.  Variable $outdegree_i$ is the number of its outgoing edges, and $outtagdegree_i$  is the number of different tags of its outgoing edges, which is equal to $|OUT_i|$. Similarly, $outtagdegree_i \leq outdegree_i$.
			
\begin{figure}[!htbp]
		\centering
		\includegraphics[width=0.8\linewidth]{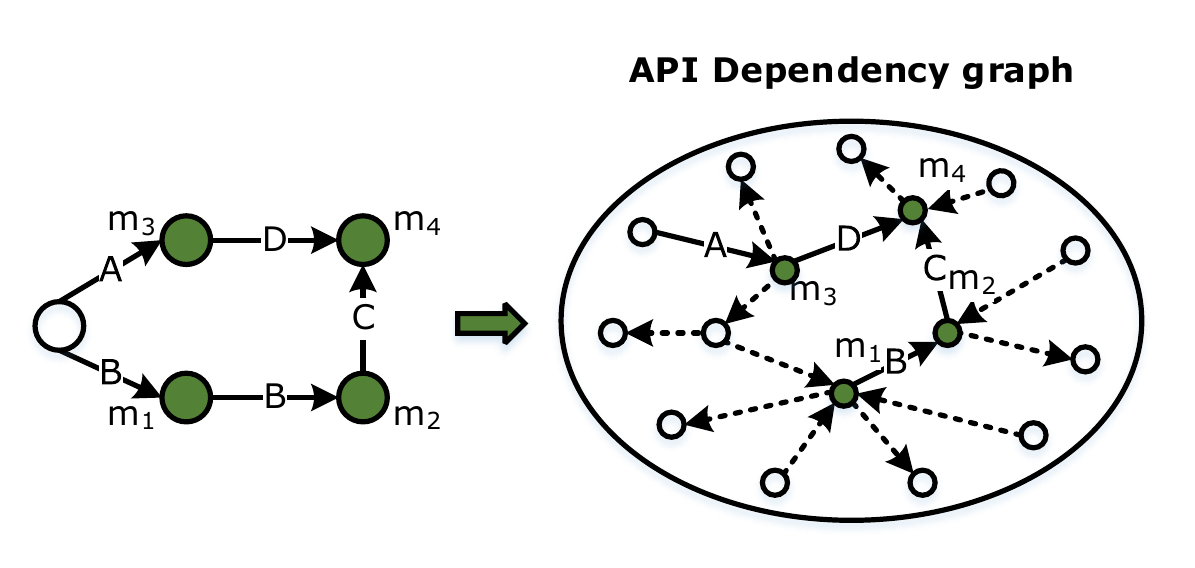}
		\caption{\textbf{Panoramic view of an ADG.} The left part is from Fig.~\ref{fig:Figure2b}. The ellipsis indicates the unlisted nodes in the ADG.}
		\label{fig:figure5}
	\end{figure}

	Note that the ADG represents the API dependencies in a codebase, not just those from the code snippet shown in Fig. \ref{fig:Figure2a}. Fig.~\ref{fig:figure5} shows a panoramic schematic of the ADG; the green nodes are from  Fig.~\ref{fig:Figure2b}. Notably, when inputting a textual program description, the API dependencies involved in the generated code snippet correspond to a part of the ADG instead of the entire ADG.
	
	\noindent \textbf{Reachability}  \quad The nodes in the ADG are different from those in a classic graph because our nodes may have multiple dependent parameters. Only when all dependent parameters are satisfied can the corresponding node/method be reachable/invoked. This special reachability of nodes is an essential property of the ADG, which is defined as follows. 
	
	\textbf{Definition 1} (Reachability) \textit{For any node $ m_i $ in the ADG, the condition is considered to be reachable if and only if the dependent parameters for $ m_i $ are all satisfied \footnote{An instance of the parameter type is provided by its predecessor method.}. In our ADG model, there might be one or more dependent parameters for invoking node $ m_i $. Only when these dependent parameters are satisfied simultaneously can node $ m_i $ be successfully reached.} 
	
	The special reachability here is different from the properties of the classic graph. A node $ v $ of a classic directed graph is reachable from another node $ u $ when there exists a path that starts at $ u $ and ends at $ v $.

	\subsubsection{Data Structures}
	\label{sec:Data Structures}
	
\begin{figure}[!htbp]
		\centering
		\includegraphics[width=0.6\linewidth]{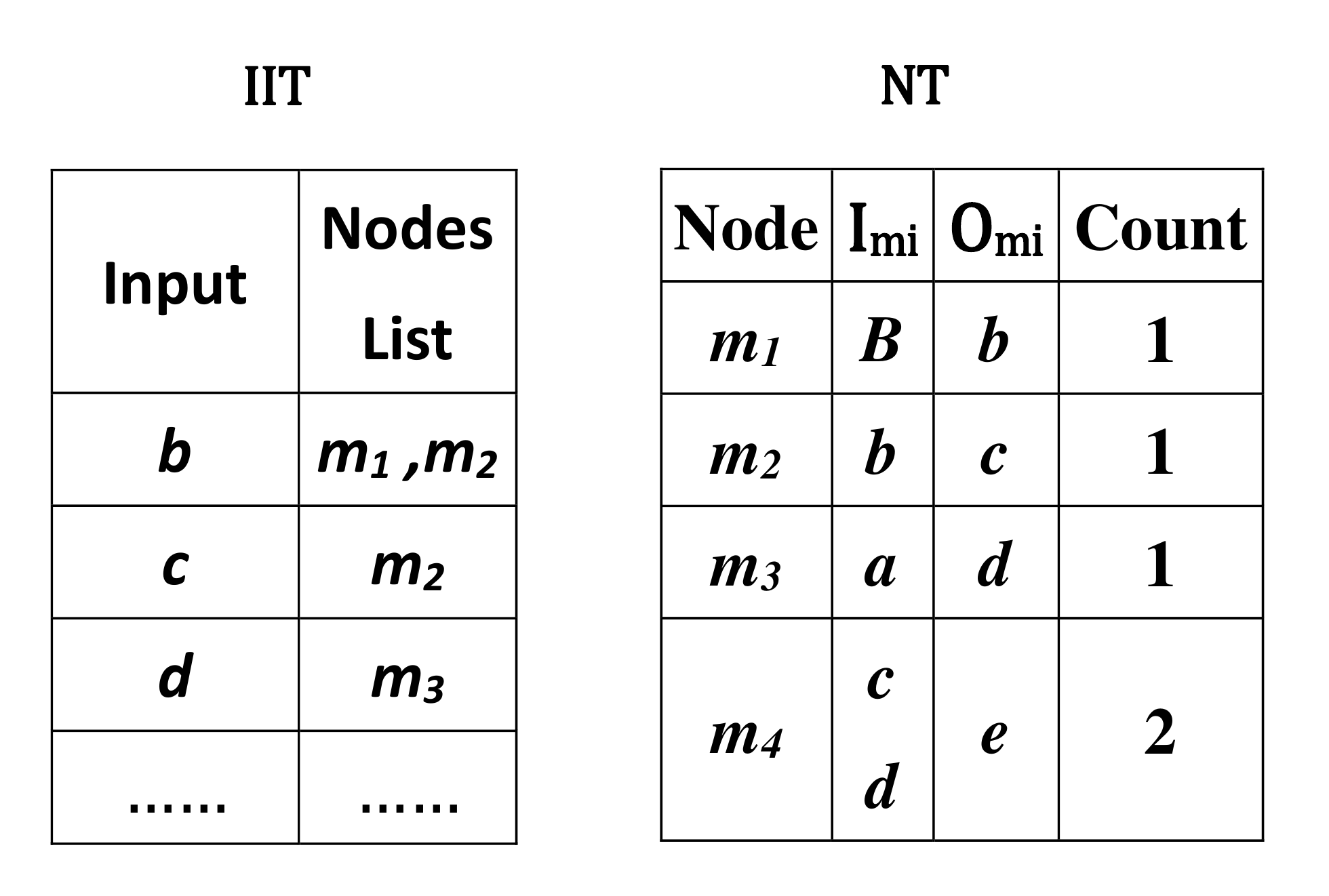}
		\caption{\textbf{Data structure of the example illustrated in Fig.~\ref{fig:figure2}} }
		\label{fig:data structureE}
	\end{figure}

	A significant challenge in our graph embedding algorithm is to efficiently compute reachability constraints (e.g., the providers of the nodes' dependent parameters) for individual nodes in the ADG. Such constraints are dependent on the graph structure. This issue will be addressed by using the specially designed data structures of our model, including the inverted index table and the nodes table. An illustrative example is shown in Fig.~\ref{fig:data structureE}.

    \textbf{Inverted Index Table (IIT)}: a table for representing and storing the ADG. The entries in the IIT are key-value pairs (parameter, nodes list), where the key is a parameter with the value being the list of nodes that need the parameter as an input. The use of the IIT can avoid the traditional graph storage structure, such as adjacency tables and adjacency matrices, which cause high computational complexity, and thus can effectively improve the computational efficiency of reachability constraints. For example, in the IIT the time complexity of retrieving all provider nodes of a dependent/input parameter is only O(1). In contrast, if an adjacency table or an adjacency matrix is used, all nodes (n in total) must be iterated over to determine their input parameters with the time complexity of O(n).

    \textbf{Nodes Table (NT)}: a table for storing nodes and related information.  The NT uses a quadruple to represent nodes in the form of \{$m_i$, $I_{m_i}$ , $O_{m_i}$, $Count$\}, where $m_i$ is the unique identifier of the node, and $I_{m_i}$ and $O_{m_i}$ are two lists that store {\tt IN,OUT} of the node. We use a counting mechanism to judge whether a node is reachable. In the NT, each node has a counter initialized to the number of its dependent parameters/inputs. Whenever a provider of some input parameter is retrieved, the count of the node that needs that parameter as an input is reduced by one. Thus, if the counter equals zero, then all dependent parameters of that node have been satisfied and it becomes a reachable node.

    It should be noted that since we use the IIT to construct the ADG, the following notation is used: $c$ is the average number of API input parameters, $p$ is the average number of individual input parameter providers, and $n$ is the total number of nodes in the ADG. Therefore, the time complexity of constructing the ADG is $O(n\times c\times p)$.
    	
	\subsection{ADG Embedding}
	\label{ADG}
\begin{figure}[!htbp]
		\centering
		\subfigure[\textbf{Example of a dependent parameter.} Node $ m_{4} $ is not satisfied, so this subgraph of the ADG is not reachable.]{
			\includegraphics[width=0.60\linewidth]{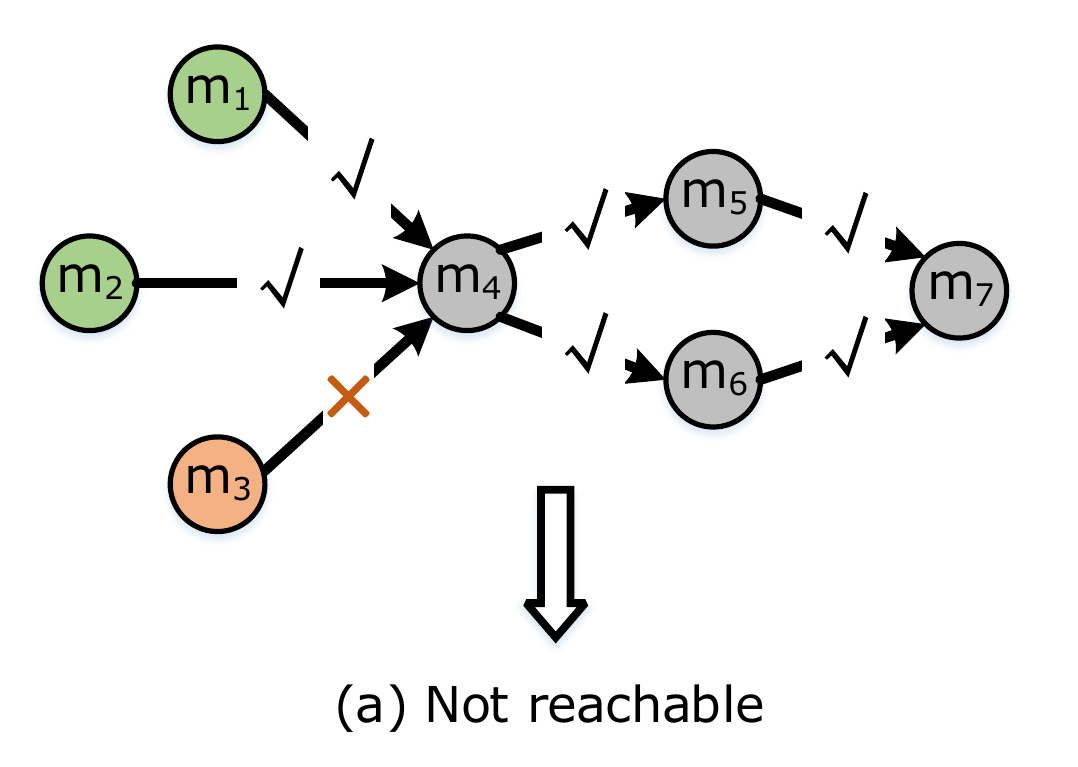}
		}
		
		\subfigure[\textbf{Example of each dependent parameter.} The dependent parameters for node $ m_{4} $ are satisfied, but for node $ m_{7} $, a dependent parameter is not satisfied. Accordingly, we can obtain a partially ordered set from nodes $ m_{1},m_{2},m_{3} $ to node $ m_{5} $, but not for nodes $ m_{6} $ and $ m_{7} $.]{
			\includegraphics[width=0.65\linewidth]{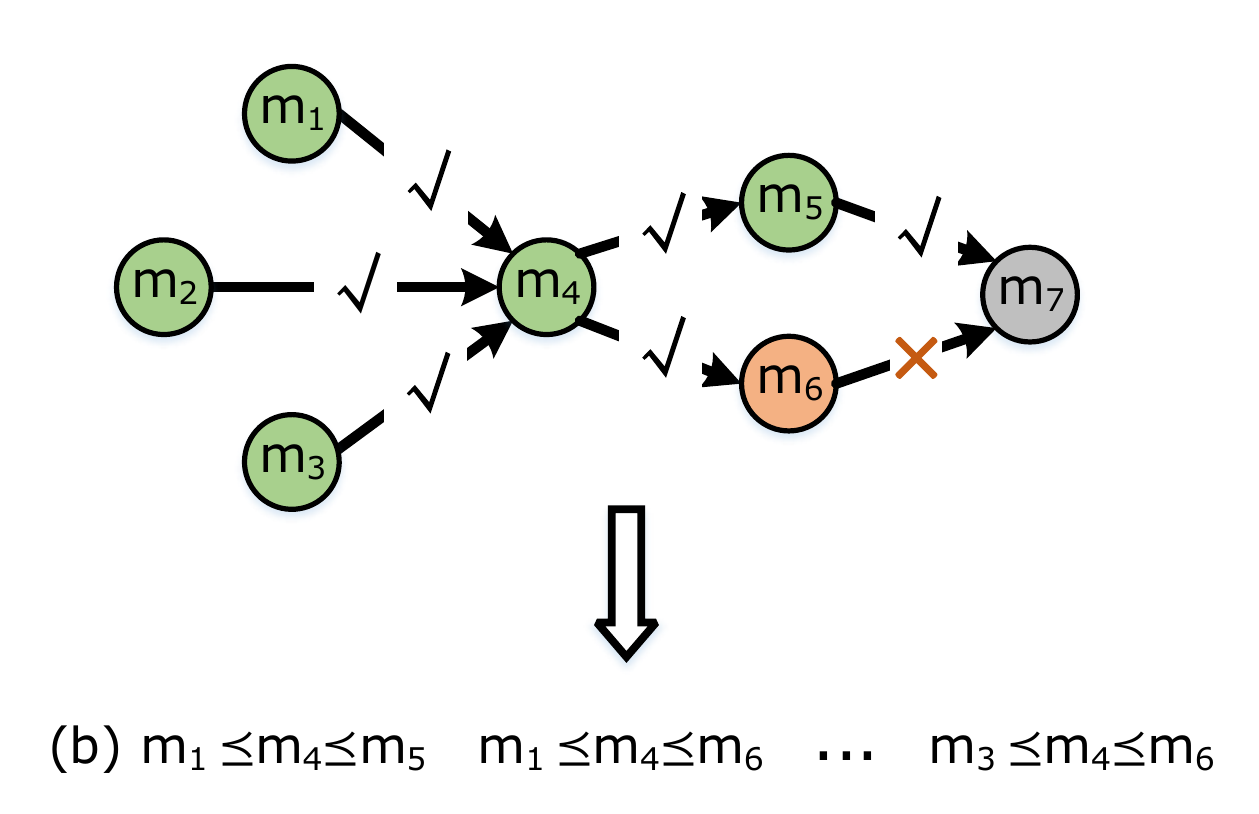}
		}
		\caption{\textbf{Motivations of the ADG embedding algorithm.}}
		\label{fig:figure6}
	\end{figure}

	After the construction of the ADG,	the next task is to produce a vectorized representation of nodes through graph embedding. Since the distinctive ADG is directed and tagged, in this paper, we propose a new graph embedding algorithm called \textbf{ADG} \textbf{embedding} \textbf{algorithm} (Algorithm~\ref{alg:r2p}), which extracts the features and embeds them into a low-dimensional space to retain the global structure information.
	Our algorithm is inspired by Hamilton et al. \cite{HamiltonW}. More specifically, it is based on the two motivations shown in Fig.~\ref{fig:figure6}:

	\begin{itemize}
		
		\item The special reachability of each node of the ADG is considered: if the dependent parameters of one node are not satisfied completely, then that node is unreachable. Considering this property helps avoid incorrect API invocations. 
		
		\item The invocation order among APIs is considered: the API invocation sequence is ordered. This constraint restricts the generated sequence to be in the correct order.
		
	\end{itemize}
	
\begin{figure*}[!htbp]
		\centering
		\includegraphics[width=1.1\linewidth]{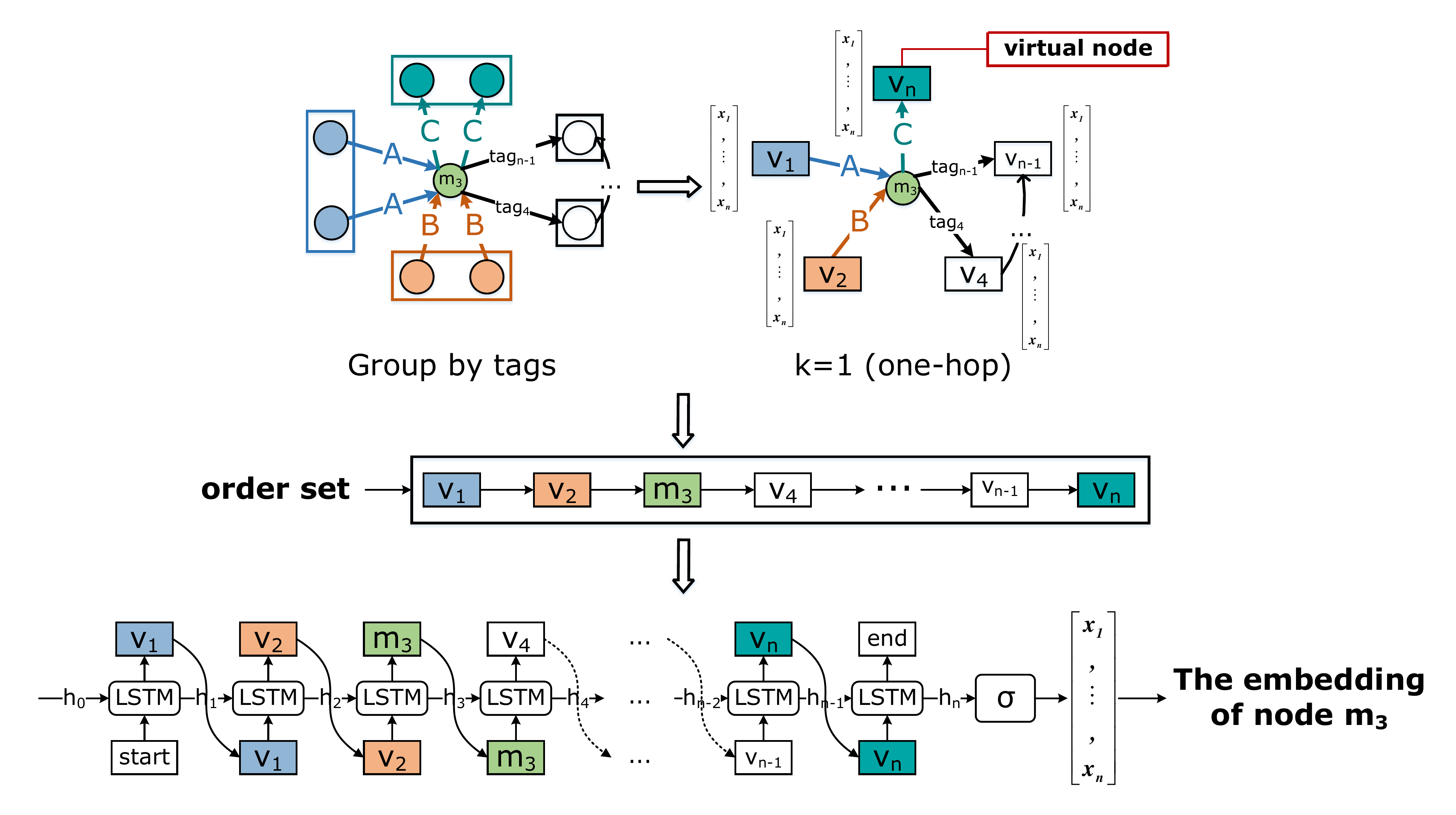}
		\caption{\textbf{Visual illustration of the process of the ADG embedding algorithm.} $ m_{1} $ to $ m_{n} $ represent virtual nodes, and the ordered set reflects the timing sequence. }
		\label{fig:figure7}
	\end{figure*}

	For a graph $ \mathcal{G}(\mathcal{M},\mathcal{E}) $, the ADG embedding process is shown in Fig.~\ref{fig:figure7}, and the detailed steps of Algorithm~\ref{alg:r2p} are illustrated on Lines 2-12. The process is described in detail below.
	
	\begin{algorithm*}[!htbp]
		\caption{ADG embedding algorithm}
		\LinesNumbered
		\label{alg:r2p}
		\KwIn{Input graph $ \mathcal{G}(\mathcal{M},\mathcal{E}) $; hops $ K $; initial input features of nodes $ \bm{x}_{m}, \forall m \in \mathcal{M} $; weight matrices $ {\bf W}^{k},\forall k \in \{1,2,...,K\} $; nonlinearity $ \sigma $;  virtualization functions $ \mathrm{VIRTUALIZED}^{\vdash}_{k},\mathrm{VIRTUALIZED}^{tag \dashv}_{k} , \forall k \in \{1,2,...,K\} $; neighbourhood functions $ \mathcal{N}_{\vdash},\mathcal{N}_{tag\dashv} $; aggregator function $ \mathrm{LSTM\_ AGGREGATE}_{k}, \forall k \in \{1,2,...,K\} $}
		
		\KwOut{Vector representations $ \bm{z}_{m} $ for all $ m \in \mathcal{M} $}
		
		$ \bm{h}^{0}_{m}\gets \bm{x}_{m},\forall m\in \mathcal{M} $
		
		\For{$ k=1,...K $}
		{
			\For{$ m\in \mathcal{M} $}
			{
			    $ \bm{h}^{k}_{\mathcal{N}_{\vdash}(m)} \gets \mathrm{VIRTUALIZED}^{\vdash}_{k}(\{ \bm{h}^{k-1}_{n_{\vdash}},\forall n \in \mathcal{N}_{\vdash}(m)\}) $;
			    
				$ \bm{h}^{k}_{\mathcal{N}_{tag\dashv}(m)} \gets \mathrm{VIRTUALIZED}^{tag\dashv}_{k}(\{ \bm{h}^{k-1}_{n_{tag\dashv}},\forall n \in \mathcal{N}_{tag\dashv}(m)\}) $;

				$ \bm{ordered\;set} $ $ \gets \{\bm{h}^{k}_{\mathcal{N}_{\vdash}(m)}
				,	            \bm{h}^{k}_{\mathcal{N}_{tag\dashv}(m)}\} $;
				
				\For {$ u $ in $ \bm{ordered\;set} $}
				{
					$ \widetilde{\bm{h}^{k}_{m}}\gets \mathrm{LSTM\_AGGREGATE}_{k}(\{\bm{h}^{k-1}_{u},\forall u \in \bm{ordered\;set}\}) $
				}

				$ \bm{h}^{k}_{m}\gets \sigma ({\bf W}^{k}\cdot \widetilde{\bm{h}^{k}_{m}}) $
			}    
		}
		
		$ \bm{z}_{m}\gets \bm{h}^{K}_{m},\forall m \in \mathcal{M} $
	\end{algorithm*}
	
	\begin{itemize}
		
		\item We define the initial input features $ \bm{x}_{m}, \forall m \in \mathcal{M} $ of each node. By training the weight matrices $ {\bf W}^{k} $, the vector of each node is constantly updated. Note that each node represents an API method, and the initial value $ \bm{x}_{m} $ is treated as a one-hot vector.
		
		\item We categorize the tagged neighbours $ m $ into tagged forward neighbours $\mathcal{N}_{\vdash}(m) $ and backward neighbours $ \mathcal{N}_{tag\dashv}(m) $. Specifically, $\mathcal{N}_{\vdash}(m) $ represents the nodes that direct to $ m $ and provide the same input parameters of $ m $, and $ \mathcal{N}_{tag\dashv}(m) $ represents the nodes that are directed to by $ m $ and accept a single output parameter of $m$.
		
		\item Note that on Lines 4 and 5, a function denoted by $\mathrm{VIRTUALIZED}$ is used to mean\_aggregate or concatenate the features of a group of identical tagged neighbours into a single feature. The resulting feature is treated as a virtual node feature of the nodes in that group. Specifically, we virtualize the forward neighbours 
		$ \{\bm{h}^{k-1}_{n_{\vdash}},\forall n \in \mathcal{N}_{\vdash}(m)\} $
		of $m$ by aggregating or concatenating to vectors 
		$ \bm{h}^{k}_{\mathcal{N}_{\vdash}(m)} $. We then virtualize the tagged backward neighbours 
		$ \{\bm{h}^{k-1}_{n_{tag\dashv}},\forall n \in \mathcal{N}_{tag\dashv}(m)\} $
		of $m$ by aggregating or concatenating to vectors 
		$ \bm{h}^{k}_{\mathcal{N}_{tag\dashv}(m)} $, where $ k \in\{1,...,K\} $ is the iteration index.

		\item Line 6 corresponds to the formation process of the $ \bm{ordered\;set} $. We order the virtualized forward and backward neighbours of $ m $ into the $ \bm{ordered\;set} $ via topological sorting, which maintains the invocation sequence relationships between methods in each code snippet.
		
		\item To fuse the sequenced relationships, on Lines 7 and 8, we leverage LSTM to aggregate the nodes in the $ \bm{ordered\;set} $ since LSTM processes inputs in a sequential manner. 
		Specifically, the LSTM aggregator adopts LSTM to encode the characteristics of the neighbours of a node. The aggregator considers the actual call order between neighbours, enters the node and its neighbours into the LSTM in that order, and then accesses the neighbours behind the current node in turn and aggregates their information. In particular, the sequential $ \bm{ordered\;set} $ of $m$ is aggregated as $ \widetilde{\bm{h}^{k}_{m}} $ in one hop size.
		
		\item Based on the previous step's result $ \widetilde{\bm{h}^{k}_{m}} $ and a fully connected network with nonlinear activity function $ \sigma $, we update the forward representation of $ m $ and the tagged backward representation of $ m $, 
		$ \bm{h}^{k}_{\mathcal{N}_{\vdash}(m)} $, $ \bm{h}^{k}_{\mathcal{N}_{tag\dashv}(m)} $ for the next iteration; this process is done in Line 10.
		
	\end{itemize}
	
	By utilizing the ADG embedding, our approach fuses node feature information with both global (structural dependencies) and sequential (method invocation constraints) API dependencies implied in the ADG. Therefore, the decoder of our approach can utilize structural information to achieve substantial improvements over the existing results.

	\section{Experimental Setting}
		\label{sec:set}%
	\begin{figure}[htbp]
		\centering
		\subfigure[\textbf{HearthStone (HS).}]{
			\includegraphics[width=1.15\linewidth]{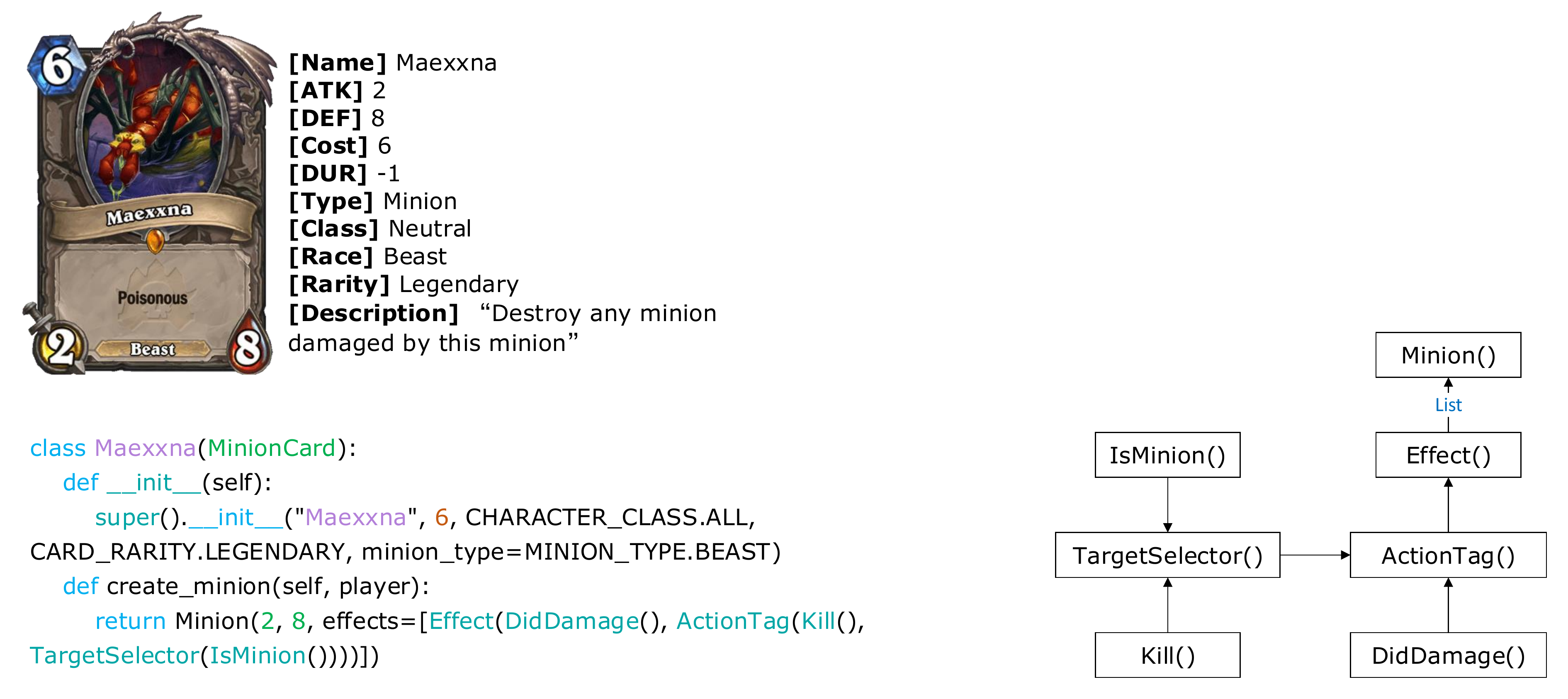}
			\label{fig:figure8a}
		}
		
		\subfigure[\textbf{Magic The Gathering (MTG).}]{
			\includegraphics[width=1.15\linewidth]{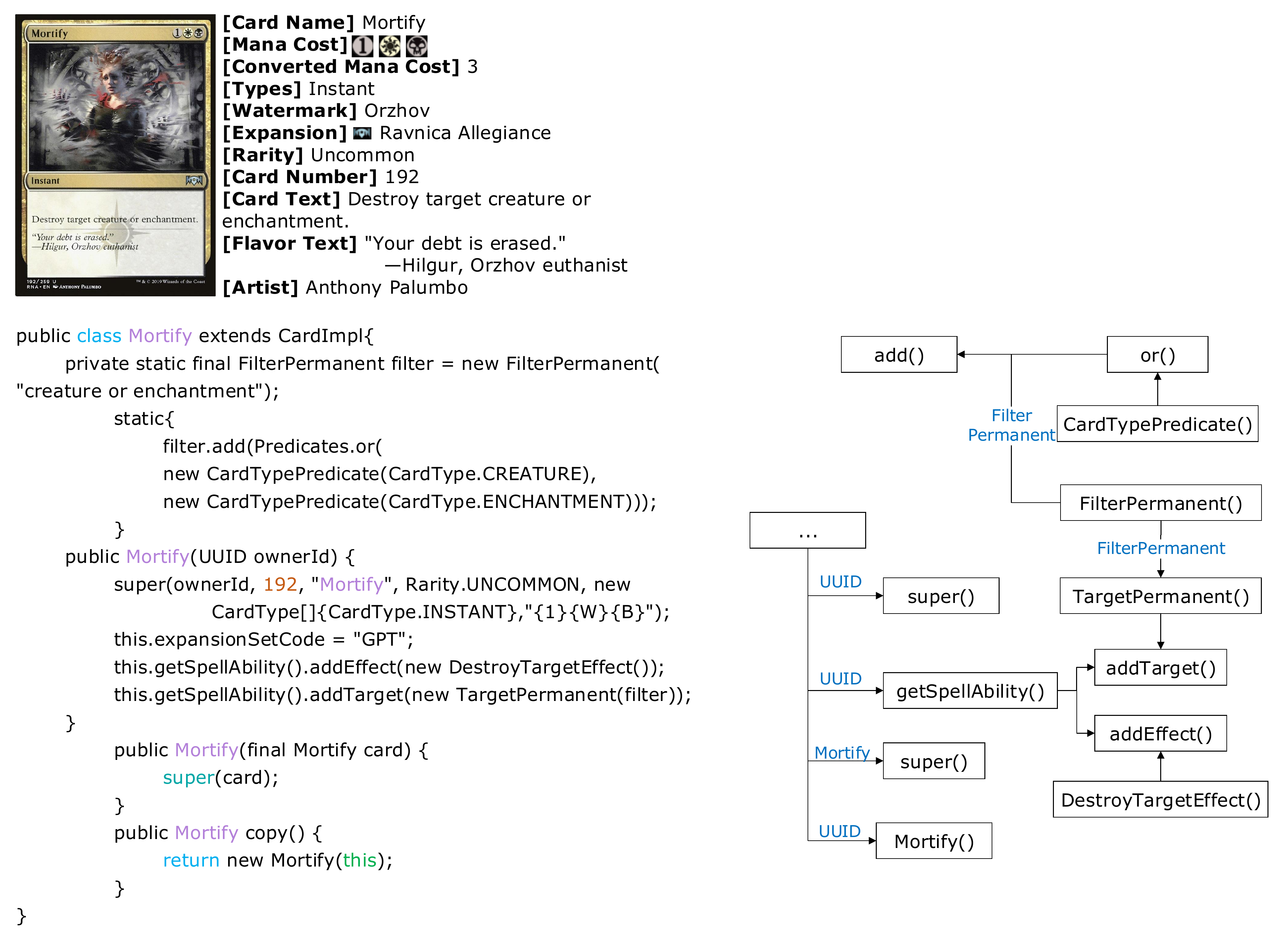}
			\label{fig:figure8b}
		}
		
		\subfigure[\textbf{Eclipse's JDT (E-JDT).}]{
			\includegraphics[width=1.15\linewidth]{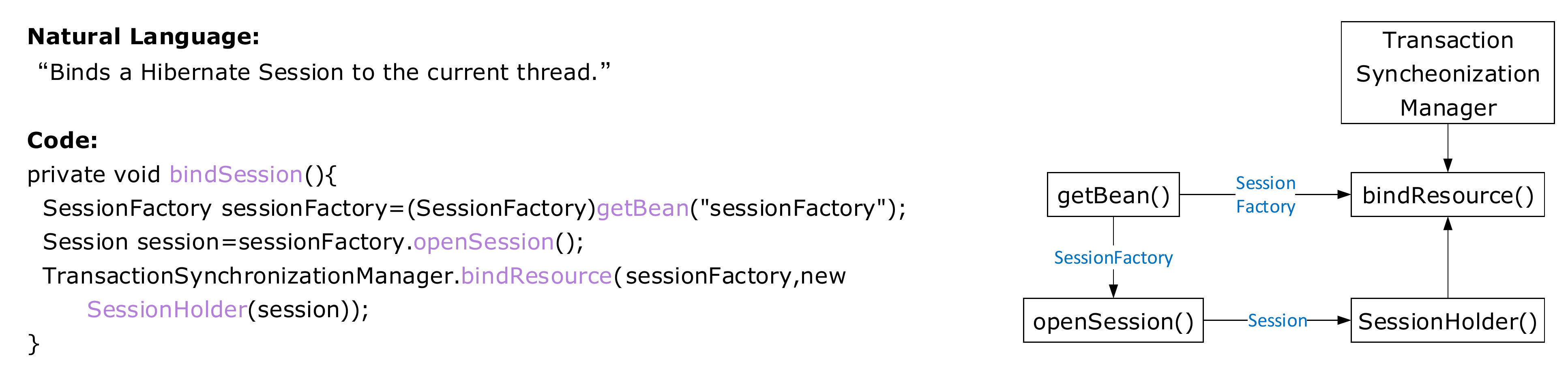}
			\label{fig:figure8c}
		}
		\caption{\textbf{Example input and output of the three datasets.} For (a) HS and (b) MTG, the upper part of each example is the input description, and the lower part is the corresponding output program. For (c) E-JDT, the figure shows a natural-language description and its corresponding code.}
		\label{fig:figure9}
	\end{figure}

	\subsection{Datasets}

	\noindent{\bf HearthStone (HS)} \cite{LingW} is an established benchmark dataset that collects Python classes from a card game. \textit{Hearthstone} is a two-player versus card game produced by the Blizzard company. Both players seek to achieve victory by using different decks, utilizing different skills and changing various digital values in the game (such as blood volume). Each deck has several attributes (attack, health, rarity, type, race, class, and cost and durability) and is described by a simple text in the text box below. Each attribute of a deck is bounded by a well-formed code, such as ``\texttt{super().\_\_init\_\_("Maexxna", 6, $\cdots$)}" in the illustration in Fig.~\ref{fig:figure8a}, which corresponds to the name and the cost of the deck. The code ``\texttt{return Minion(2, 8)}" means that the corresponding attack value is 2 and defence value is 8.
	We obtain the data from the open source implementation of TCGs\footnote{\url{github.com/danielyule/hearthbreaker/}}. In this corpus, each card is implemented in a separate class, and the imports and comments are stripped. An example of such a card and its corresponding code is shown in Fig.~\ref{fig:figure8a}.
	
	\noindent{\bf Magic the Gathering (MTG)} \cite{LingW} is a collection of Java classes that demonstrate functions of the execution cards for the game \textit{Magic: The Gathering}. This game is a digital collectable card game created by Richard Garfield. The MTG dataset is similar to HS. Players use different decks to inflict damage on the enemy through thought and strategies in the game to achieve victory. As shown in Fig.~\ref{fig:figure8b}, an MTG deck contains nine attributes (card name, attack, defence, type, cost, ID, health, rarity and suit) with some simple descriptions of the deck's functions. MTG's code adheres strictly to a standard. For example,  the code ``\texttt{super(ownerId, 192, "Mortify", Rarity.UNCOMMON, new CardType[]\{CardType.INSTANT\}, "\{1\}\{W\}
	\{B\}");}" specifies that the card number is 192, card name is ``Mortify", rarity is uncommon, and the mana cost is ``1", ``white", and ``black". We also collect MTG data from an open source implementation\footnote{\url{github.com/magefree/mage/}} that contains data from 13,279 different cards.
	We establish an input sequence with a set of fields in each card. The output sequence is the target code snippet, which represents complex class structures, as depicted in Fig.~\ref{fig:figure8b}.
	
	\noindent{\bf Eclipse Java Development Tools (E-JDT)} \cite{HuX} is a source code dataset collected from the Eclipse Java Development Tools compiler. This dataset comprises 69,708 different Java methods and related comments. In contrast to HS and MTG, the comment sentences appearing in the Javadocs are used as input sequences, while the Java code obtained from the Javadoc guidance is used as output sequences. An example of this dataset is given in Fig.~\ref{fig:figure8c}.

    As illustrated in Fig.~\ref{fig:figure9}, we show the API dependencies on the right side of each example, which are encoded by the ADG embedding algorithm. Taking the code of E-JDT as an example, the invocation constraint for ``\texttt{bindResource()}" is that the input parameter and a container must be provided by calling ``\texttt{SessionHolder()}” and ``\texttt{getBean()}”, respectively, before ``\texttt{bindResource ()}” is called. In addition, the API sequences contained in the code must follow the API invocation order, i.e., ``\texttt{getBean()$\rightarrow$openSession()$\rightarrow$ SessionHolder()$\rightarrow$bindResource()}”, subject to the invocation constraints described above. 
    
	The statistics of the HS, MTG, and E-JDT datasets are listed in Table~\ref{tab:table1}. For HS and E-JDT, we use 80\% for training, 10\% for validation, and 10\% for testing. For MTG, we use 90\% for training, 5\% for validation, and 5\% for testing. 
	
	\begin{table}[!htbp]
		\centering
		\caption{\textbf{Statistics of the datasets.} }
		\label{tab:table1}
		\begin{tabular}{lrrr}
			\toprule
			Dataset & HS    & MTG   & E-JDT \\
			\midrule
			Training & 533   & 11,969 & 470,486 \\
			Development & 66    & 664   & 58,811 \\
			Validation  & 66    & 664   & 58,811 \\
			\midrule
			Avg.words in description & 26.31  & 50.00  & 14.72  \\
			Max.words in description & 38    & 147   & 1873 \\
			Avg.tokens in code & 38.09  & 117.85  & 57.47  \\
			Max.tokens in code & 197   & 885   & 8472 \\
			Avg.methods in code & 3.53  & 11.96  & 5.40  \\
			Max.methods in code & 26    & 126   & 1729 \\
			\bottomrule
		\end{tabular}%
		
		\begin{tablenotes}
			\raggedright
			\footnotesize
			\item \textit{This table shows the number of training, development, and testing items in the three datasets. Words and tokens represent the number of characters in each description and code item, respectively.}
		\end{tablenotes}
	\end{table}      
		
	\subsection{Metrics}
	Our approach is evaluated in terms of eight metrics: \textit{Acc} \cite{SunZ}, \textit{BLEU} \cite{bleu}, \textit{F1}, \textit{CIDEr} \cite{cider}, \textit{ROUGE-L}, \textit{ROUGE-1}, \textit{ROUGE-2} \cite{rouge} and \textit{RIBES} \cite{ribes}. 
	
	\begin{itemize}
		\item \textbf{Acc} \quad 
		
		Following Ling et al. \cite{LingW}, Yin et al. \cite{YinP} and Sun et al. \cite{SunZ}, we calculate the accuracy (denoted by \textit{Acc}) based on a string match: 
		$$
          Acc = \frac{N_{correct}}{N_{sample}}
		$$
	    where $N_{correct}$ denotes the number of correctly generated examples, and $N_{sample}$ denotes the number of all samples.

		\item \textbf{BLEU} \quad The token-level \textit{BLEU} \cite{bleu} is selected as our second metric since it calculates the precision of the match between the generated code snippets and reference code snippets, which is a reasonable approach to evaluating the quality of the generated code snippets. This score is computed as:
		$$
		BLEU=BP\times exp(\Sigma ^{N}_{n=1}w_{n}logp_{n})
		$$
		$$ BP=\begin{cases}
		1& \text{if c $ > $ r}\\
		e^{(1-r)/c}& \text{if c $ \le $ r}
		\end{cases} $$
		where $ p_{n} $ is the ratio of length $ n $ of the tokens' subsequences in the generated code. 
		
		\item \textbf{CIDEr} \quad CIDEr \cite{cider} calculates the cosine angle of the TF-IDF vector of the candidate text to obtain the similarity with the reference text:
		$$ CIDEr_n(c_i, s_i) = \frac{1}{M}\times  \sum_{j=1}^{M} \frac{g^n(c_i)\times g^n(s_{ij})}{|| g^n(c_i) ||\times  || g^n(s_{ij}) ||} $$
		$$ g_k(c_i) = TF(k)\times  IDF(k) $$
		$$ TF(k) = \frac{h_k(c_i)}{\sum h_l(c_i)} $$
		$$ IDF = log(\frac{N}{\sum_{1}^{N}min(1, \sum_{1}^{M} h_k(c_i))}) $$
		where $g_k(c_i)$ denotes the TF-IDF weight of n-gram $\omega_k$, $h_k(c_i)$ denotes the occurrence time of $\omega_k$ in sentence $c_i$, $\sum h_l(c_i)$ denotes the sum of occurrence times of all n-grams occurring in sentence $c_i$, and $g^n(c_i)$ denotes the TF-IDF weight vector of n-gram $\omega_k$ in sentence $c_i$. The main goal of the CIDEr metric is to determine whether the model has captured critical information to evaluate whether the generated code implements the major requirements of the textual program description.
	
		\item \textbf{ROUGE} \quad ROUGE \cite{rouge} stands for ``recall-oriented understudy for gisting evaluation." ROUGE-N is based on n-grams. For any $n$, we count the total number of n-grams across all reference code and determine how many are present in the candidate code. ROUGE-L is based on the longest common subsequence (LCS):
		$$ ROUGE-N = \frac{\sum Count_{match}(gram_n)}{ \sum Count(gram_n)} $$
		$$ P = \frac{LCS(A, B)}{m} $$
		$$ R = \frac{LCS(A, B)}{n} $$
		$$ ROUGE-L = F = \frac{(1+b^2)RP}{R+b^2P} $$
		where $LCS(A, B)$ denotes the longest common subsequence between sentences A and B, and $m$ and $n$ denote the lengths of reference code and generated code, respectively.
				
		\item \textbf{RIBES} \quad RIBES \cite{ribes} is an automatic evaluation method based on rank correlation that considers the distance between the generation result and the token order of the reference code. The Pearson's correlation coefficient $\rho$ is used to measure the differences in rank: 
		$$ NSR = \frac{\rho +1}{2} $$
		$$ \rho = 1-\frac{\sum_id_i^2}{_{n+1}C_3} $$
		where NSR denotes the normalized Spearman's $\rho$, and $d_i$ indicates the difference in the rank of the $i$-th element. The RIBES metric considers the token order to assess whether the method invocation order in the generated code is correct.
		
		\item \textbf{F1} \quad To evaluate the performance of different algorithms comprehensively, the F1 value is used to assess both precision and recall. The F1 score is calculated as
		$$ F1 = \frac{2\times Precision\times Recall}{Precision+Recall} $$
		$$ Precision = BLEU $$
		$$ Recall = ROUGE-1$$
	\end{itemize}

	\subsection{Setup}
	\noindent\textbf{Parameter Settings} \quad The method was implemented in Python under the PyTorch 1.5.0 framework. 
	Experiments were performed on a machine with an i9-9900K CPU and 2 $\times$ RTX-2080ti GPUs.
	During the training process, parameters were set as follows:
	
	\begin{itemize}
				
		\item The word embedding module was used to obtain a continuous vector for inputs to the encoder. Code vocabulary sizes of HS, MTG, and E-JDT were 1570, 3162 and 123,975, respectively. Description vocabulary sizes for HS, MTG, and E-JDT were counted as 1301, 2354, and 20,038, respectively. 
		For HS and MTG, the dimension of word embedding was set to 100 following \cite{wang2017dynamic}, and for E-JDT, the dimension was set to 256 following the setting in \cite{hu2020deep} since its vocabulary size was much larger.

		\item For HS, MTG and E-JDT, we set the dimensions of the hidden layer to 256.
		
		\item The characteristics of the constructed ADG contain the numbers of nodes, edges, and tags, the total in-degree and out-degree, and the average in-degree and out-degree. These statistics are listed in Table~\ref{tab:table2}.
			
		\item  We set the training epochs on HS, MTG, and E-JDT to 25,000, 400, and 15, respectively. 
		
		\item Early stopping was performed by calculating the \textit{BLEU} score with the validation set. For HS, the \textit{BLEU} score was calculated every 1000 iterations, and if there were no increments for more than 10 times, processing was terminated. For MTG and E-JDT, the stopping time was set to 10 and 3 respectively.

		\item For the decoder, we used the beam search with a size of 5 during prediction.
		
		\item To prevent overfitting, we applied dropout \cite{SrivastavaN} to the output of each sub-layer. In all cases, we set $P_{drop}=0.1$
		
		\item Glorot initialization \cite{GlorotX} was used to initialize all parameters randomly.

	\end{itemize}

	\begin{table}[!htbp]
		\centering
		\caption{\textbf{Statistics of the ADGs.} }
		\begin{tabular}{lrrr}
			\toprule
			& \multicolumn{1}{l}{\textbf{HS}} & \multicolumn{1}{l}{\textbf{MTG}} & \multicolumn{1}{l}{\textbf{E-JDT}} \\
			\midrule
			Nodes & 1,204  & 117,246  & 2,162,968  \\
			Edges & 3,726  & 2,452,088  & 18,048,043  \\
			Max.in & 48    & 215   & 290  \\
			Avg.in & 3.03  & 28.11  & 10.25  \\
			Max.out & 22    & 201   & 264  \\
			Avg.out & 3.16  & 13.71  & 6.44  \\
			\bottomrule
		\end{tabular}%
		\begin{tablenotes}
			\raggedright
			\footnotesize
			{\item \textit{This table shows the quantities of nodes and edges, and the in-degree and out-degree values of nodes.}}
		\end{tablenotes}
		\label{tab:table2}%
	\end{table}%
		
	\section{Research Questions}
	\label{sec:rq}%

	In the experiments, we evaluated the performance of various approaches on generating Python and Java code to answer the following three research questions.

	\textit{\textbf{RQ1: Is the performance of ADG-Seq2Seq better than that of the compared approaches?}}
	
	The objective of this RQ is to evaluate the effectiveness of our method.
	Compared to previous approaches concentrating on grammar rules, we considered more aspects and used the ADG to generate code. The superiority of our approach is highlighted by comparisons with several state-of-the-art models.

	\textbf{\textit{RQ2: Does the proposed ADG-based embedder contribute to code generation?}}

	The purpose of  this RQ is to  evaluate the benefits of ADG embedding.
	Extensive ablation tests were performed to evaluate the contribution of each component. To this end, we removed or substituted a single component each time. Specifically, the components we evaluated were as follows. 
	
	\begin{itemize}[leftmargin=*]
		
		\item\textbf{Graph embedding.} We removed the ADG embedding from the decoder; i.e., the method used only the Seq2Seq model with attention to vectorize the target code.
		
		\item\textbf{Different graph embedding strategies.} We replaced our ADG embedding with other algorithms, including GraphSAGE \cite{HamiltonW} and GCN \cite{kipf2016semi}.
		
		\item\textbf{Special reachability.} We did not consider the special reachability of the ADG and did not classify the forward and backward nodes according to different tags.
		
		\item\textbf{Directed edges.} We neglected the directions of edges between nodes to evaluate the impact of such directions.
		
		\item\textbf{Labelled edges.} We removed the labels of edges in the ADG, namely the tags, to evaluate the influence of edge labels.
		
		\item\textbf{Different Hop Sizes.} We compared ADG embeddings with hop sizes of one and two.
		
		\item\textbf{Different Aggregators.} We compared different aggregators, such as mean, pooling\footnote{The pooling we used followed the method of GraphSAGE [27], namely, max pooling.}, and LSTM aggregators, in the ADG embedding.
	\end{itemize}

	\textbf{\textit{RQ3: How do various aspects of the datasets affect the performance of our approach?}}

	This RQ is explored to evaluate the impact of each aspect of the datasets on the performance of our approach.
	For all existing solutions, several aspects of datasets might affect the performance of code generation. 
	Since the embedder was first introduced into code generation, we devoted a considerable amount of effort to the evaluation of the code structure. Several new experiments were designed to explore the influences of various aspects. Specifically, we conducted two main experiments to evaluate the performance of the model at different description lengths and different code lengths. In addition, other aspects of datasets, such as various forms of descriptions (semi-structured or unstructured), different code implementation languages (Python or Java), and various graph complexities (sparse or dense), are also discussed and analysed in detail.

	\section{Results}
	\label{Results}
	\subsection{Answer to RQ1}
	\label{ARQ1}

	To answer RQ1, the proposed ADG-Seq2Seq method was compared with benchmark methods reported in the literature: attention-based Seq2Seq, Transformer, SNM, ASN, GB-CNN, and TreeGen. Attention-based Seq2Seq and Transformer are classic models that are widely used in neural machine translation, while the other approaches are designed for code generation. These approaches continuously promote and effectively improve the performance of code generation. The differences between these methods are described in the following. Details of the implementations of these methods are available in Appendix A.

	\begin{itemize}
		\item \textbf{Attention-based Seq2Seq.} The attention-based Seq2Seq model is a typical neural machine translation model \cite{NeubigG} that adopts an encoder-decoder architecture and leverages an attention mechanism to better map different tokens.
		
		\item \textbf{Transformer.} Transformer, proposed by Vaswani et al. \cite{VaswaniA}, is also based on the encoder-decoder architecture but integrates a more advanced multihead attention mechanism. Sun et al. \cite{SunZ2} trained a transformer-based code generation model using plain code tokens and comments. In our experiments, we also compare our model with this transformer-based code generation model.

		\item \textbf{SNM.} The syntactic neural model \cite{YinP} is a grammar model that uses the AST to capture the syntax of source code snippets.
		
		\item \textbf{ASN.} The abstract syntax network \cite{RabinovichM} develops a specifically designed decoder with a dynamically determined modular structure paralleling the structure of the output tree, such that its output is represented by an AST.
		
		\item \textbf{GB-CNN.} Grammar-based CNNs \cite{SunZ} use a grammar-based structural CNN decoder for code generation, where the underlying grammar information in an AST is parsed and absorbed through a three-layer CNN.
		
		\item \textbf{TreeGen.} TreeGen \cite{SunZ2} leverages the Transformer's multihead attention mechanism  to alleviate the problem of a long-range dependency and introduces a new AST encoder to incorporate syntax rules and AST structures into the model.
	\end{itemize}

		\begin{table*}[htbp]
		\small
		\centering
		\caption{\textbf{Experimental results obtained by our approach and the compared methods.}}

		\label{tab:table3}
		\resizebox{\textwidth}{40mm}{
			\begin{threeparttable}
				\begin{tabular}{clcccccccc}
					\toprule
					&  & Acc   & Bleu  & F1    & CIDEr & RougeL & Rouge1 & Rouge2 & RIBES \\
					\midrule
					\multirow{7}[2]{*}{\begin{sideways}HS\end{sideways}} & Attn-Seq2Seq & { { 1.5}}   & 60.4  &  62.8   & 0.56  & 64.0  & 65.5  & 32.3  & 52.1  \\
					& Transformer & 10.6  & 68.0  & 70.8  & 1.35  & 72.8  & 73.8  & 36.5  & 63.7  \\
					& SNM   & 16.2  & 75.8  & 77.3  & 1.57  & 77.5  & 78.9  & 39.0  & 73.4  \\
					& ASN   & 18.2  & 77.6  & 78.7  & 1.56  & 77.0  & 79.8  & 39.2  & 73.3  \\
					& GB-CNN & 27.3  & 79.6  & 81.4  & 1.62  & 82.8  & 83.2  & 39.7  & 76.2  \\
					& TreeGen & \textbf{31.8} & \textbf{80.8} & 82.2 & 1.88  & 82.9  & 83.6  & 40.2  & 75.9  \\
					& ADG-Seq2Seq & 27.3  & 78.1  & \textbf{82.5}  & \textbf{2.02} & \textbf{87.4} & \textbf{87.5} & \textbf{43.5} & \textbf{80.4} \\
					\midrule
					\multirow{7}[2]{*}{\begin{sideways}MTG\end{sideways}} & Attn-Seq2Seq & { { 1.4}}   & 54.8  & 59.0   & 0.33  & 63.1  & 63.9  & 31.4  & 51.4  \\
					& Transformer & 10.1   & 58.1  & 63.6  & 0.94  & 69.9  & 70.3  & 34.7  & 62.0  \\
					& SNM   & 19.3  & 62.3  & 70.0  & 1.28  & 79.4  & 79.8  & 38.8  & 72.2  \\
					& ASN   & 21.2  & 63.7  & 70.3  & 1.23  & 78.3  & 78.4  & 38.6  & 72.0  \\
					& GB-CNN & 25.0  & 65.9  & 72.1  & 1.35  & 79.8  & 79.6  & 37.9  & 75.8  \\
					& TreeGen & 26.7  & 65.3  & 71.9  & 1.48  & 78.4  & 80.0  & 38.3  & 75.1  \\
					& ADG-Seq2Seq & \textbf{29.4} & \textbf{69.2} & \textbf{76.3} & \textbf{1.69} & \textbf{85.4} & \textbf{85.1} & \textbf{42.2} & \textbf{79.7} \\
					\midrule
					\multirow{7}[2]{*}{\begin{sideways}E-JDT\end{sideways}} & Attn-Seq2Seq & { { 0.8}}   & 50.1  & 56.6   & 0.31  & 64.8  & 65.1  & 32.5  & 50.9  \\
					& Transformer & { { 8.4}}   & 54.2  & 60.7  & 0.53  & 68.5  & 68.9  & 33.9  & 60.8  \\
					& SNM   & 19.2  & 59.9  & 65.9  & 1.58  & 73.3  & 73.1  & 36.5  & 70.4  \\
					& ASN   & 18.9  & 58.5  & 65.6  & 1.54  & 74.1  & 74.7  & 36.7  & 69.3  \\
					& GB-CNN & 26.1  & 60.8  & 68.8  & 1.67  & 77.6  & 79.4  & 37.9  & 70.2  \\
					& TreeGen & 26.3  & 61.1  & 69.0  & 1.73  & 76.0  & 79.4  & 37.8  & 71.9  \\
					& ADG-Seq2Seq & \textbf{28.6} & \textbf{65.3} & \textbf{73.8} & \textbf{1.85} & \textbf{81.7} & \textbf{84.7} & \textbf{41.8} & \textbf{78.8} \\
					\bottomrule
				\end{tabular}%
				
				\begin{tablenotes}
					\raggedright
					\footnotesize
					{
					\item [1] \textit{Attn-Seq2Seq denotes the attention-based Seq2Seq model. }
					\item [2] \textit{For models with existing experimental data, we use the experimental results presented in the original paper, and for those without, we retrain the model with the default parameters provided in the original paper and calculate the experimental results under various metrics. Detailed information is provided in Appendix A.}
					}
				\end{tablenotes}
			\end{threeparttable}
		}
	\end{table*}

\begin{figure*}[!htbp]
		\centering
		\subfigure[HS: Acc]{
			\includegraphics[width=0.22\linewidth]{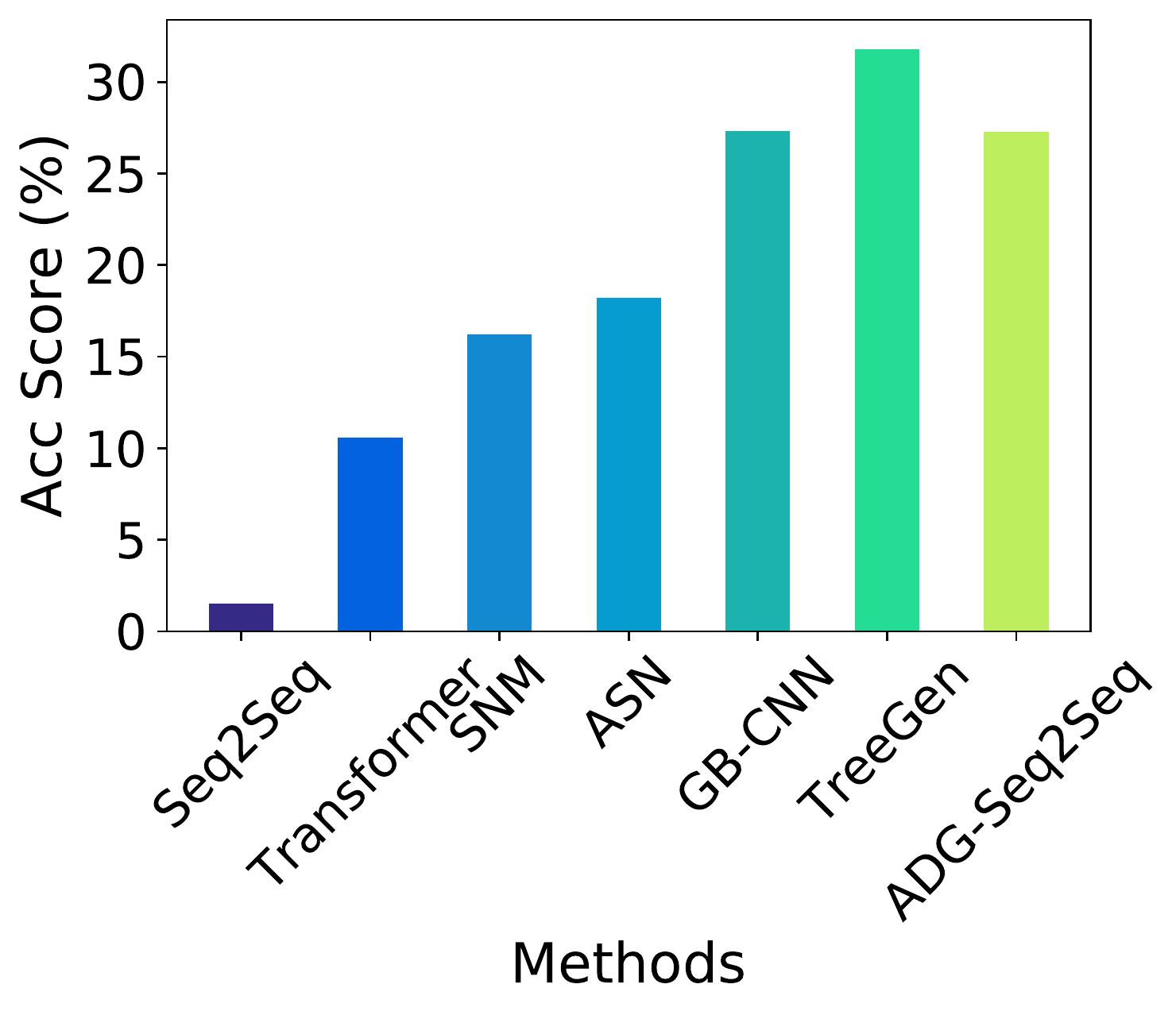}
			\label{HS:Acc}
		}
		\subfigure[HS: BLEU]{
			\includegraphics[width=0.22\linewidth]{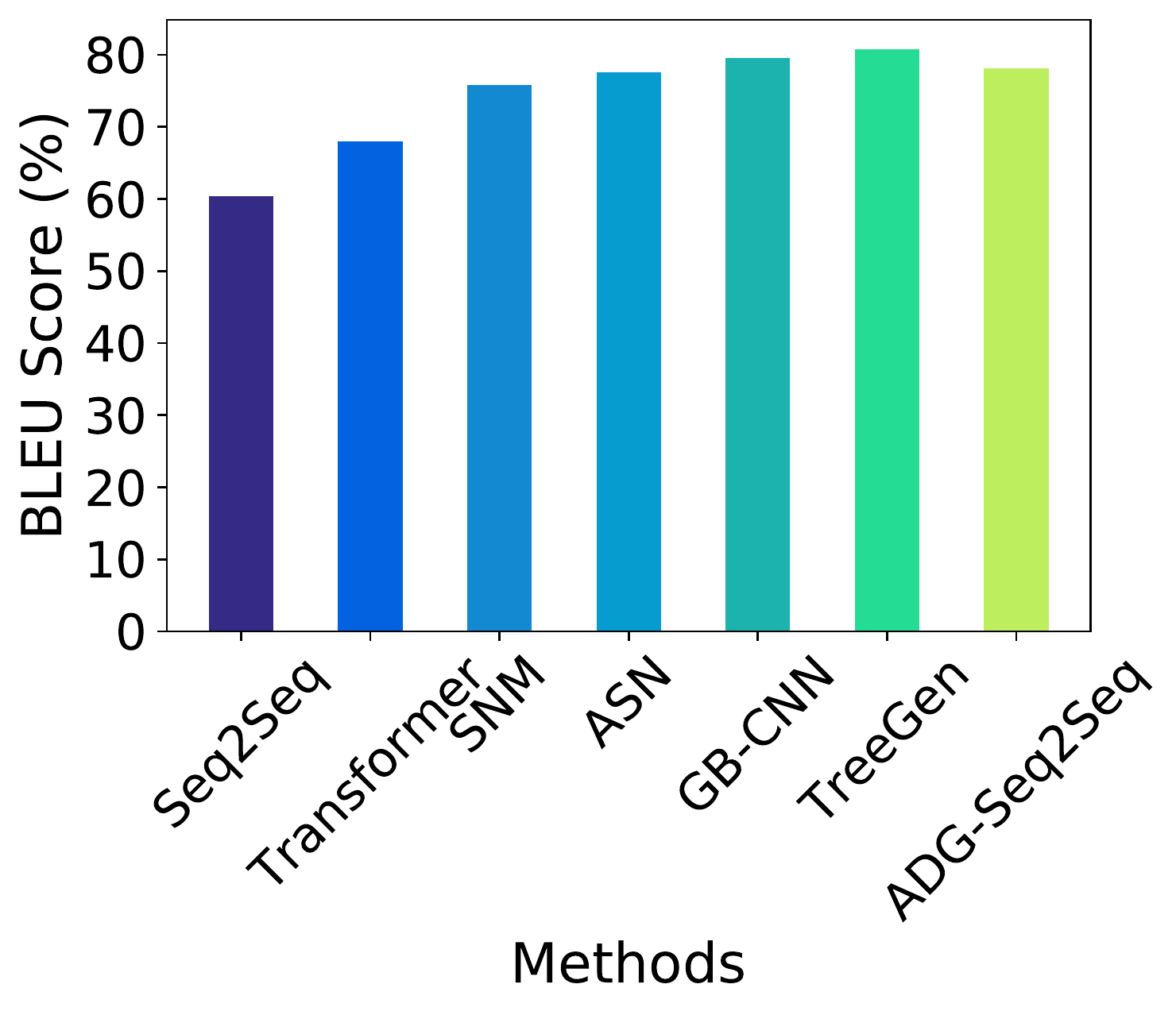}
			\label{HS:BLEU}
		}
		\subfigure[HS: F1]{
			\includegraphics[width=0.22\linewidth]{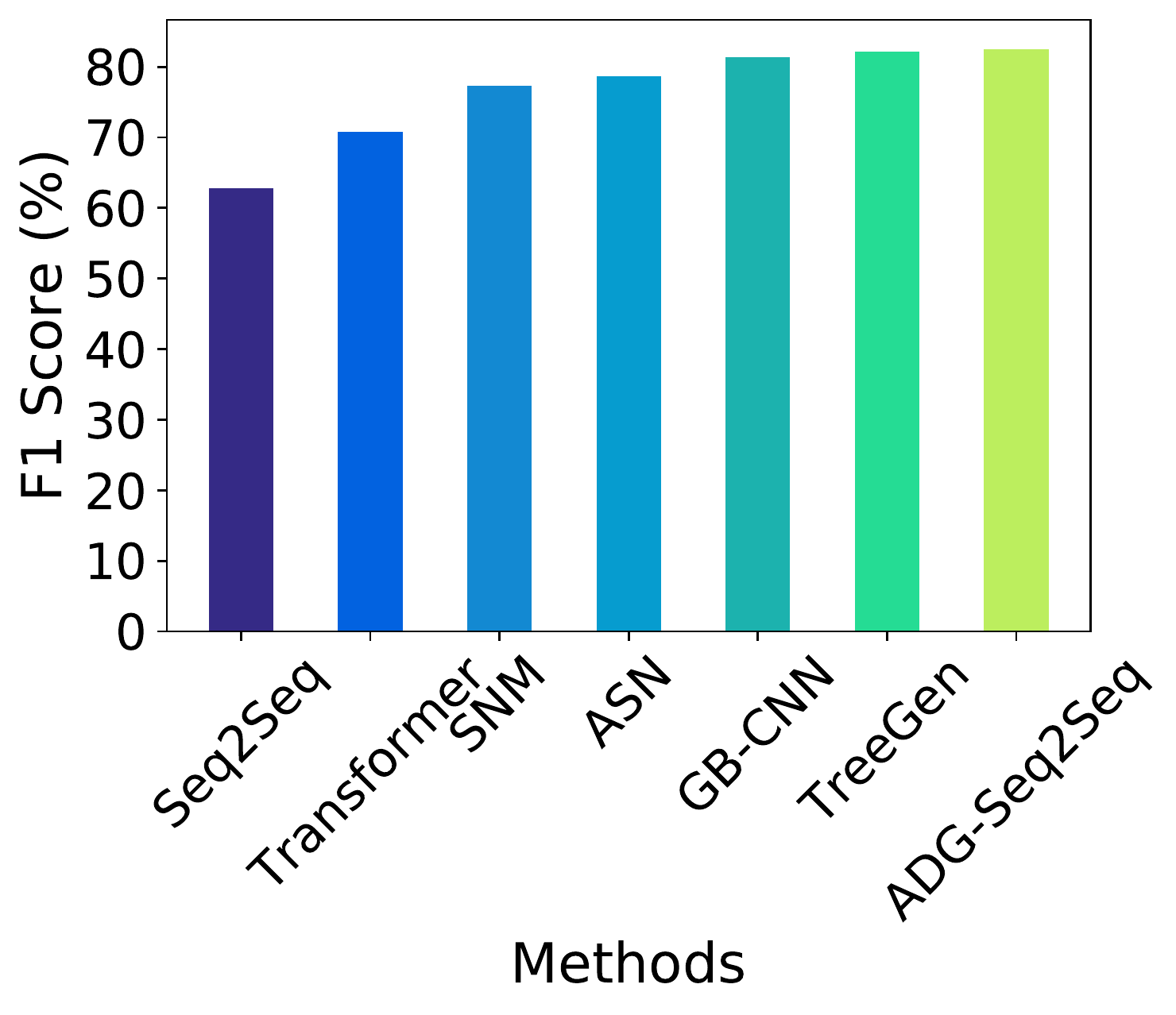}
			\label{HS:F1}
		}
		\subfigure[HS: CIDEr]{
			\includegraphics[width=0.22\linewidth]{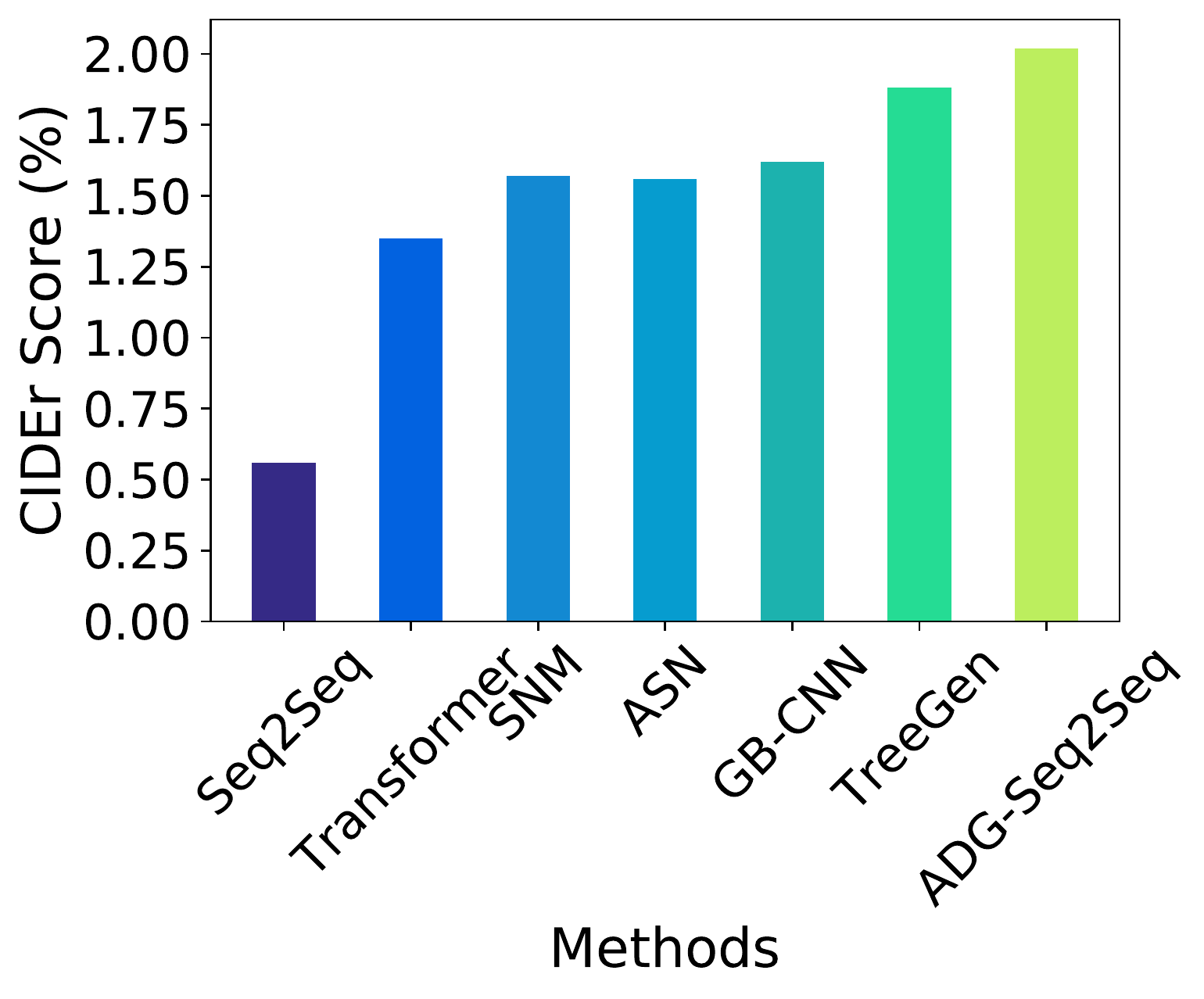}
			\label{HS:CIDEr}
		}
		
		\subfigure[HS: ROUGE-L]{
			\includegraphics[width=0.22\linewidth]{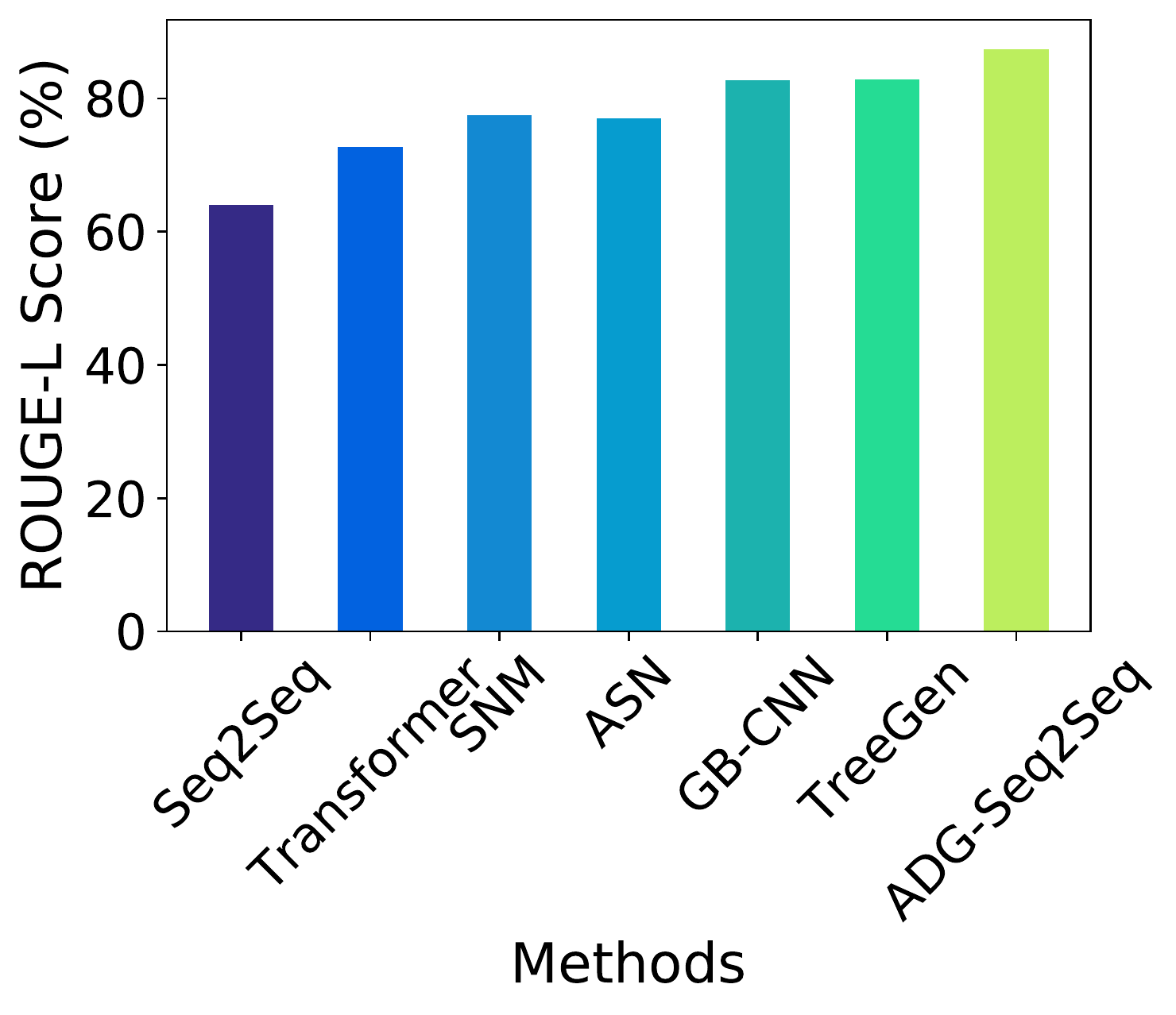}
			\label{HS:ROUGE-L}
		}
		\subfigure[HS: ROUGE-1]{
			\includegraphics[width=0.22\linewidth]{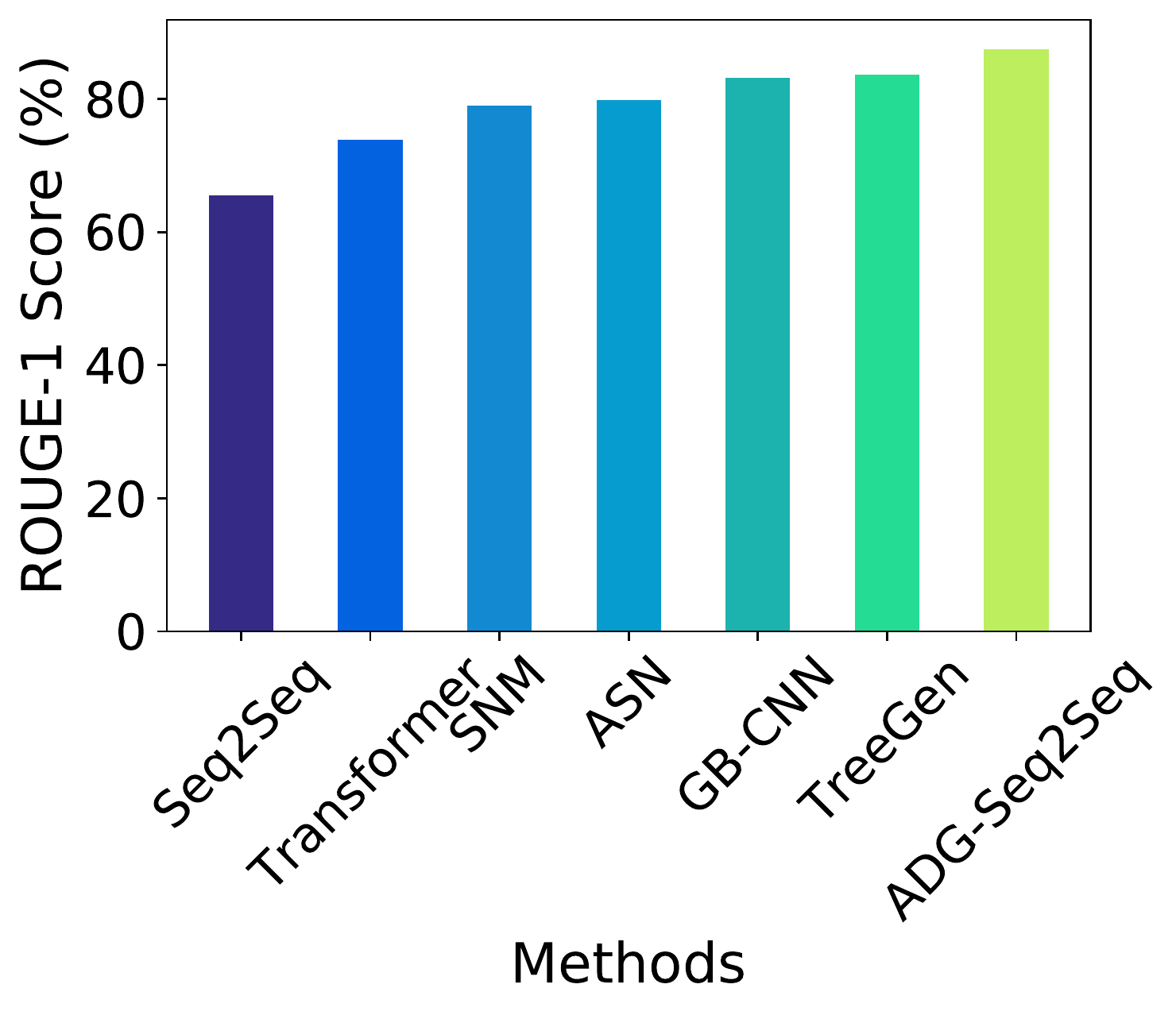}
			\label{HS:ROUGE-1}
		}
		\subfigure[HS: ROUGE-2]{
			\includegraphics[width=0.22\linewidth]{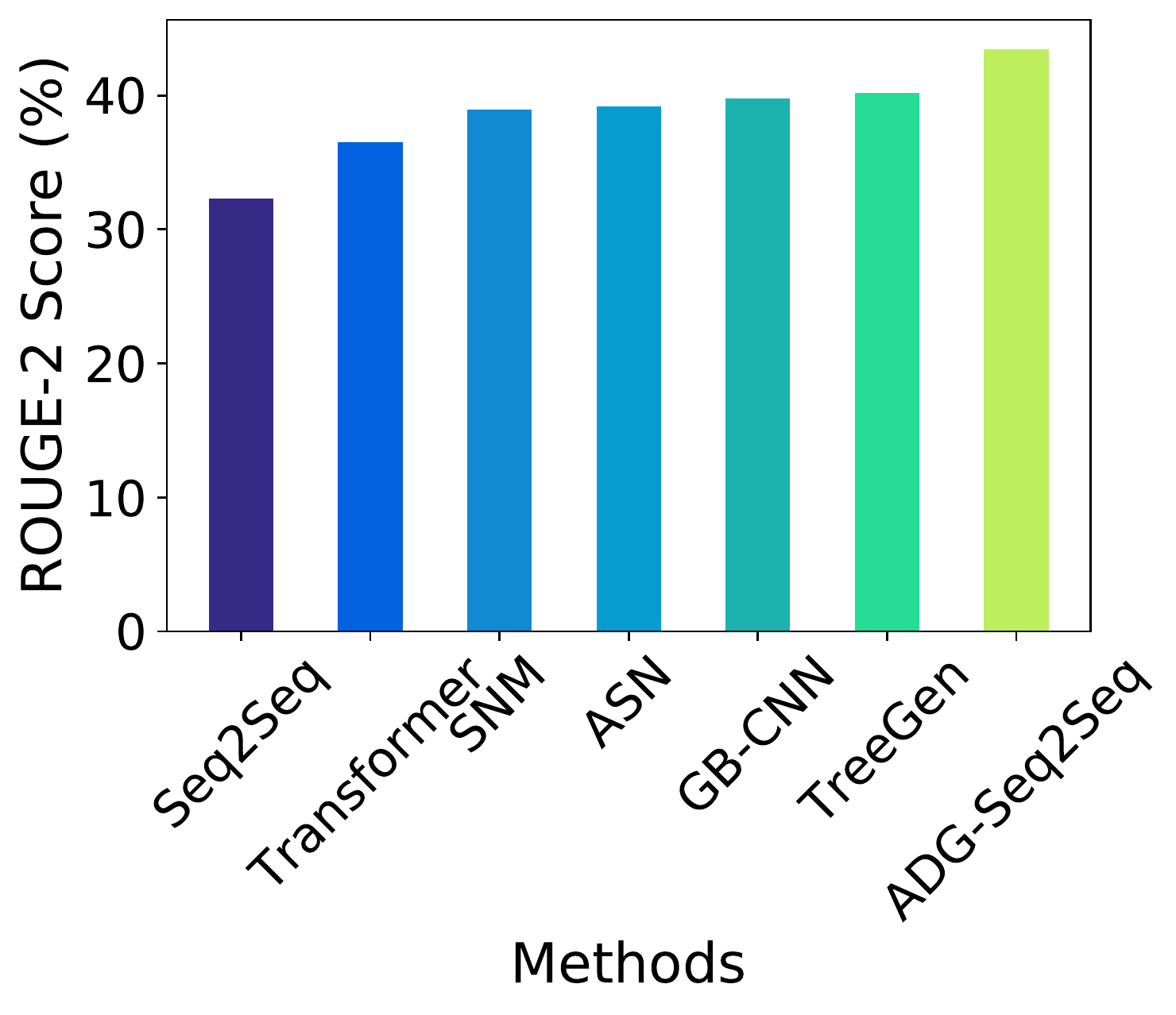}
			\label{HS:ROUGE-2}
		}
		\subfigure[HS: RIBES]{
			\includegraphics[width=0.22\linewidth]{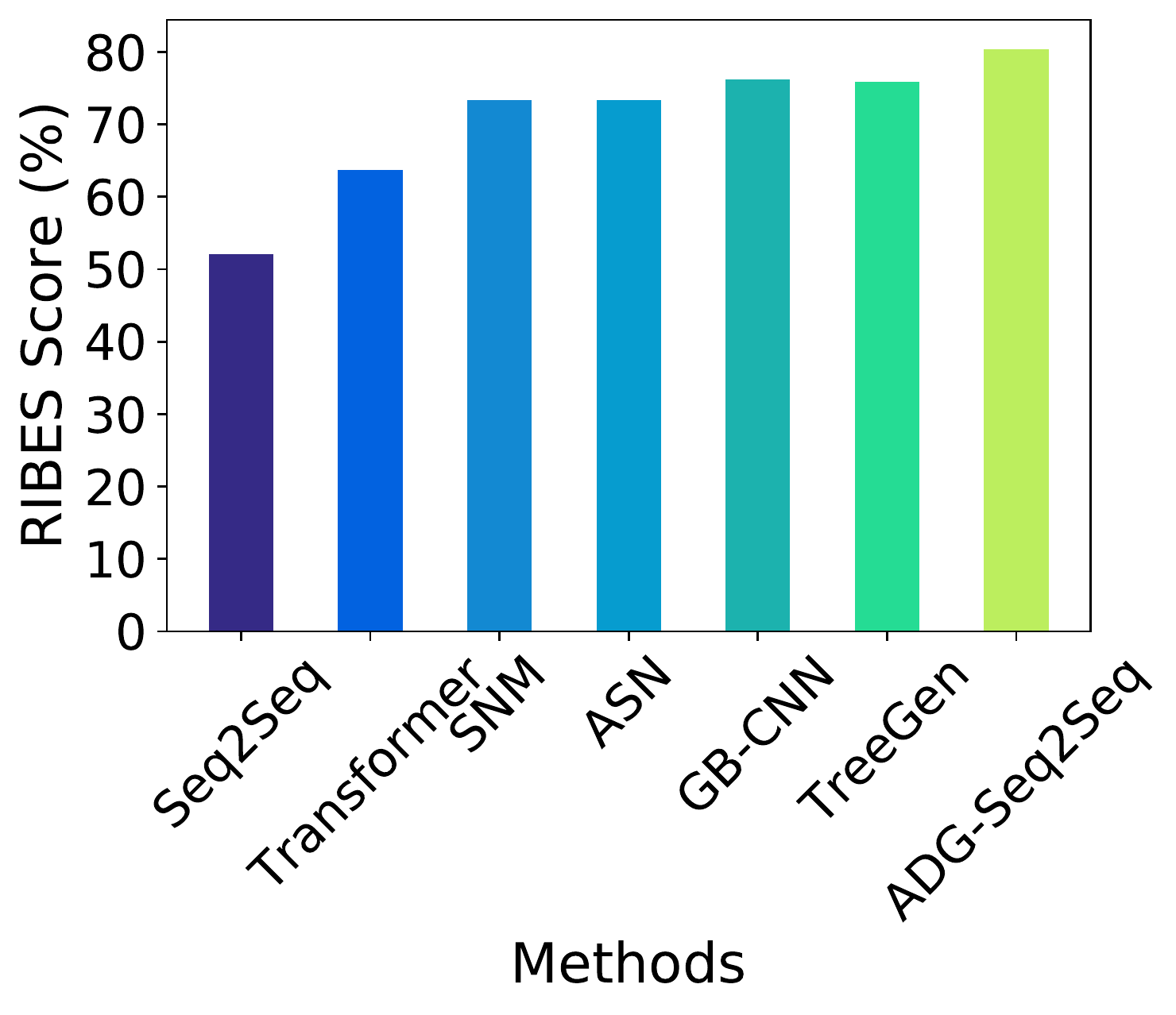}
			\label{HS:RIBES}
		}
		
		\subfigure[MTG: Acc]{
			\includegraphics[width=0.22\linewidth]{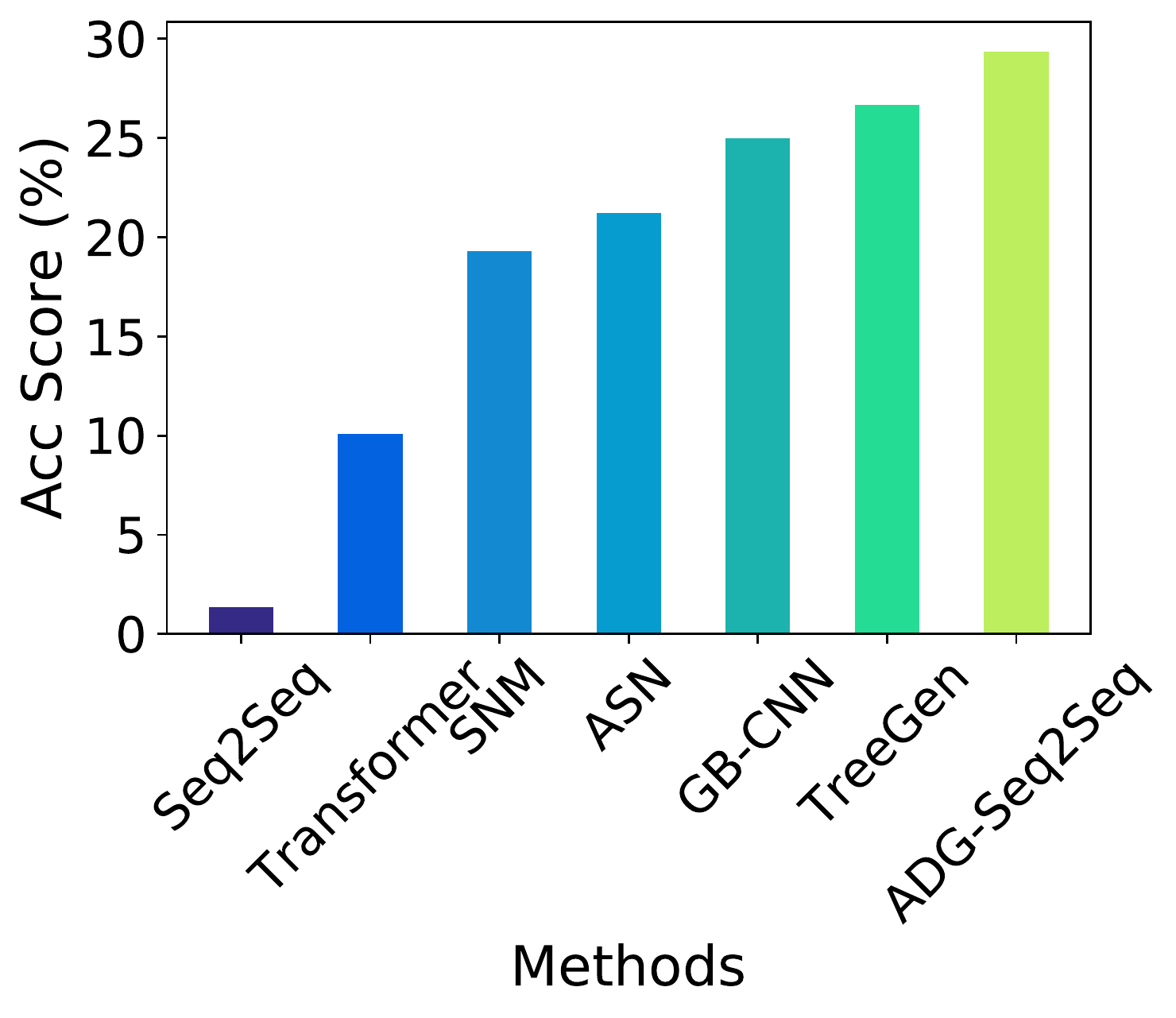}
			\label{MTG:Acc}
		}
		\subfigure[MTG: BLEU]{
			\includegraphics[width=0.22\linewidth]{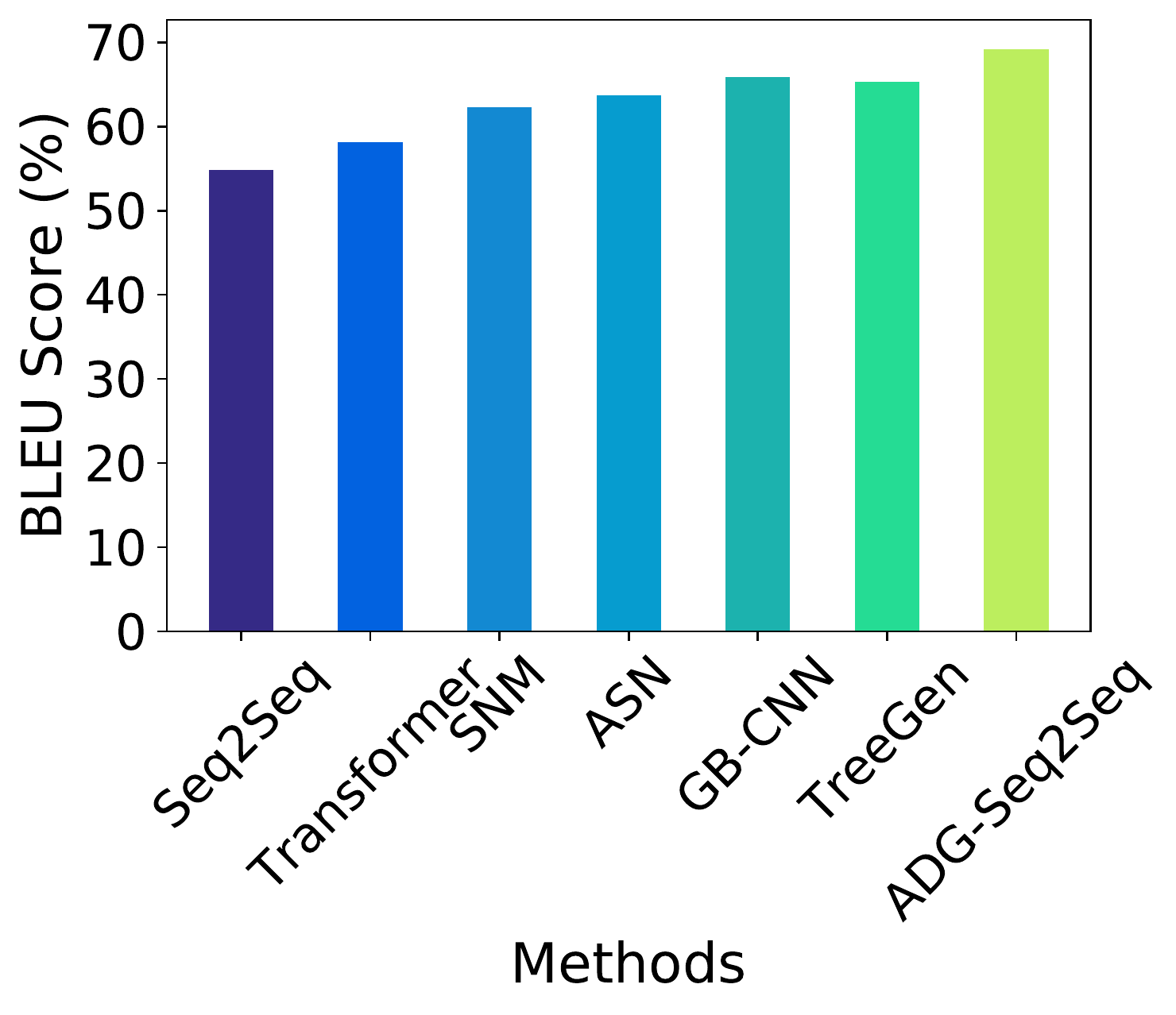}
			\label{MTG:BLEU}
		}
		\subfigure[MTG: F1]{
			\includegraphics[width=0.22\linewidth]{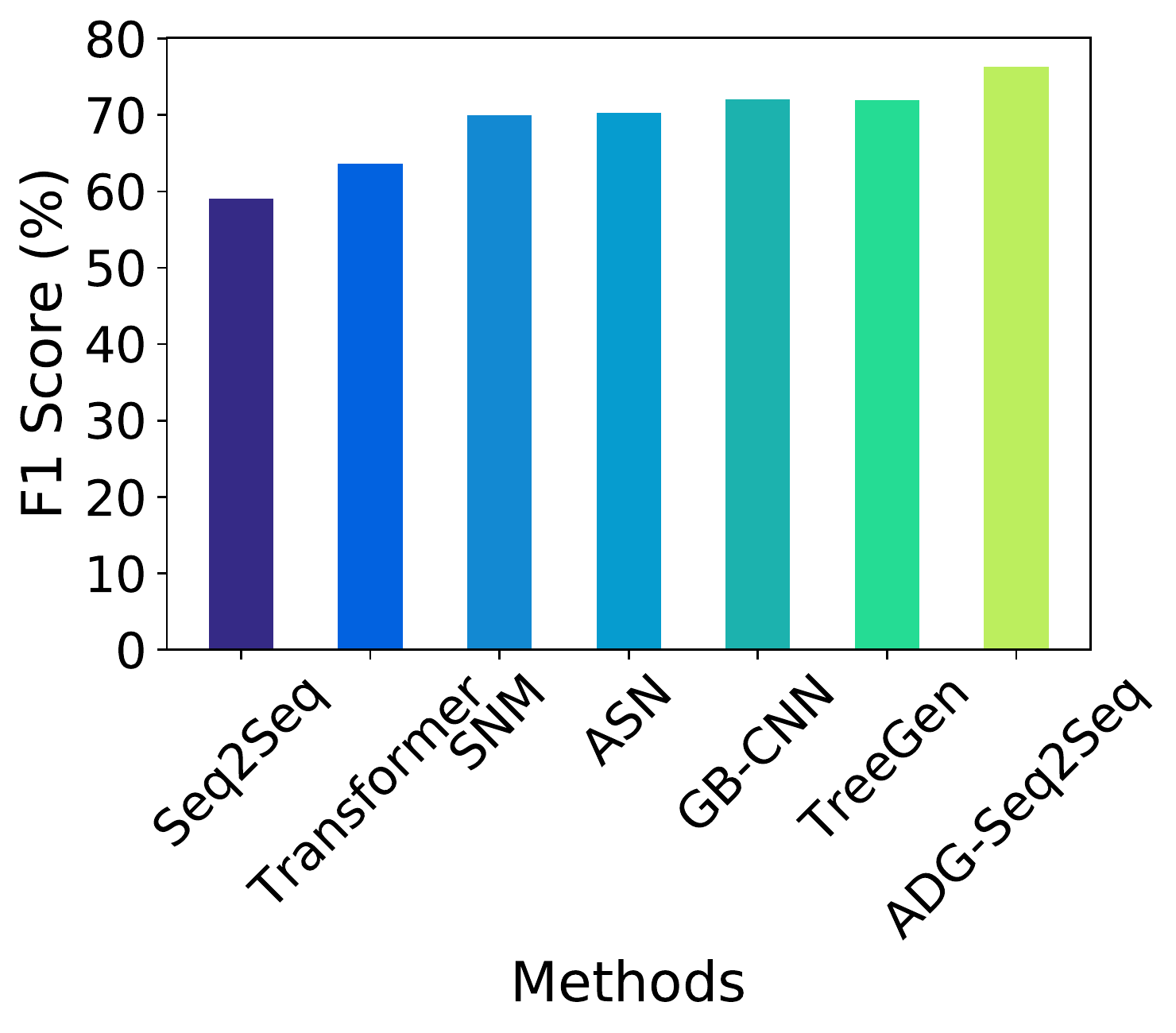}
			\label{MTG:F1}
		}
		\subfigure[MTG: CIDEr]{
			\includegraphics[width=0.22\linewidth]{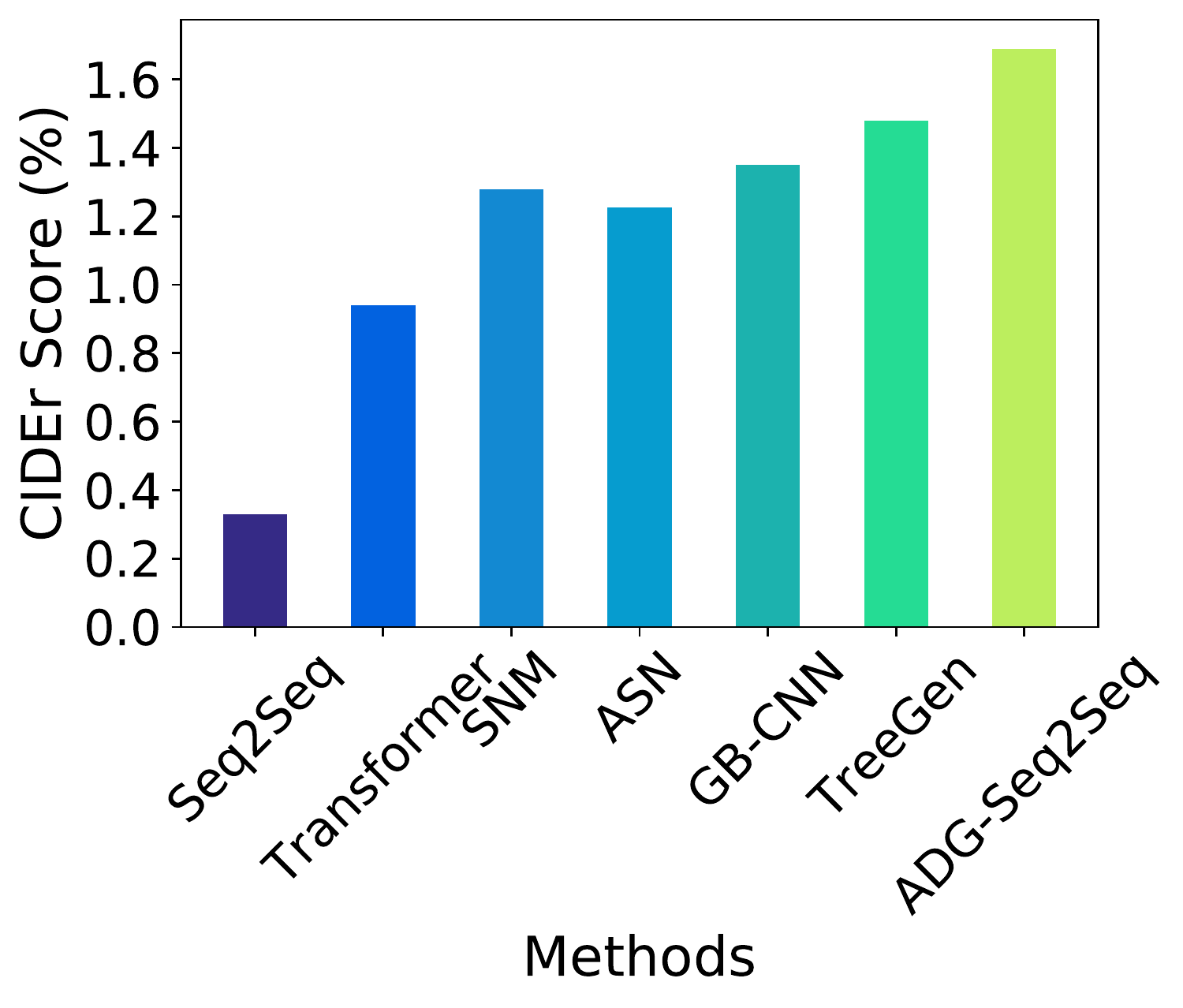}
			\label{MTG:CIDEr}
		}	
		
		\subfigure[MTG: ROUGE-L]{
			\includegraphics[width=0.22\linewidth]{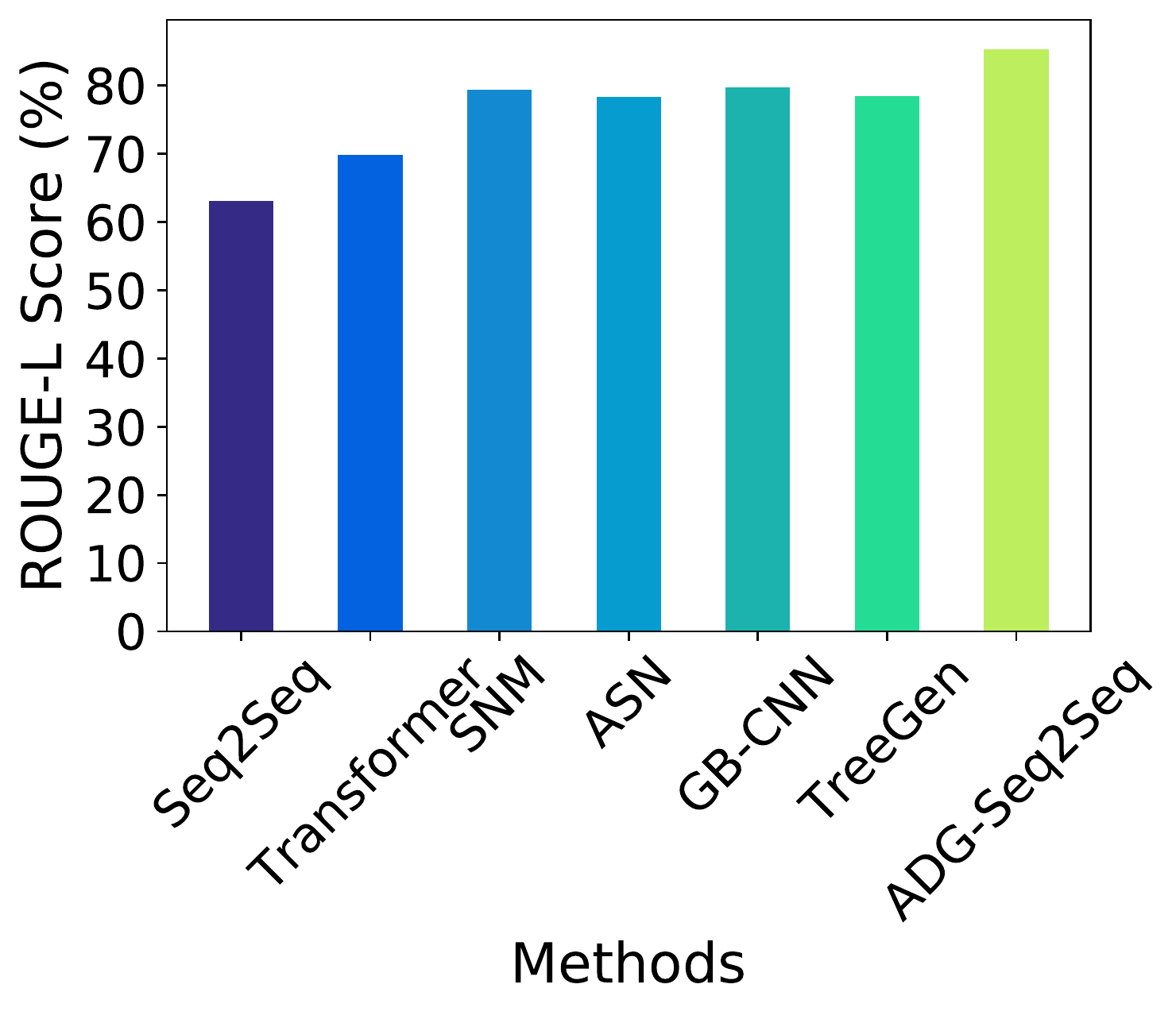}
			\label{MTG:ROUGE-L}
		}
		\subfigure[MTG: ROUGE-1]{
			\includegraphics[width=0.22\linewidth]{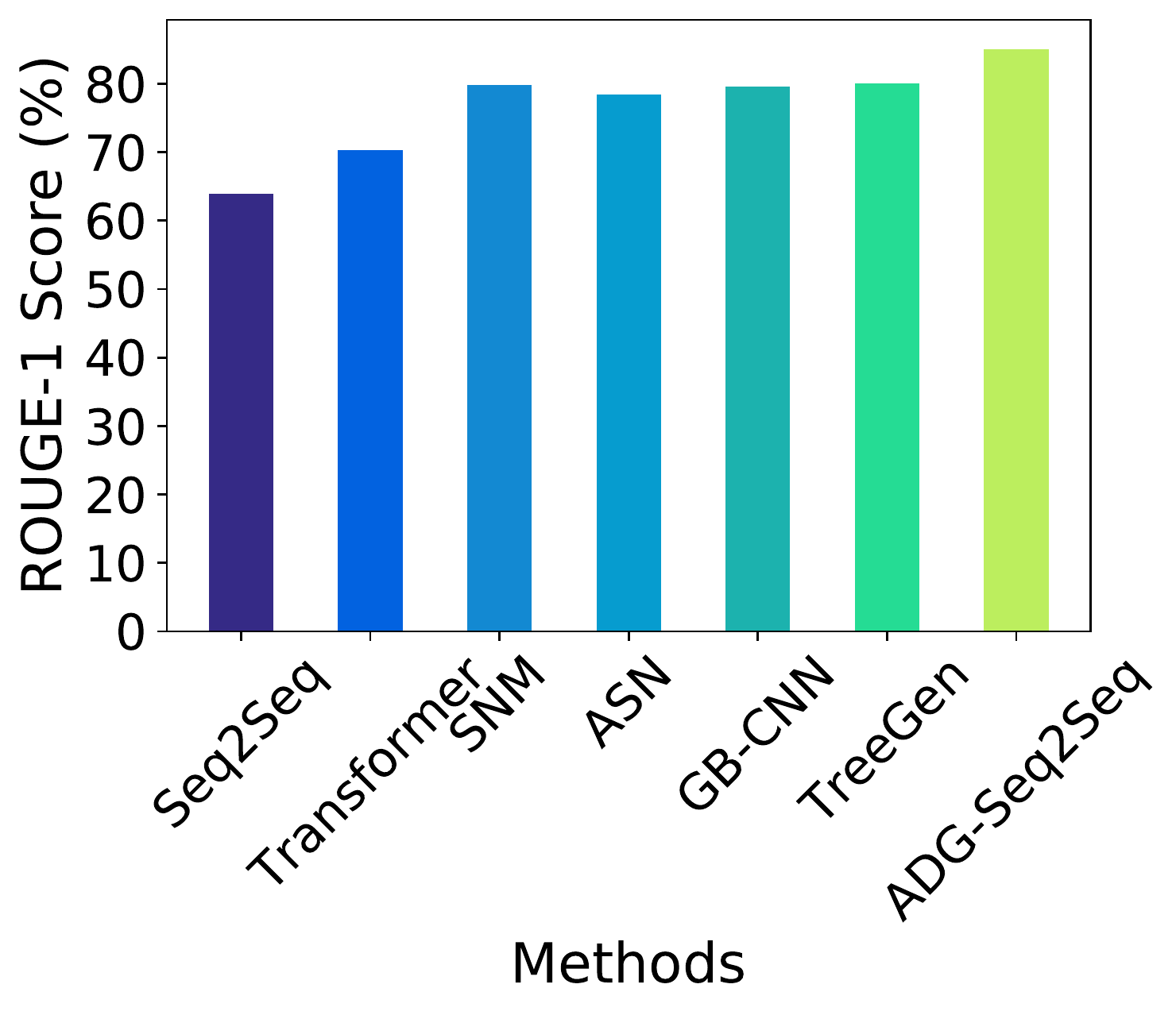}
			\label{MTG:ROUGE-1}
		}
		\subfigure[MTG: ROUGE-2]{
			\includegraphics[width=0.22\linewidth]{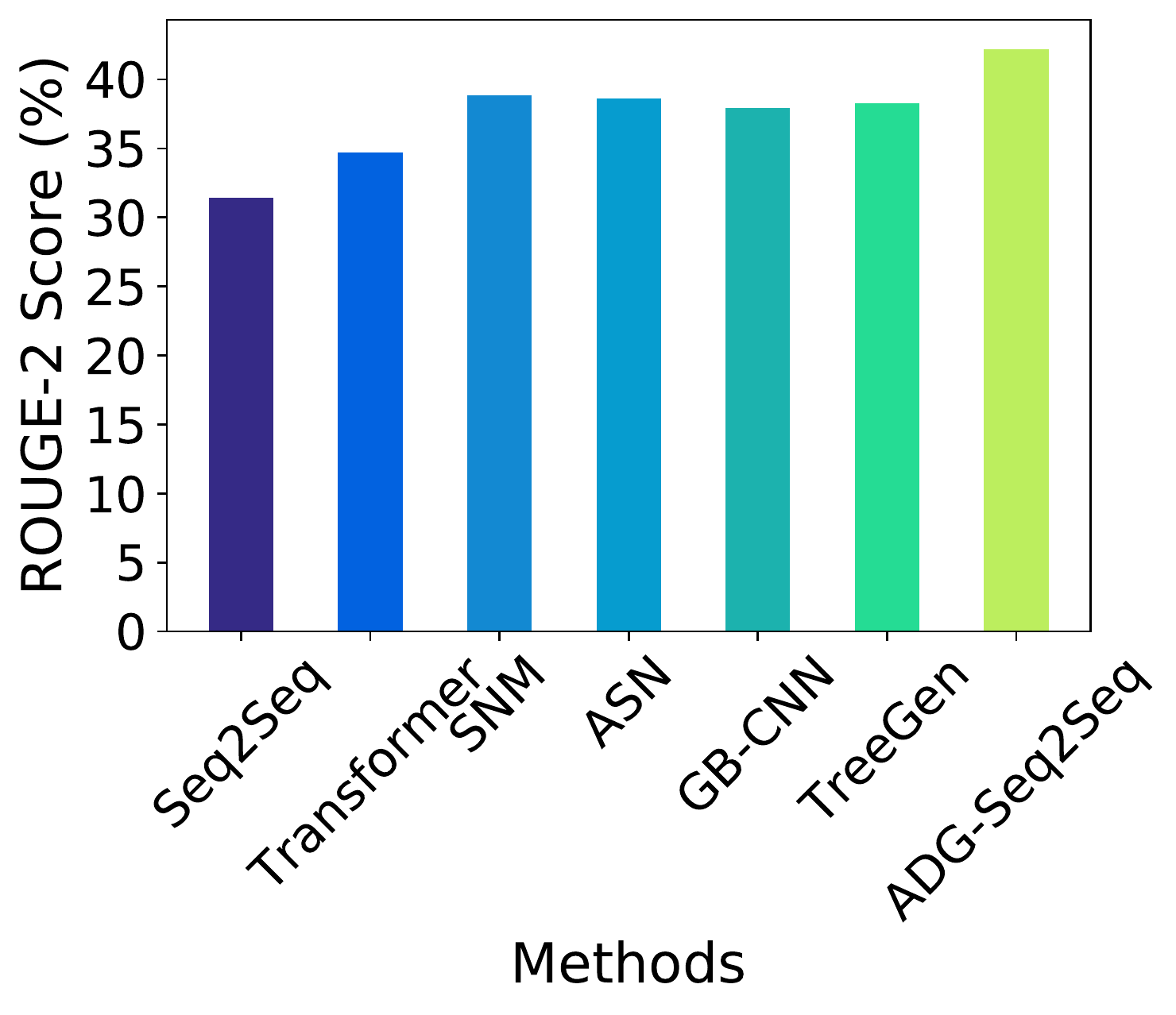}
			\label{MTG:ROUGE-2}
		}
		\subfigure[MTG: RIBES]{
			\includegraphics[width=0.22\linewidth]{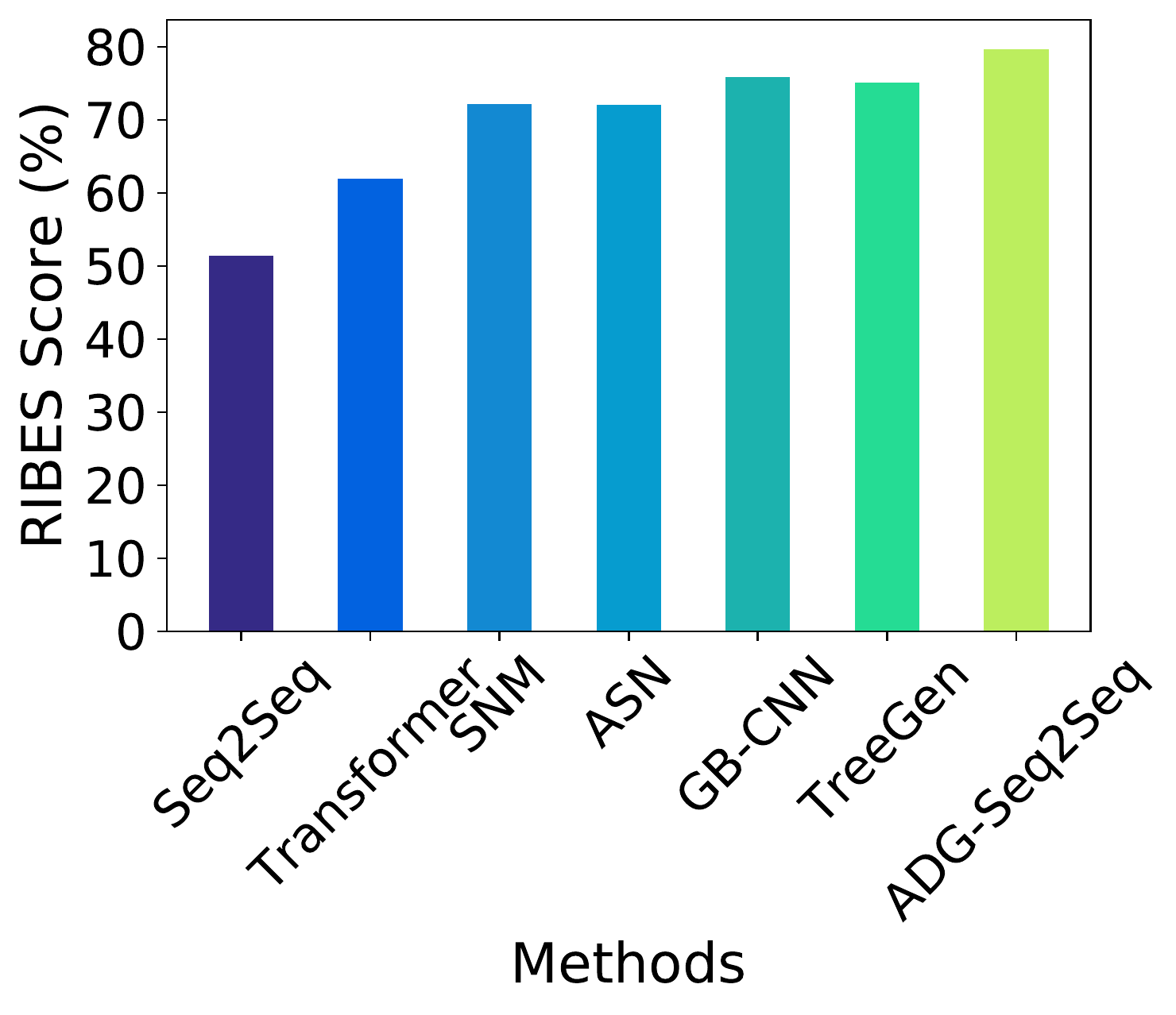}
			\label{MTG:RIBES}
		}
		
		\subfigure[EJDT: Acc]{
			\includegraphics[width=0.22\linewidth]{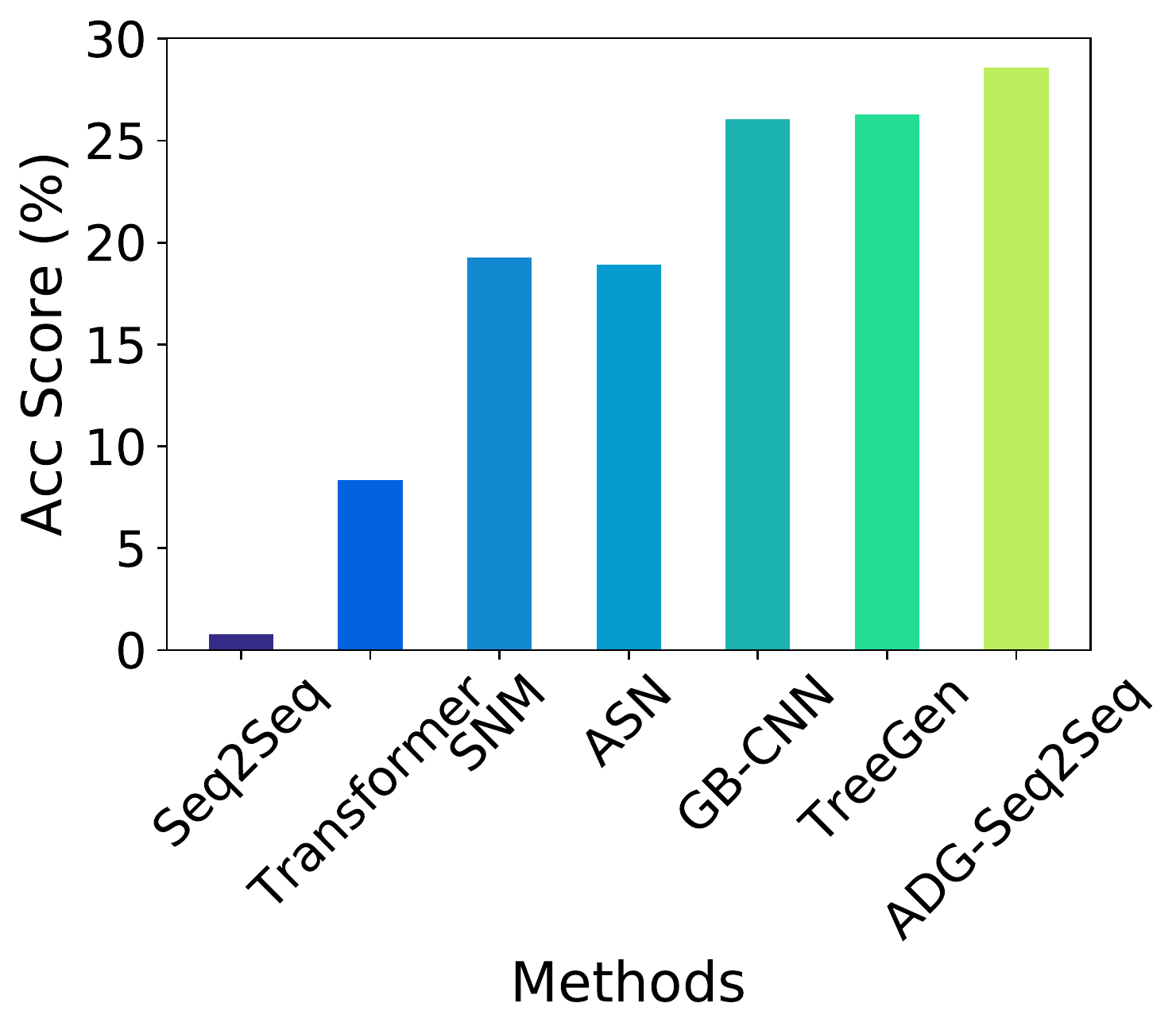}
			\label{E-JDT:Acc}
		}
		\subfigure[EJDT: BLEU]{
			\includegraphics[width=0.22\linewidth]{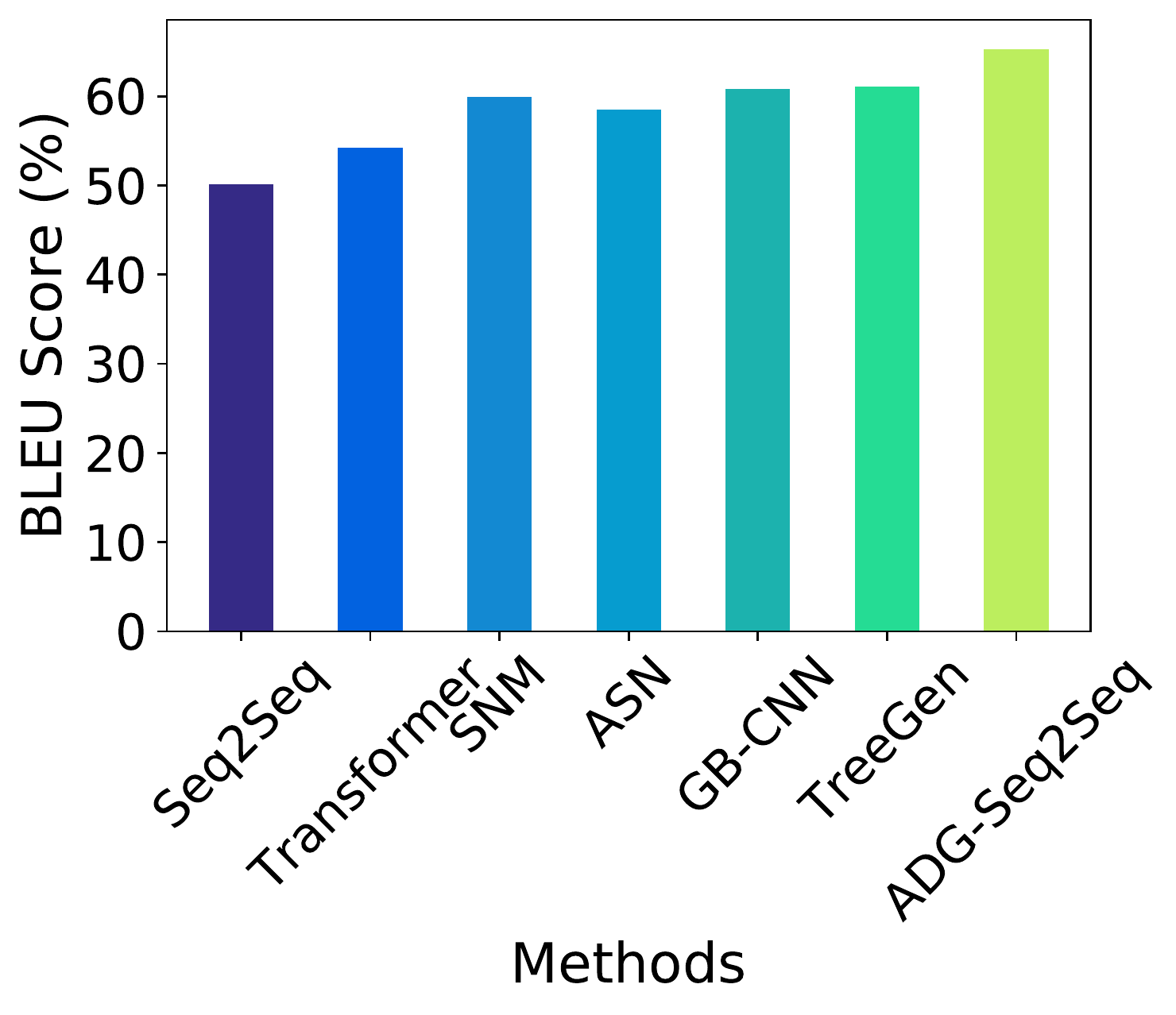}
			\label{E-JDT:BLEU}
		}
		\subfigure[EJDT: F1]{
			\includegraphics[width=0.22\linewidth]{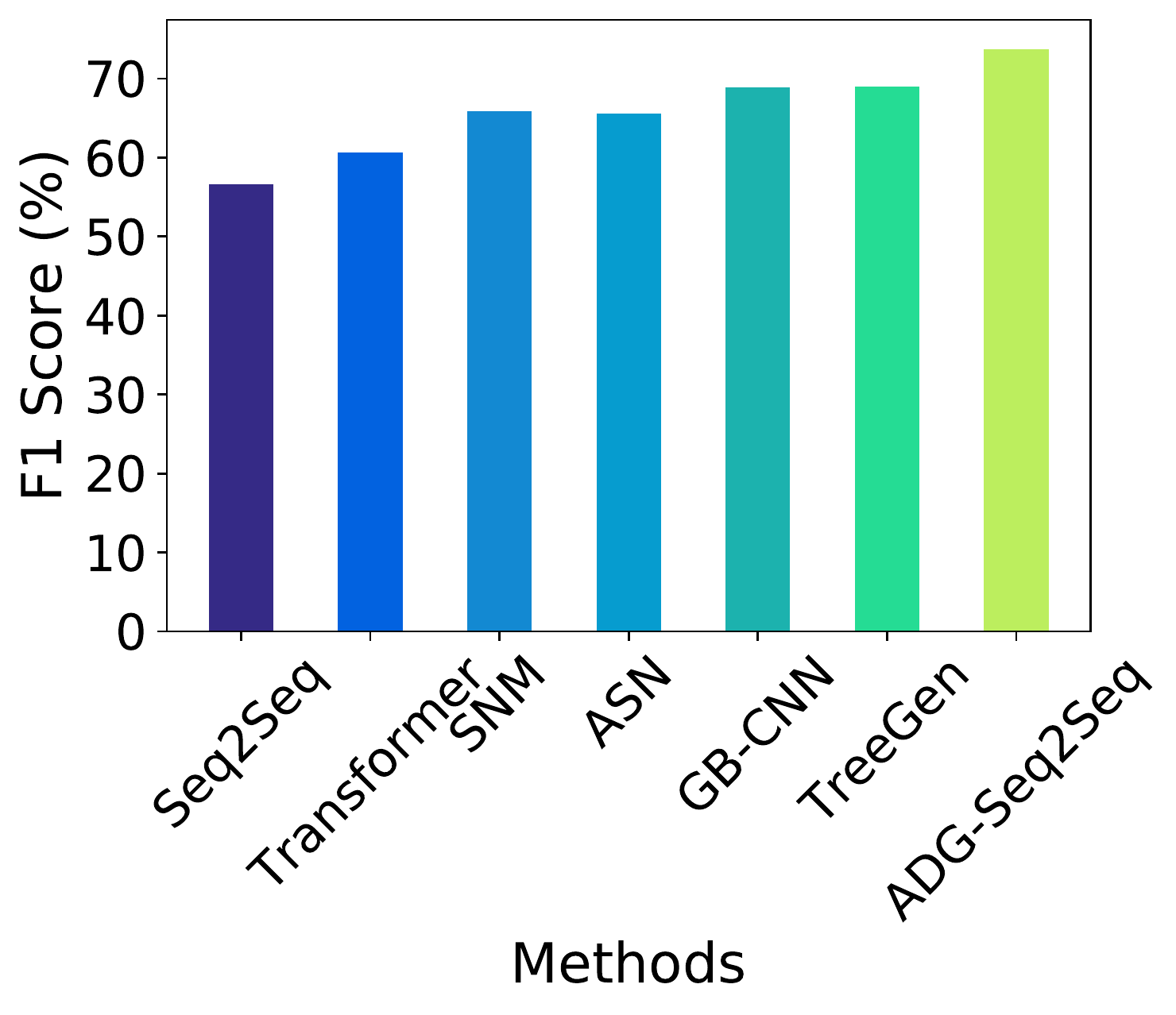}
			\label{E-JDT:F1}
		}
		\subfigure[EJDT: CIDEr]{
			\includegraphics[width=0.21\linewidth]{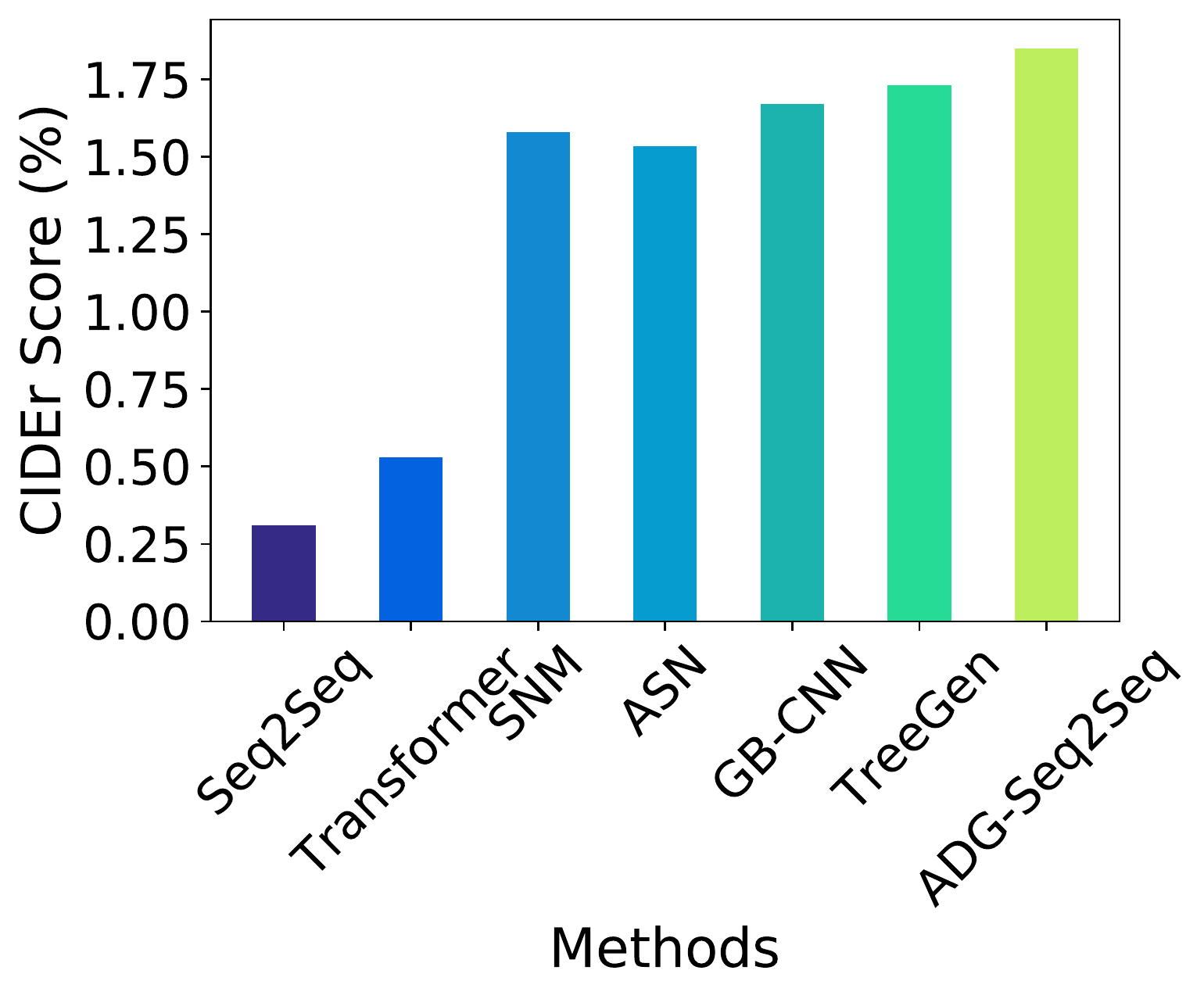}
			\label{E-JDT:CIDEr}
		}
		
		\subfigure[EJDT: ROUGE-L]{
			\includegraphics[width=0.22\linewidth]{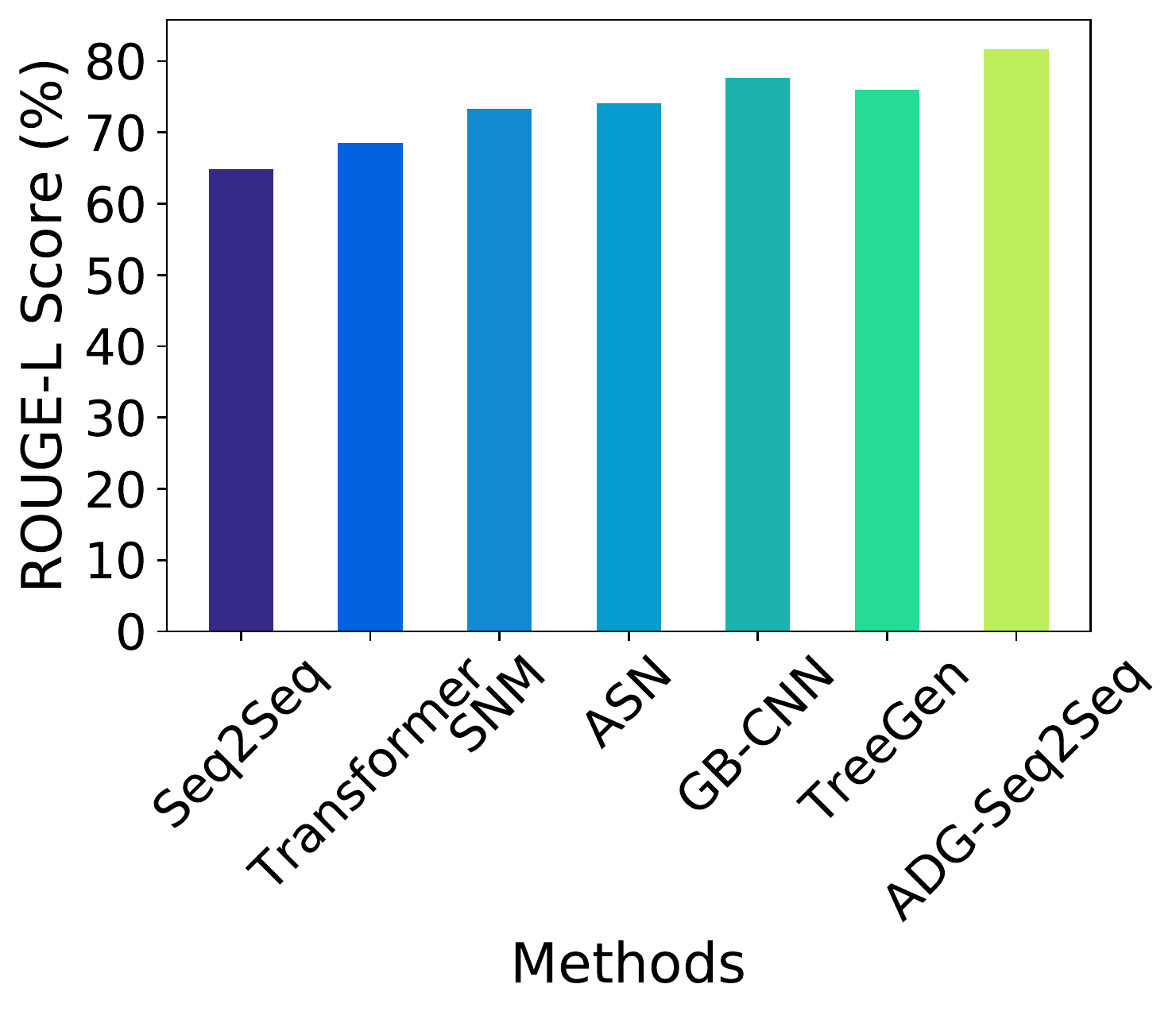}
			\label{E-JDT:ROUGE-L}
		}
		\subfigure[EJDT: ROUGE-1]{
			\includegraphics[width=0.22\linewidth]{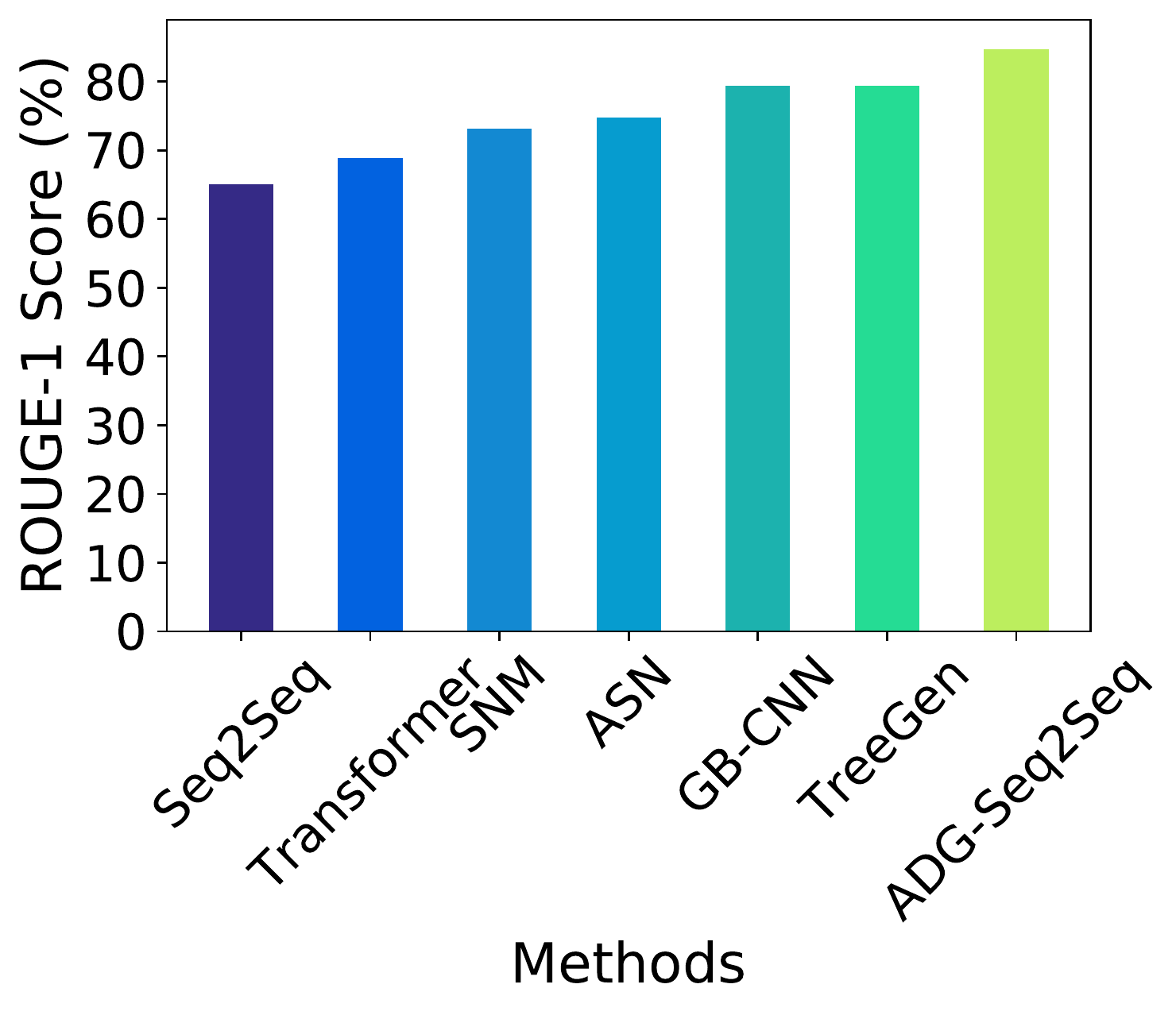}
			\label{E-JDT:ROUGE-1}
		}
		\subfigure[EJDT: ROUGE-2]{
			\includegraphics[width=0.22\linewidth]{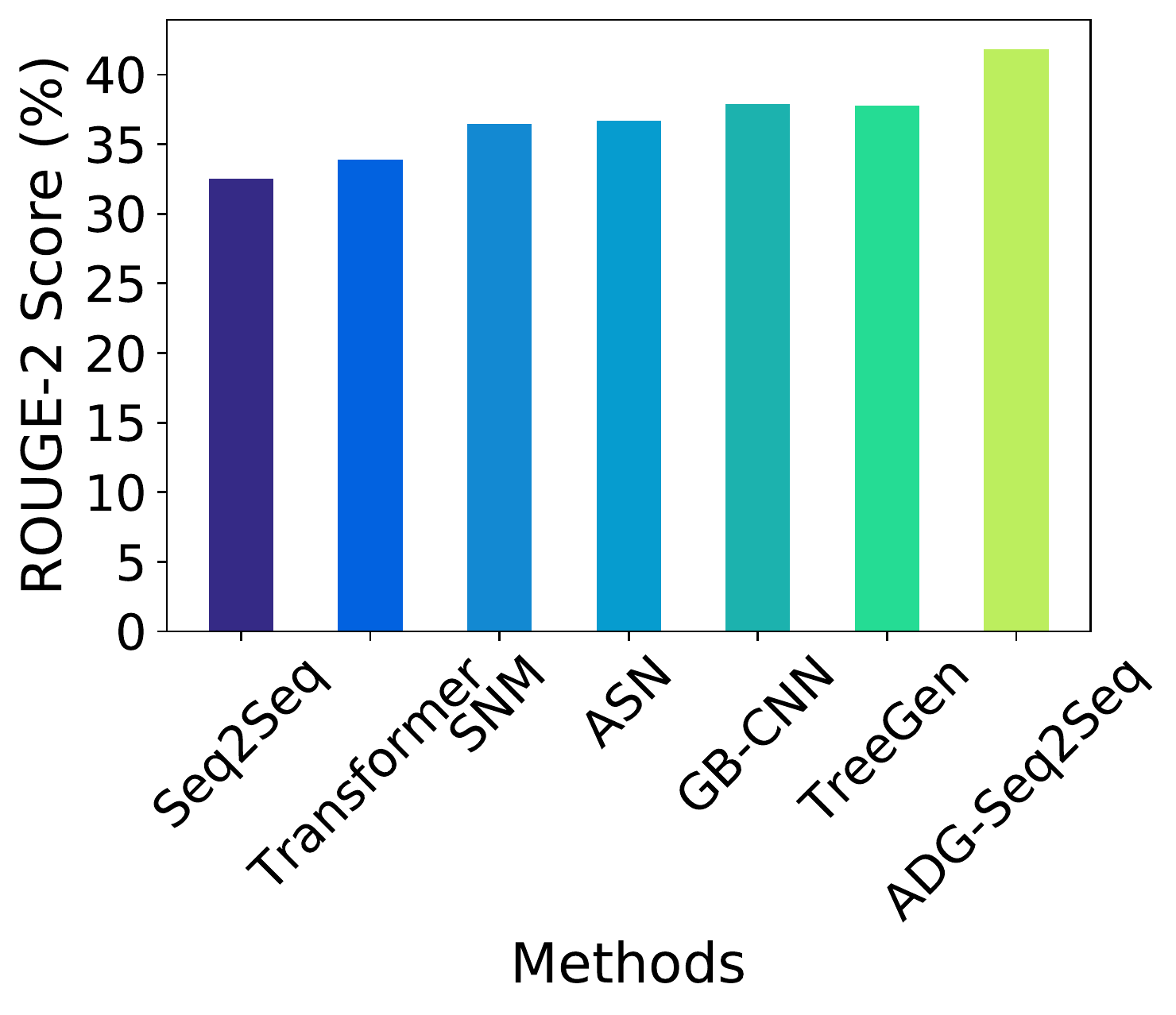}
			\label{E-JDT:ROUGE-2}
		}
		\subfigure[EJDT: RIBES]{
			\includegraphics[width=0.22\linewidth]{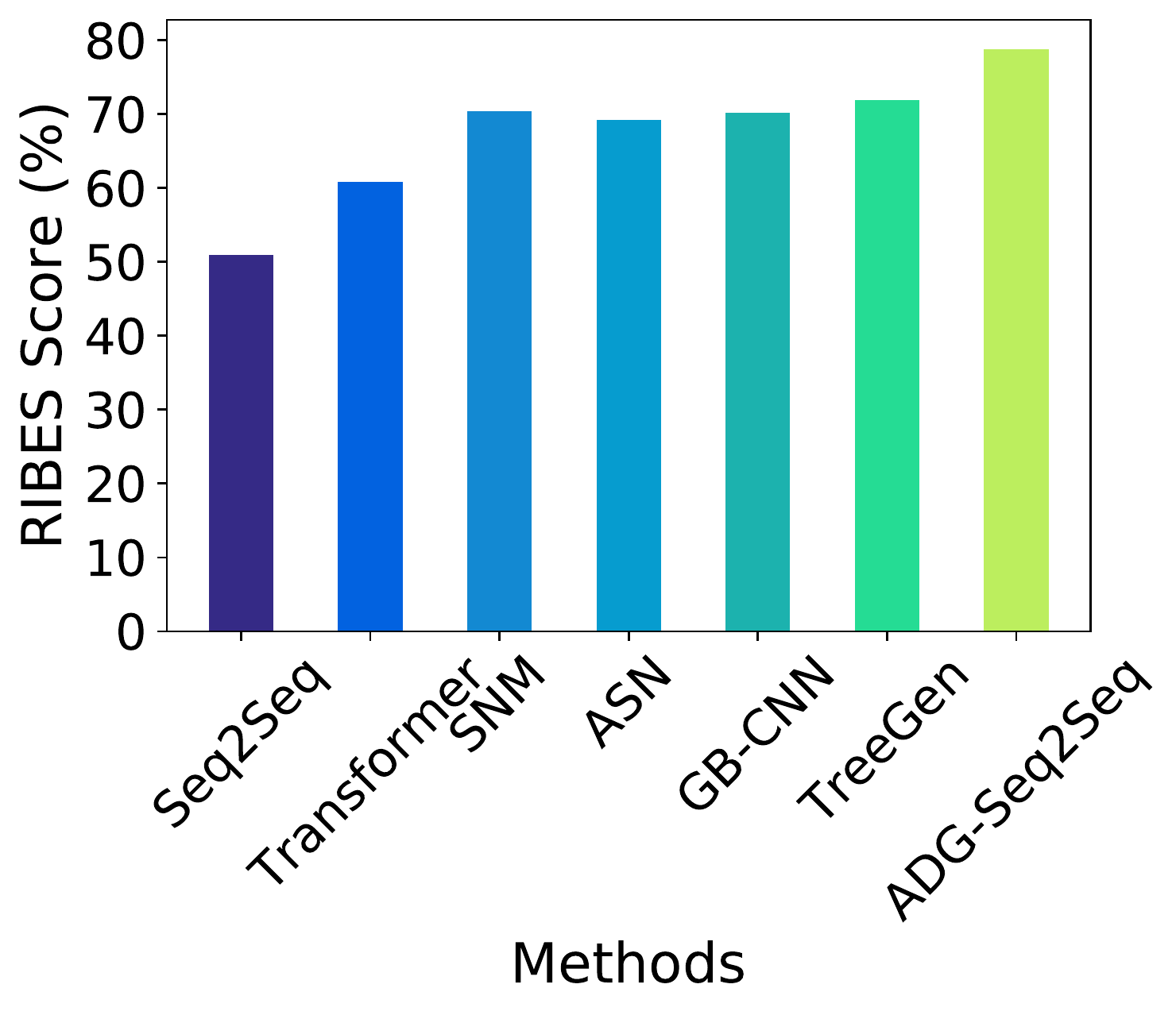}
			\label{E-JDT:RIBES}
		}
		\caption{\textbf{Various metric scores obtained in the experiments.} }
		\label{fig:result}
	\end{figure*}

	Comprehensive comparison results of all three datasets are provided in Table~\ref{tab:table3} and illustrated in Fig.~\ref{fig:result}.
	
	\textit{\textbf{HS}}: Fig.~\ref{fig:result} (a) - (h) shows the results of various methods. TreeGen achieves the best Acc and BLEU scores, while the proposed method outperforms the others in terms of F1, CIDEr, ROUGE, and RIBES. TreeGen uses the AST-based Transformer to learn syntactical information within API methods and exhibits a better capacity for addressing long-range dependency, thus increasing the related scores. Compared to TreeGen, ADG-Seq2Seq pays more attention to the global and sequential structure of the program and designs an ADG embedding model to learn the underlying structure and dependencies among APIs. The learned global and sequential information is conducive to generating more comprehensive code. Therefore, the performance of ADG-Seq2Seq on other metrics, such as ROUGE, which considers the recall rate between tokens, is the best, regardless of whether ROUGE-1, ROUGE-2, or ROUGE-L is considered. Beyond ROUGE, the highest scores achieved in terms of CIDEr reflect the enormous potential of our method, which demonstrates the importance of the global structural information for capturing important representative tokens in code snippets. Regarding RIBES, which focuses on the token order, ADG-Seq2Seq outperforms other Seq2Seq models, indicating that embedding the call constraint information helps the model more accurately predict the order in which API methods are called.
	
	\textit{\textbf{MTG}}: We further evaluate the proposed method on the MTG dataset, as shown in Fig.~\ref{fig:result} (i) - (p). The performance of various methods drops slightly in comparison to the performance on the HS dataset. Compared to HS, the programming language in MTG changes from Python to Java, and the data complexity increases significantly, as illustrated in Table~\ref{tab:table1}. Therefore, it is more difficult to predict the target code as the code length and description length grow significantly. ADG-Seq2Seq outperforms other methods on all metrics, which demonstrates that the use of ADG to represent complex API dependencies and the program-oriented embedder in our method can help address such changes in the code generation task.

	\textit{\textbf{E-JDT}}: In contrast to the other two datasets, E-JDT is characterized by unstructured descriptions and increased uncertainty with a broad range of code lengths in terms of both tokens and methods, as shown in Table~\ref{tab:table1}. The performance comparison in Fig.~\ref{fig:result} (q) - (x) demonstrates that ADG-Seq2Seq can produce more comprehensive and accurate code from natural language statements than other state-of-the-art methods on this dataset. 
	
	The above results show that the SOTA models and our model outperform the naive NMT models (e.g., Seq2Seq) and the Transformer in terms of all metrics by a substantial margin, demonstrating the importance of the integrated structural information obtained from AST or ADG. 
	Among these methods, GB-CNN, TreeGen, and ADG-Seq2Seq achieve relatively good performance, but each has its weakness. In particular, although GB-CNN \cite{SunZ} and TreeGen \cite{SunZ2} perform well on HS in terms of Acc and BLEU, they perform poorly on HS in terms of the other six metrics and perform poorly in terms of all metrics on MTG and E-JDT. GB-CNN and TreeGen show improvements over previous approaches by learning the grammar rules in ASTs and handling the long-range dependency problem in sequence-to-sequence translation. However, an AST represents the structural information within the API method from a local perspective.  Hence, the information used for learning confined in each individual AST lacks connections and interactions with that in other ASTs. In contrast, our model represents the API dependencies in the whole libraries or the entire project from a global perspective and connects them in the ADG, thereby contributing to more comprehensive and accurate code generation.

	Additionally, our method performs slightly worse on HS in terms of Acc and BLEU. We argue that the reason is that our joint learning model requires large training samples and adequate connections between nodes in the ADG. Hence, the effectiveness of our approach, especially as measured by precision-related indicators, degrades if the quantities of sample data and connections are insufficient, as shown in Tables~\ref{tab:table1} - \ref{tab:table2}. ADG-Seq2Seq has the best performance on MTG and E-JDT, where the quantities of training samples and data connections in the constructed ADG are typically greater than those for HS by several orders of magnitude.
	
	In addition, we observe that the recall-based ROUGE-L exhibits more improvements than the precision-based BLEU 
	on all three datasets. The precision-based metric BLEU-N excels at assessing the precision of generated sentences and their linguistic fluency. ROUGE-L, in contrast, is primarily recall-based and therefore tends to reward long sentences with high recall \cite{cider}. To obtain further insights, we have verified the experimental results and analysed them in detail. The experimental results show that on HS, TreeGen generates more ``structured" code with more stable quality because of its ability to batch input and output data. ADG-Seq2Seq, in contrast, relies more substantially on sequential information, which, coupled with the API dependencies, tends to generate longer results than TreeGen. Based on the above observations, the results of TreeGen are better in terms of BLEU, while ADG-Seq2Seq performs better under the recall-based metrics. Note that these results are for the HS dataset; the improvements could vary across datasets.

	An evaluation metric called the percentage of valid code (PoV) was proposed by Wei et al. \cite{Dual} to evaluate the performance of code generation based on the percentage of the generated code that could be parsed into an AST (i.e., compilable code). We also compute the PoV value to assess the quality of the generated code. For ADG-Seq2Seq, the PoVs tested on HS, MTG and E-JDT are 68.2\%, 55.1\% and 53.0\%, respectively. The PoVs produced by the attention-based Seq2Seq are much lower (22.7\%, 20.8\%, and 19.3\%, respectively), demonstrating the ability of our model to 
    increase the percentage of valid code.

	In summary, ADG-Seq2Seq achieves the best performance on 22 indicators (except for two indicators in HS), which demonstrates the superiority of our method for neural code generation.
			
	\subsection{Answer to RQ2}
	\label{ARQ2}
	To validate the benefits of the ADG embedding, we evaluate different 
	variants of embedders and different design options of the ADG.

	\subsubsection{Examination of Variations}
		
	In this paper, we develop an ADG embedding algorithm to learn the node features from the ADG. The graph embedding algorithms GCN and GraphSAGE are known for their efficiency and good performance in many areas. To explore the advantages of our methods, we further compare four variations using different graph embedding strategies, namely, GCN, GraphSAGE, ADG embedding, and the model without graph embedding. We refer to the reported results \cite{HamiltonW,zhou2020effective} and set the number of GCN layers to two and the number of hops for GraphSAGE and ADG embedding to two.
	
	A quantitative comparison of the final code generation results (in terms of BLEU on E-JDT) based on these approaches is shown in Table~\ref{tab:table4}. The model without graph embedding produces the worst results. The model using GraphSAGE outperforms GCN, while the results based on the ADG embedding are the best. These comparisons show that the models with graph embedding are better than those without. This result confirms the positive effect of the graph embedding algorithm, notably, the ADG embedding algorithm designed explicitly for program structure, on the quality of code generation results.

	\begin{table}[!htbp]
		\centering
		\caption{\textbf{Results of various graph embedding modules.} The scores represent the BLEU metric.}
		\begin{threeparttable}
			\begin{tabular}{p{120px}rrr}
				\toprule
				\textbf{Graph Embedding} & \textbf{HS} & \textbf{MTG}   & \textbf{E-JDT} \\
				\midrule
				$-$ADG Embedding\tnote{1} & 60.4  & 54.8  & 50.1  \\
				$+$ADG Embedding\tnote{2} & \textbf{78.1} & \textbf{69.2} & \textbf{65.3} \\
				ours $\rightarrow$ GCN\tnote{3} & 69.3  & 60.8  & 55.2  \\
				ours $\rightarrow$ GraphSAGE\tnote{4} & 70.8  & 63.4  & 58.5  \\
				\bottomrule
			\end{tabular}%
			\begin{tablenotes}
				\raggedright
				\footnotesize
				{\item [1] \textit{Remove the graph embedding module, namely, an NMT model.}
					\item [2] \textit{A model with our proposed module.}
					\item [3] \textit{Replace the ADG embedding module with GCN, a GCN-based Seq2Seq model.}
					\item [4] \textit{Replace the ADG embedding module with GraphSAGE, a GraphSAGE-based Seq2Seq model.}}
			\end{tablenotes}
		\end{threeparttable}
		\label{tab:table4}%
	\end{table}%

	\begin{figure}[!htbp]
		\centering
		\includegraphics[width=0.65\linewidth]{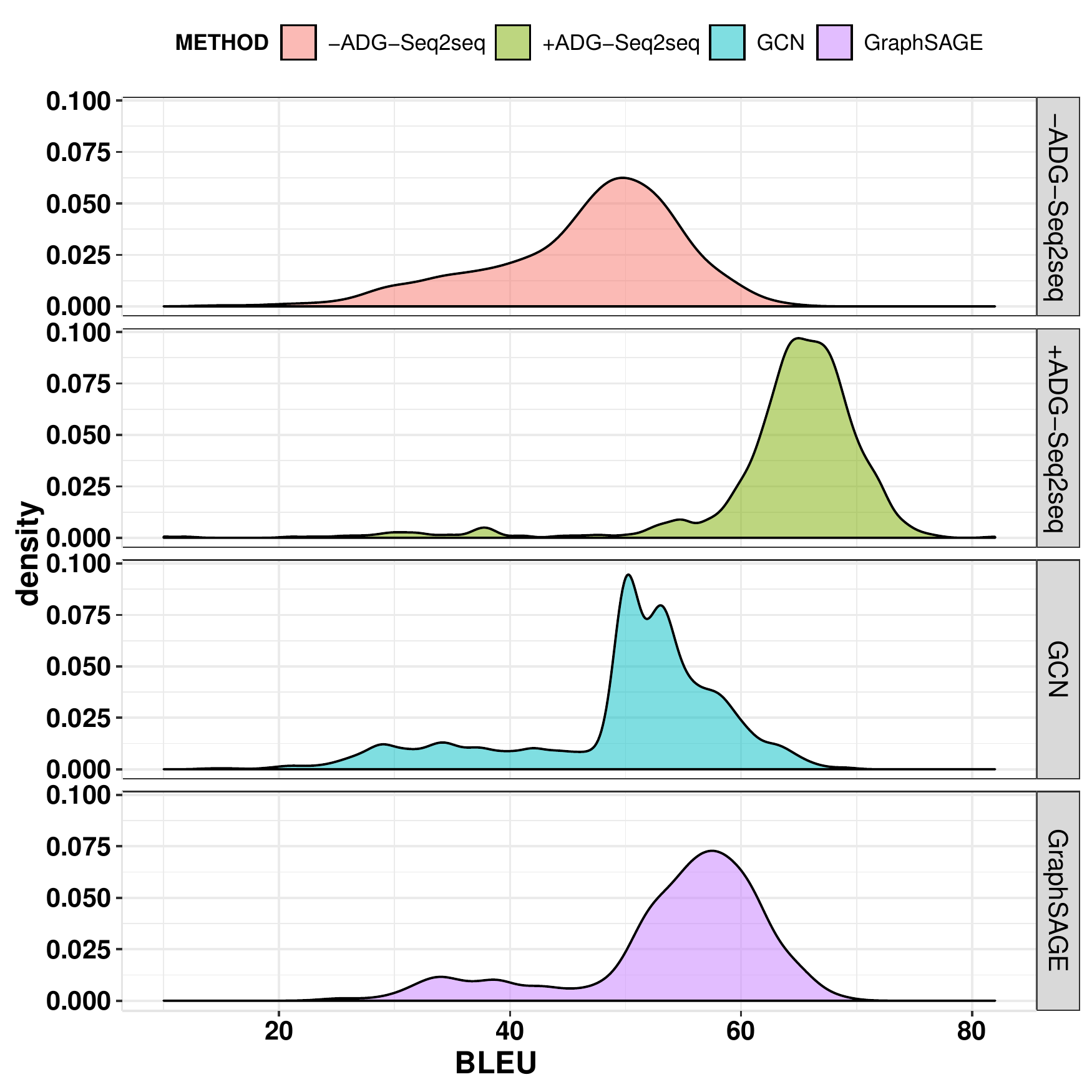}
		\caption{\textbf{\textit{BLEU} distribution density of each item in testing.} Various graph embedding algorithms applied to the E-JDT dataset generate different distributions.}
		\label{fig:density}
	\end{figure}
	
	Another observation is that ADG-Seq2Seq appears to be the most stable model. A visual distribution comparison of graph embedding approaches is shown in Fig.\ref{fig:density}. The method without graph embedding obtains the lowest values at the peak, mean, and median points. The results produced by the models based on GCN and GraphSAGE are better, but their left tails appear to be much thicker, which indicates that there are more abnormally weak scores. The distribution generated by ADG-Seq2Seq is approximately symmetric, with the highest peak, mean, and median. Moreover, the shape of the distribution tends to be tighter, so most data are centred around the peak, which indicates that more results of our method achieve better ratings. The above results reveal the significance of our innovative design for learning the program graph structure considering the special reachability and invocation order between APIs. This approach helps preserve the calling order and restricts invalid invocations in the predictions. As a result, our approach outperforms the competitors.
	
	In summary, the results presented in Table~\ref{tab:table4} and Fig.~\ref{fig:density} show that our ADG embedding mechanism has advantages over other variations and is effective for neural code generation.
	
	\subsubsection{Examination of Design Options}
	\label{ablation test}
	We examine different design options of the ADG and the corresponding algorithm, as shown in Table \ref{tab:LPers}. The detailed results are as follows. 

	\begin{table*}[htbp]
		\small
		\centering
		\caption{\textbf{Experimental results of our model with different ADG parameter settings. }} 
		\resizebox{\textwidth}{45mm}{
		\begin{tabular}{clcccccccc}
			\toprule
			&       & \multicolumn{1}{c}{Acc} & \multicolumn{1}{c}{Bleu} & \multicolumn{1}{c}{F1} & \multicolumn{1}{c}{CIDEr} & \multicolumn{1}{c}{RougeL} & \multicolumn{1}{c}{Rouge1} & \multicolumn{1}{c}{Rouge2} & \multicolumn{1}{c}{RIBES} \\
			\midrule
			\multirow{8}[2]{*}{\begin{sideways}HS\end{sideways}} & Full model & \textbf{27.3} & \textbf{78.1} & \textbf{82.5} & \textbf{2.02} & \textbf{87.4} & \textbf{87.5} & \textbf{43.5} & \textbf{80.4} \\
			& -- directed edges & 21.2  & 71.5  & 77.4  & 1.85  & 84.3  & 84.3  & 41.9  & 77.8  \\
			& -- labelled edges & 24.2  & 72.2  & 77.9  & 1.87  & 84.6  & 84.7  & 42.1  & 78.1  \\
			& One-hop size & 24.2  & 70.1  & 76.3  & 1.81  & 83.6  & 83.7  & 41.6  & 77.3  \\
			& Two-hop size & 27.3  & 78.1  & 82.5  & 2.02  & 87.4  & 87.5  & 43.5  & 80.4  \\
			& Mean aggregator & 19.7  & 70.0  & 76.2  & 1.81  & 83.6  & 83.6  & 41.6  & 77.2  \\
			& Pooling aggregator & 21.2  & 71.4  & 77.3  & 1.85  & 84.2  & 84.3  & 41.9  & 77.8  \\
			& LSTM aggregator & 27.3  & 78.1  & 82.5  & 2.02  & 87.4  & 87.5  & 43.5  & 80.4  \\
			\midrule
			\multirow{8}[2]{*}{\begin{sideways}MTG\end{sideways}} & Full model & \textbf{29.4} & \textbf{69.2} & \textbf{76.3} & \textbf{1.69} & \textbf{85.4} & \textbf{85.1} & \textbf{42.2} & \textbf{79.7} \\
			& -- directed edges & 17.3  & 59.3  & 67.6  & 1.53  & 78.4  & 78.5  & 39.0  & 73.1  \\
			& -- labelled edges & 17.8  & 62.3  & 70.0  & 1.61  & 79.7  & 79.9  & 39.7  & 74.2  \\
			& One-hop size & 17.3  & 61.1  & 69.0  & 1.58  & 79.3  & 79.3  & 39.4  & 73.8  \\
			& Two-hop size & 29.4  & 69.2  & 76.3  & 1.69  & 85.4  & 85.1  & 42.2  & 79.7  \\
			& Mean aggregator & 17.5  & 59.5  & 67.7  & 1.54  & 78.5  & 78.6  & 39.1  & 73.2  \\
			& Pooling aggregator & 17.9  & 61.8  & 69.6  & 1.60  & 79.6  & 79.7  & 39.6  & 74.1  \\
			& LSTM aggregator  & 29.4  & 69.2  & 76.3  & 1.69  & 85.4  & 85.1  & 42.2  & 79.7  \\
			\midrule
			\multirow{8}[2]{*}{\begin{sideways}E-JDT\end{sideways}} & Full model & \textbf{28.6} & \textbf{65.3} & \textbf{73.8} & \textbf{1.85} & \textbf{81.7} & \textbf{84.7} & \textbf{41.8} & \textbf{78.8} \\
			& -- directed edges & 14.5  & 53.3  & 62.5  & 1.38  & 75.6  & 75.6  & 37.6  & 70.7  \\
			& -- labelled edges & 16.1  & 54.5  & 63.7  & 1.50  & 76.5  & 76.7  & 38.1  & 72.3  \\
			& One-hop size & 15.8  & 54.4  & 63.5  & 1.41  & 76.1  & 76.1  & 37.9  & 71.2  \\
			& Two-hop size & 28.6  & 65.3  & 73.8  & 1.85  & 81.7  & 84.7  & 41.8  & 78.8  \\
			& Mean aggregator & 17.7  & 56.2  & 65.0  & 1.45  & 76.9  & 77.0  & 38.3  & 71.9  \\
			& Pooling aggregator & 18.4  & 56.9  & 65.6  & 1.47  & 77.3  & 77.3  & 38.5  & 72.1  \\
			& LSTM aggregator & 28.6  & 65.3  & 73.8  & 1.85  & 81.7  & 84.7  & 41.8  & 78.8   \\
			\bottomrule
			
		\end{tabular}}
		
		\label{tab:LPers}
	\end{table*}%
		
	\begin{itemize}
		\item \textbf{Undirected Edges.} We first examine the option without edge direction. The rows marked with $-directed$ $edges$ illustrate the performance of our method learning on a graph with undirected edges. As shown by the results in these rows, the original full model outperforms this approach. According to our algorithm, dropping the direction information means ignoring the API invocation order in the learning process, which thereby leads to poor performance. In particular, the average BLEU score of this option is $12$ points lower than that of the full model on E-JDT. The RIBES score focuses on the evaluation of sequencial order, and more comprehensively reflects the influence caused by the lack of direction information. Compared with the scores of the full model, the RIBES scores of this approach decrease by large margins on all datasets. Based on the above observations, the use of edge direction in our generation framework contributes to the overall performance. Note that this experiment is equivalent to testing the case where the ADG embedding algorithm does not encode the API invocation order.
		
		\item \textbf{No Labelled Edges.} We further examine the performance when using unlabelled edges, as shown in rows $-labelled$ $edge$. In the ADG, the various labels (tags) on incoming or outgoing edges of each node represent the different input or output parameter types of the corresponding API method. The full model can learn valid dependencies among APIs by fusing these tags/types and their relationships into our generation model. Therefore, the full model outperforms the option without labels on all metrics. Note that this experiment is equivalent to testing the case where the ADG embedding algorithm does not encode the API invocation constraints.
		
		\item \textbf{Different Hop Sizes.} Since parameter K in Algorithm 1 can be specified as 1 or 2, we examine the performance of embedders using one- or two-hop sizes. The results using different hop sizes are shown in the rows $one-hop$ $size$ and $two-hop$ $size$. One-hop size means that the embedder gathers node information for only the first layer, namely, direct neighbours' information. Two-hop modelling gathers information about a node's two-layer neighbours, namely, both immediate neighbours and immediate neighbours' neighbours, the number of which is almost a square of the number of neighbours in one-hop size.. Meanwhile, the API dependency information between these neighbours is also included. Code-related features, such as API invocation order and API invocation constraints, are embedded in this information. 
		More comprehensive API dependency knowledge can be learned from ADG using a two-hop size. Therefore, the embedder using the two-hop size always achieves higher scores than that using the one-hop size, e.g., by approximately 12\% to 80\% in terms of Acc on the three datasets. This is why we use the two-hop size in the full model.
		
		\item \textbf{Different Aggregators.} Finally, we compare different aggregators used in the ADG embedding algorithm, as shown in Table~\ref{tab:LPers}. For all metrics, the results of LSTM aggregators are generally 3.35\% to 63.8\% higher than the results of the pooling aggregators, and 4.08\% to 68.10\% higher than the results of mean aggregators. These results indicate that different aggregators produce different effects. The mean aggregator ignores the invocation order of nodes as well as some of the collected information, such as the direction of edges. The pooling aggregator obtains the most crucial information among the order set and is more efficient but has the same defect as the mean aggregator. LSTM is not as efficient as the pooling aggregator but can hold complete order information in the order set, which may be the reason why LSTM achieves the highest scores.
	\end{itemize}
	
	In summary, all these results demonstrate that ADG embedding is of great value and is effective for the code generation task.

	\subsection{Answer to RQ3}
	For most NMT tasks, the lengths of input and output sequences are the primary factors affecting generation performance. In this section, we investigate external factors, such as dataset parameters, to provide the answer to RQ3.
		
	\subsubsection{Impact of Description Length}

\begin{figure}[h]
	\centering
	\subfigure[Description length (HS)]{
		\includegraphics[width=0.30\linewidth]{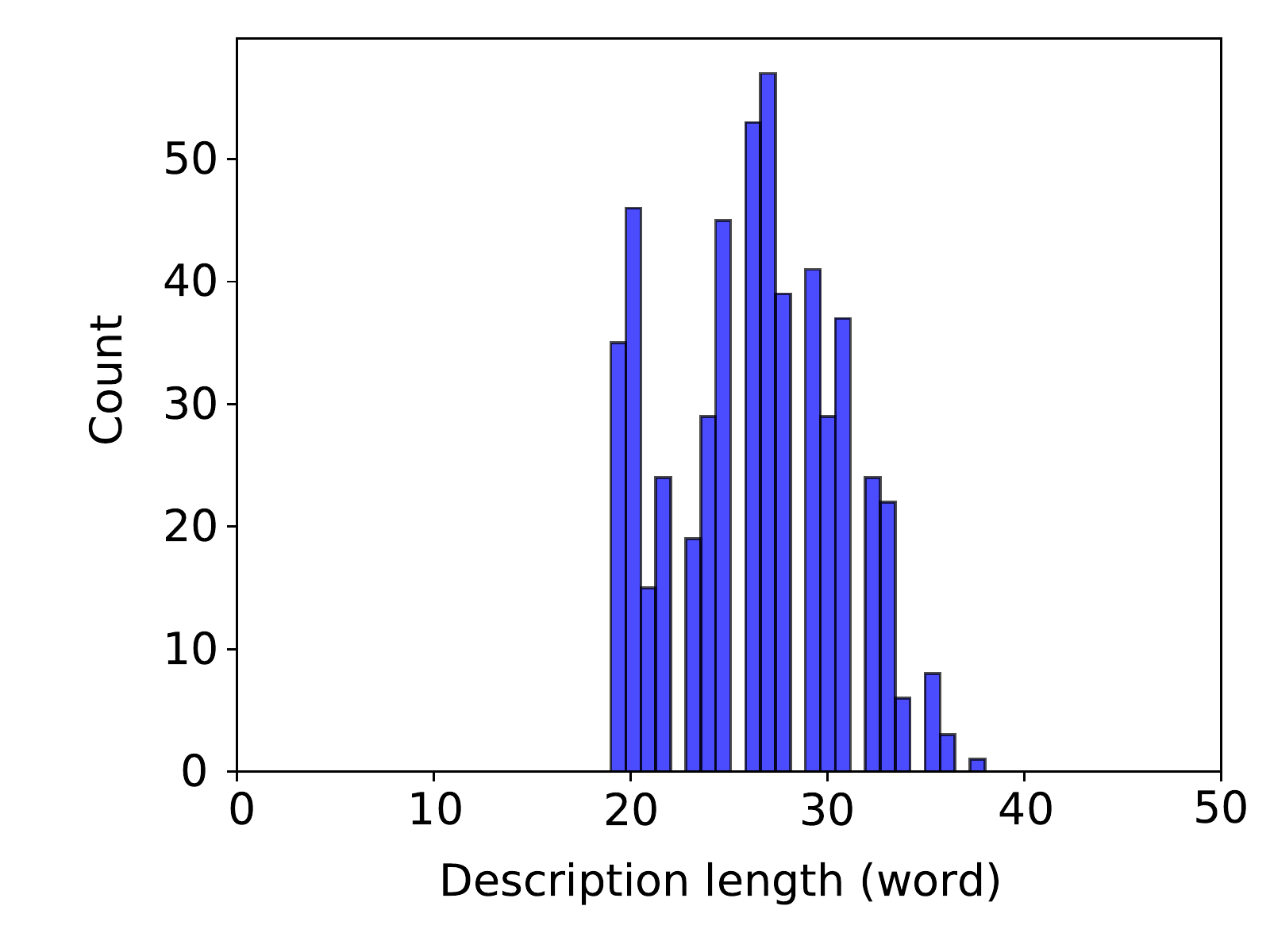}
		\label{description-HS-distribute}
	}
	\subfigure[Description length (MTG)]{
		\includegraphics[width=0.31\linewidth]{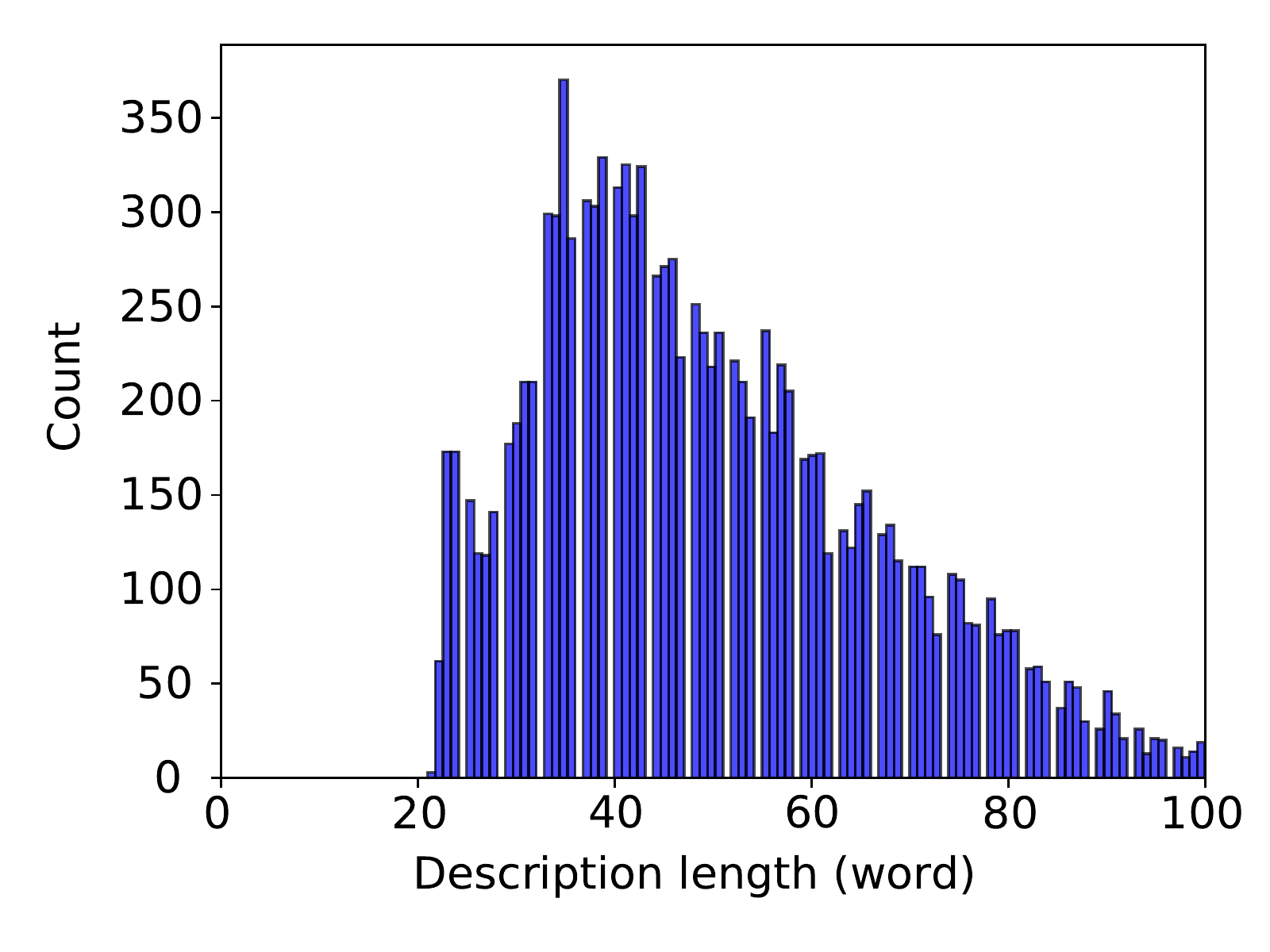}
		\label{description-MTG-distribute}
	}
	\subfigure[Description length (EJDT)]{
		\includegraphics[width=0.31\linewidth]{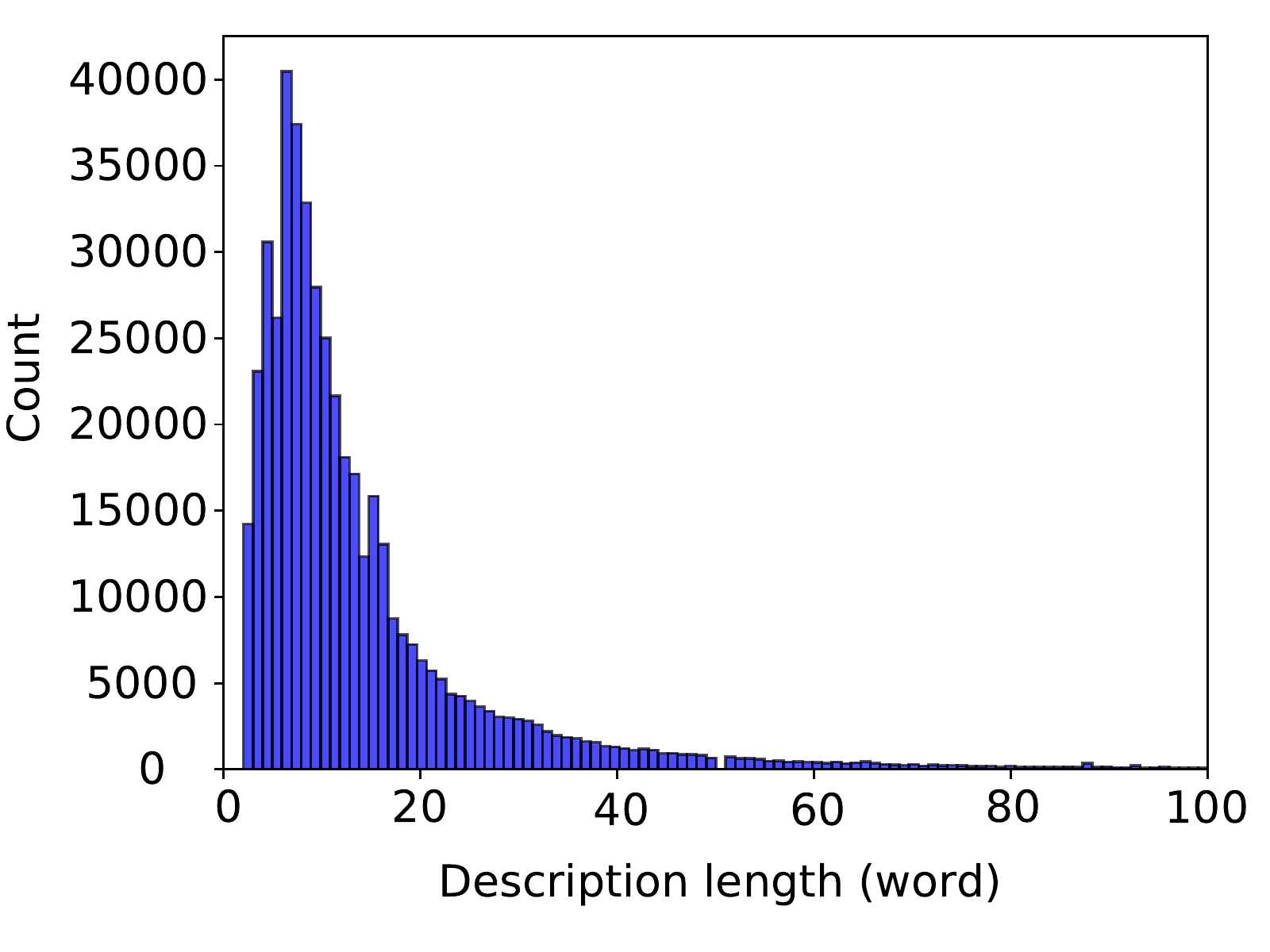}
		\label{description-EJDT-distribute}
	}
	
	\subfigure[Code length (HS)]{
		\includegraphics[width=0.30\linewidth]{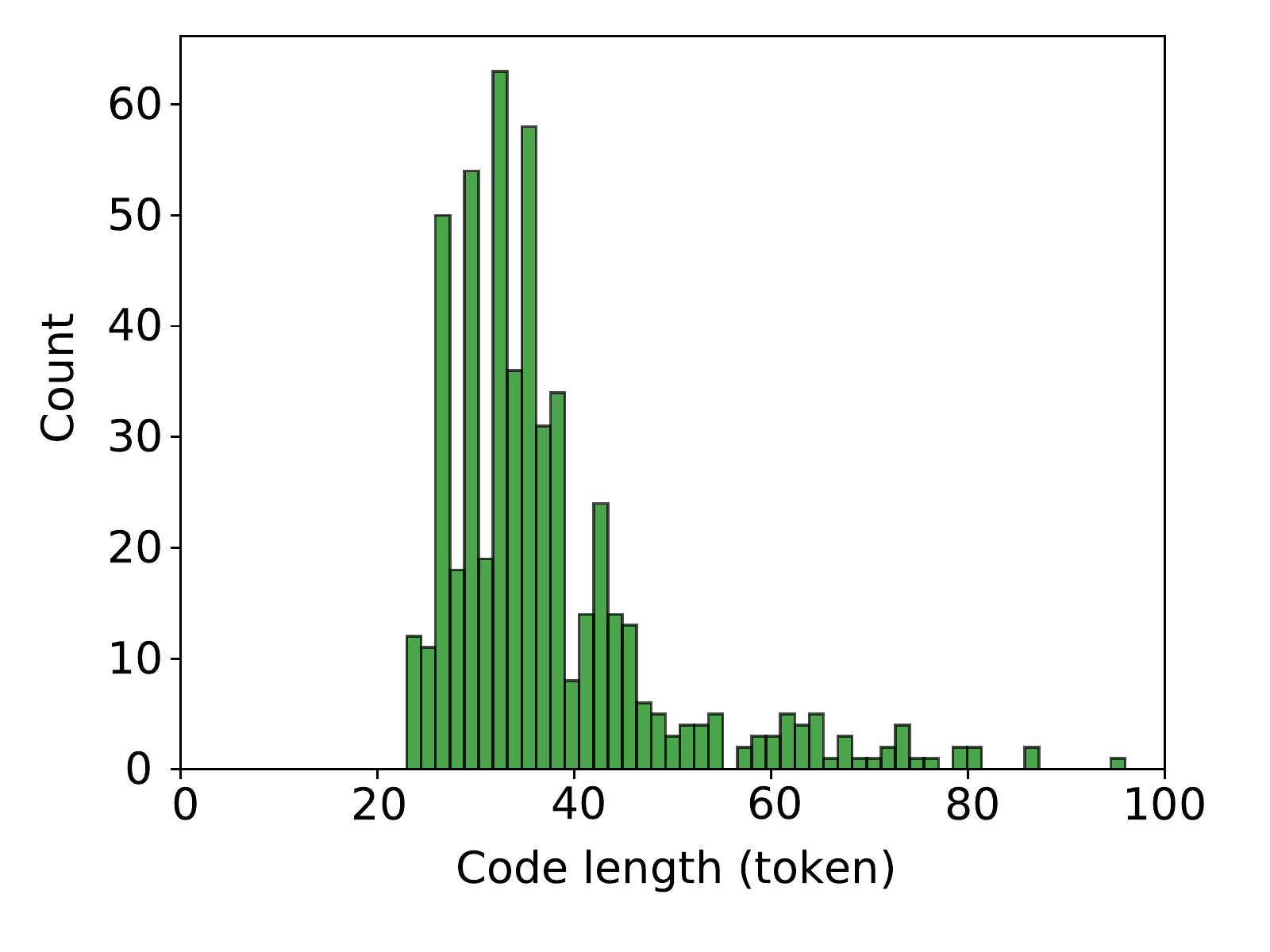}
		\label{code-HS-distribute}
	}
	\subfigure[Code length (MTG)]{
		\includegraphics[width=0.30\linewidth]{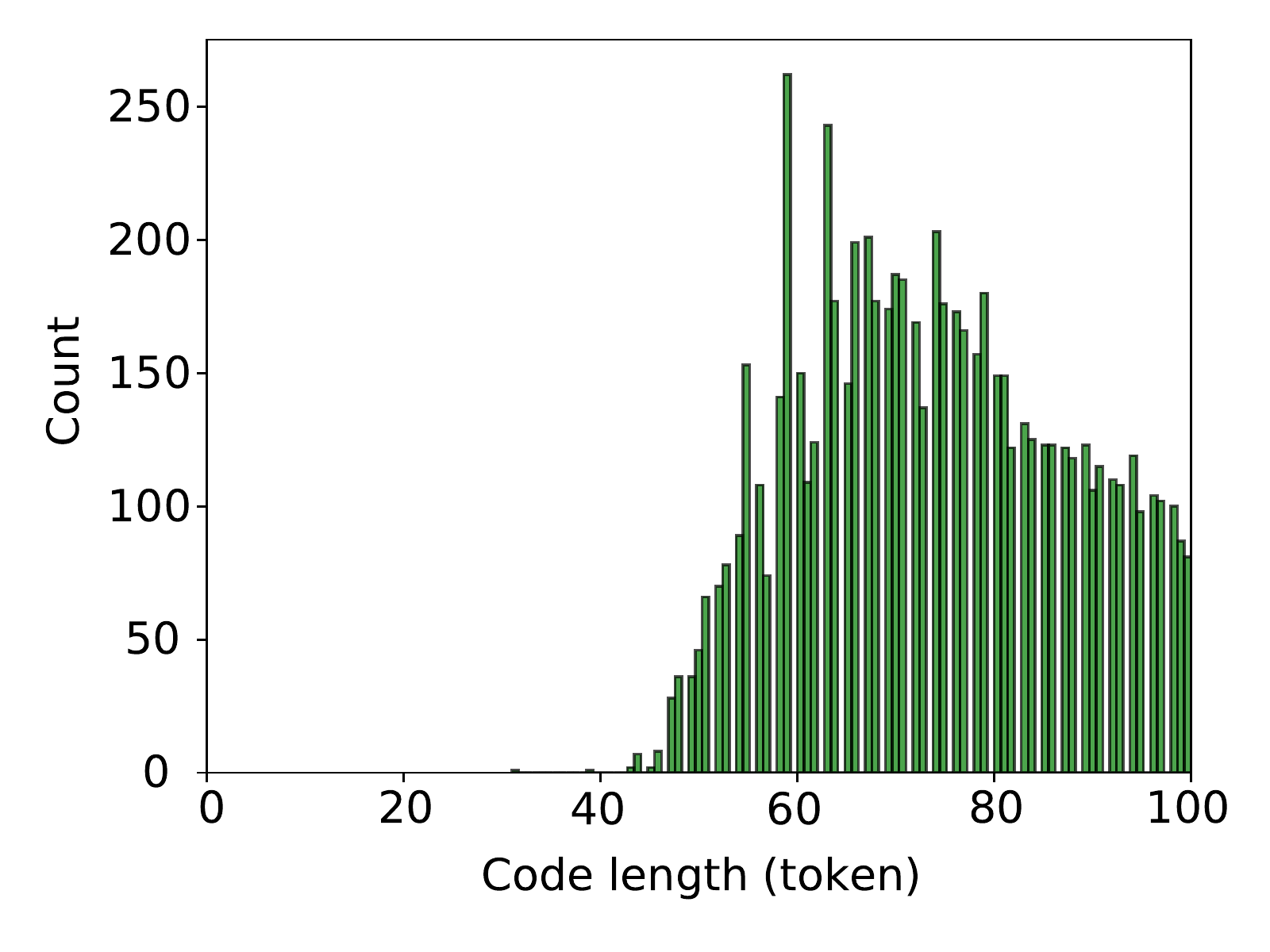}
		\label{code-MTG-distribute}
	}
	\subfigure[Code length (EJDT)]{
		\includegraphics[width=0.30\linewidth]{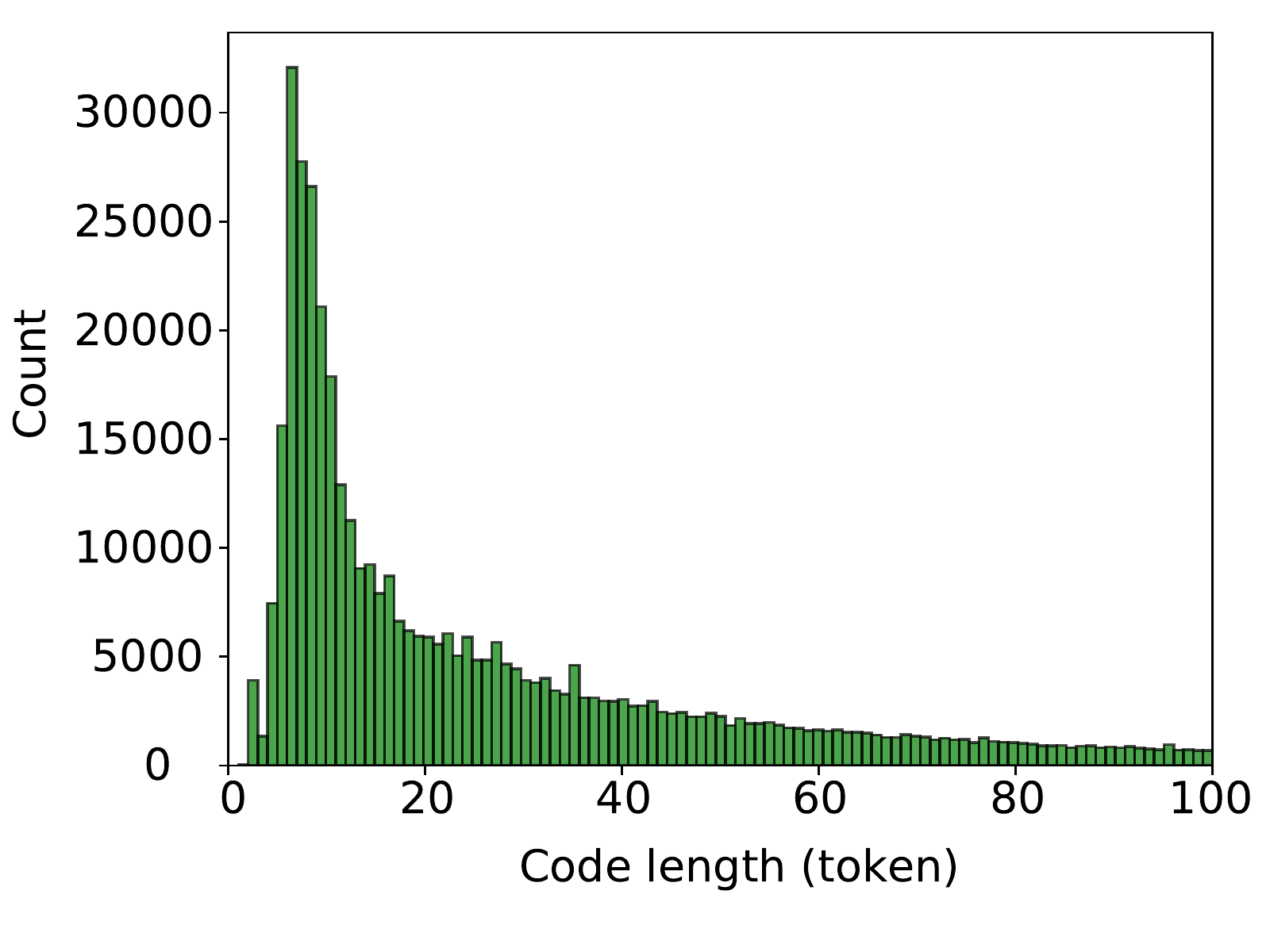}
		\label{code-EJDT-distribute}
	}
	\caption{\textbf{Distribution of data with different code/description lengths.} }
	\label{fig:dataset_statistics}
\end{figure}

	A textual program description states the purpose of the code, which is important for the establishment of learning models. Description lengths of HS are small and no more than 40, as shown in  Fig.\ref{fig:dataset_statistics}(a); MTG has a large range of description lengths, concentrating in the range of 20 to 100, as shown in Fig.\ref{fig:dataset_statistics}(b). The description lengths of E-JDT have a wide range, but most are less than 40, as shown in Fig.\ref{fig:dataset_statistics}(c). To explore the influence of varying description lengths on different metrics, we first split datasets according to length distributions given by Fig.\ref{fig:dataset_statistics}.  The statistics of the new split datasets are shown in Table~\ref{tab:deslen}.
	
\begin{figure*}[!htbp]
		\centering
		\subfigure[HS: Acc]{
			\includegraphics[width=0.22\linewidth]{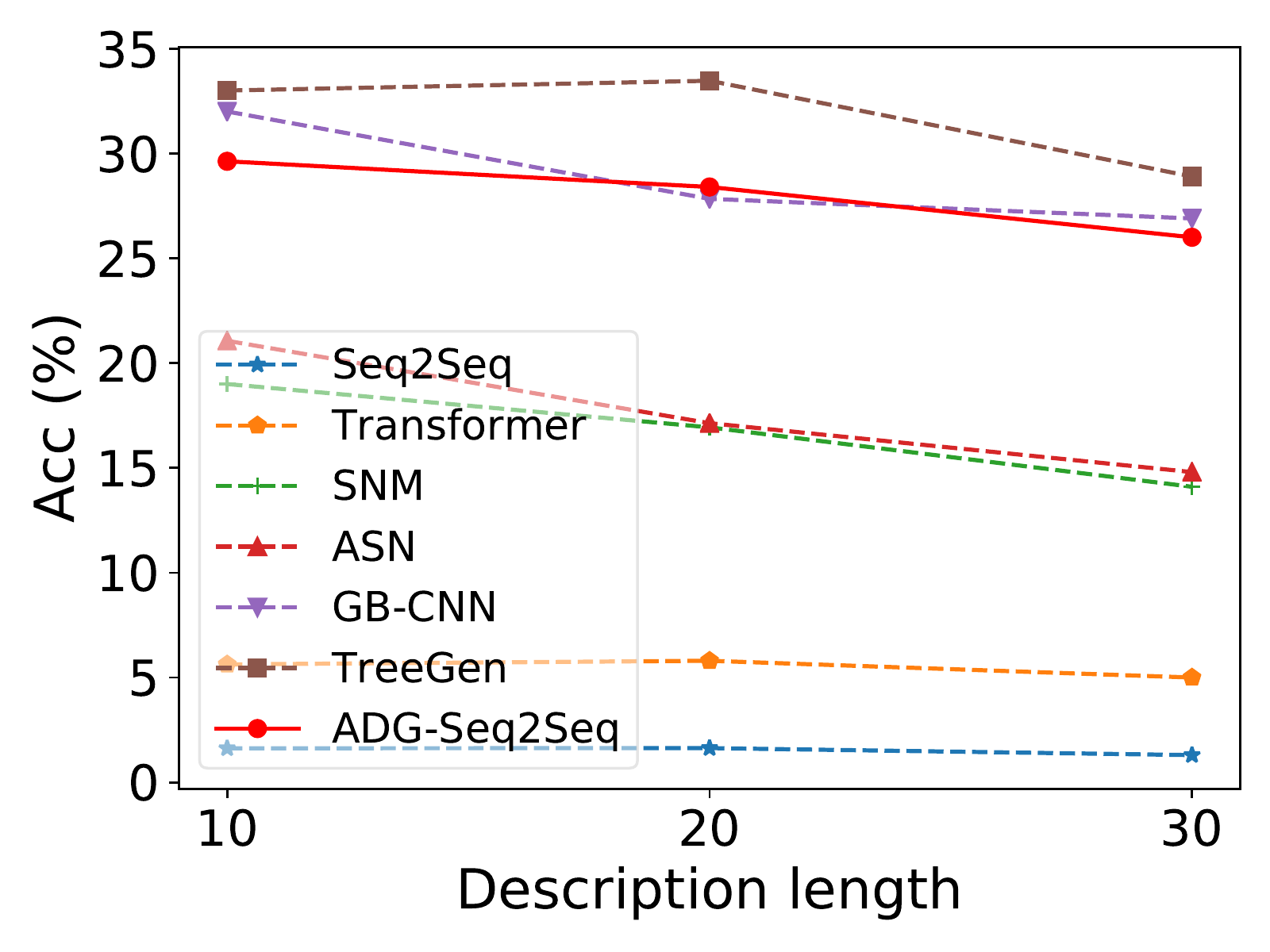}
			\label{DesLen1_HS_Acc}
		}
		\subfigure[HS: BLEU]{
			\includegraphics[width=0.22\linewidth]{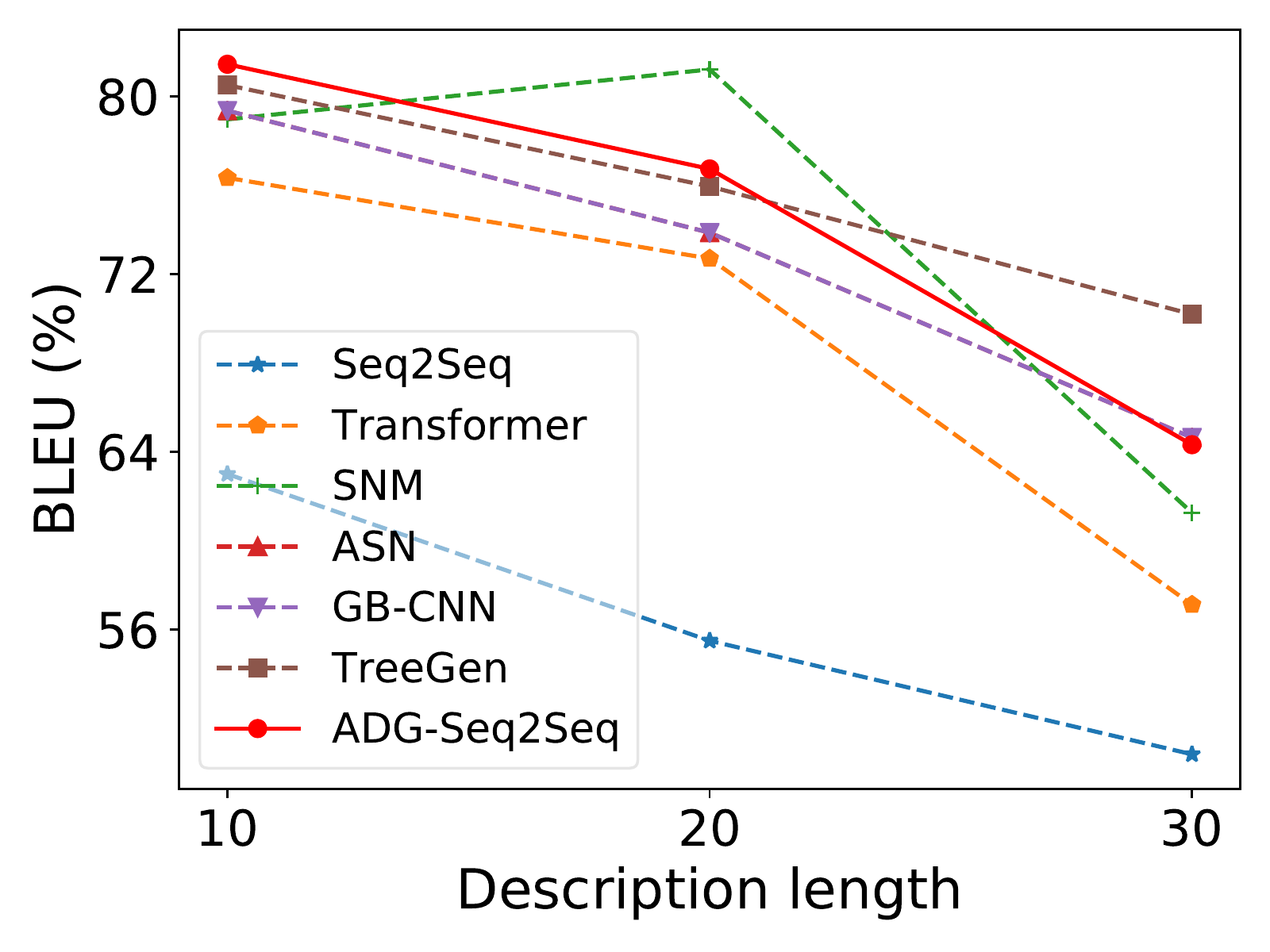}
			\label{DesLen2_HS_BLEU}
		}
		\subfigure[HS: F1]{
			\includegraphics[width=0.22\linewidth]{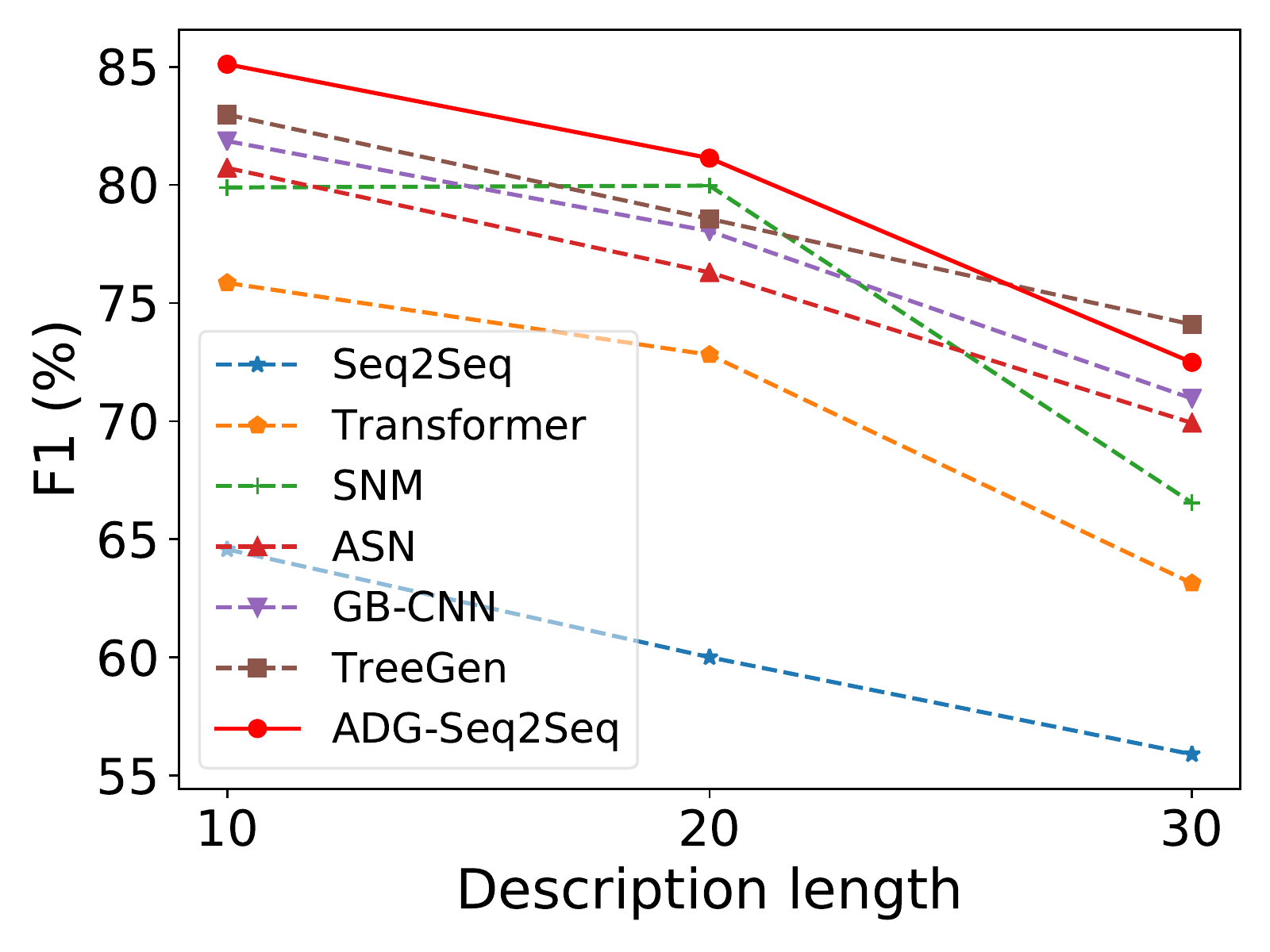}
			\label{DesLen3_HS_F1}
		}
		\subfigure[HS: CIDEr]{
			\includegraphics[width=0.22\linewidth]{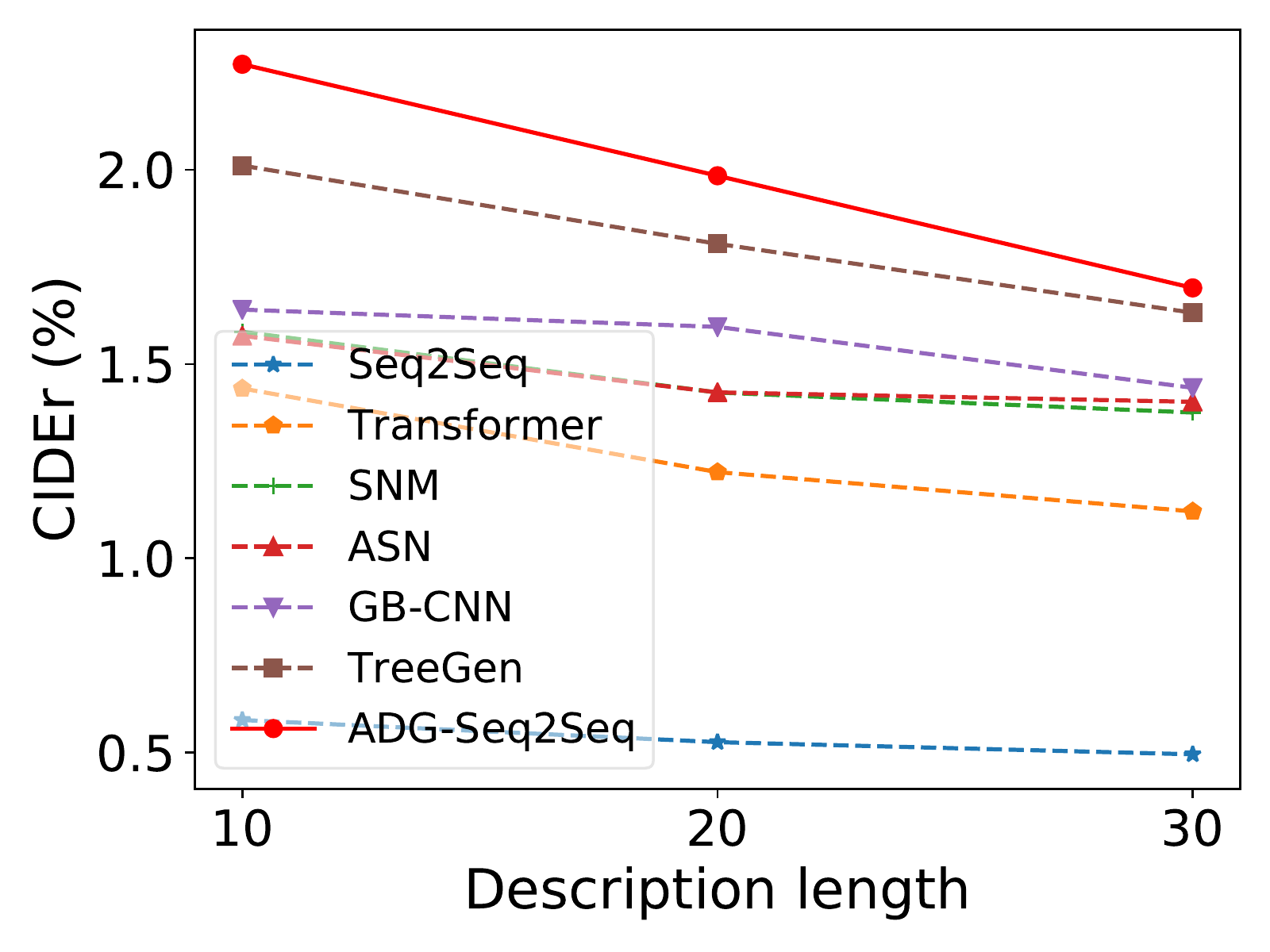}
			\label{DesLen4_HS_CIDEr}
		}
		
		\subfigure[HS: ROUGE]{
			\includegraphics[width=0.22\linewidth]{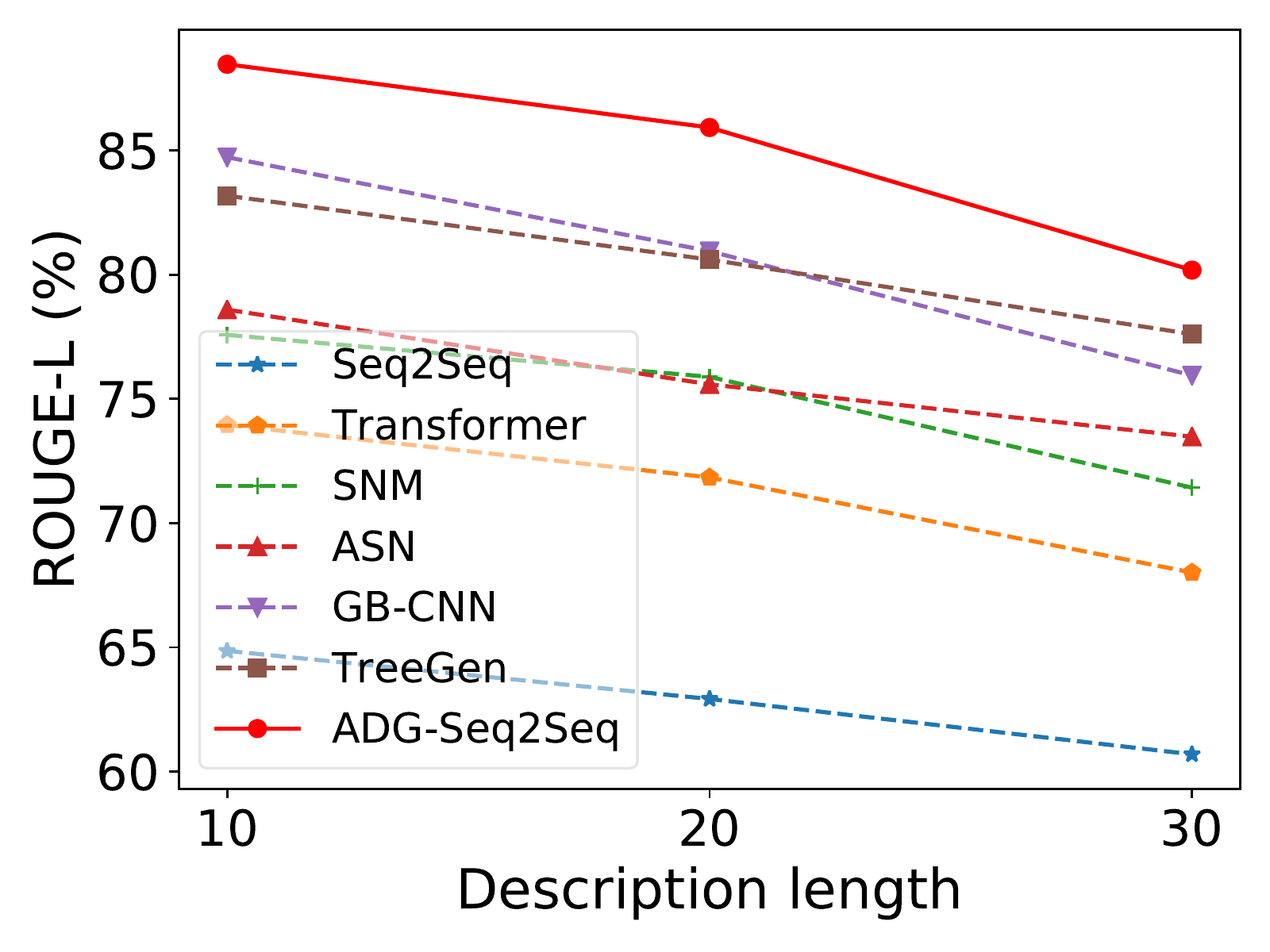}
			\label{DesLen5_HS_ROUGE-L}
		}
		\subfigure[HS: ROUGE-1]{
			\includegraphics[width=0.22\linewidth]{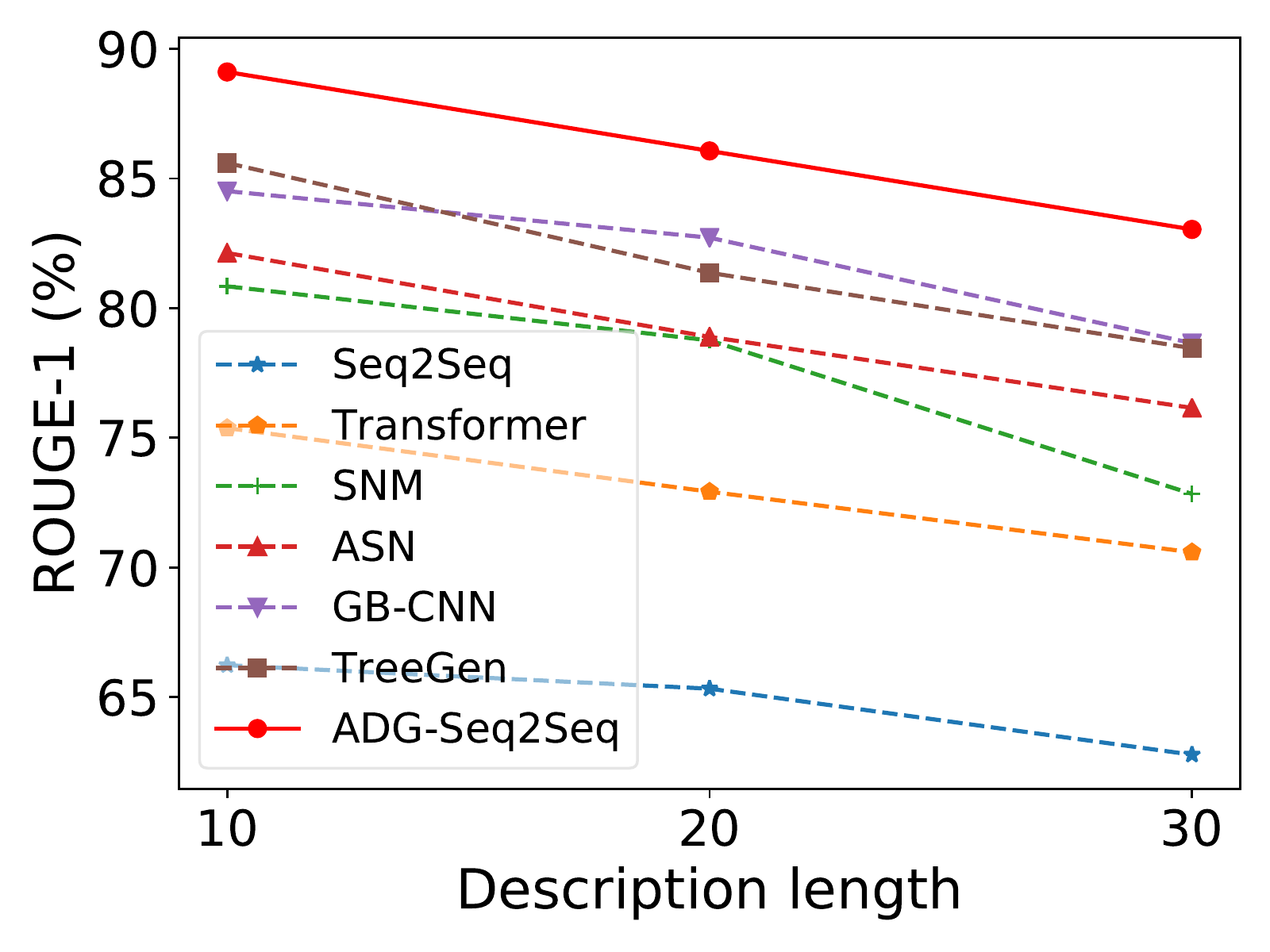}
			\label{HS-ROUGE-1}
		}
		\subfigure[HS: ROUGE-2]{
			\includegraphics[width=0.22\linewidth]{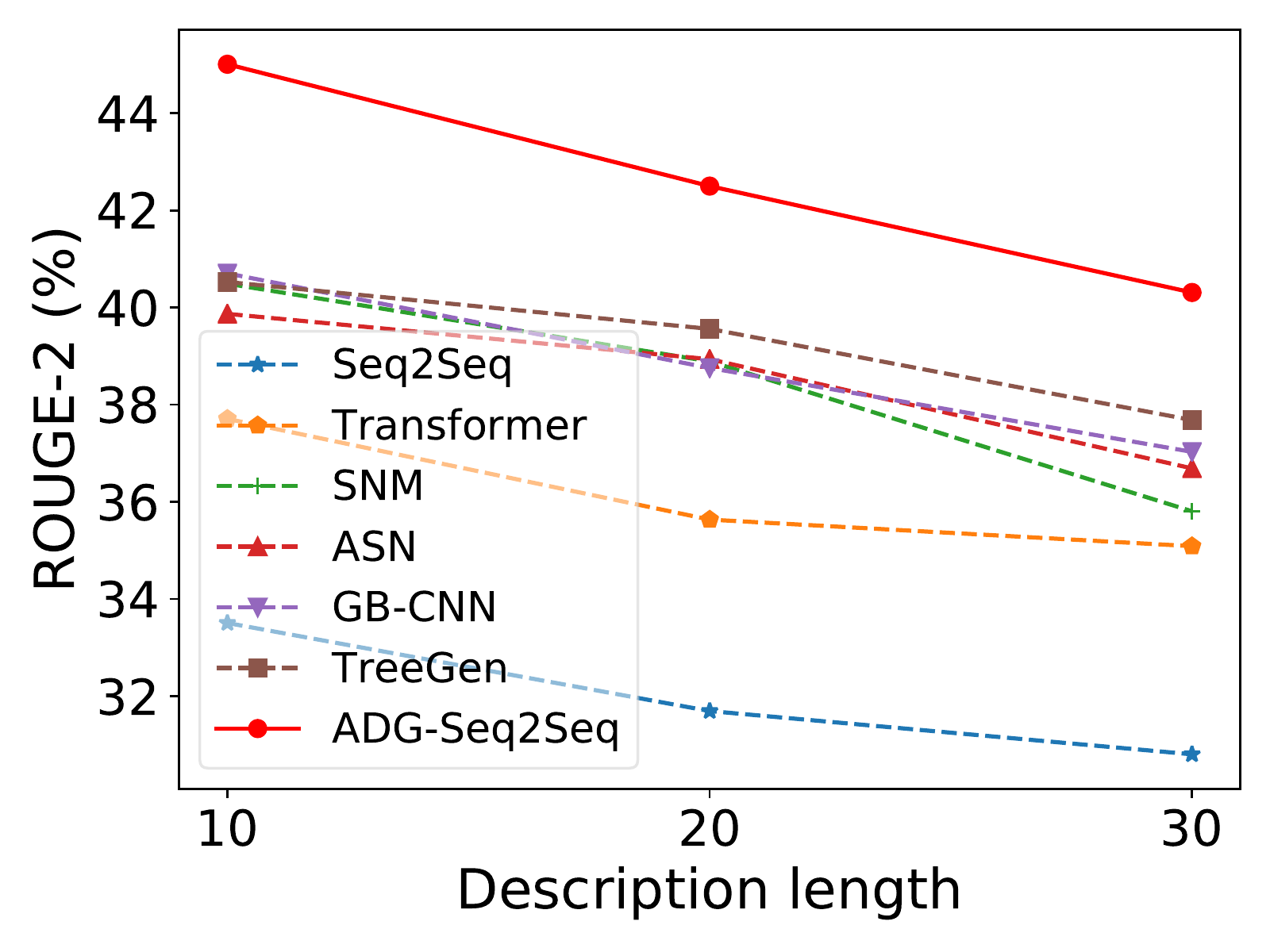}
			\label{HS-ROUGE-2}
		}
		\subfigure[HS: RIBES]{
			\includegraphics[width=0.22\linewidth]{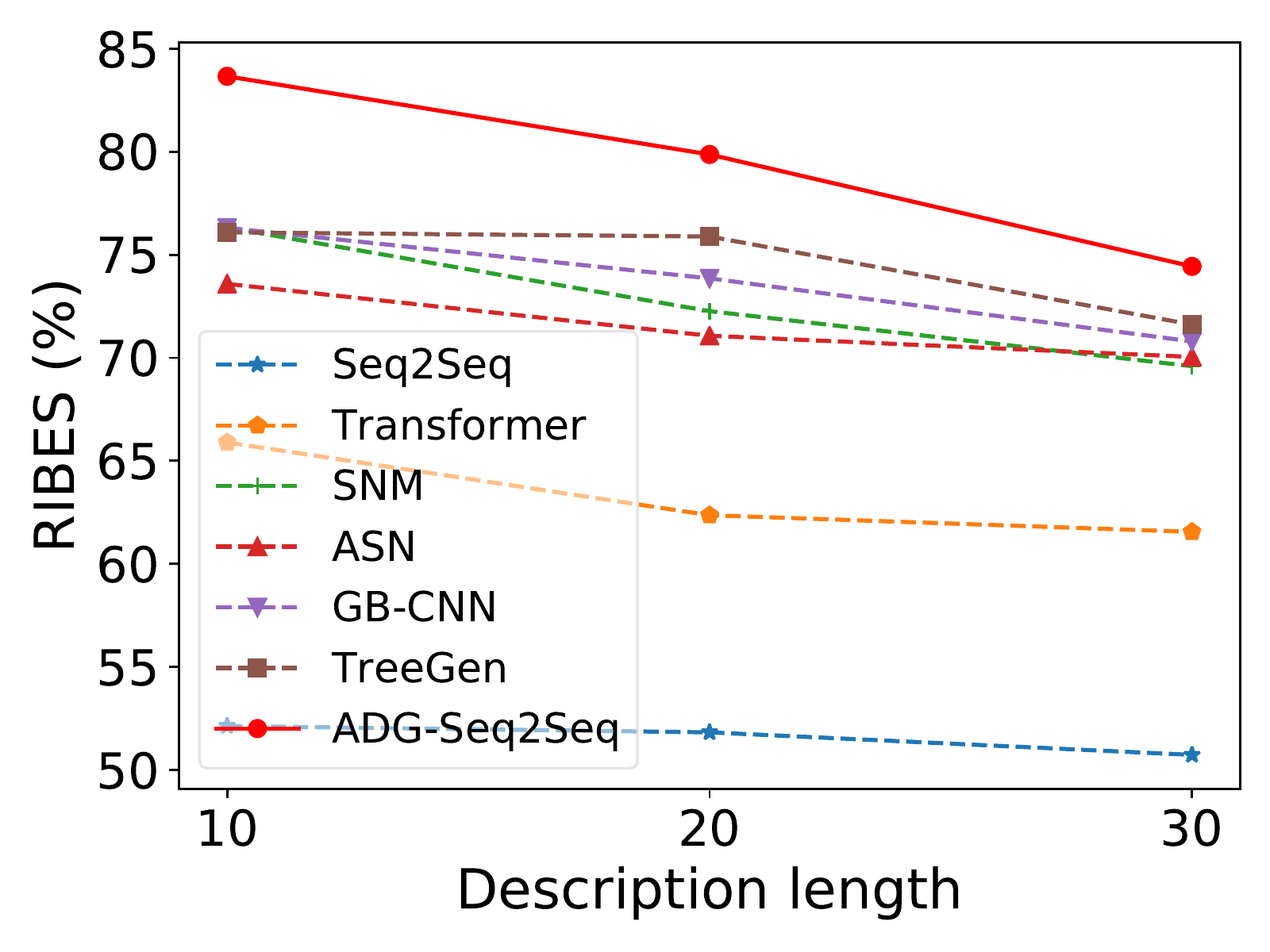}
			\label{DesLen8_HS_RIBES}
		}
		
		\subfigure[MTG: Acc]{
			\includegraphics[width=0.22\linewidth]{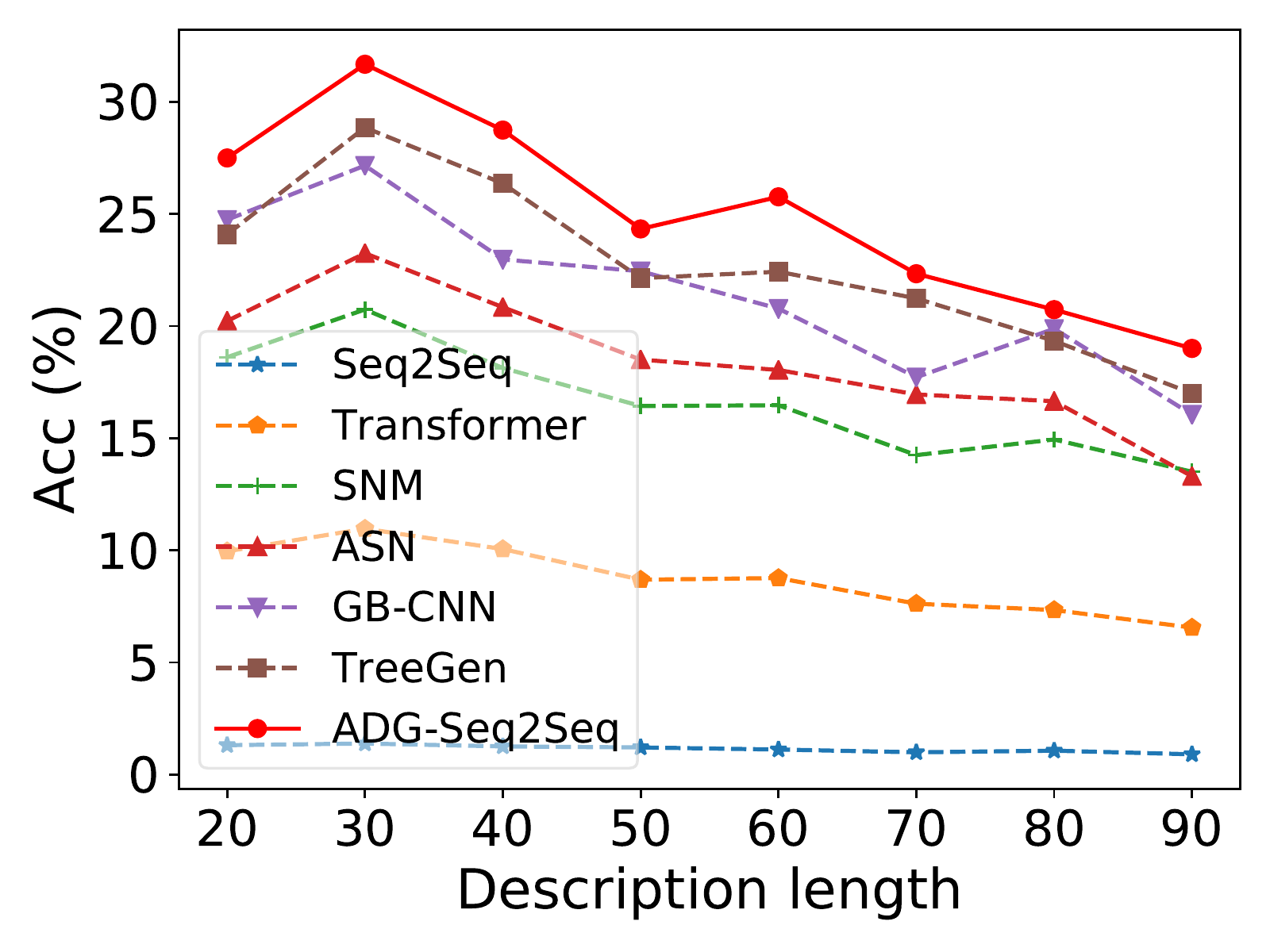}
			\label{DesLen9_MTG_Acc}
		}
		\subfigure[MTG: BLEU]{
			\includegraphics[width=0.22\linewidth]{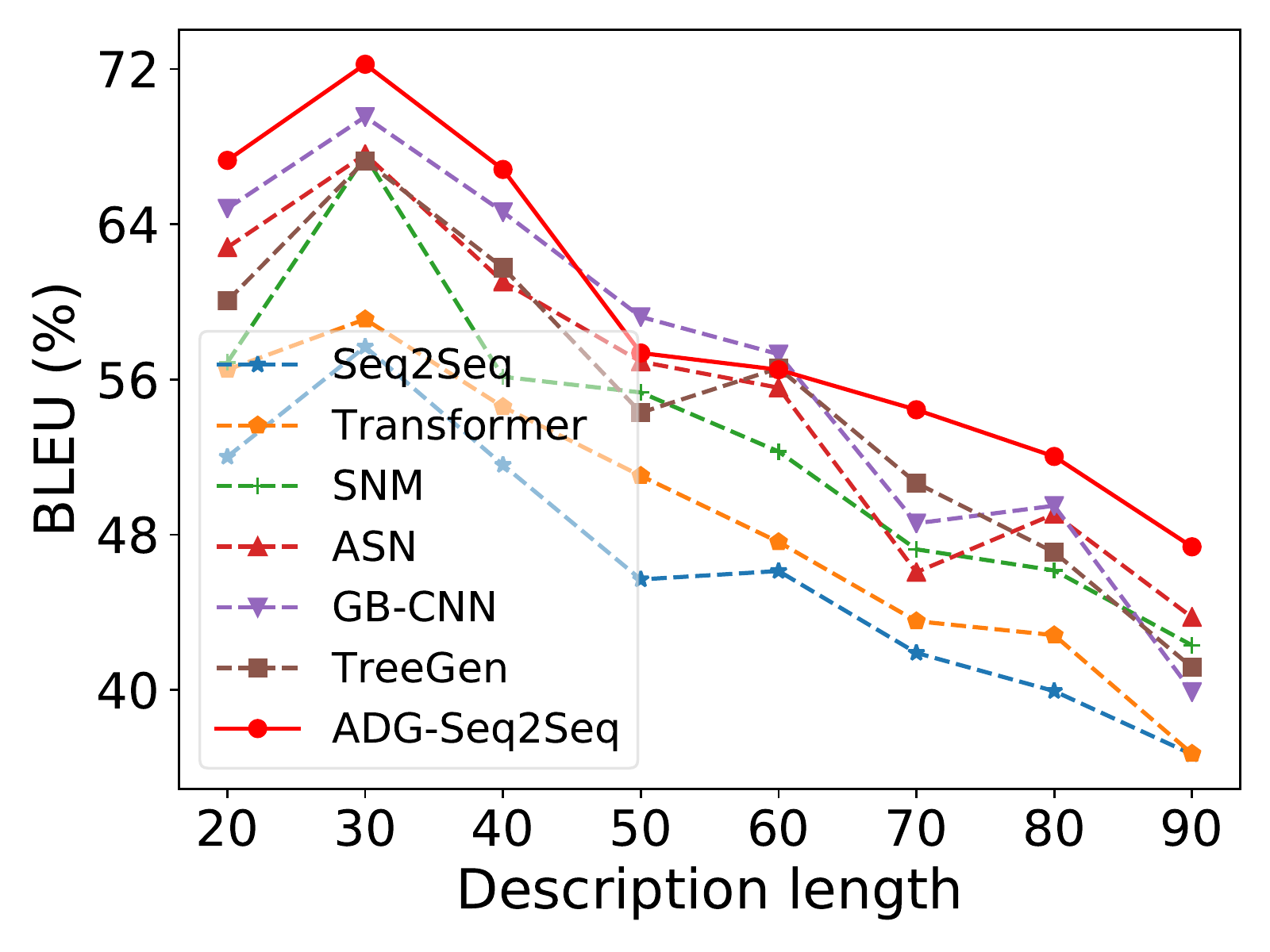}
			\label{DesLen10_MTG_BLEU}
		}
		\subfigure[MTG: F1]{
			\includegraphics[width=0.22\linewidth]{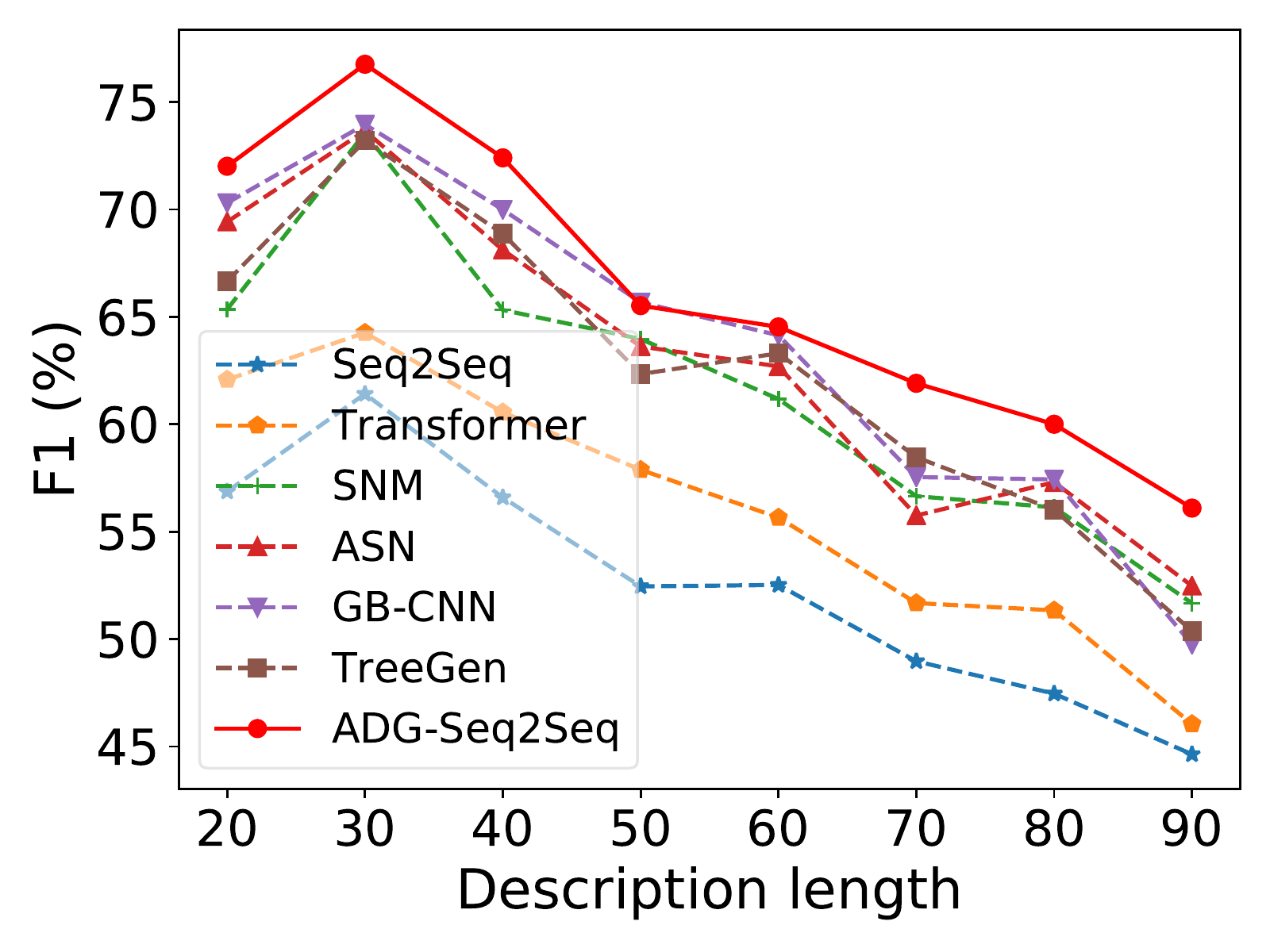}
			\label{DesLen11_MTG_F1}
		}
		\subfigure[MTG: CIDEr]{
			\includegraphics[width=0.22\linewidth]{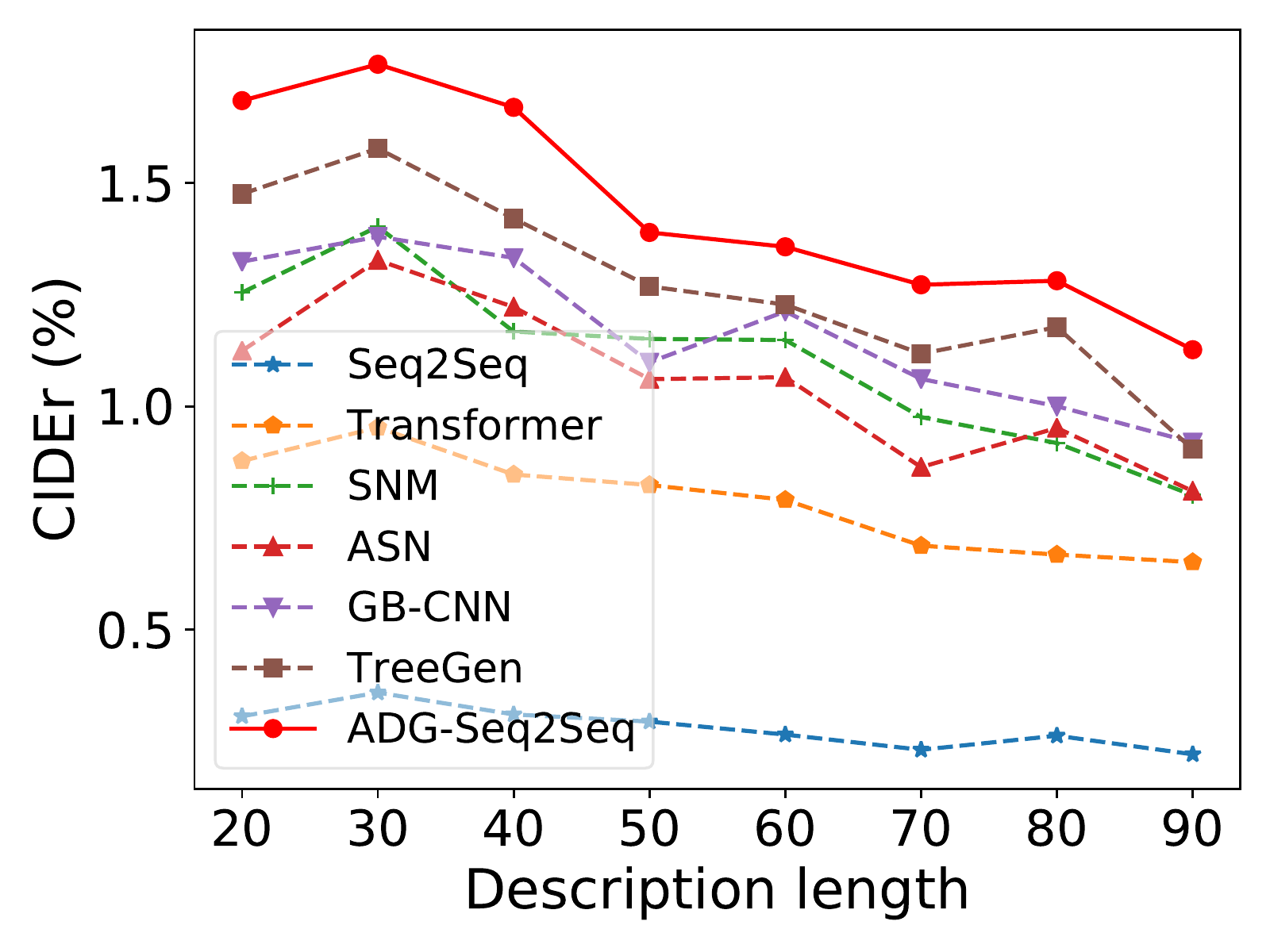}
			\label{DesLen12_MTG_CIDEr}
		}
		
		\subfigure[MTG: ROUGE-L]{
			\includegraphics[width=0.22\linewidth]{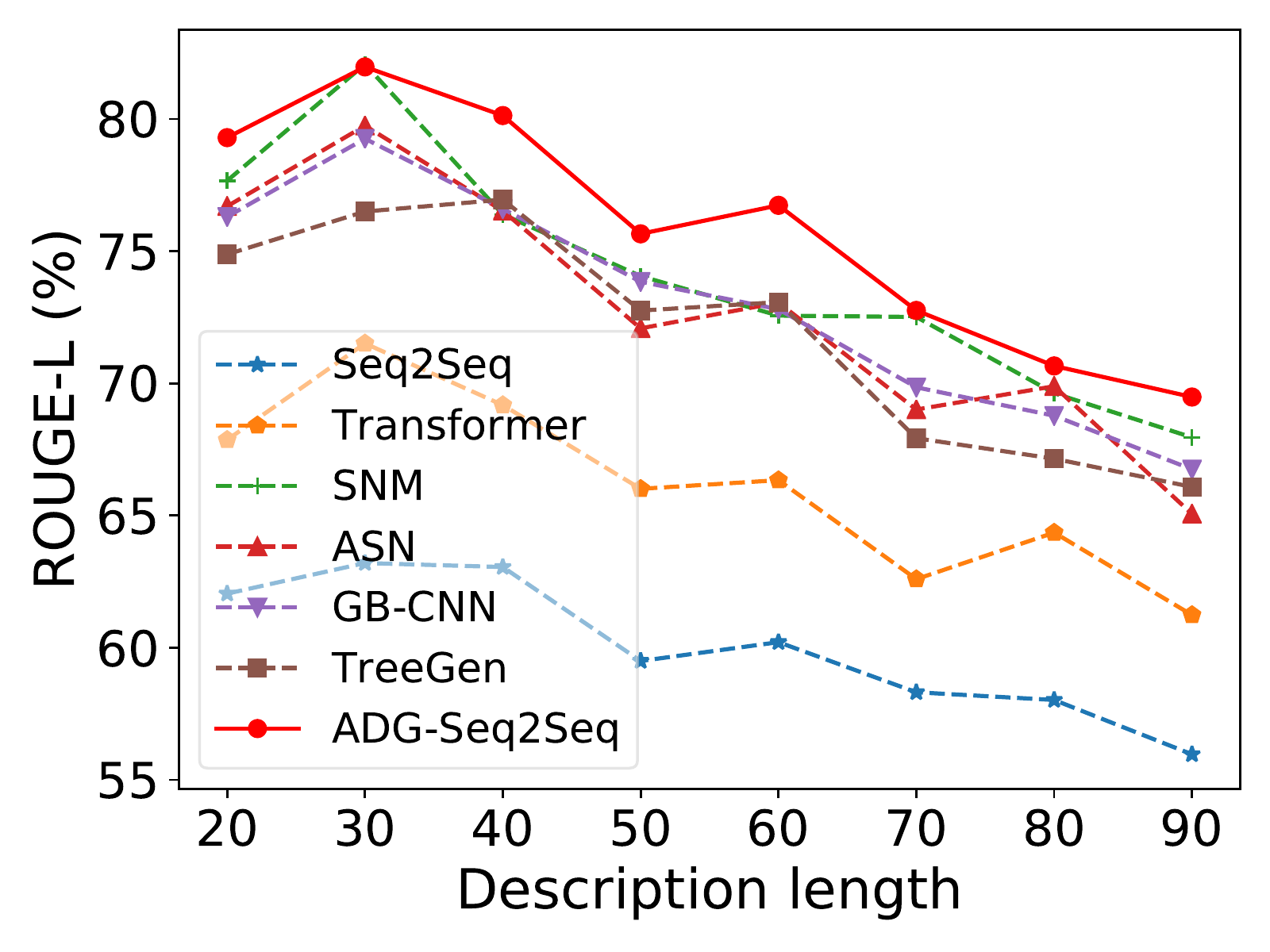}
			\label{DesLen13_MTG_ROUGE-L}
		}
		\subfigure[MTG: ROUGE-1]{
			\includegraphics[width=0.22\linewidth]{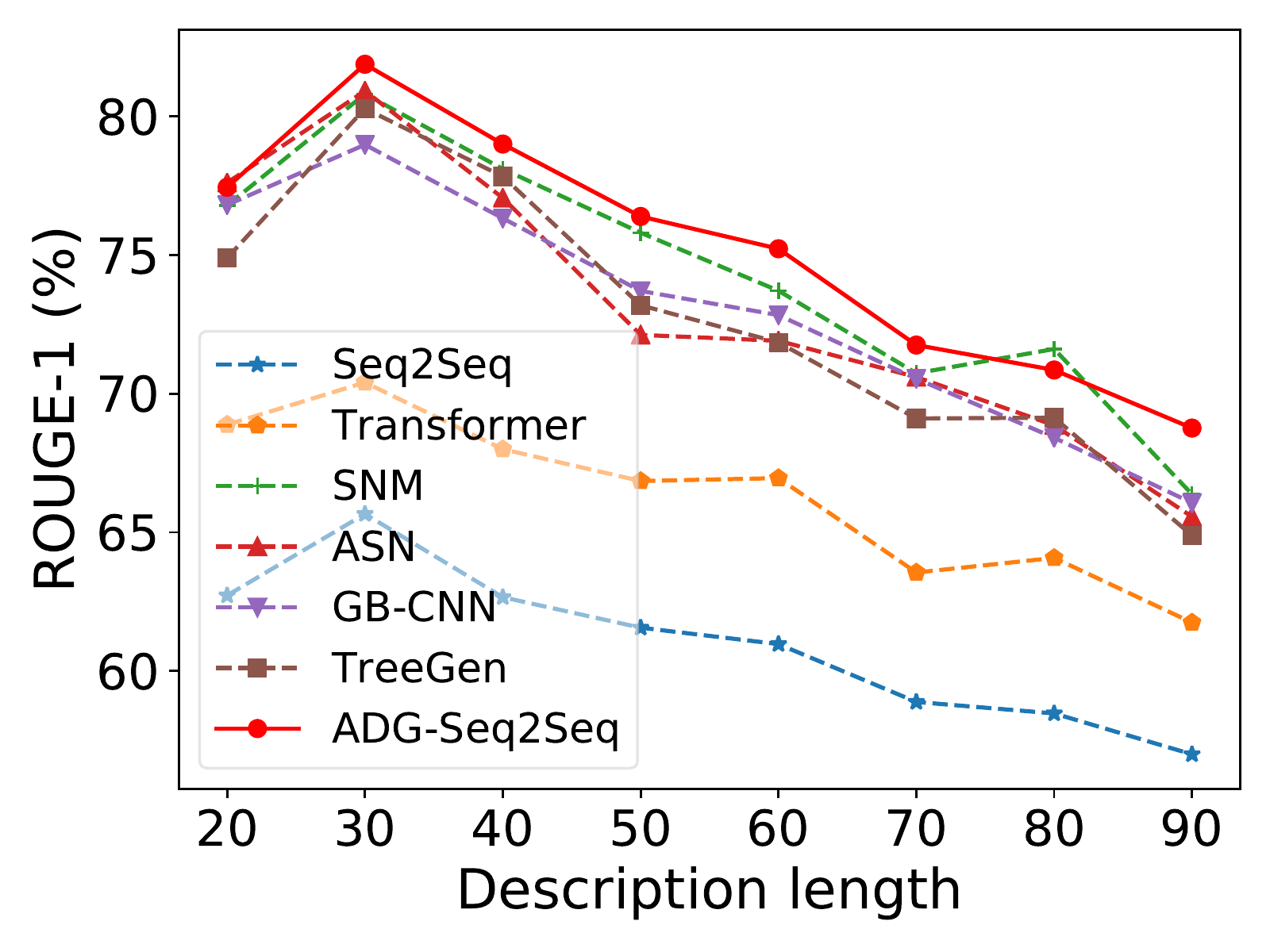}
			\label{DesLen14_MTG_ROUGE-1}
		}
		\subfigure[MTG: ROUGE-2]{
			\includegraphics[width=0.22\linewidth]{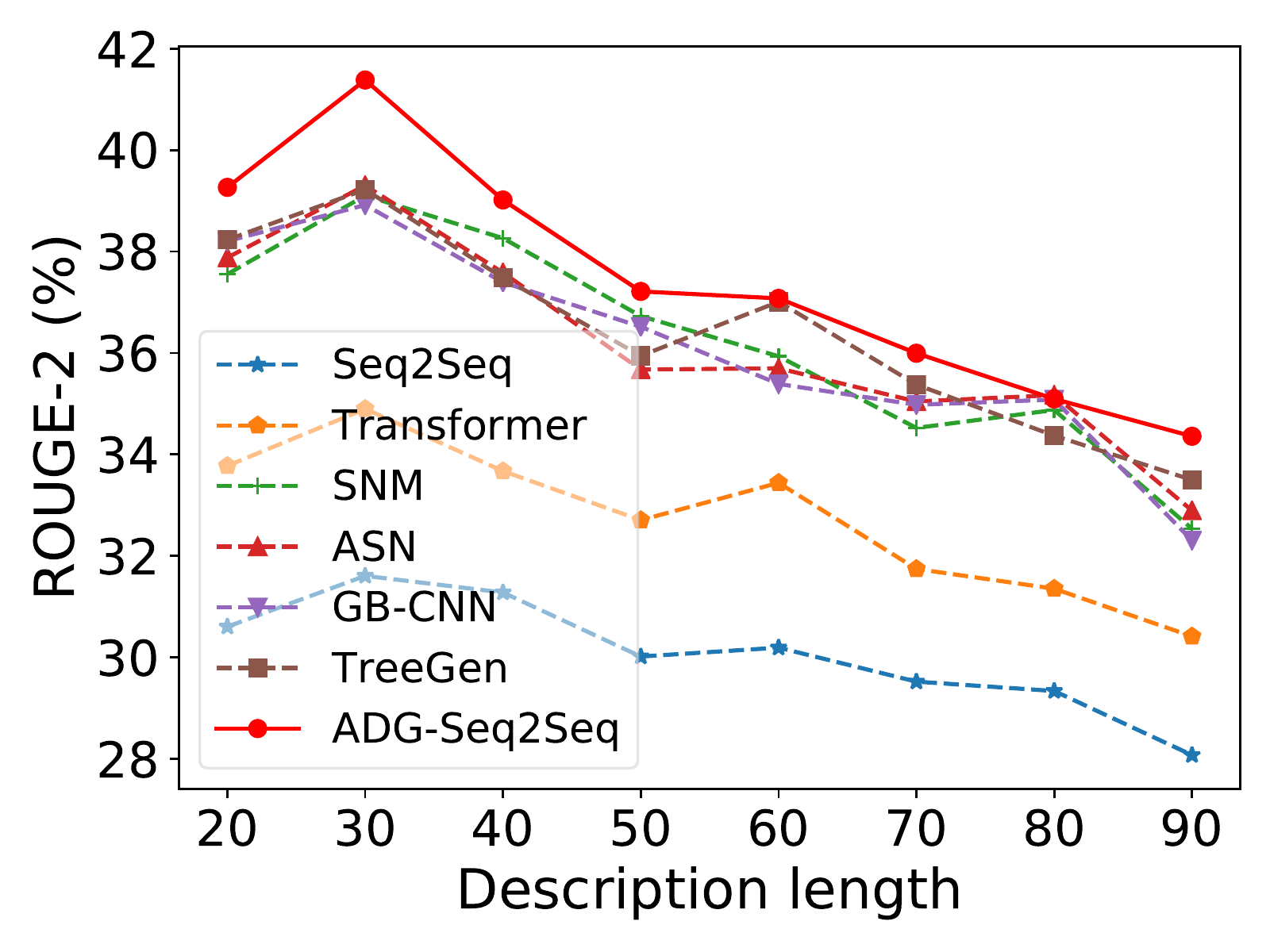}
			\label{DesLen15_MTG_ROUGE-2}
		}
		\subfigure[MTG: RIBES]{
			\includegraphics[width=0.22\linewidth]{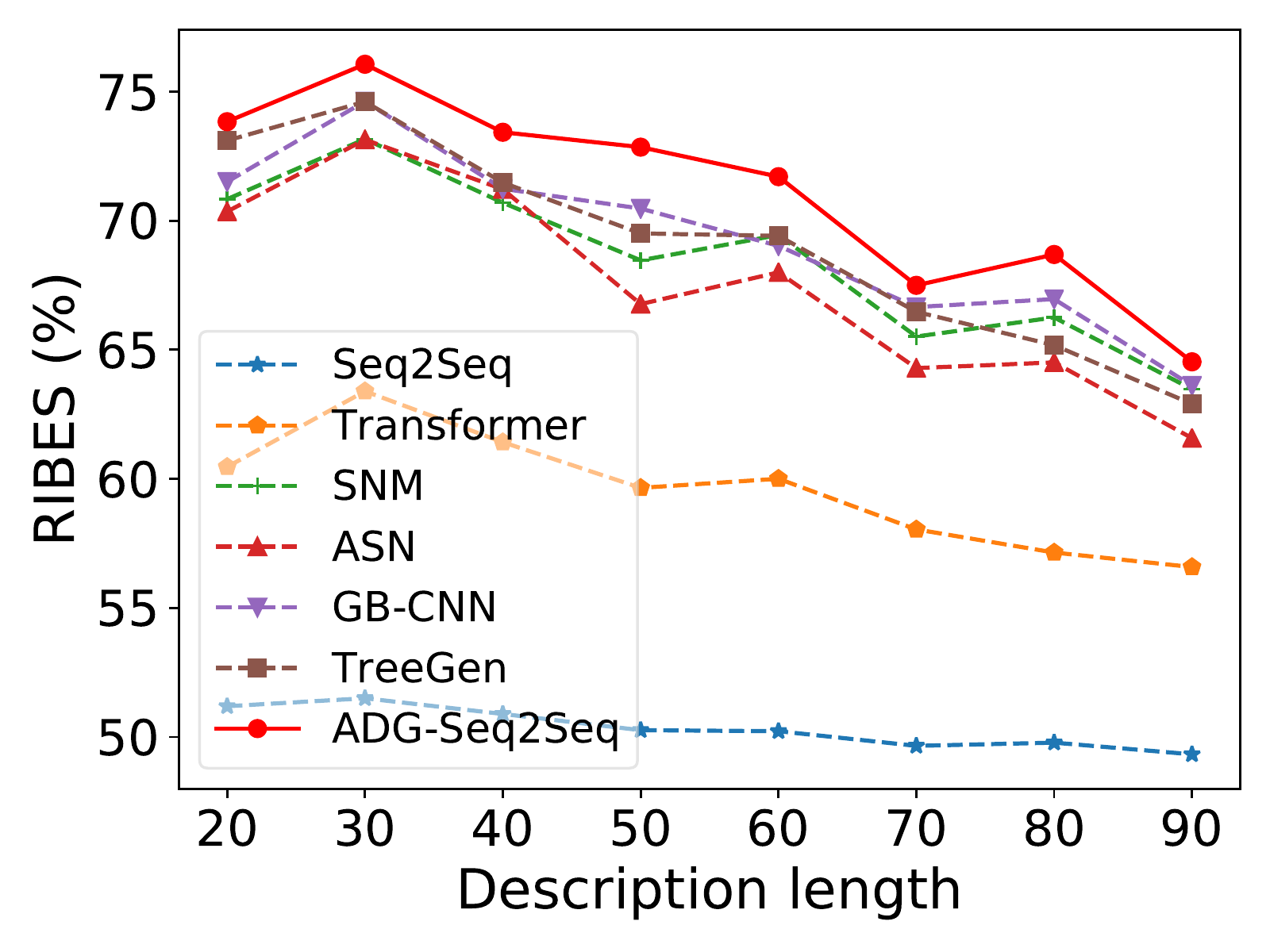}
			\label{DesLen16_MTG_RIBES}
		}
		
		\subfigure[EJDT: Acc]{
			\includegraphics[width=0.22\linewidth]{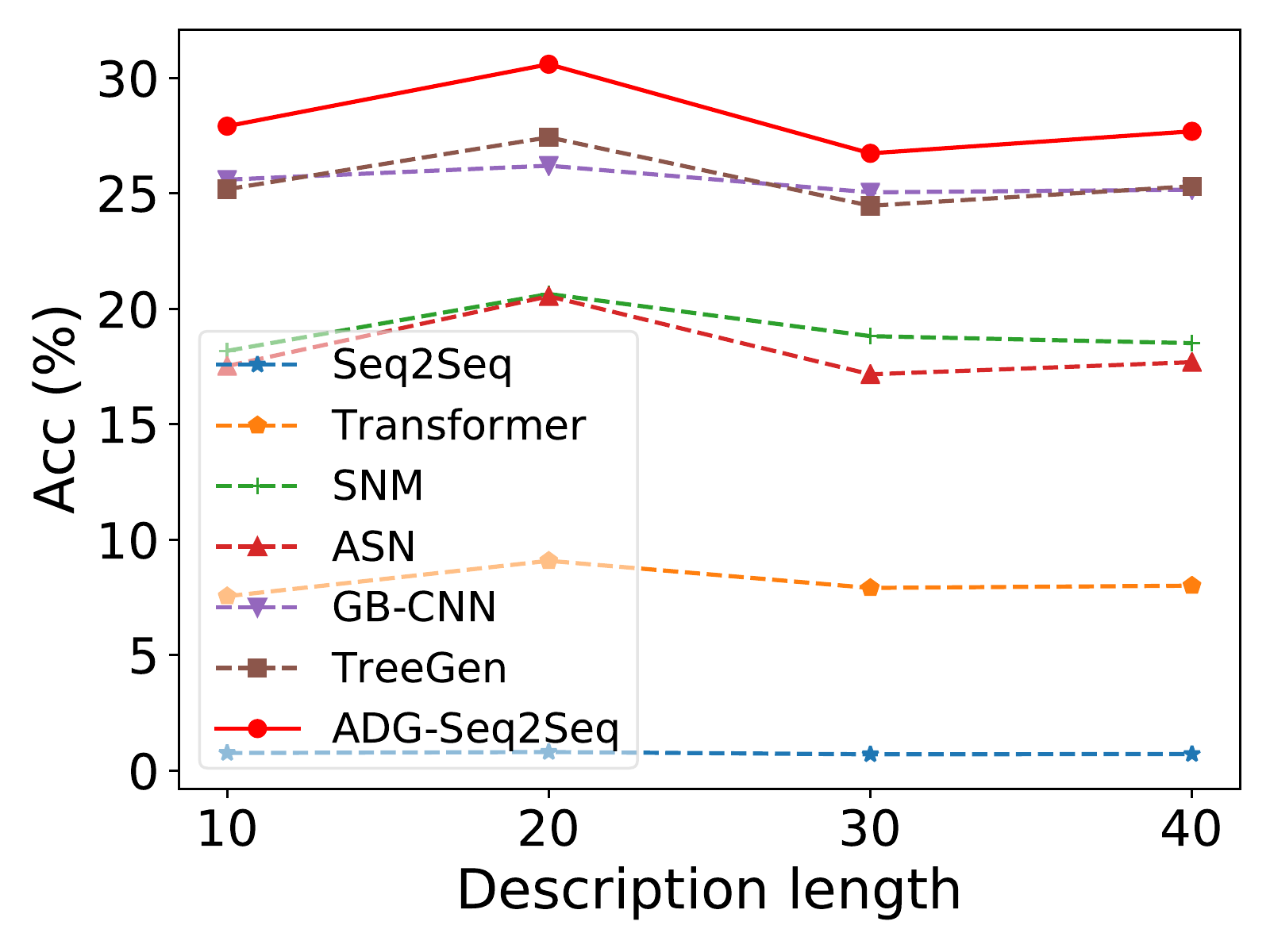}
			\label{DesLen17_E-JDT_Acc}
		}
		\subfigure[EJDT: BLEU]{
			\includegraphics[width=0.22\linewidth]{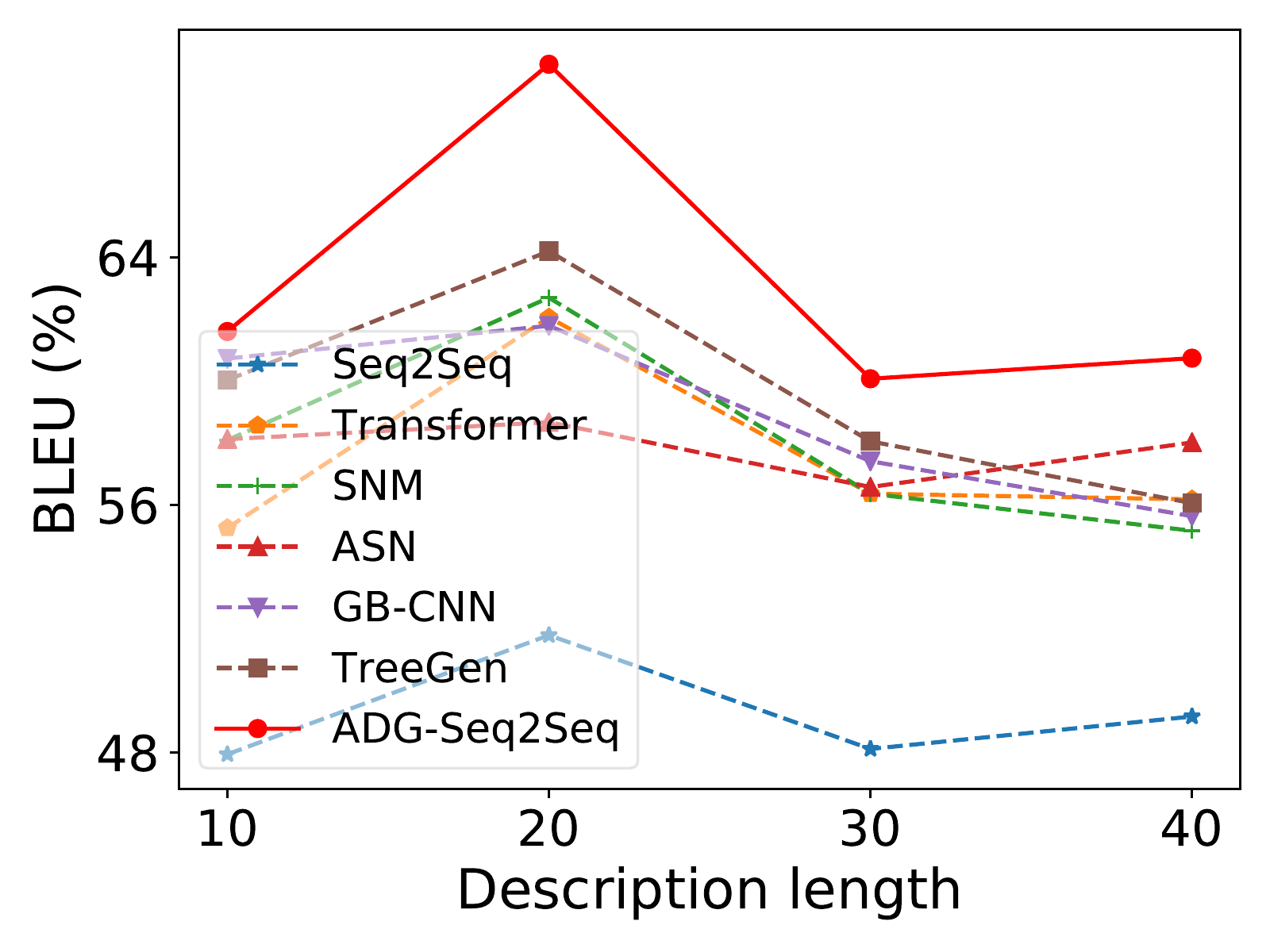}
			\label{E-JDT-BLEU}
		}
		\subfigure[EJDT: F1]{
			\includegraphics[width=0.22\linewidth]{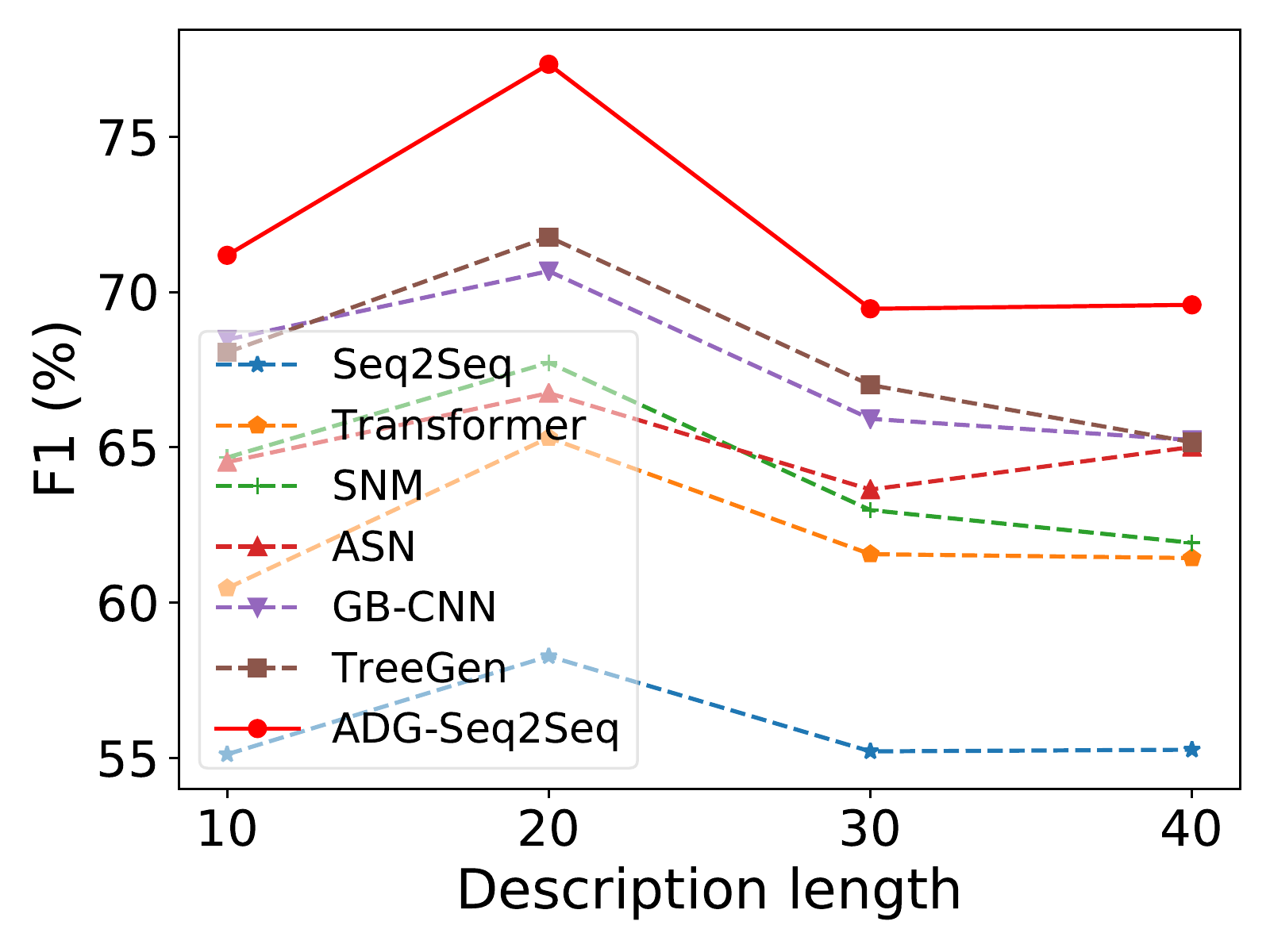}
			\label{E-JDT-F1}
		}
		\subfigure[EJDT: CIDEr]{
			\includegraphics[width=0.22\linewidth]{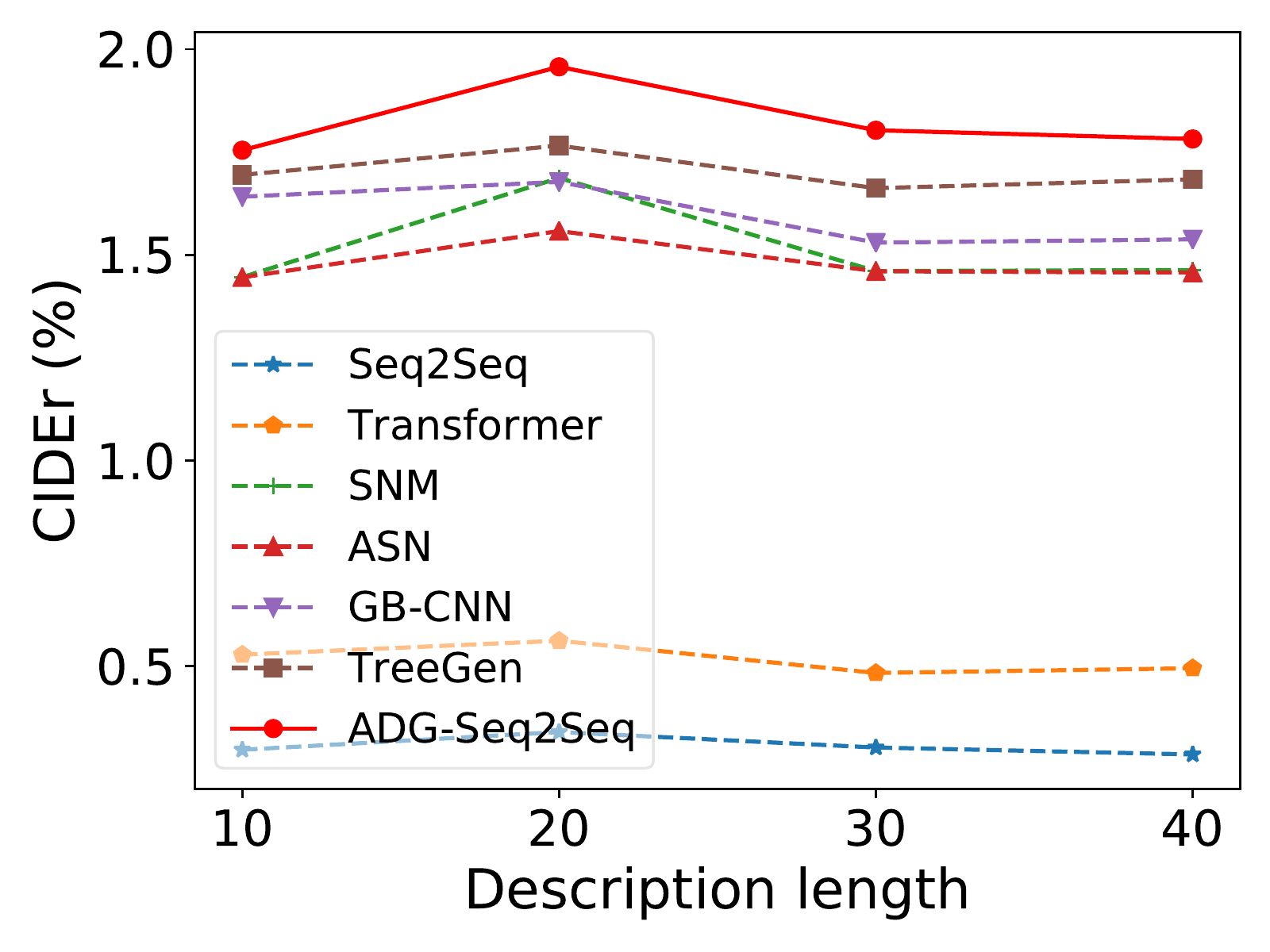}
			\label{E-JDT-CIDEr}
		}
		
		\subfigure[EJDT: ROUGE-L]{
			\includegraphics[width=0.22\linewidth]{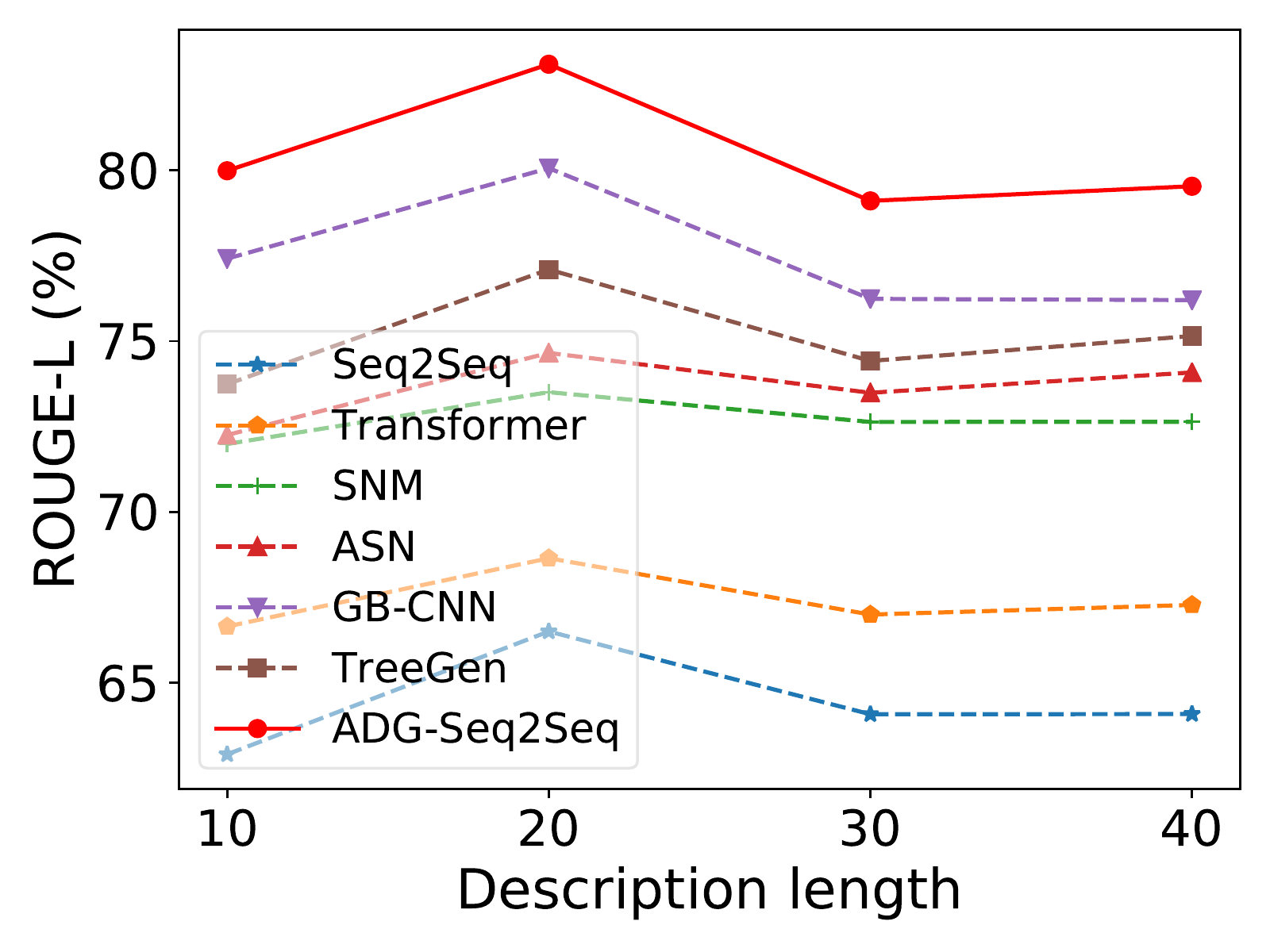}
			\label{E-JDT-ROUGE-L}
		}
		\subfigure[EJDT: ROUGE-1]{
			\includegraphics[width=0.22\linewidth]{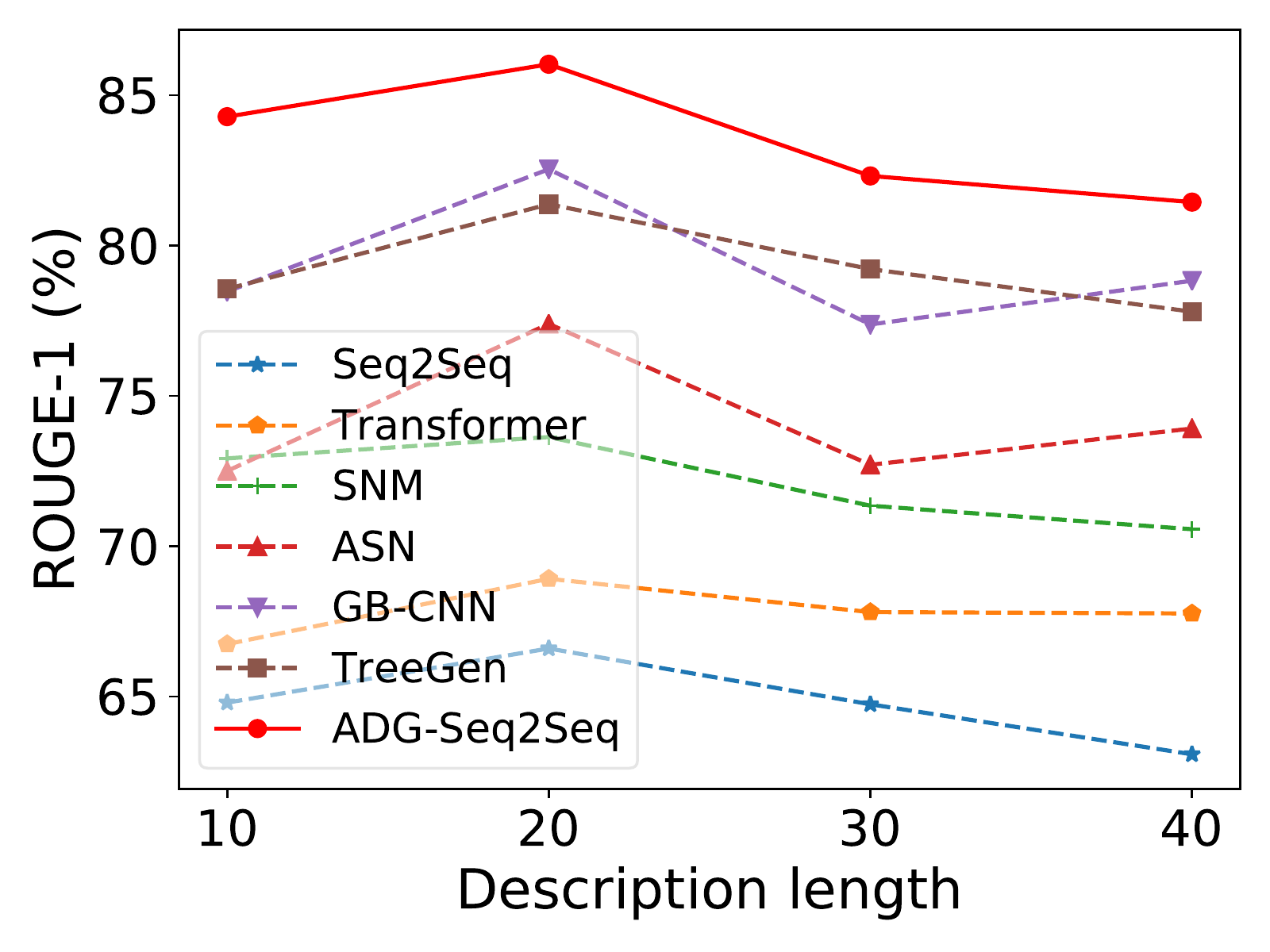}
			\label{E-JDT-ROUGE-1}
		}
		\subfigure[EJDT: ROUGE-2]{
			\includegraphics[width=0.22\linewidth]{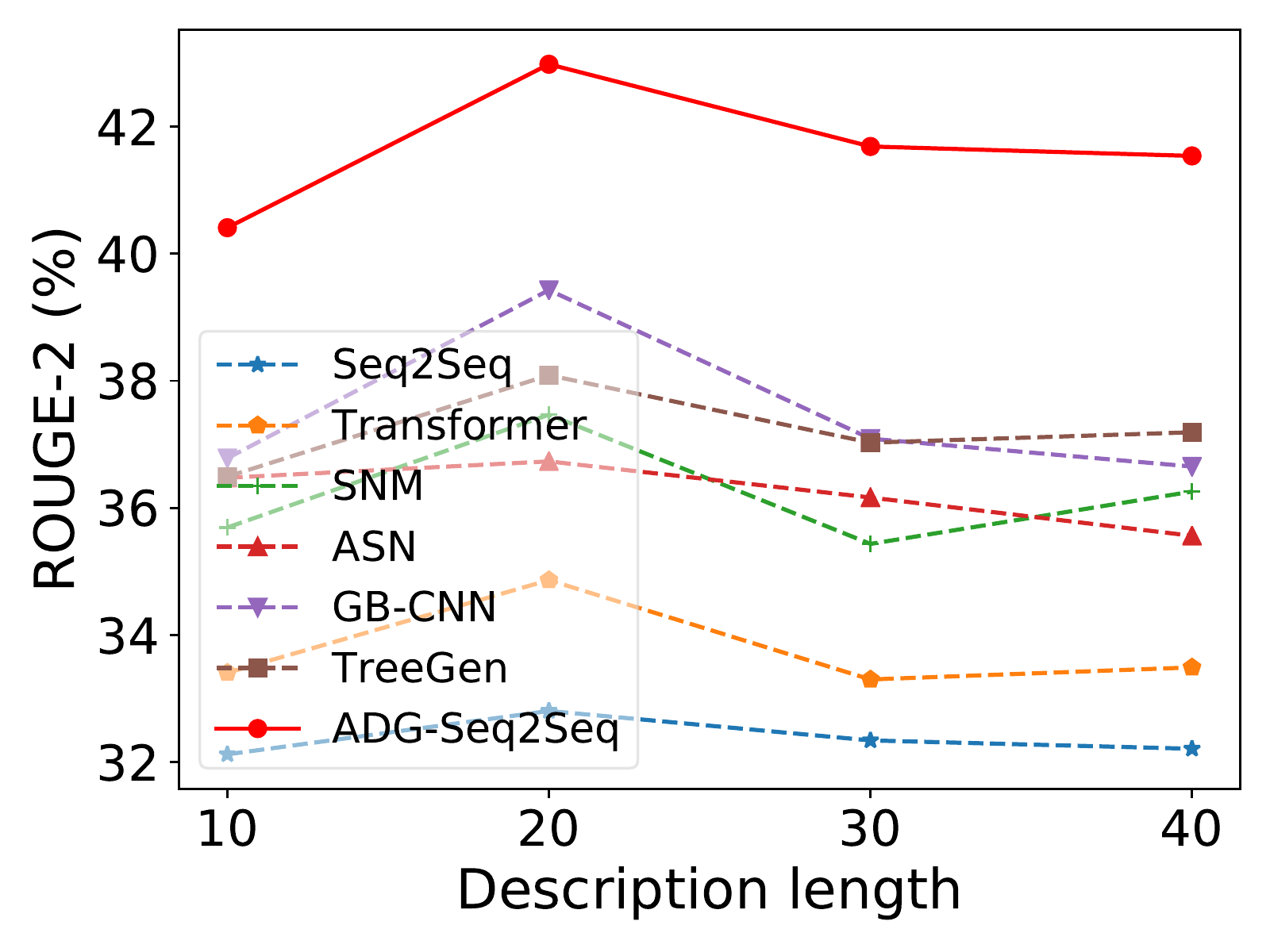}
			\label{E-JDT-ROUGE-2}
		}
		\subfigure[EJDT: RIBES]{
			\includegraphics[width=0.22\linewidth]{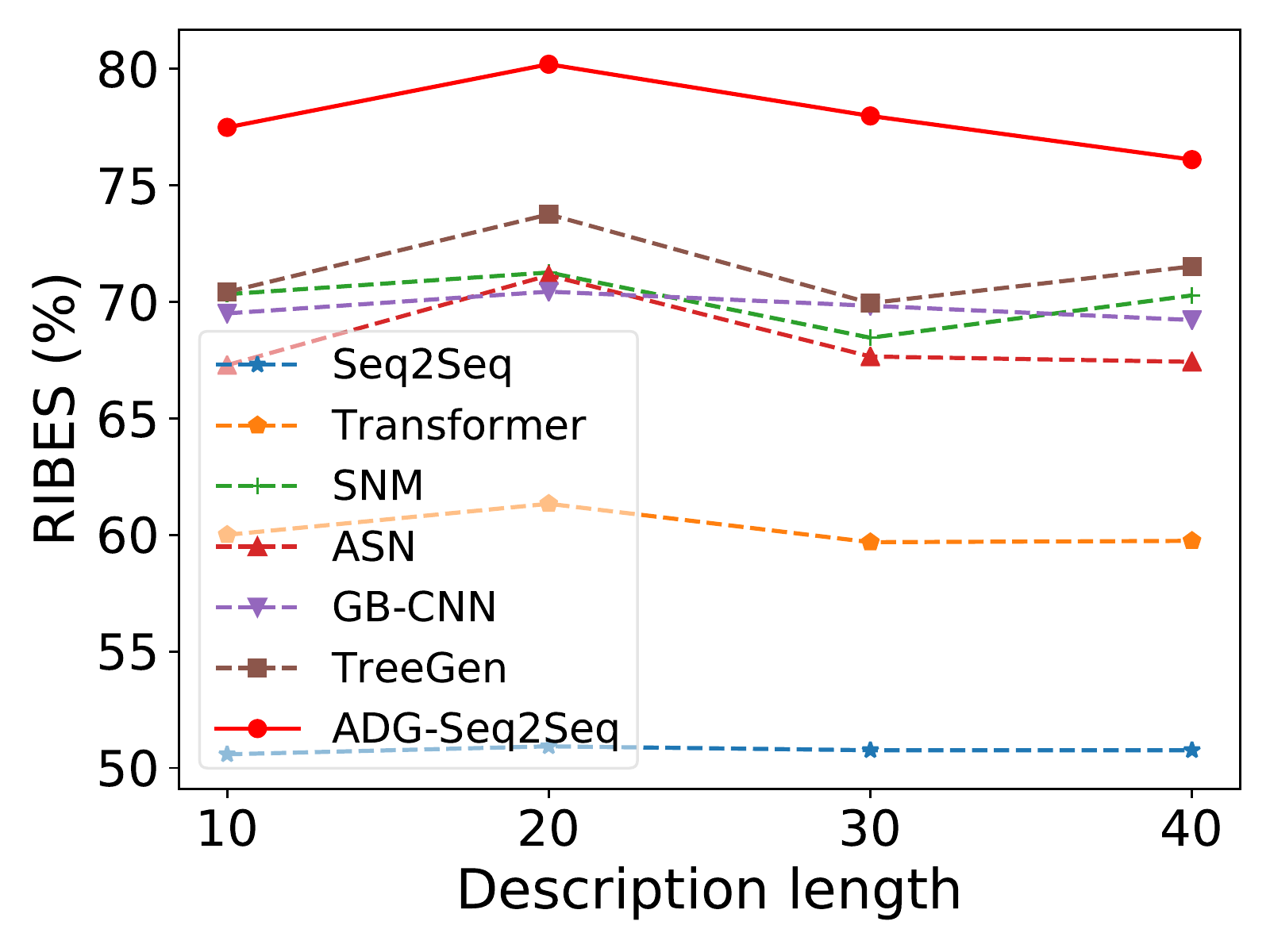}
			\label{E-JDT-RIBES}
		}
		
		\caption{\textbf{Experimental results of the comparison of state-of-the-art methods with varying description lengths.} }
		\label{des_length}
	\end{figure*}

	The experimental results for these new split datasets are illustrated in Fig.\ref{des_length}. Among the 24 indicators, except for some cases of MTG and E-JDT, which increase first and then decrease as the description length increases, all indicators exhibit a downward trend. However, the decreases occur relatively slowly.  

		\begin{table}[h]
		\centering
		\small
		\begin{threeparttable}
			\caption{\textbf{Statistics of datasets split according to description length.}}
			\label{tab:deslen}
			\begin{tabular}{p{12px}lp{25px}<{\raggedleft}p{30px}<{\raggedleft}p{25px}<{\raggedleft}r}
				\toprule
				& \textbf{LEN} & \textbf{QTY} & \multicolumn{1}{l}{\textbf{Avg.w}} & \multicolumn{1}{l}{\textbf{Avg.t}} & \multicolumn{1}{l}{\textbf{Avg.m}} \\
				\midrule
				\multirow{4}[2]{*}{\begin{sideways}HS\end{sideways}} & [0~, 10] & 0     &   -    &    -   & -  \\
				& [10, 20] & 81    & 19.6  & 27.1  & 2.0  \\
				& [20, 30] & 0.3k  & 26.1  & 37.3  & 3.5  \\
				& [30, 40] & 0.1k  & 32.4  & 49.6  & 4.9  \\
				\midrule
				\multirow{8}[2]{*}{\begin{sideways}MTG\end{sideways}} & [20, 30] & 1.3k  & 26.3  & 68.2  & 5.8  \\
				& [30, 40] & 2.9k  & 35.8  & 93.0  & 8.9  \\
				& [40, 50] & 2.7k  & 45.1  & 113.9  & 11.4  \\
				& [50, 60] & 2.0k  & 55.3  & 129.5  & 13.5  \\
				& [60, 70] & 1.3k  & 65.3  & 146.7  & 15.7  \\
				& [70, 80] & 0.9k  & 75.2  & 157.4  & 16.9  \\
				& [80, 90] & 0.5k  & 84.8  & 169.8  & 18.2  \\
				& [90, 100] & 0.2k  & 94.9  & 186.9  & 19.9  \\
				\midrule
				\multirow{4}[2]{*}{\begin{sideways}E-JDT\end{sideways}} & [0~, 10] & 257.8k & 6.3   & 52.4  & 5.1  \\
				& [10, 20] & 128.2k & 14.4  & 53.9  & 5.0  \\
				& [20, 30] & 39.5k & 24.9  & 67.7  & 6.1  \\
				& [30, 40] & 19.1k & 34.8  & 73.6  & 6.4  \\
				\bottomrule
				
			\end{tabular}
			\begin{tablenotes}
				\raggedright
				\footnotesize{
					\item [1] \textit{Avg.w means the average number of words in the descriptions.}
					\item [2] \textit{Avg.t means the average number of tokens in code.}
					\item [3] \textit{Avg.m means the average number of methods in code.}
					\item [4] \textit{QTY means the quantity of data in the subset.}}
			\end{tablenotes}    
		\end{threeparttable}
	\end{table}

	For HS, the highest scores are obtained if the description length is approximately 20, which indicates that the learning models can capture complete semantic information from short descriptions. Based on the discussion in Section~\ref{ARQ1}, ADG-Seq2Seq can learn more comprehensive knowledge through global view modelling, thereby contributing to the prediction of complete code and correct call sequences. Therefore, our approach outperforms other models in terms of ROUGE, CIDEr, and RIBES. Limited by the fewer samples, the scores of precision-related metrics (including ACC, BLEU, and F1) are lower than those of TreeGen and GB-CNN. However, this scenario does not represent the upper bound of our model. For example, when sufficient number of samples are available (see Table~\ref{tab:deslen}) on MTG and E-JDT, ADG-Seq2Seq outperforms the competitors regardless of the description length. Moreover, the degradation of performance of our model on HS is faster than that on  MTG and E-JDT. One explanation is that as description length increases in HS, fewer samples increase, which might negatively affect the performance of our method.

	\subsubsection{The Impact of Code Length}
	
	\begin{table}[htbp]
		\small
		\centering
		\caption{\textbf{Statistics of the datasets with different code lengths.}} 
		\label{tab:LPer}%
		\resizebox{\textwidth}{28mm}{
		\begin{threeparttable}
			\begin{tabular}{p{2px}p{4px}rrrrrrrrrrrr}
				\toprule
				&       & \multicolumn{1}{c}{\begin{scriptsize}QTY\end{scriptsize}} & \multicolumn{1}{c}{\begin{scriptsize}Avg.w\end{scriptsize}} & \multicolumn{1}{c}{\begin{scriptsize}Max.w\end{scriptsize}} & \multicolumn{1}{c}{\begin{scriptsize}Avg.t\end{scriptsize}} & \multicolumn{1}{c}{\begin{scriptsize}Avg.m\end{scriptsize}} & \multicolumn{1}{c}{\begin{scriptsize}Max.m\end{scriptsize}} & \multicolumn{1}{c}{\begin{scriptsize}Avg.e\end{scriptsize}} & \multicolumn{1}{c}{\begin{scriptsize}Max.e\end{scriptsize}} & \multicolumn{1}{c}{\begin{scriptsize}Avg.in\end{scriptsize}} & \multicolumn{1}{c}{\begin{scriptsize}Max.in\end{scriptsize}} & \multicolumn{1}{c}{\begin{scriptsize}Avg.out\end{scriptsize}} & \multicolumn{1}{c}{\begin{scriptsize}Max.out\end{scriptsize}} \\
				\midrule
				\multirow{4}[2]{*}{\begin{sideways}HS\end{sideways}} & 30    & 0.15k & 22.2  & 32    & 27.5  & 2.0   & 4     & 5.6   & 7     & 2.7   & 10    & 2.8   & 5  \\
          & 40    & 0.25k & 27.1  & 36    & 34.9  & 2.8   & 6     & 7.6   & 9     & 3.7   & 14    & 3.9   & 6  \\
          & 50    & 0.08k & 28.7  & 35    & 44.0  & 4.7   & 8     & 12.8  & 15    & 6.3   & 23    & 6.6   & 11  \\
          & 60    & 0.02k & 27.8  & 38    & 55.3  & 7.0   & 10    & 19.3  & 23    & 9.4   & 35    & 9.9   & 16  \\
				\midrule
				\multirow{5}[2]{*}{\begin{sideways}MTG\end{sideways}} & 100   & 6.89k & 45.1  & 118   & 74.4  & 6.7   & 23    & 28.7  & 51    & 19.3  & 38    & 9.4   & 36  \\
          & 150   & 2.19k & 53.4  & 133   & 120.4  & 11.8  & 35    & 50.3  & 89    & 33.8  & 67    & 16.5  & 63  \\
          & 200   & 1.26k & 54.7  & 134   & 173.8  & 17.8  & 37    & 75.9  & 134   & 51.0  & 101   & 24.9  & 95  \\
          & 250   & 0.70k  & 61.8  & 142   & 222.3  & 24.7  & 60    & 105.5  & 186   & 70.9  & 141   & 34.6  & 131  \\
          & 300   & 0.37k & 67.0  & 129   & 272.9  & 31.7  & 51    & 115.1  & 215   & 70.7  & 159   & 44.4  & 144  \\
				\midrule
				\multirow{5}[2]{*}{\begin{sideways}EJDT\end{sideways}} & 20    & 237.56k & 13.0  & 1702  & 9.8   & 0.9   & 168   & 3.4   & 6     & 2.1   & 5     & 1.3   & 4  \\
          & 40    & 85.34k & 14.4  & 1692  & 29.2  & 3.2   & 255   & 132.0  & 247   & 7.2   & 16    & 4.5   & 15  \\
          & 60    & 43.26k & 15.4  & 1389  & 49.5  & 5.4   & 169   & 19.5  & 36    & 12.0  & 27    & 7.5   & 24  \\
          & 80    & 26.27k & 16.6  & 630   & 69.8  & 7.5   & 108   & 27.2  & 51    & 16.7  & 37    & 10.5  & 34  \\
          & 100   & 16.79k & 17.5  & 806   & 89.9  & 9.4   & 165   & 34.2  & 64    & 21.0  & 47    & 13.2  & 43  \\
				\bottomrule
			\end{tabular}%
			\begin{tablenotes}
				\raggedright
				\footnotesize
				{\item \noindent \scriptsize \textit{\textbf{w} is the number of words in a description, \textbf{t} is the number of tokens in code, \textbf{m} is the number of methods in code, \\ \textbf{e} is the number of edges in the graph for the code, \textbf{in} and \textbf{out} are the numbers of neighbours directly connected to the node.}}
			\end{tablenotes}
		\end{threeparttable}}
	\end{table}%
	
	To further analyse the impact of code length, we split the three datasets according to their code length distributions, as shown in Fig.~\ref{fig:dataset_statistics}(d), (e), and (f). The code length levels of HS were set to -30, -40, -50, and -60; those of MTG were set to -100, -150, -200, -250, and -300; and those of E-JDT were set to -20, -40, -60, -80, and -100. The statistics of these new fourteen split datasets are illustrated in Table~\ref{tab:LPer}.\footnote{As presented in the experimental results, we concatenate the name of each dataset and the code length level to indicate a split dataset. Taking two items of HS as examples, HS-30 consists of the data with code length between 0 and 30, while the code length of HS-60 ranges from 50 to 60. }
		
\begin{figure*}[!htbp]
		\centering
		\subfigure[HS: Acc]{
			\includegraphics[width=0.225\linewidth]{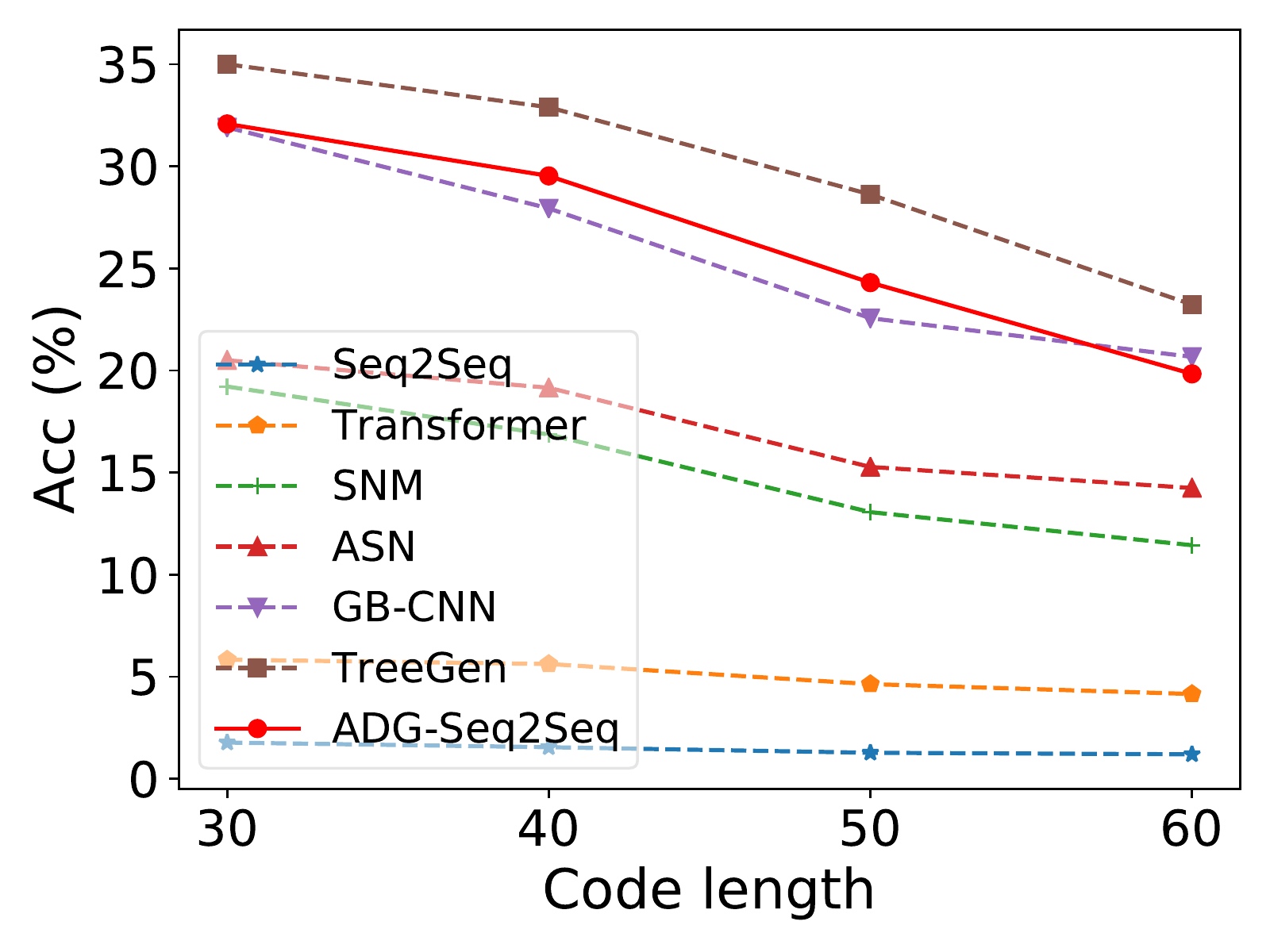}
			\label{CodeLen1_HS_Acc}
		}
		\subfigure[HS: BLEU]{
			\includegraphics[width=0.225\linewidth]{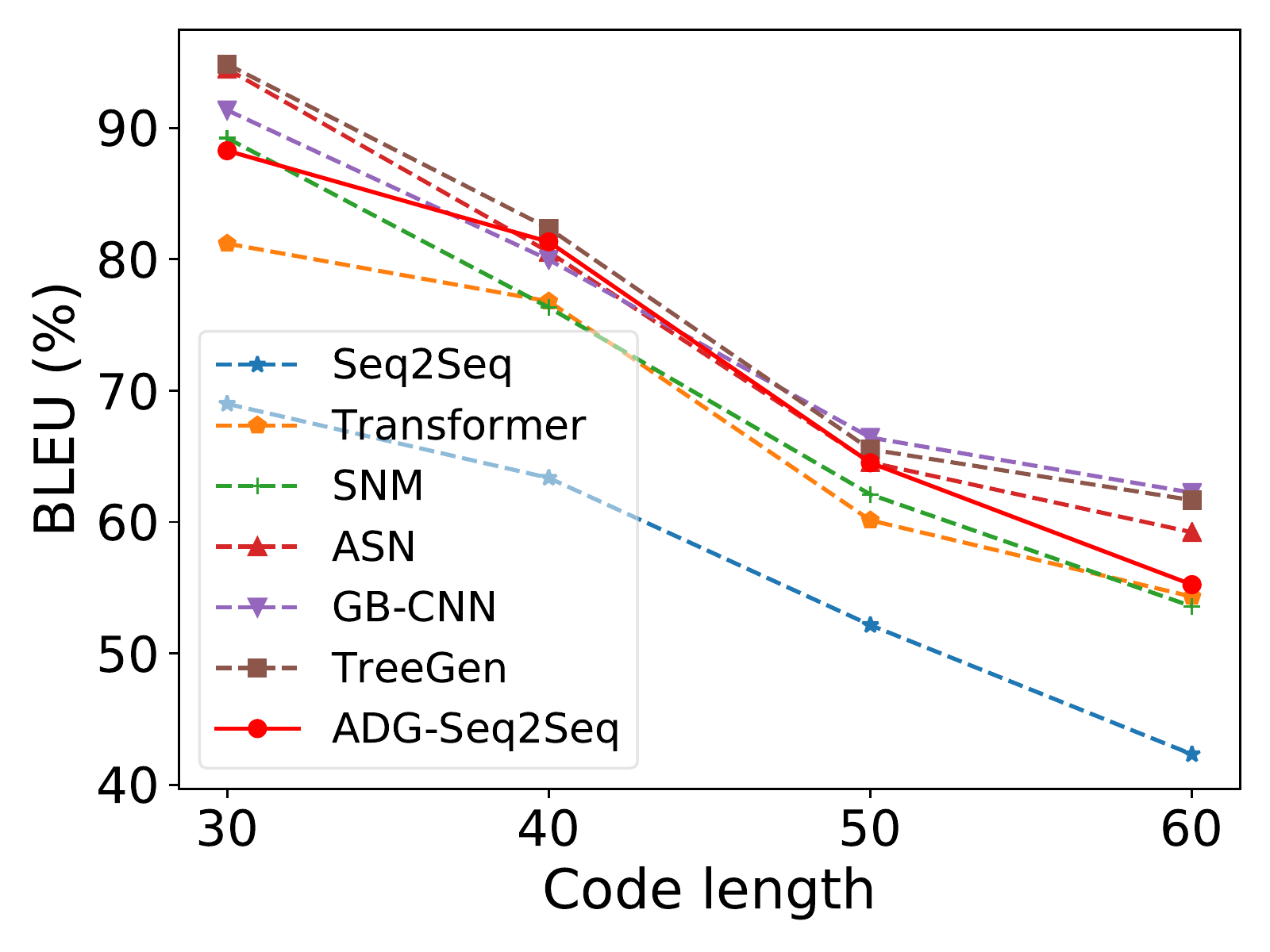}
			\label{CodeLen2_HS_BLEU}
		}
		\subfigure[HS: F1]{
			\includegraphics[width=0.225\linewidth]{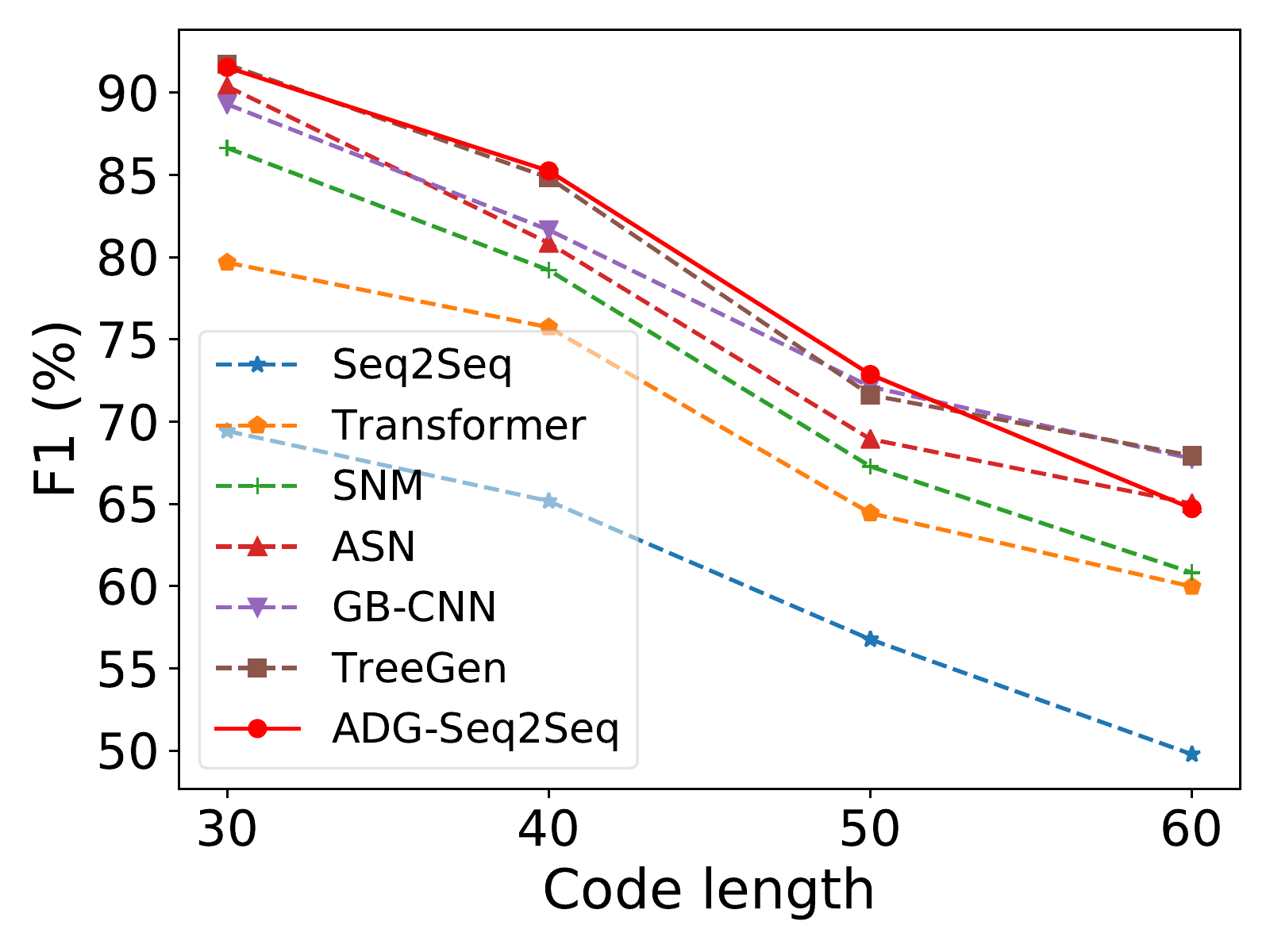}
			\label{CodeLen3_HS_F1}
		}
		\subfigure[HS: CIDEr]{
			\includegraphics[width=0.225\linewidth]{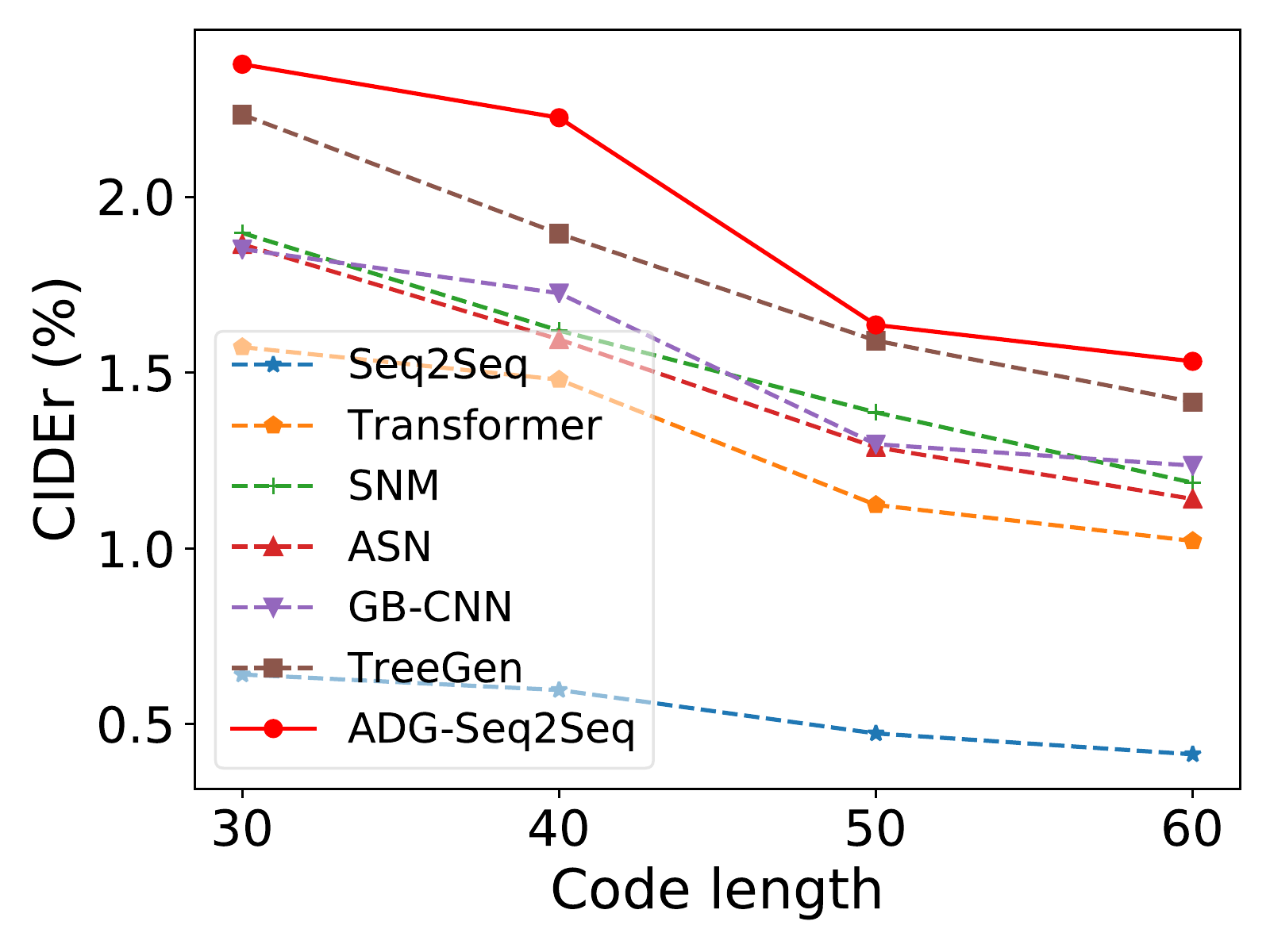}
			\label{CodeLen4_HS_CIDEr}
		}
		
		\subfigure[HS\_ROUGE-L]{
			\includegraphics[width=0.225\linewidth]{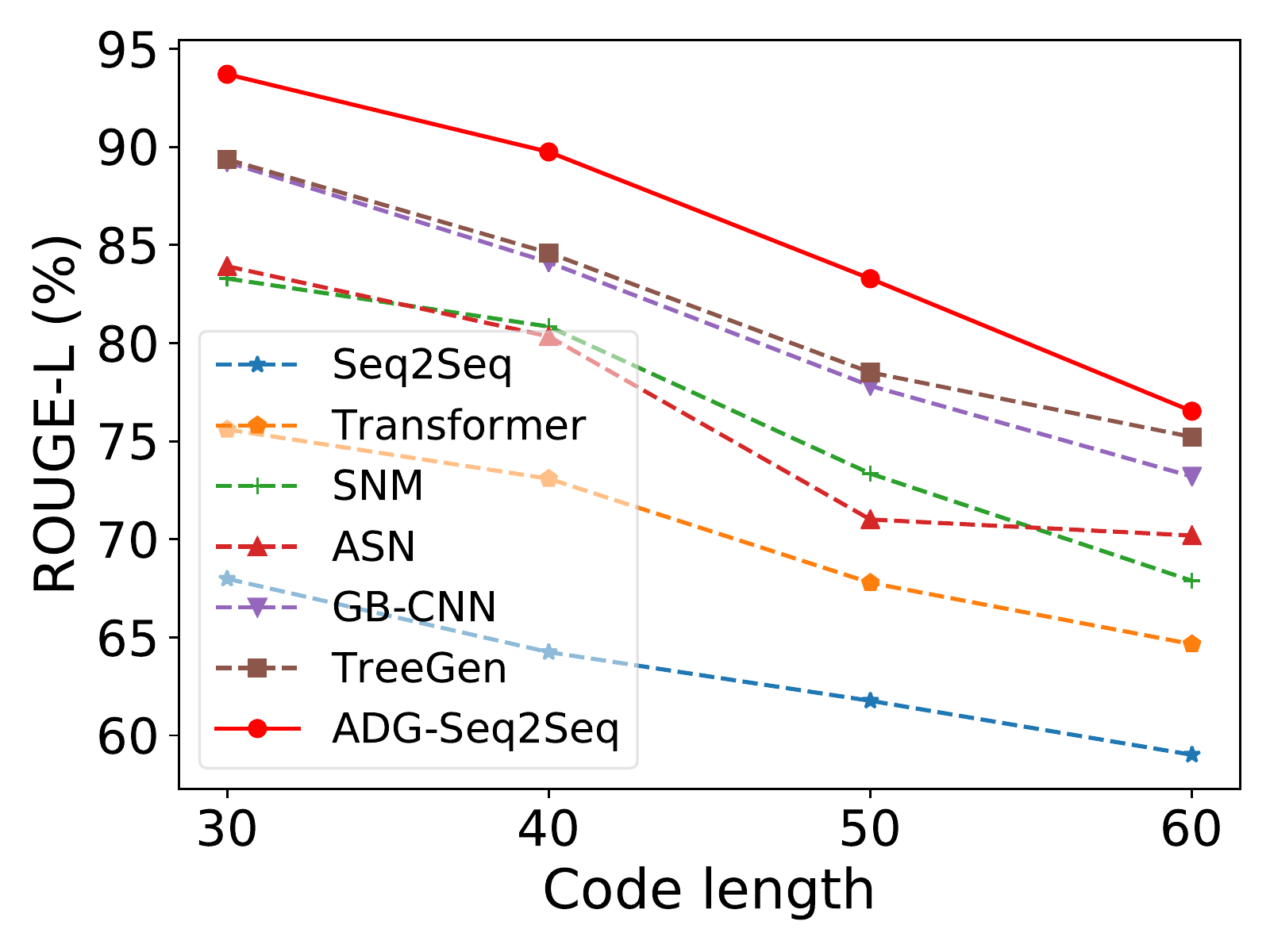}
			\label{CodeLen5_HS_ROUGE-L
			}
		}
		\subfigure[HS: ROUGE-1]{
			\includegraphics[width=0.225\linewidth]{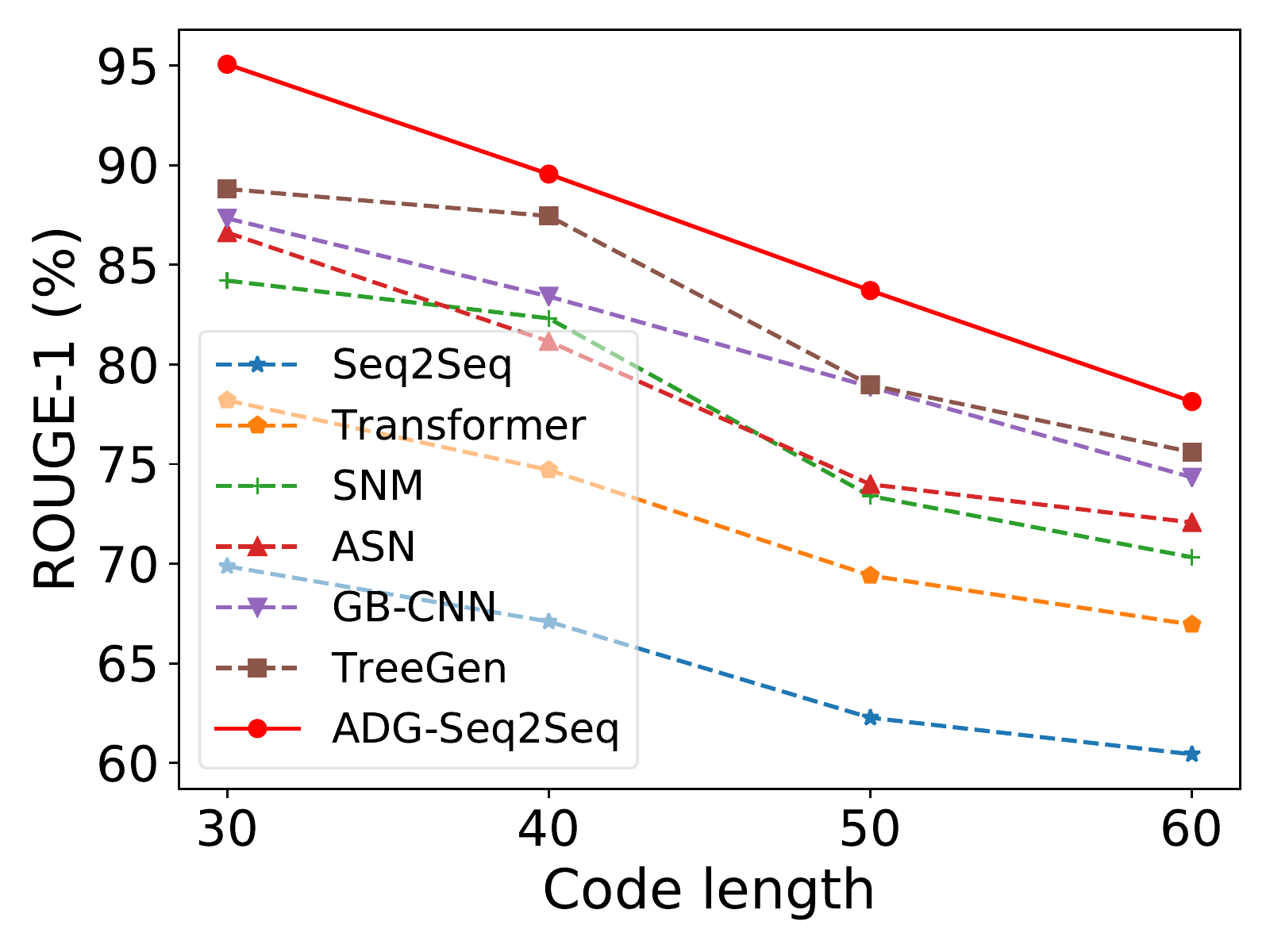}
			\label{CodeLen6_HS_ROUGE-1}
		}
		\subfigure[HS: ROUGE-2]{
			\includegraphics[width=0.225\linewidth]{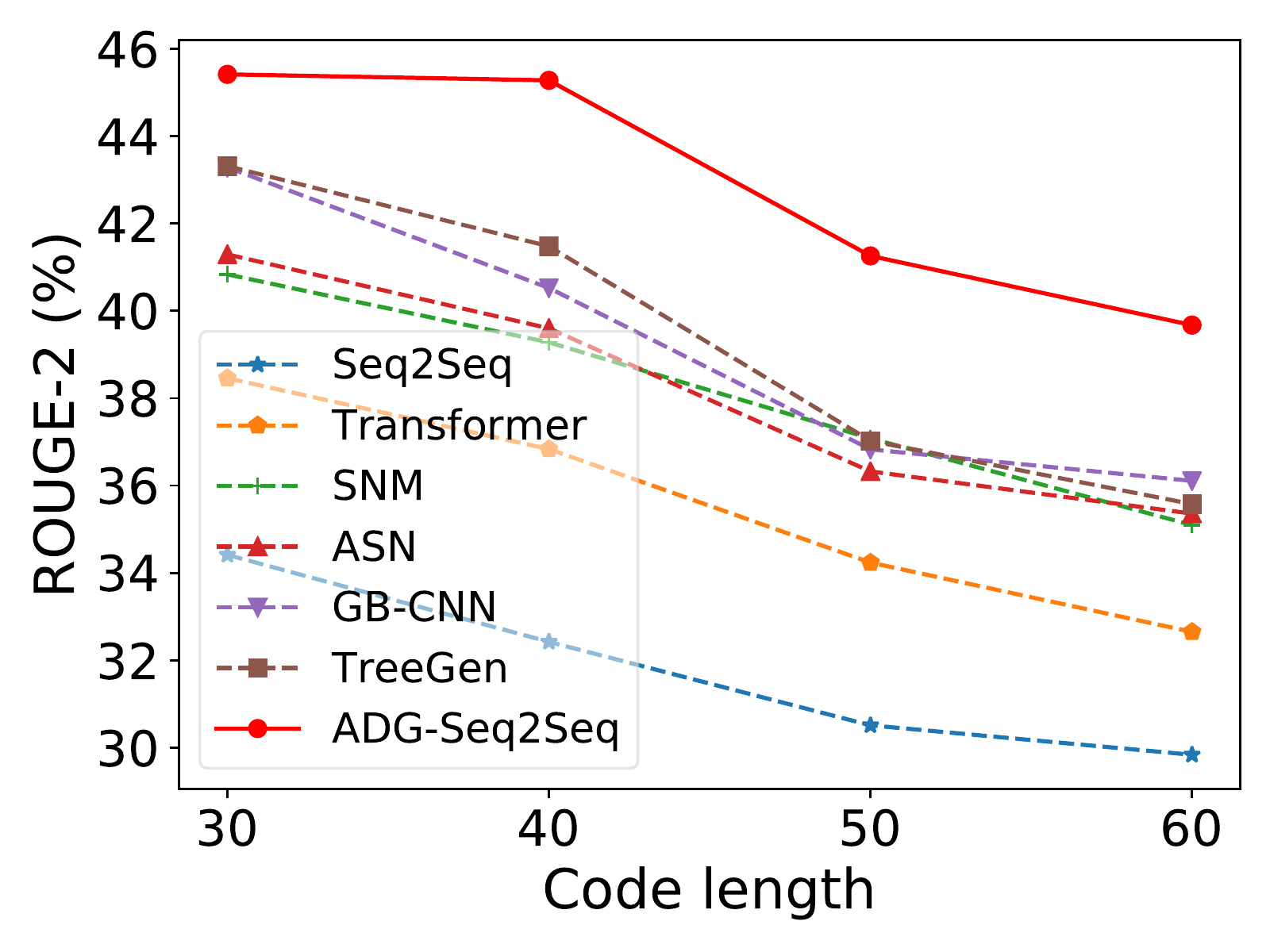}
			\label{CodeLen7_HS_ROUGE-2}
		}
		\subfigure[HS: RIBES]{
			\includegraphics[width=0.225\linewidth]{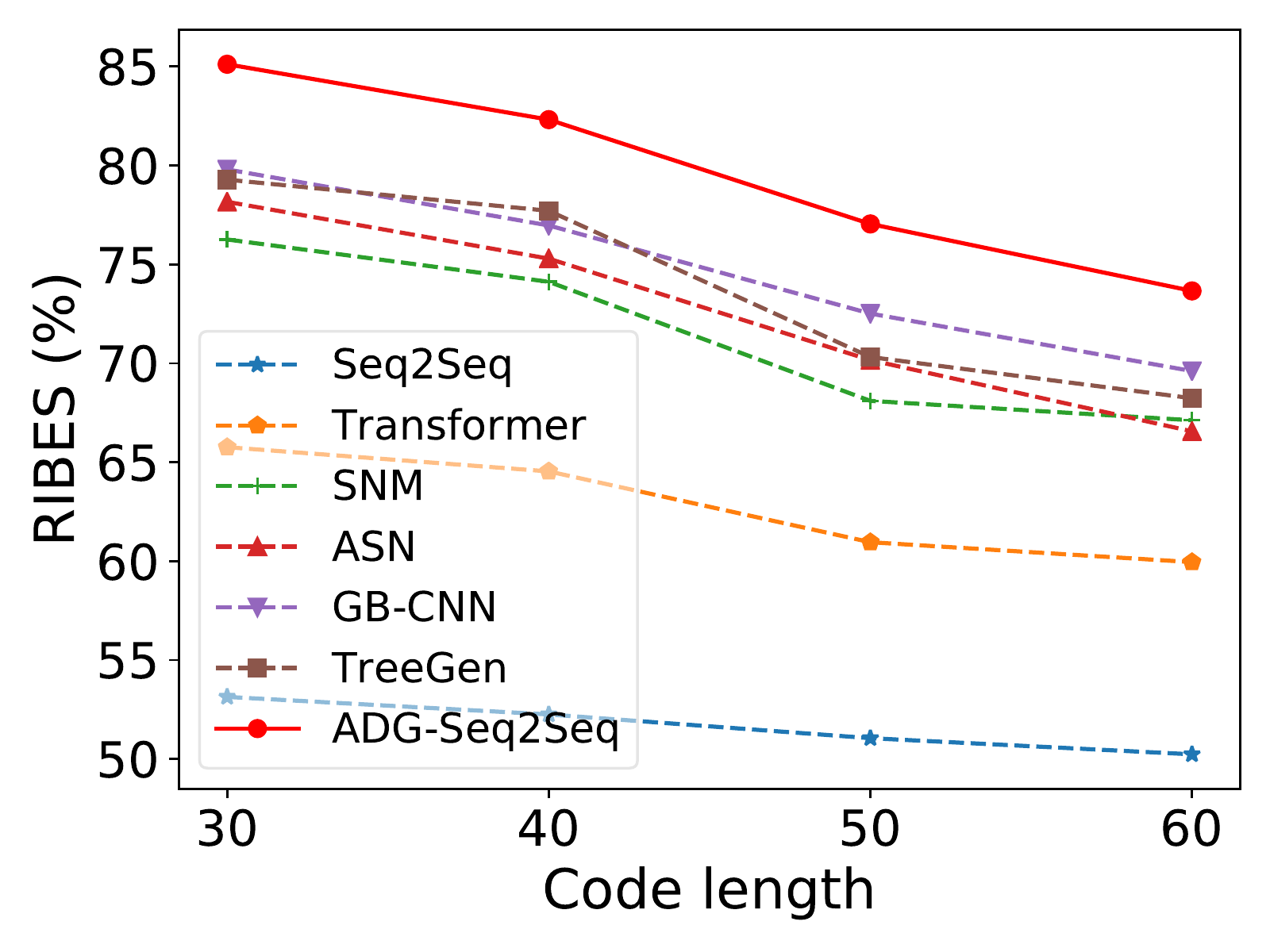}
			\label{CodeLen8_HS_RIBES}
		}
		
		\subfigure[MTG: Acc]{
			\includegraphics[width=0.225\linewidth]{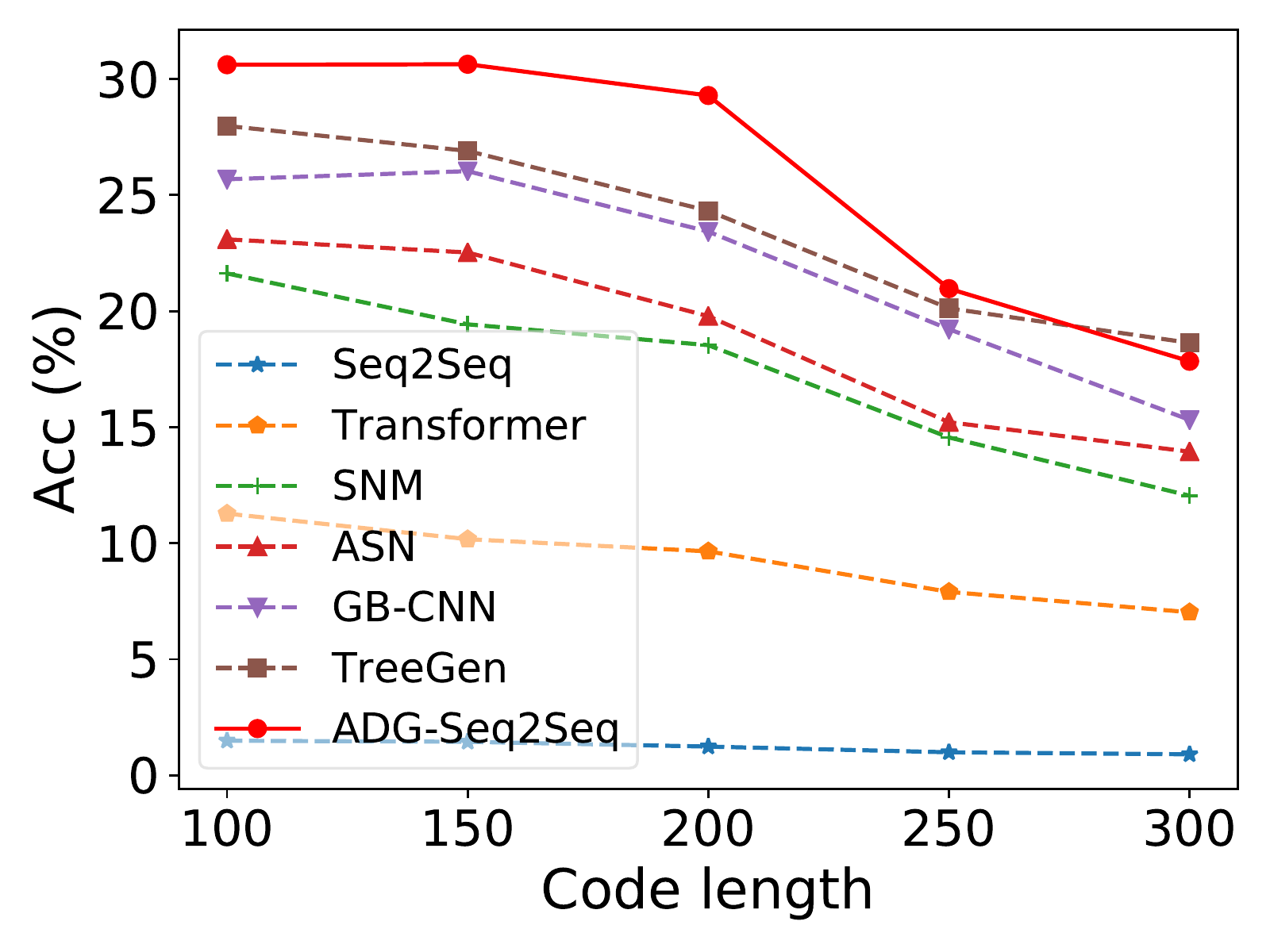}
			\label{CodeLen9_MTG_Acc}
		}
		\subfigure[MTG: BLEU]{
			\includegraphics[width=0.225\linewidth]{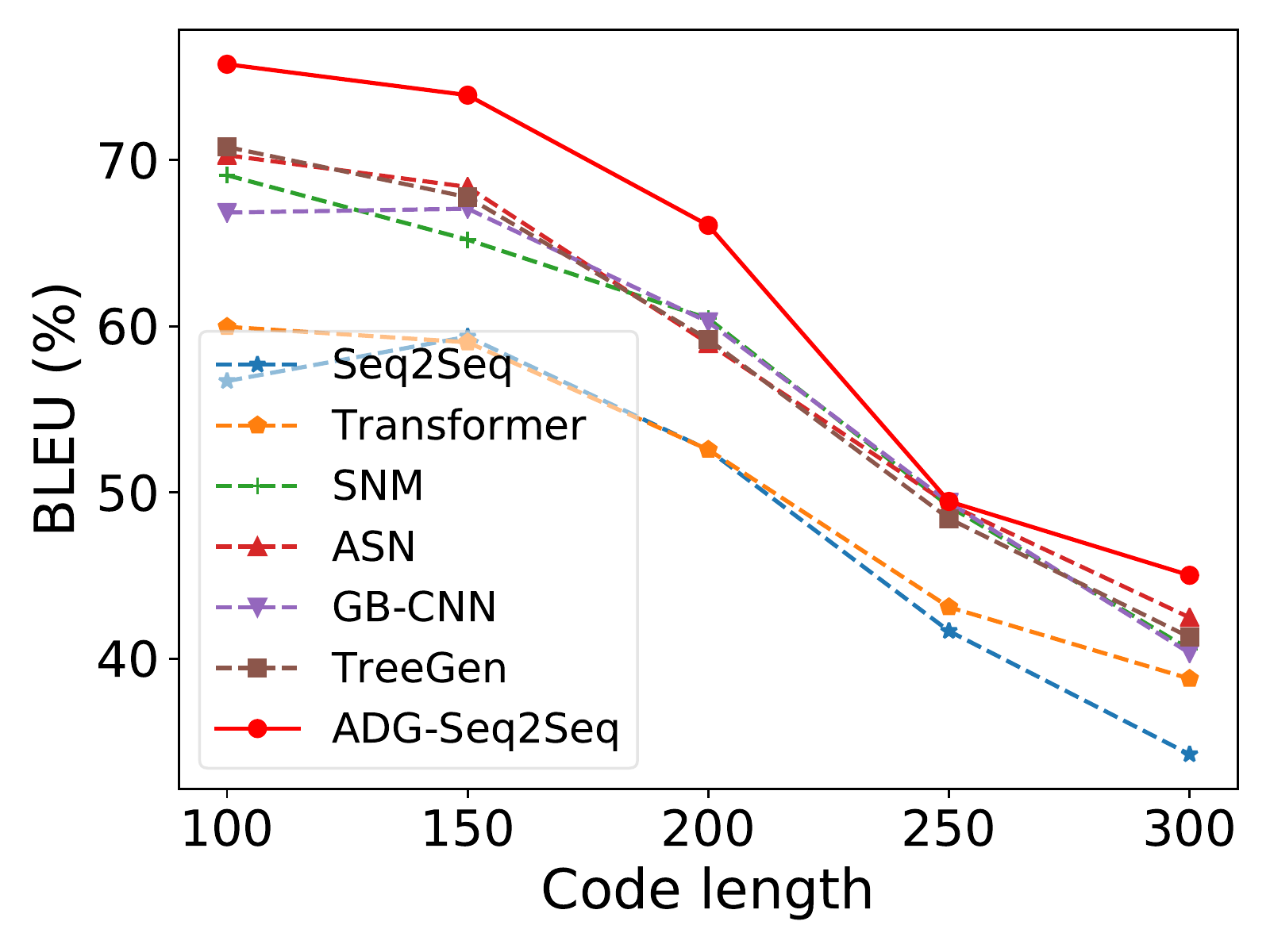}
			\label{CodeLen10_MTG_BLEU}
		}
		\subfigure[MTG: F1]{
			\includegraphics[width=0.225\linewidth]{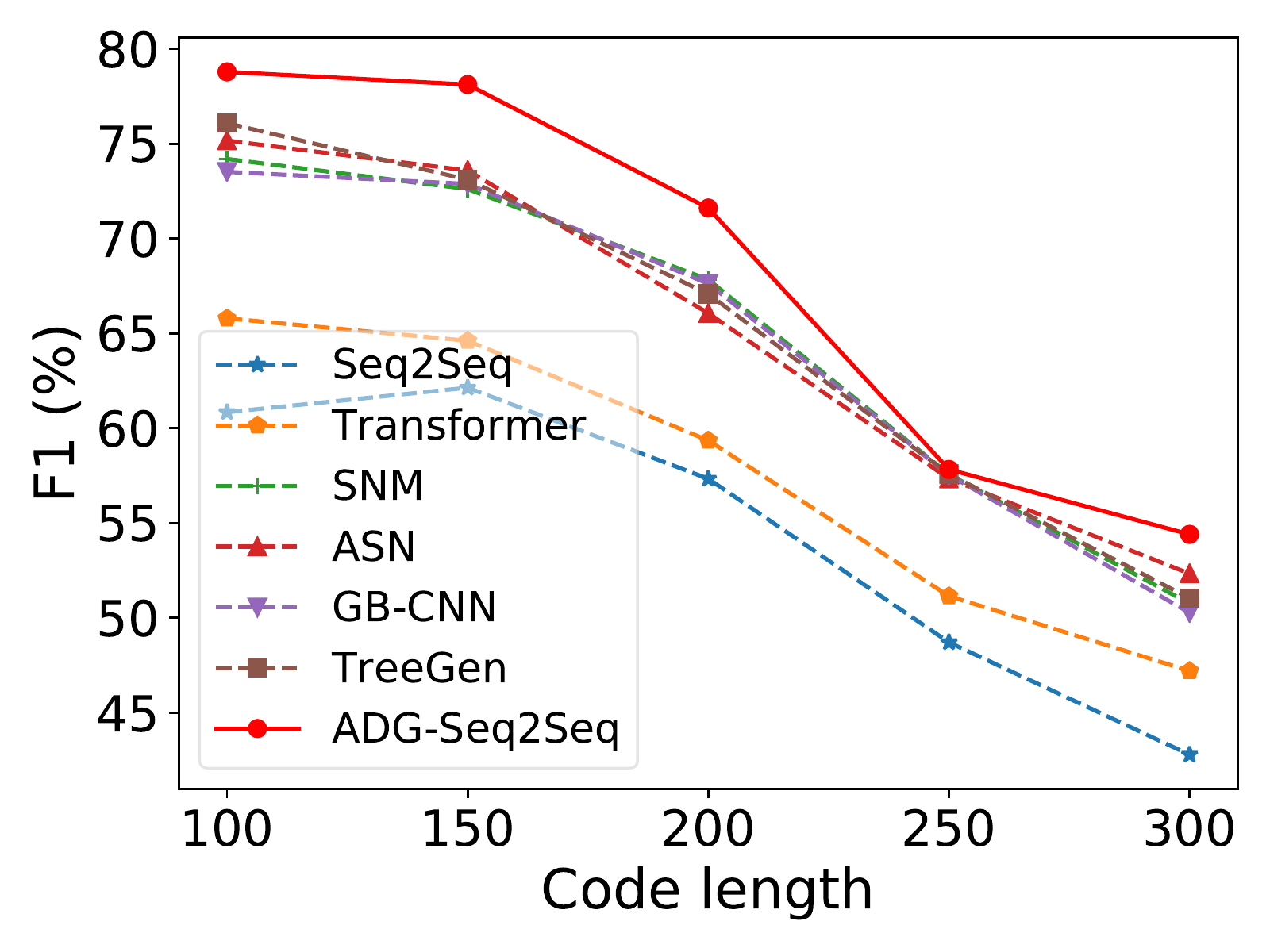}
			\label{CodeLen11_MTG_F1}
		}
		\subfigure[MTG: CIDEr]{
			\includegraphics[width=0.225\linewidth]{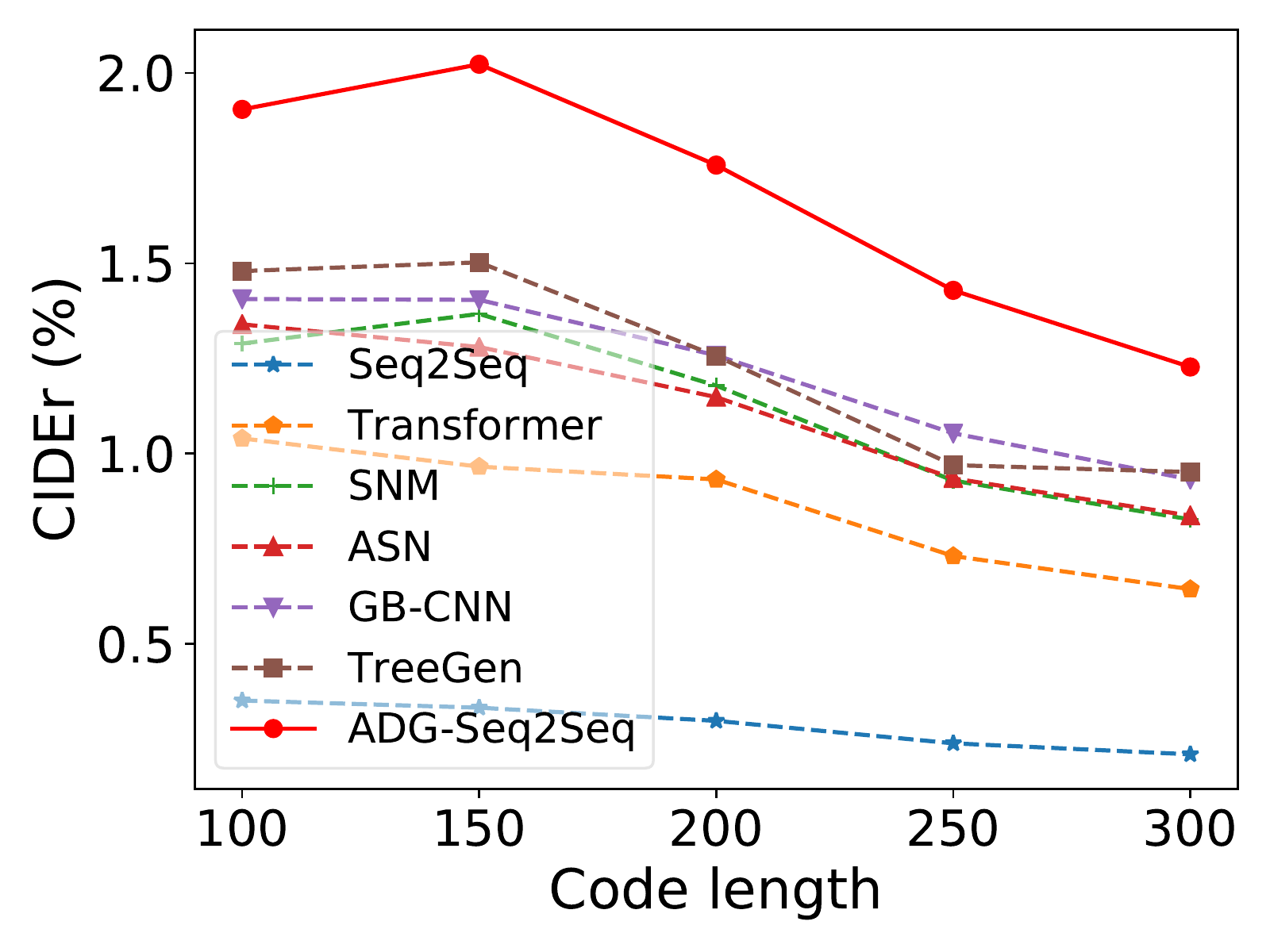}
			\label{CodeLen12_MTG_CIDEr}
		}
		
		\subfigure[MTG: ROUGE-L]{
			\includegraphics[width=0.225\linewidth]{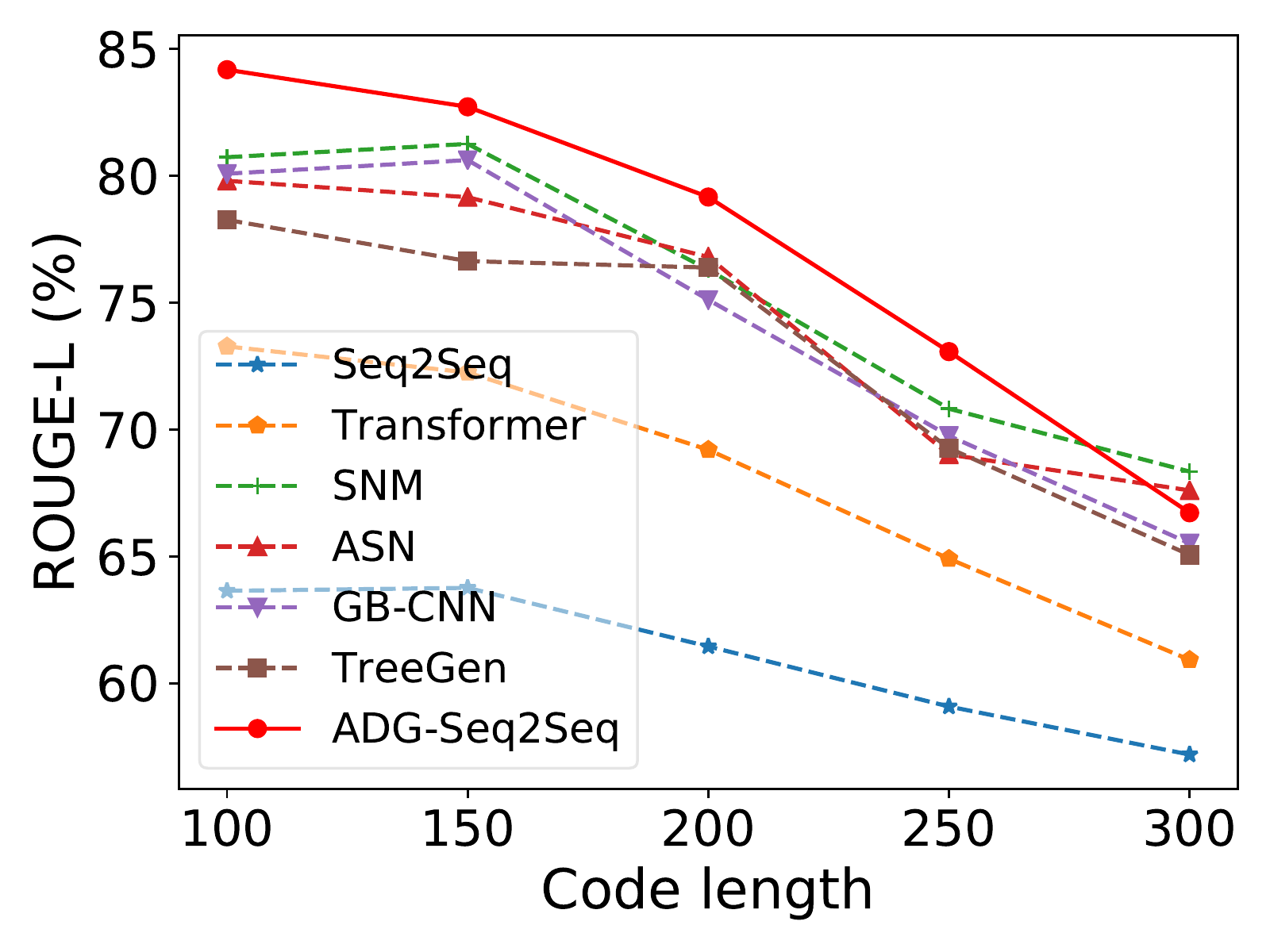}
			\label{CodeLen13_MTG_ROUGE-L}
		}
		\subfigure[MTG: ROUGE-1]{
			\includegraphics[width=0.225\linewidth]{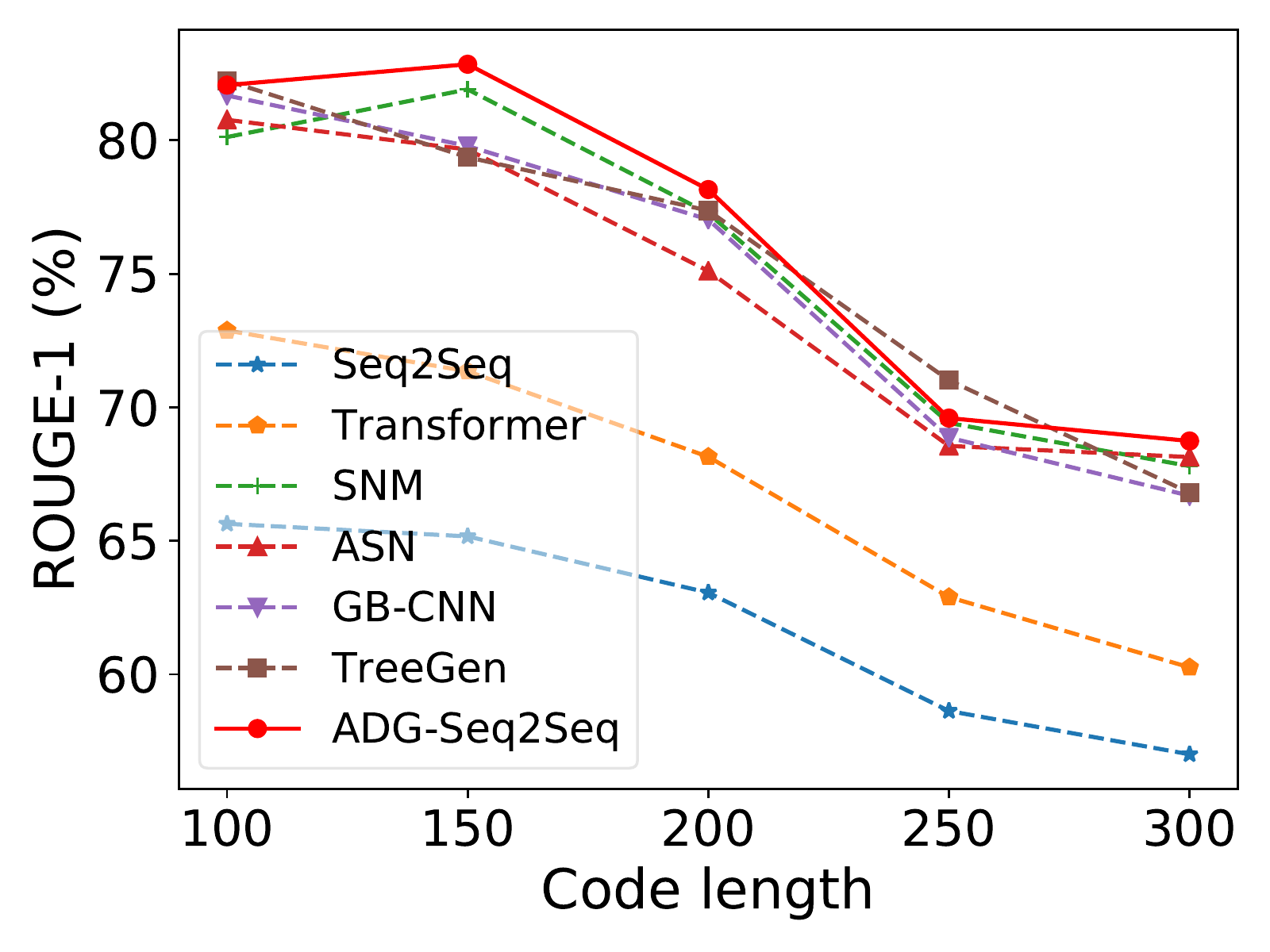}
			\label{CodeLen14_MTG_ROUGE-1}
		}
		\subfigure[MTG: ROUGE-2]{
			\includegraphics[width=0.225\linewidth]{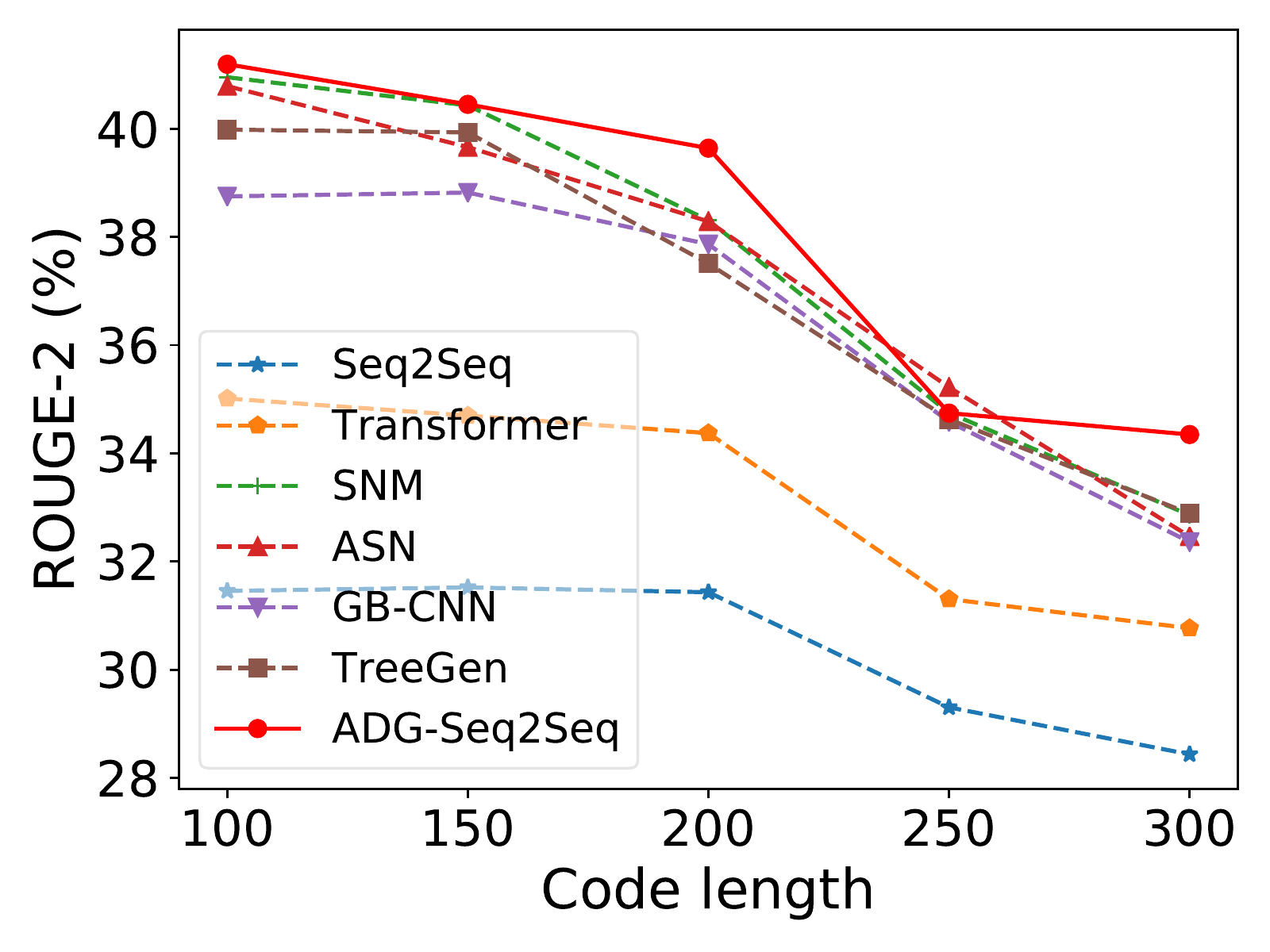}
			\label{CodeLen15_MTG_ROUGE-2}
		}
		\subfigure[MTG: RIBES]{
			\includegraphics[width=0.225\linewidth]{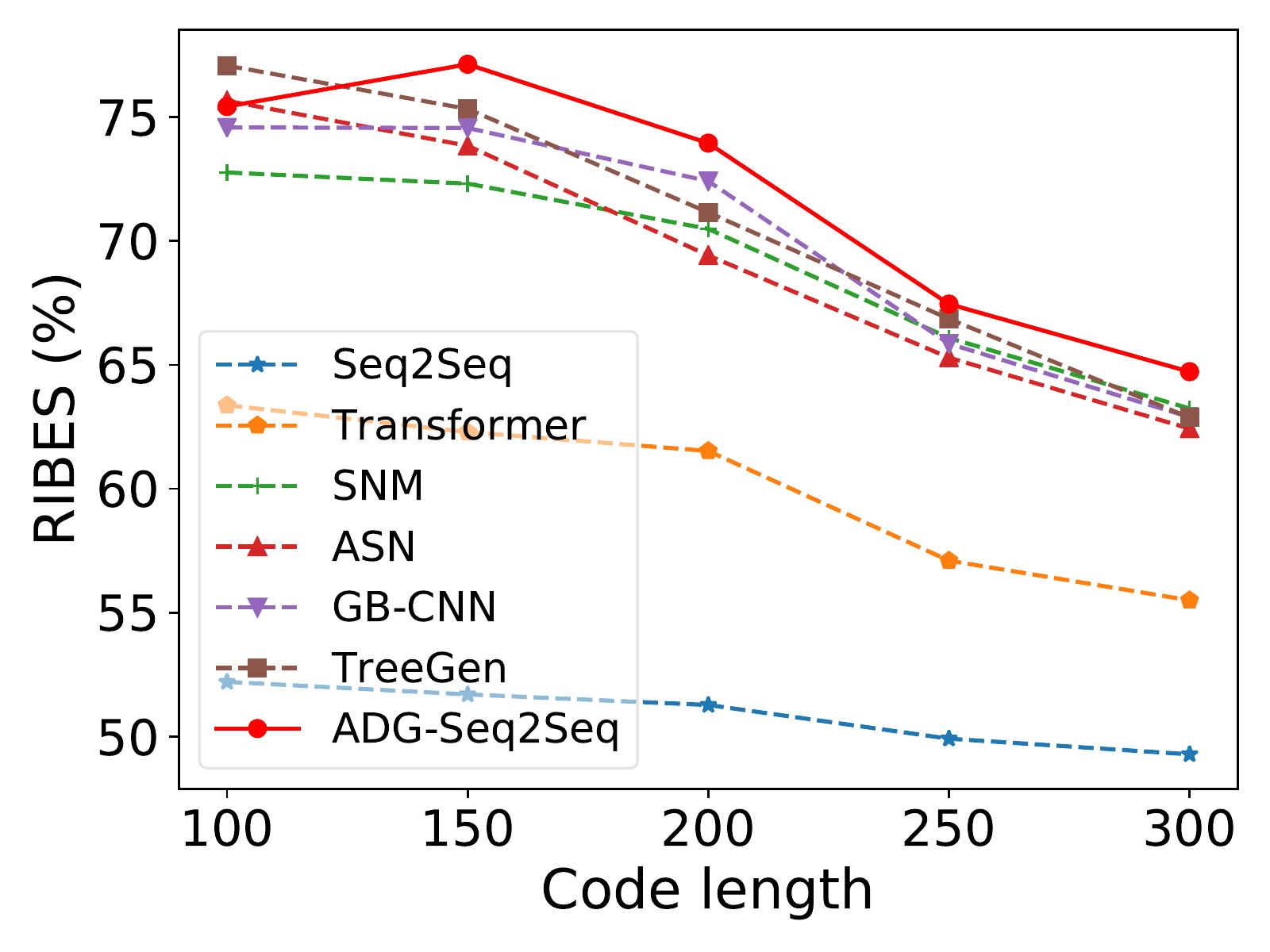}
			\label{CodeLen16_MTG_RIBES}
		}
		
		\subfigure[E-JDT: Acc]{
			\includegraphics[width=0.225\linewidth]{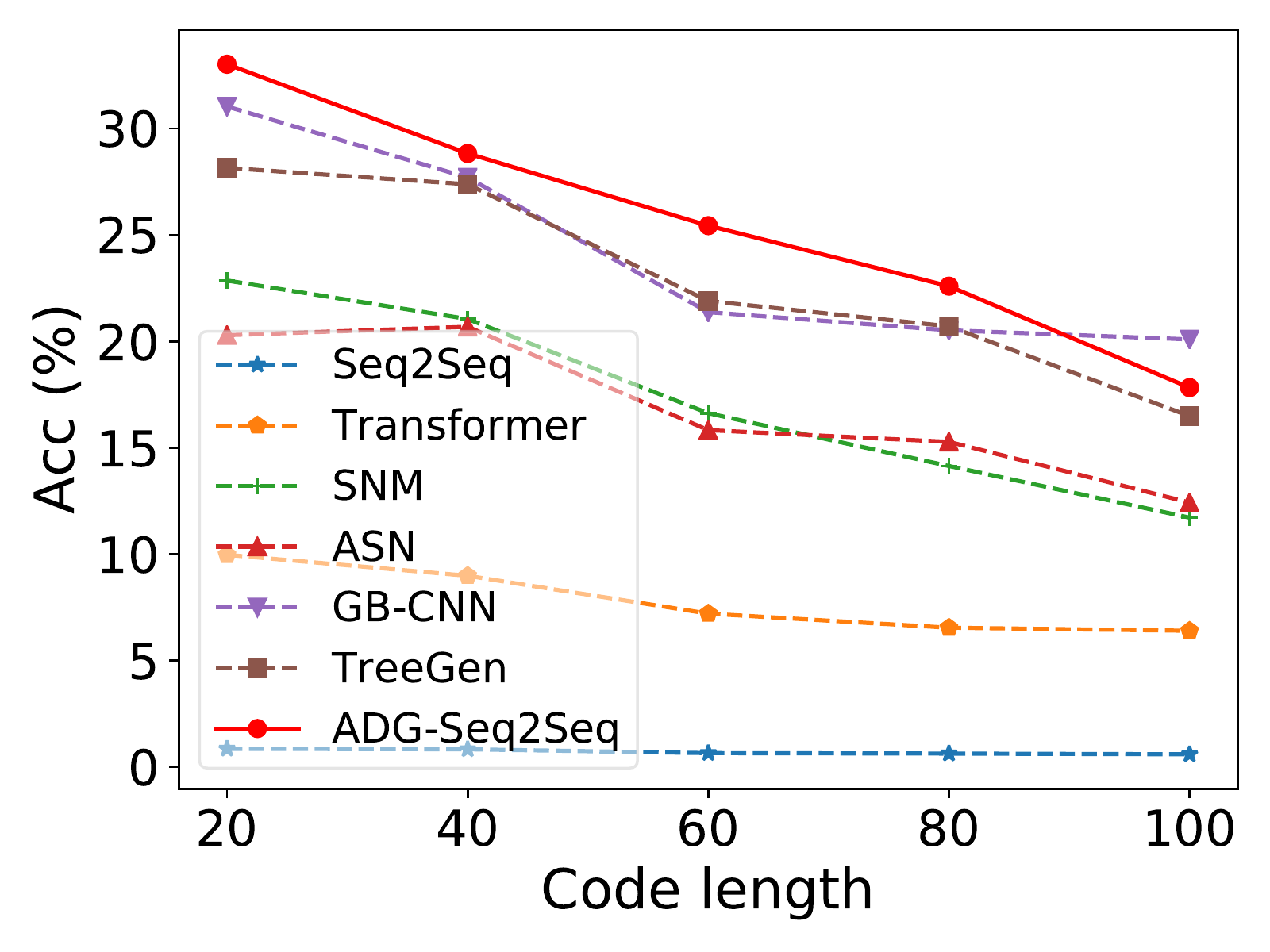}
			\label{CodeLen17_E-JDT_Acc}
		}
		\subfigure[E-JDT: BLEU]{
			\includegraphics[width=0.225\linewidth]{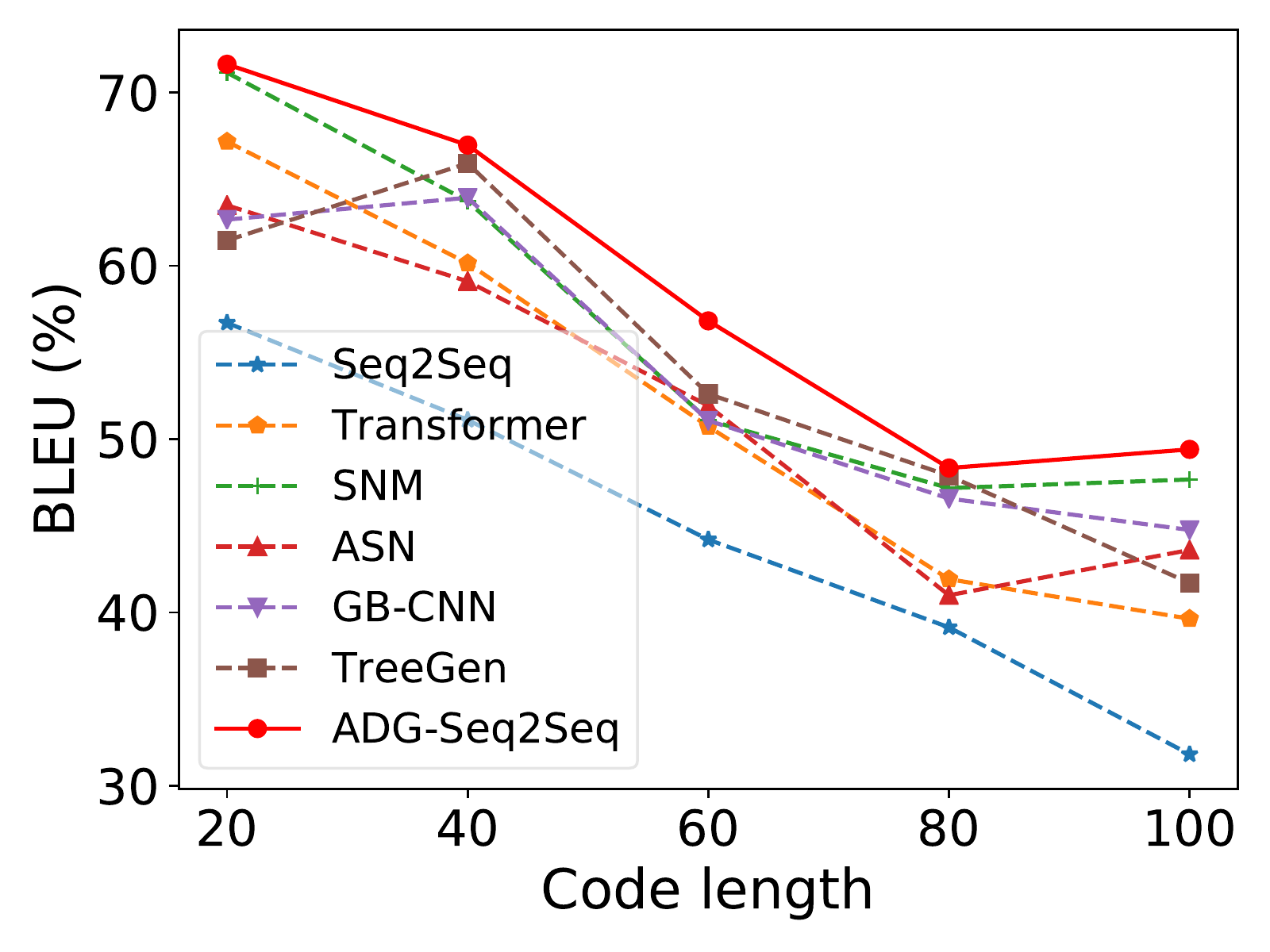}
			\label{CodeLen18_E-JDT_BLEU}
		}
		\subfigure[E-JDT: F1]{
			\includegraphics[width=0.225\linewidth]{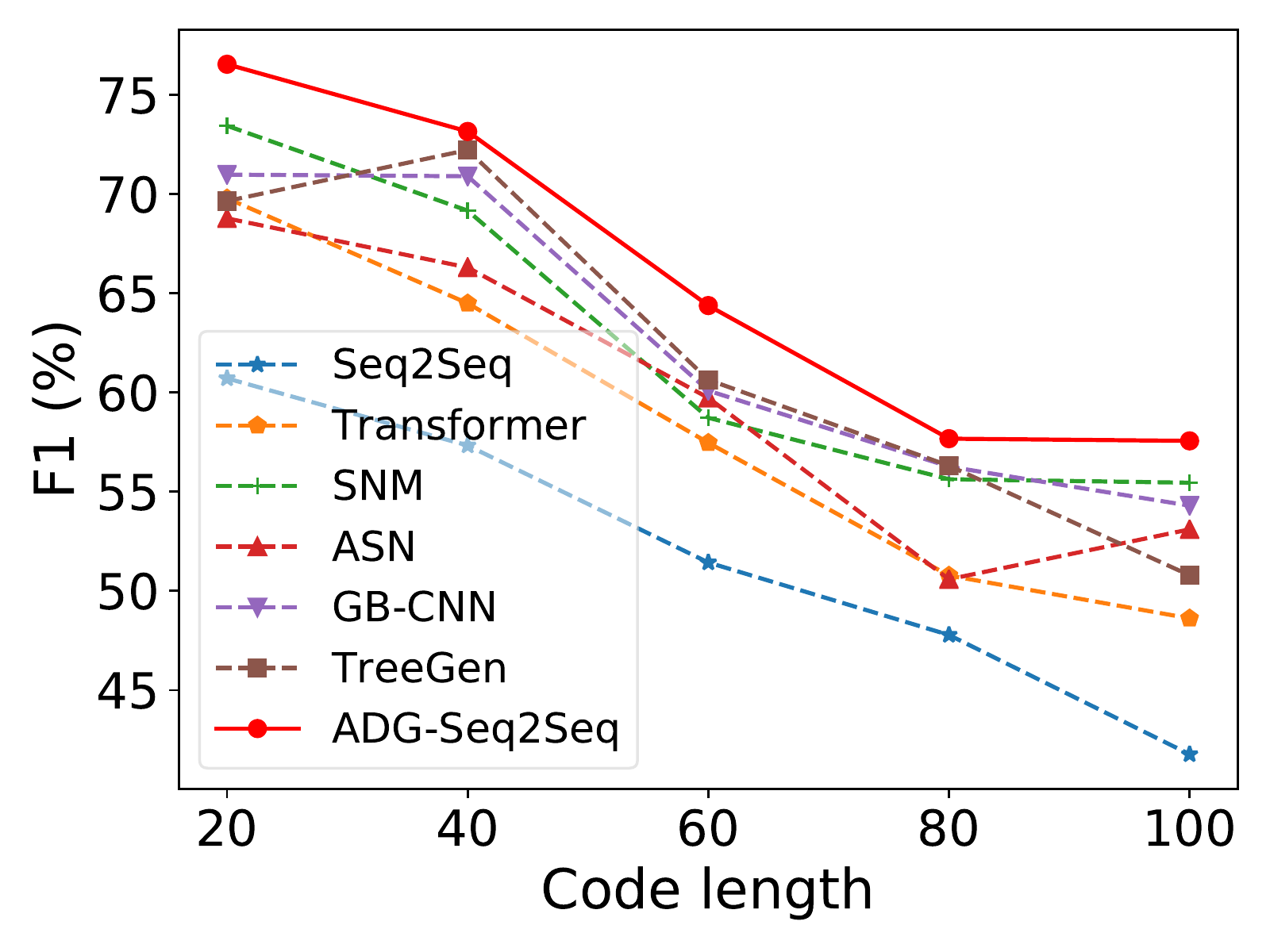}
			\label{CodeLen19_E-JDT_F1}
		}
		\subfigure[E-JDT: CIDEr]{
			\includegraphics[width=0.225\linewidth]{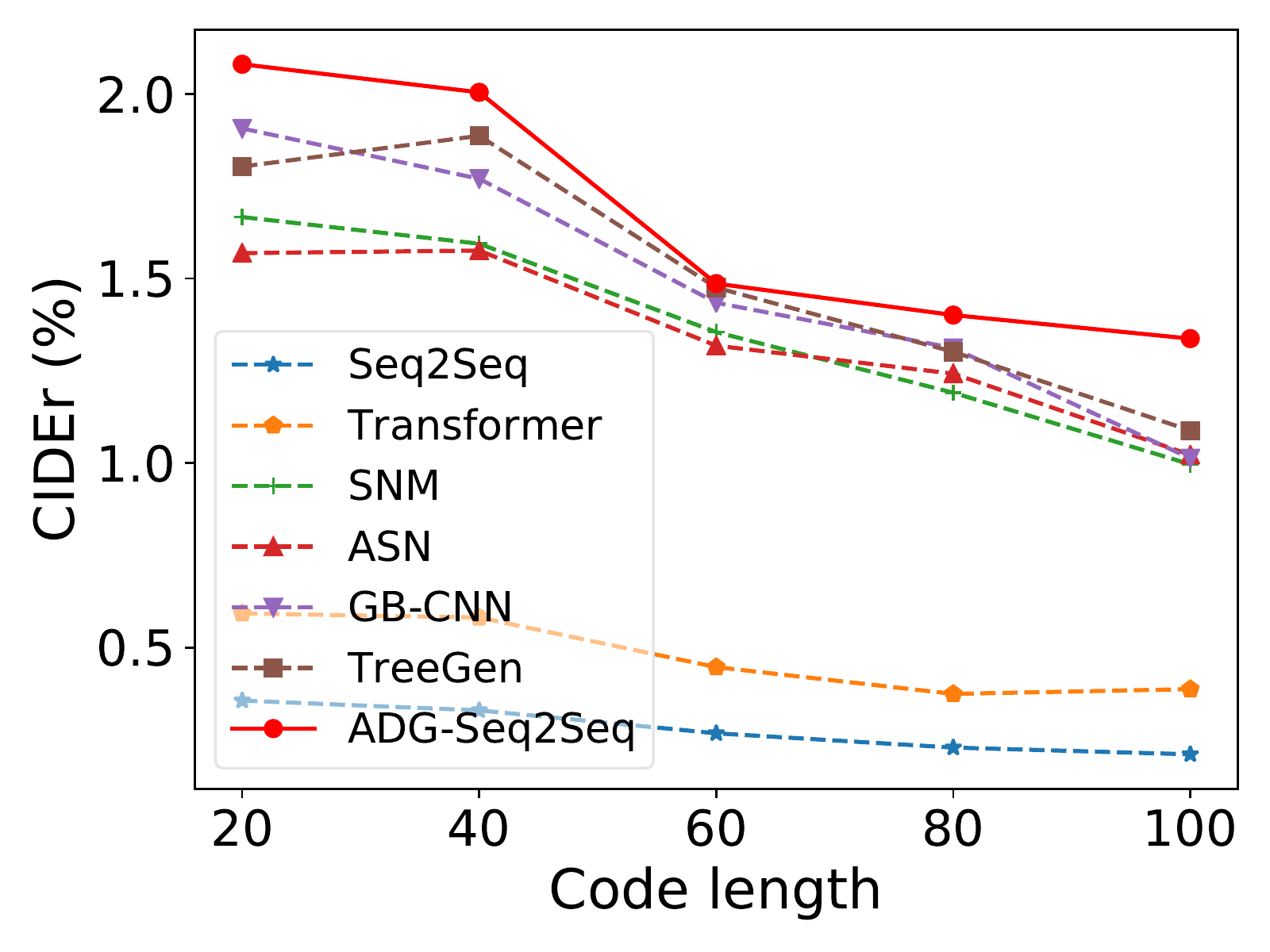}
			\label{CodeLen20_E-JDT_CIDEr}
		}
		
		\subfigure[E-JDT: ROUGE-L]{
			\includegraphics[width=0.225\linewidth]{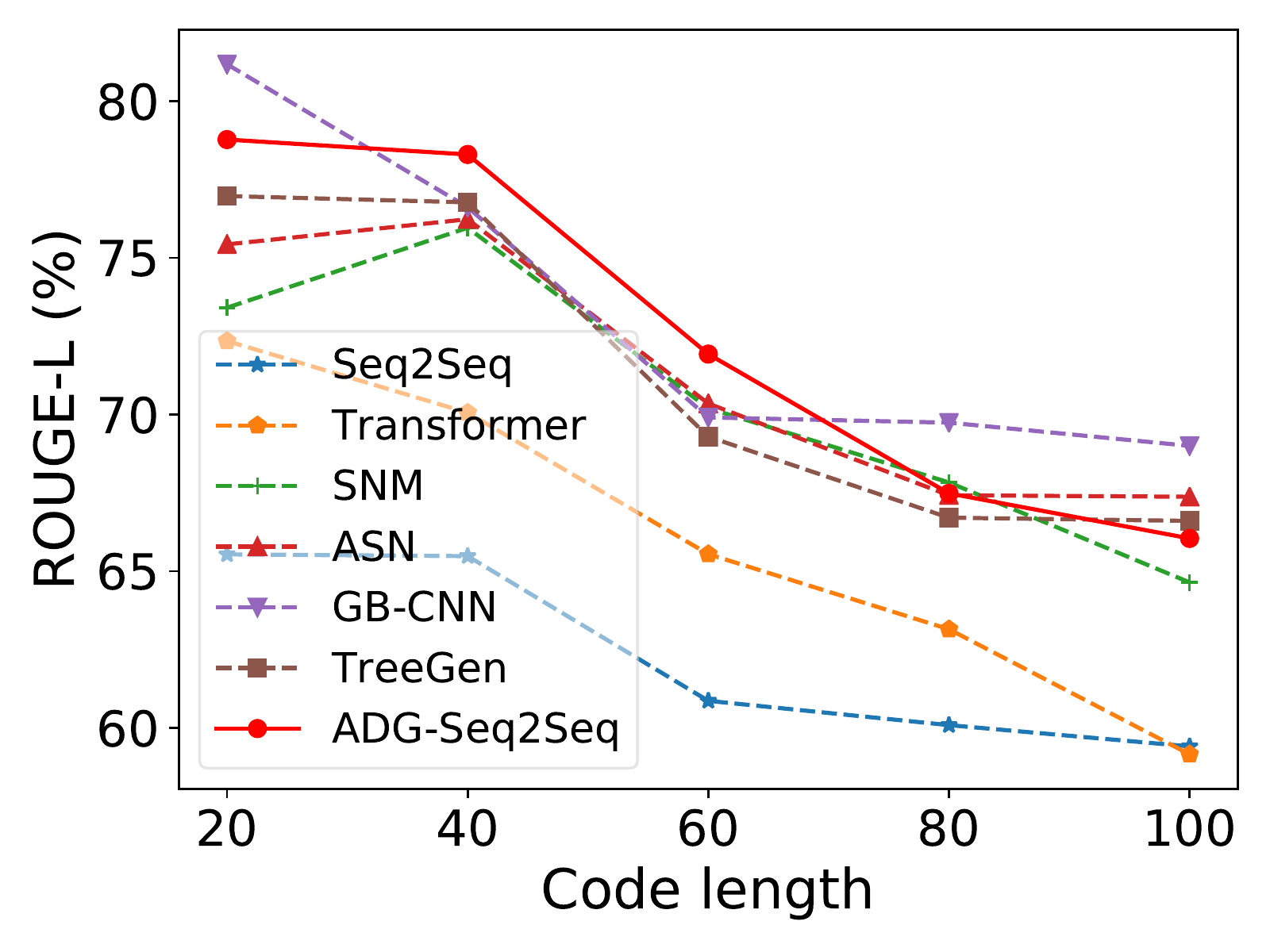}
			\label{CodeLen21_E-JDT_ROUGE-L}
		}
		\subfigure[E-JDT: ROUGE-1]{
			\includegraphics[width=0.225\linewidth]{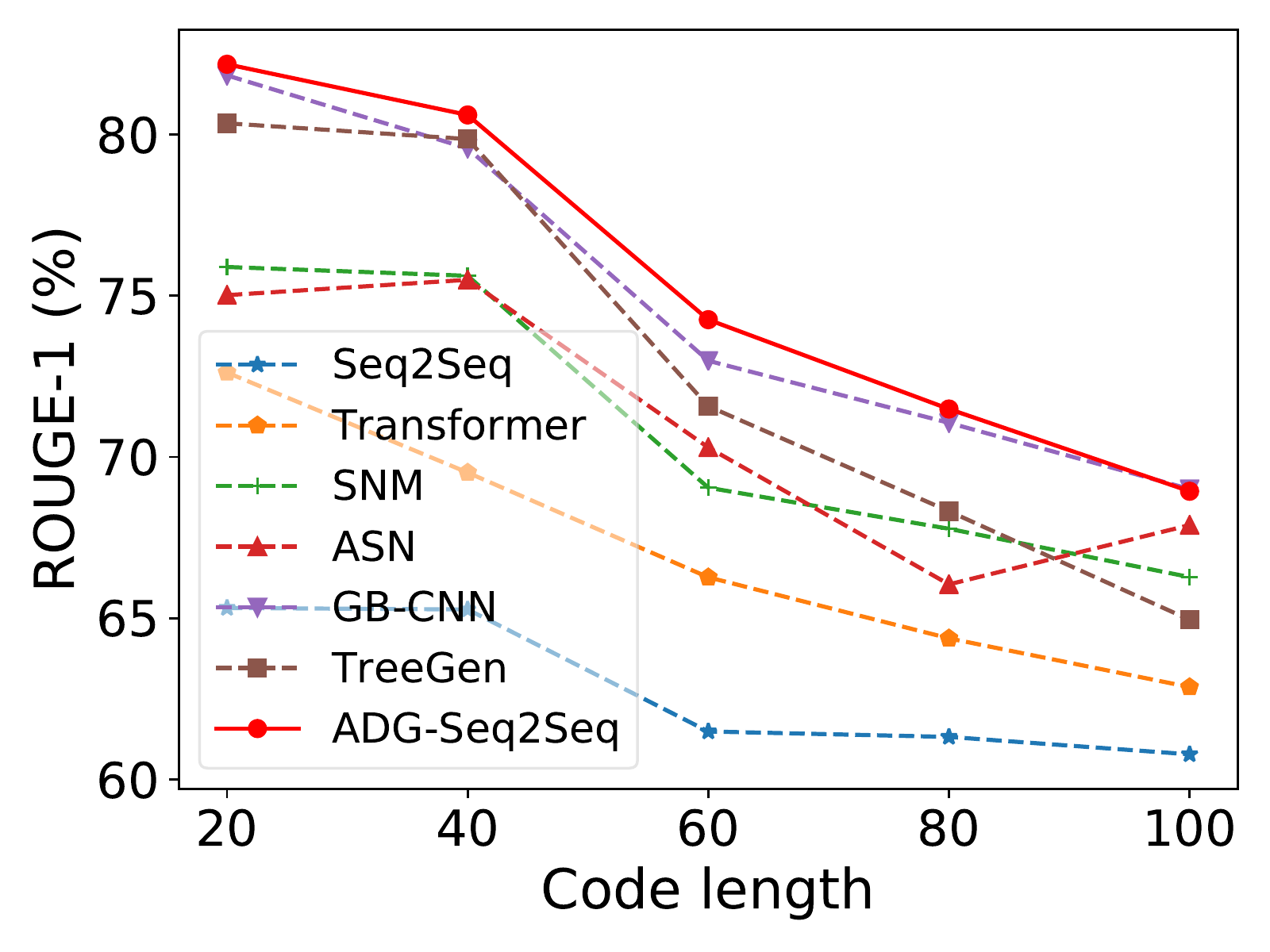}
			\label{CodeLen22_E-JDT_ROUGE-1}
		}
		\subfigure[E-JDT: ROUGE-2]{
			\includegraphics[width=0.225\linewidth]{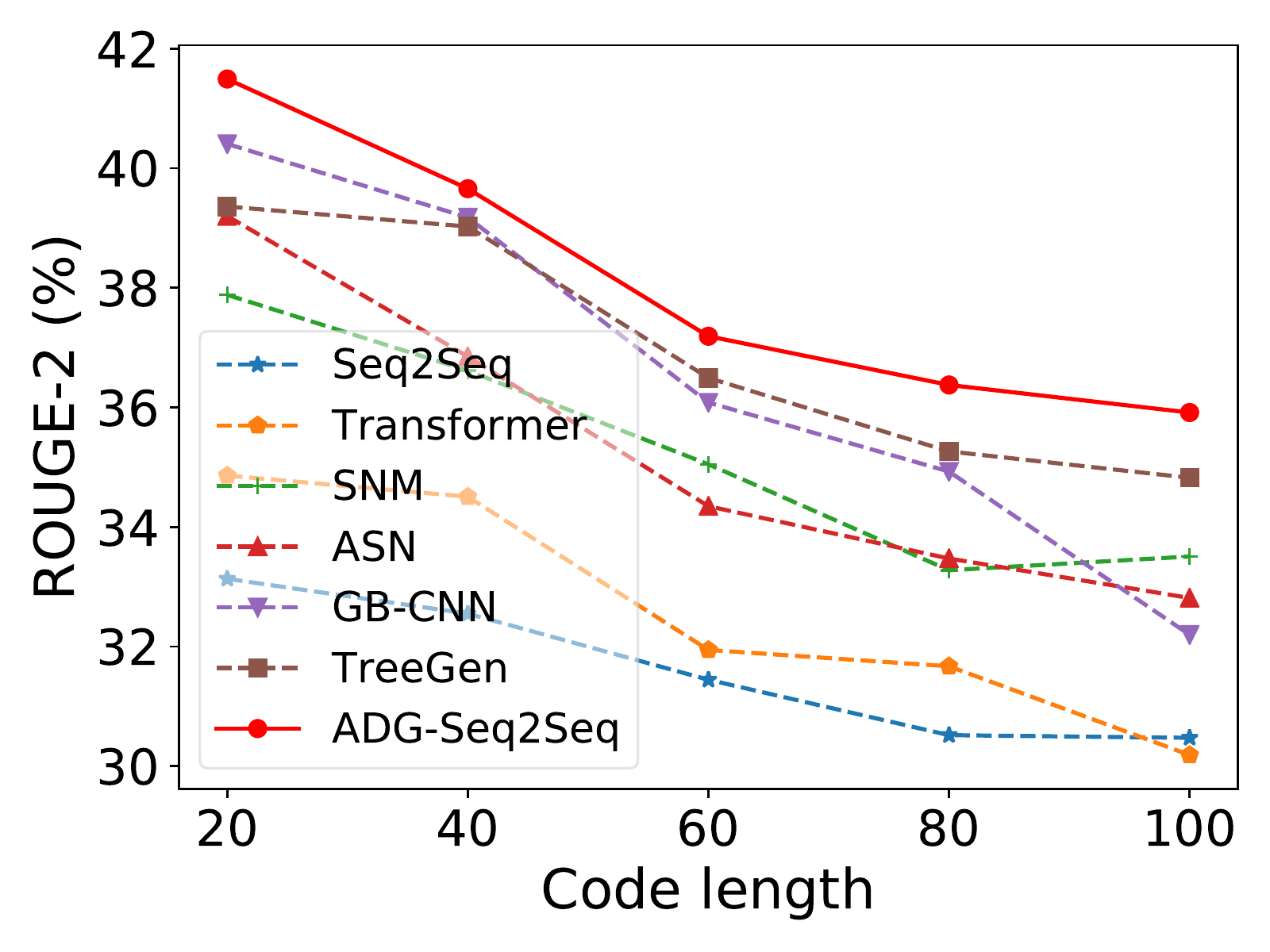}
			\label{CodeLen23_E-JDT_ROUGE-2}
		}
		\subfigure[E-JDT: RIBES]{
			\includegraphics[width=0.225\linewidth]{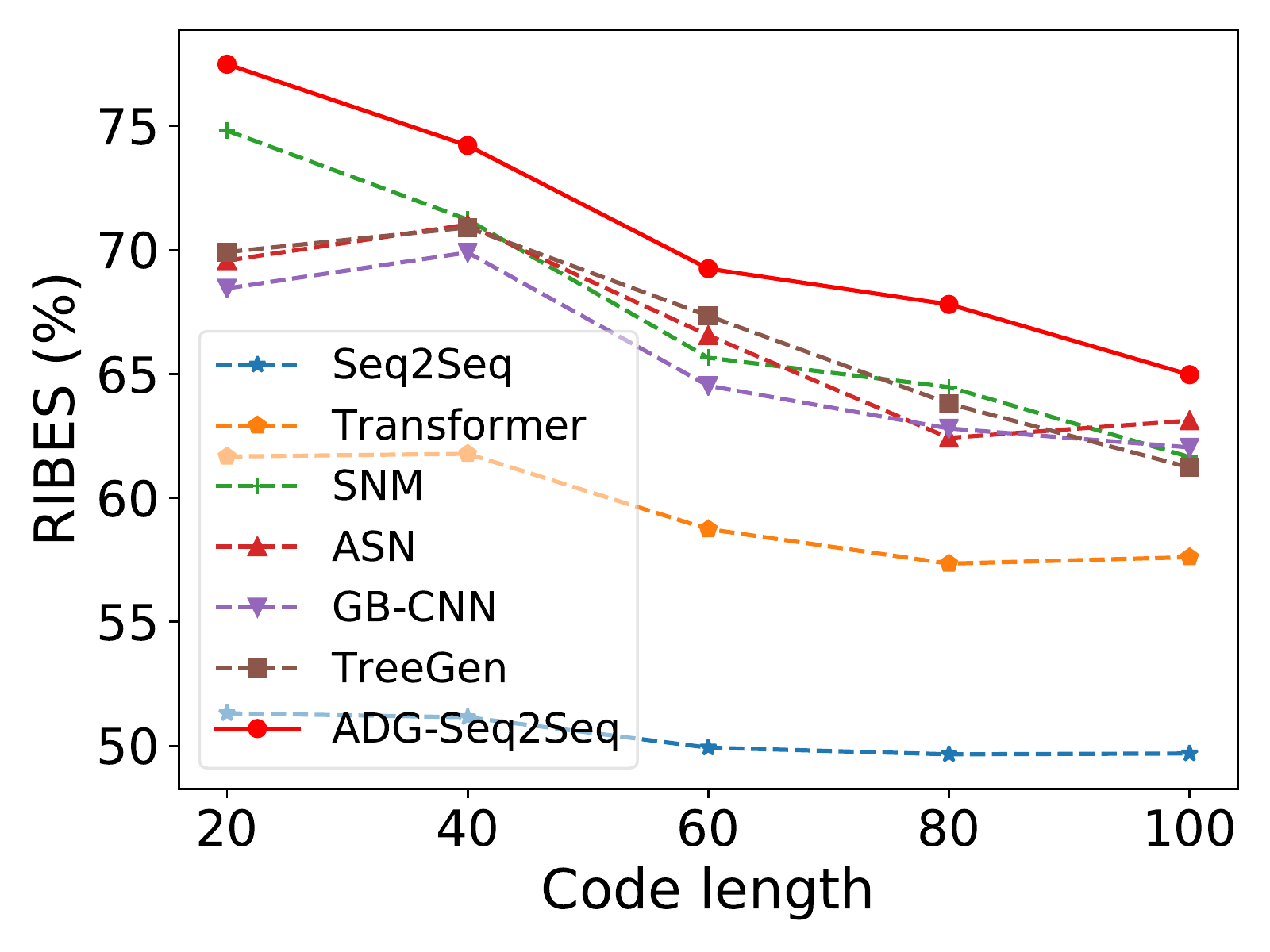}
			\label{CodeLen24_E-JDT_RIBES}
		}
		
		\caption{\textbf{Experimental results of the comparison of state-of-the-art methods with varying code lengths.}}
		\label{fig:code_len}
	\end{figure*}

	Fig.~\ref{fig:code_len} provides the comprehensive comparison results. The scores on all metrics tend to be lower for greater code lengths, which is unsurprising because various factors, e.g., the average counts of tokens, methods, edges, incomings, and outgoings increase accordingly, as shown in Table~\ref{tab:LPer}, making predictions more difficult. In general, ADG-Seq2Seq achieves the best scores on 87.5\% of indicators. 
		
	Furthermore, TreeGen and GB-CNN perform better on the Python dataset, while ADG-Seq2Seq shows advantages on the Java datasets. In addition to considering the influence of sample size, as discussed previously, we analyse different datasets in depth from the perspective of the programming language. 
	
	Python-based HS originates from a card game, and each code snippet corresponds to a specific card. The ADG generated based on HS is sparse since the APIs developed in Python, especially those located on different cards, lack data dependencies on each other. Due to the influence of the sparse graph, the generalization and inference performance of our learning model is weakened. In contrast, AST-based models learn the code for each card from the local view and achieve better results.
	
	For MTG and E-JDT, the situation changes. MTG and E-JDT are both programmed in Java, which contains more general API methods. These APIs have rich interactive data dependencies among different code snippets. Consequently, denser ADGs  are constructed, so the performance of our model is improved. The detailed data presented in Table~\ref{tab:LPer} support our arguments.
	
	Compared to the other datasets, the target code is difficult to predict for E-JDT since all description lengths are less than 20, but the lengths of the corresponding code continue to increase, as shown in Table~\ref{tab:LPer}. ADG-Seq2Seq remains superior on E-JDT. When the code length exceeds 80, some scores drop sharply since the code length is approximately five times the description length, which poses a considerable challenge to the reasoning ability of our model. 
	
	In addition to the length of each data item, the size of the dataset could influence the experimental results. However, the primary purpose of this experiment is to compare the quality of code generation between different methods with different code lengths and description lengths. For the various methods, the comparison is fair since the same training dataset is being used. The experimental results show that ADG-Seq2Seq tends to outperform its rivals 
	when the dataset contains as few as 80 items or as many as 85,000 items. This shows that our method is stable even as the code length varies.
	
	In summary, the red curves (ADG-Seq2Seq) shown in Fig.~\ref{fig:code_len} are above the other curves, which shows that ADG-Seq2Seq outperforms the compared methods in most cases and can generate more accurate results than state-of-the-art as the code length varies.

	\section{Qualitative Analysis}
		\label{sec:dicc}%
	The promising performance of the proposed method in generating code has been verified in quantitative experiments. In this section, we provide case studies and error analyses to investigate specific situations in which our method performed well or poorly. 
	
	\subsection{Case Study}
\begin{figure}[!htbp]
    \centering
    \includegraphics[width=\linewidth]{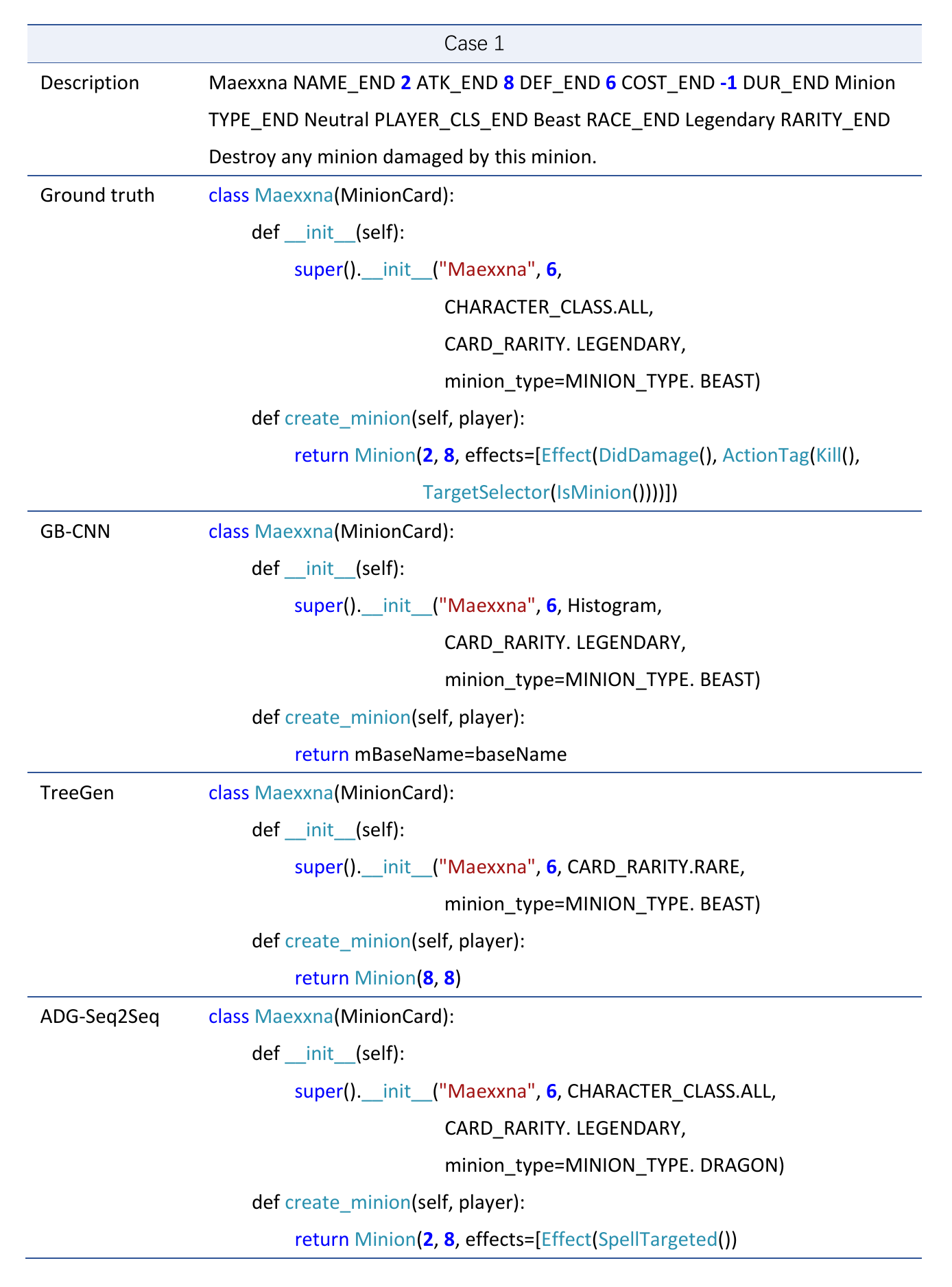}
    \caption{Code examples generated by GB-CNN, TreeGen and ADG-Seq2Seq on HS.}
    \label{fig:case1}
    \end{figure}

\begin{figure}[!htbp]
    \centering
    \includegraphics[width=\linewidth]{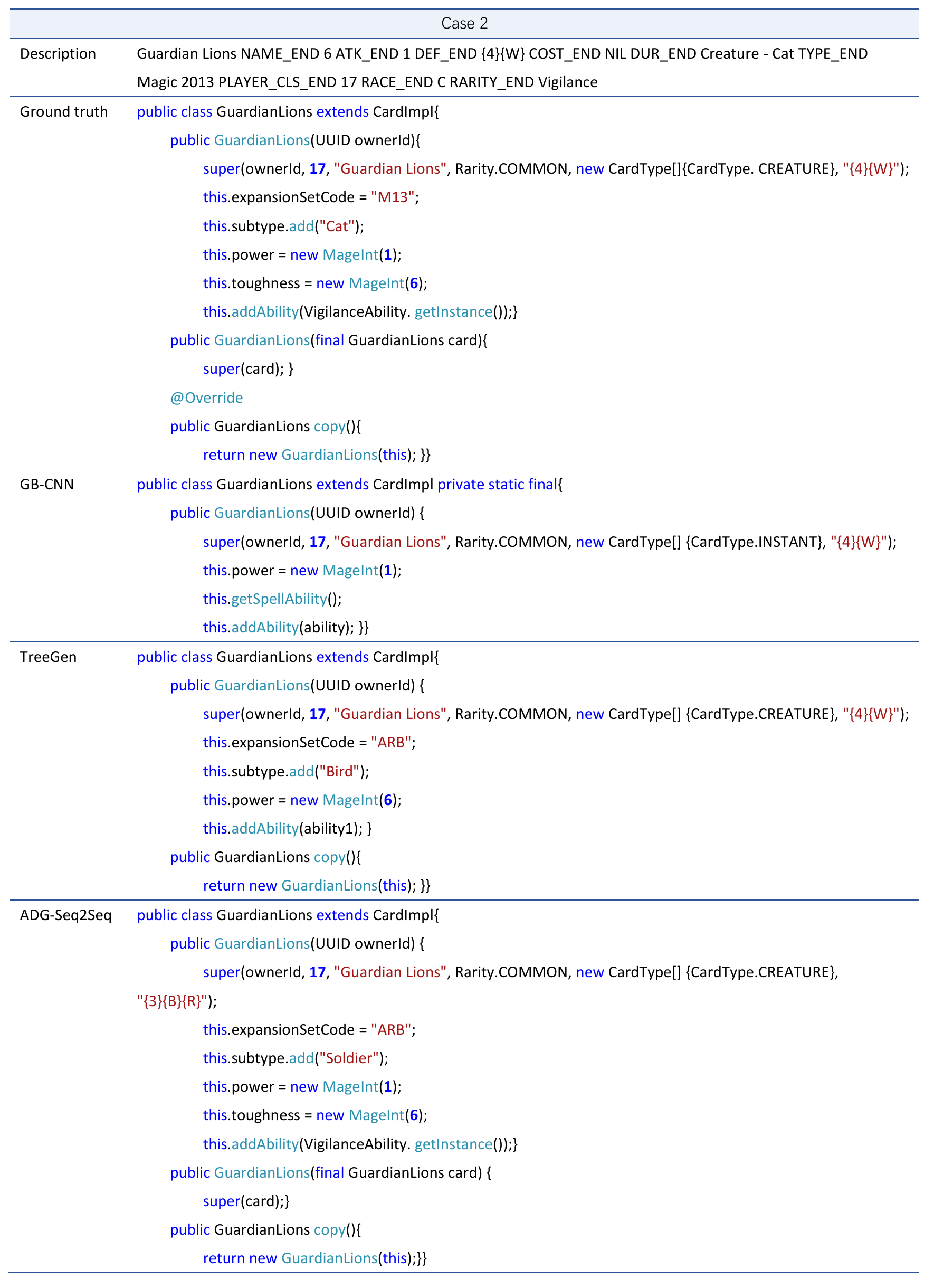}
    \caption{Code examples generated by GB-CNN, TreeGen and ADG-Seq2Seq on MTG.}
    \label{fig:case2}
    \end{figure}

\begin{figure}[!htbp]
    \centering
    \includegraphics[width=\linewidth]{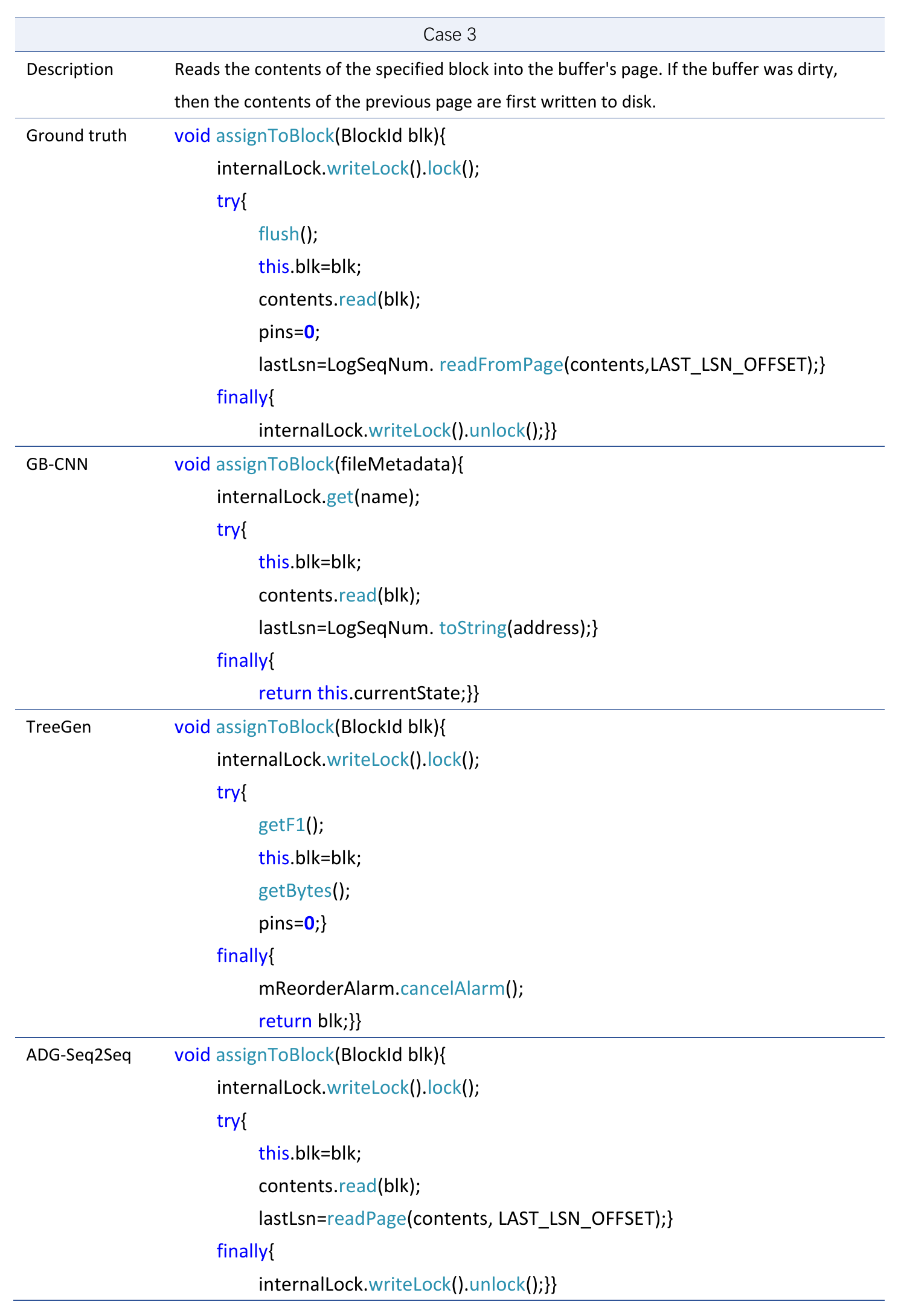}
    \caption{Code examples generated by GB-CNN, TreeGen and ADG-Seq2Seq on EJDT.}
    \label{fig:case3}
    \end{figure}

\begin{figure*}[!htbp]
		\centering
		\subfigure[HS: attention weights]{
			\includegraphics[height=0.60\linewidth]{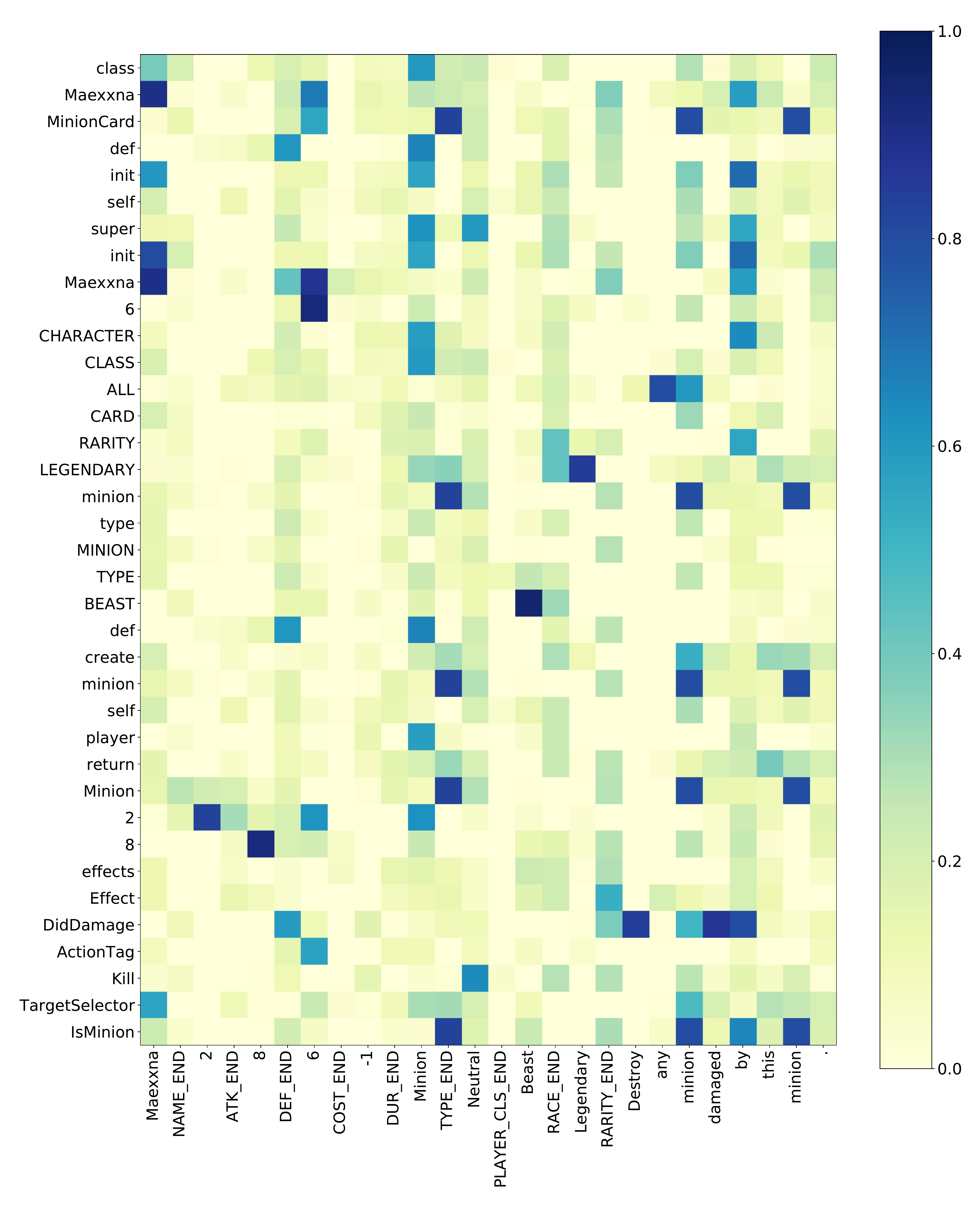}
			\label{caseattn1}
		}
		\subfigure[MTG: attention weights ]{
			\includegraphics[height=0.60\linewidth]{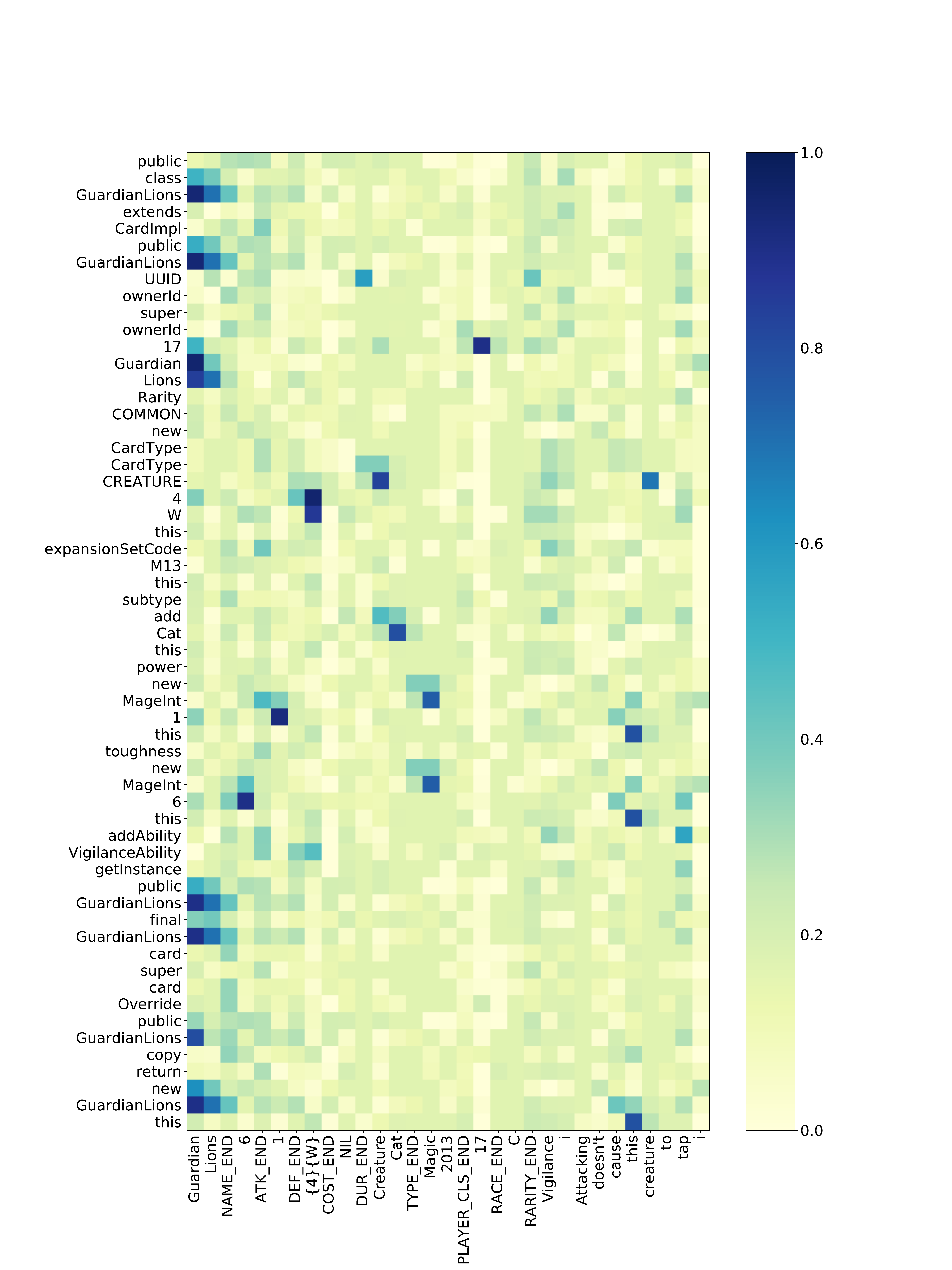}
			\label{case1attn2}
		}
		\subfigure[E-JDT: attention weights]{
			\includegraphics[height=0.60\linewidth]{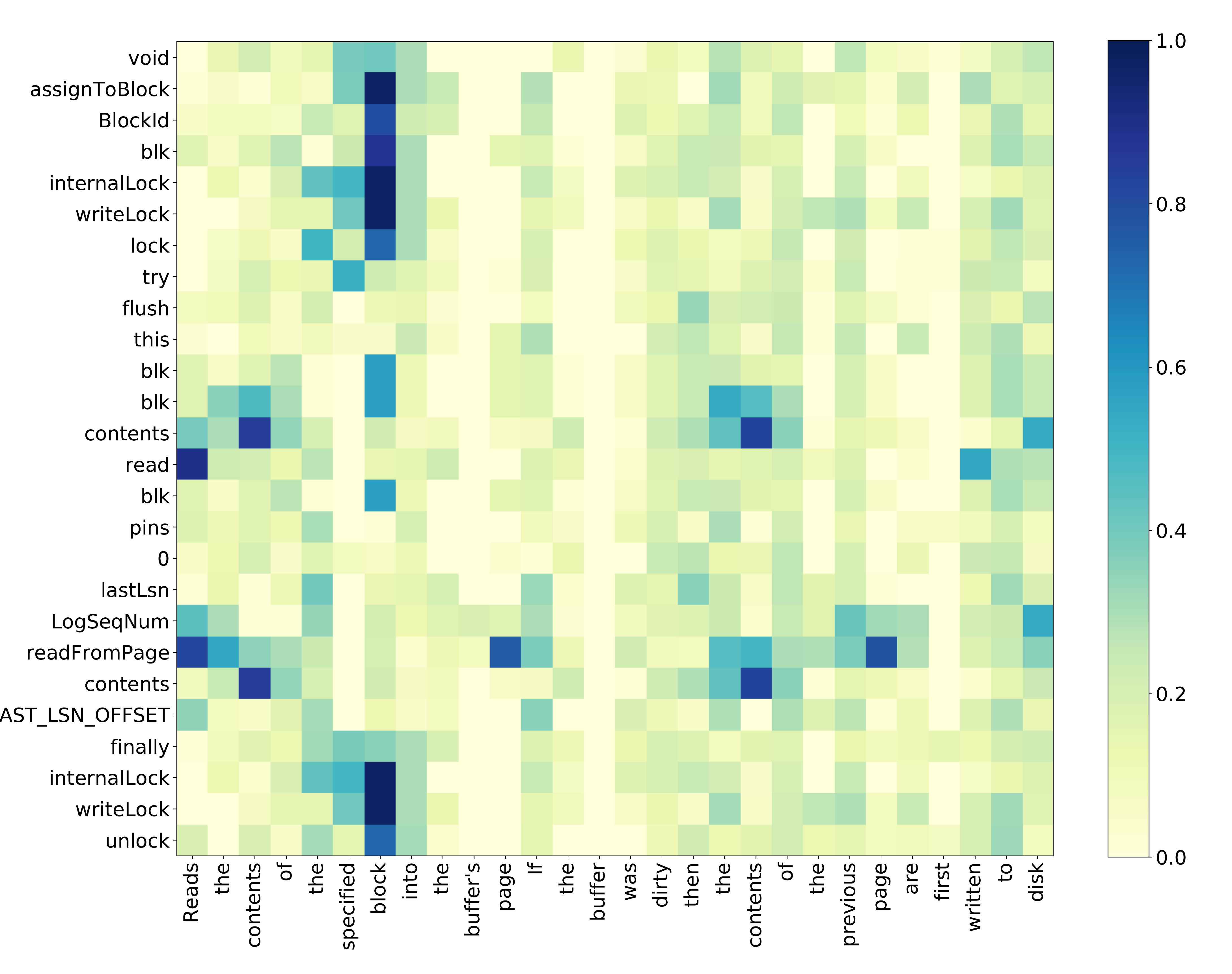}
			\label{case1attn3}
		}		
		\caption{\textbf{Attention weights of Cases 1-3 for ADG-Seq2Seq.}}
		\label{fig:case1attn}
	\end{figure*}

	We present an example for each dataset and select the results generated by GB-CNN, TreeGen, and ADG-Seq2Seq. Figs.~\ref{fig:case1}-\ref{fig:case3} show three cases of code descriptions, original code snippets regarded as the ground truth, and the corresponding predicted results. Fig.~\ref{fig:case1attn} visualizes attention in our proposed model for the target sequences. GB-CNN, TreeGen, and ADG-Seq2Seq predict reasonable and grammatical code even if they are not entirely consistent with the ground truth.

	\textit{\textbf{HS}}: The first example shown in \textit{Case 1} is a code snippet programmed in Python. The semi-structured description makes code prediction in this scenario relatively easy. The generated code is incomplete relative to the reference code but implements the required main functionality. The main problems with these prediction results include 1) omitting key functions/APIs and 2) using an incorrect number of arguments/parameters. Two key functions, ``\texttt{super().init\_}" and ``\texttt{Minion()}" are called in the reference code, but only one of them appears in the prediction results of GB-CNN, while TreeGen and ADG-Seq2Seq generate all functions. We also carefully inspect the number of arguments in each function. ADG-Seq2Seq generates the above two functions with the correct order and number of arguments, while GB-CNN and TreeGen fail in the prediction of ``\texttt{Minion()}". This demonstrates the importance of learning the invocation constraints implied in the ADG for the exact usage of the arguments. 
			
	Furthermore, the prediction of our model uses different parameter names that do not appear in the current description and uses new API combinations that cannot be found in the training samples. For example, our model generates ``\texttt{MINION\_Type\\.DRAGON}" and ``\texttt{Effect(SpellTargeted())}" instead of ``\texttt{MINION\_Type.BEAST}" and \\``\texttt{Effect(DidDamage())}" in the reference code. This indicates that the model has learned to generalize beyond simple translation. Although such a new substitution or combination reduces the accuracy of the model, it shows that the model has learned to generalize to some extent from global data dependencies among multiple semantically or structurally similar training card examples. The generalization and performance of our model are expected to improve as more training samples and tighter global data dependencies become available.
		
	\textbf{\textit{MTG}}: \textit{Case 2} is an example from the MTG dataset. Similar to the HS dataset, this card game dataset provides semi-structural descriptions but uses Java as its programming language, making the code structure more complex. For example, the number of functions and the frequency of API calls far exceed those in HS. Comparing the predicted results produced by the three models, it is clear that ADG-Seq2Seq generates more comprehensive code and covers the largest number of functions, 5 and 2 more than GB-CNN and TreeGen, respectively. Additionally, in contrast to the other two methods, ADG-Seq2Seq generates the correct invocation in a long chain of APIs, e.g., ``\texttt{this.addAbility(VigilanceAbility.getInstance());}". 
	This example shows that the model has learned the usage of common API calls through ADG embedding. 
		
	\textit{\textbf{E-JDT}}: The third dataset is E-JDT; it was extracted from the Eclipse plug-in library. As \textit{Case 3} shows, there are more data and call dependencies among the API methods than within the isolated card programs. Thus, combined with the API invocation order, our approach could predict the target code by virtue of the ADG. However, the challenge of this dataset for code generation is that the description is in the form of unstructured natural language, where function signatures cannot be learned directly from the tokens or keywords in the description, as would be possible on HS and MTG datasets. Therefore, satisfactory results are much more difficult to obtain for E-JDT. Nevertheless, our approach predicts method invocations such as ``\texttt{this.blk = blk}, \texttt{contents.read(blk)}.", by learning an applicable call sequence embedded in our model. Despite producing the incorrect class name, ADG-Seq2Seq predicts the method name in ``\texttt{lastLsn = readPage (contents, LAST\_LSN\_OFFSET);}," more reasonably than GB-CNN in ``\texttt{lastLsn = LogSeqNum.toString(address);}.". The results illustrated in Fig.~\ref{case1attn3} also show that attention weights produced by our model focus on sequences of APIs instead of variables, which exhibit a better capacity to capture the invocation structure 
	of the code.
	
	\subsection{Error Analysis}
		
	To understand the source of errors, we randomly sample 30, 100 and 100 failed cases from HS, MTG, and E-JDT datasets, respectively. We observe that errors causing these cases can be approximately categorized into four types: 1) Category I (C-I): argument-related errors, including argument dislocation, incorrect number, and incorrect name; 2) Category II (C-II): function-related errors, including incorrect function name and invalid function call; 3) Category III (C-III): variable-related errors, including incorrect assignment and incorrect variable name; and 4) Category IV (C-IV): partial implementation errors.

	Argument-related errors occur more often in E-JDT, probably because of the polysemy and ambiguity of natural language words. For function-related errors, our model sometimes recommends similar names instead of reference names, such as ``\texttt{ReadPage}" instead of its reference ``\texttt{ReadFromPage}" in \textit{Case 3} in Fig.~\ref{fig:case3}. 
	For variable-related errors, our model fails to predict the assignment statement, ``\texttt{pins = 0}", as illustrated by \textit{Case 3} in Fig.~\ref{fig:case3}. The reason is that such variables are not encoded by graph embedding, so their weights are easily weakened or even ignored in our joint learning approach. A feasible solution is to integrate the variables into the graph embedding network. We leave this step to future research. Finally, the main reason for partial implementation errors is that the missing code has no explicit requirement description. Fig.~\ref{fig:case1} provides a typical example in \textit{Case 1}: the missing code ``\texttt{actiontag()}" is not described in the reference. Notably, many partial implementation errors also stem from combined impacts of the above four types of errors.

	Table \ref{tab:error distributions} lists the error distributions of the other leading competitors, GB-CNN and TreeGen, for comparison. The overall distribution of errors for our method is generally consistent with those for the other methods. Specifically, the C-I and C-II error rates of our method are slightly lower than those of the other methods, while the C-III error rate for our method is somewhat higher than those for the other methods. The C-IV error rates for all methods are at a high level. The above statistical results indicate that our error rates are within an acceptable range compared to the other methods. Furthermore, attempting to reduce C-III and C-IV errors is an intriguing direction for future studies.
	
    \begin{table}[htbp]
      \centering
      \caption{Error distributions of GB-CNN, TreeGen and our method. }
        \begin{tabular}{clcccc}
        \toprule
              & Method      & \multicolumn{1}{l}{C-I} & \multicolumn{1}{l}{C-II} & \multicolumn{1}{l}{C-III} & \multicolumn{1}{l}{C-IV} \\
        \midrule
        \multirow{3}[2]{*}{HS} & GB-CNN & 16.7\% & 16.7\% & 3.3\% & 63.3\% \\
              & TreeGen & 16.7\% & 23.3\% & 3.3\% & 56.7\% \\
              & Ours  & 10.0\% & 20.0\% & 6.7\% & 63.3\% \\
        \midrule
        \multirow{3}[2]{*}{MTG} & GB-CNN & 16.0\% & 22.0\% & 5.0\% & 57.0\% \\
              & TreeGen & 17.0\% & 28.0\% & 4.0\% & 51.0\% \\
              & Ours  & 15.0\% & 25.0\% & 5.0\% & 55.0\% \\
        \midrule
        \multirow{3}[2]{*}{E-JDT} & GB-CNN & 27.0\% & 23.0\% & 5.0\% & 45.0\% \\
              & TreeGen & 25.0\% & 26.0\% & 5.0\% & 44.0\% \\
              & Ours  & 25.0\% & 20.0\% & 7.0\% & 48.0\% \\
        \bottomrule
        \end{tabular}%
      \label{tab:error distributions}%
    \end{table}%
	
	\subsection{Human Evaluation}
	To investigate the effectiveness of the generated code, similar to \cite{LiuZ}, we used a combination of manual verification and qualitative analysis to evaluate the quality of such code. For verification, nine evaluators assessed the results. Six were Ph.D. students in computer science at Shandong Normal University, and the other three were programmers hired from the Inspur Company\footnote{The Inspur Company is a leading cloud computing and big data service provider in China. More information is available at https://en.inspur.com/en/2488385/index.html.}. 
	The evaluators have two to five years of Python or Java programming experience.

	\subsubsection{Survey Procedure}
	We randomly selected 50, 100 and 100 results from the test sets of HS, MTG, and E-JDT, respectively. Each set of results was evaluated by two Ph.D. students and a senior programmer. The test evaluators were required to rate each generated code on a scale of 0 to 10 according to the code description. The ground truth code was also provided for reference. Specifically, a score of (0) indicates that the generated code is worthless, while a full score (10) means that the code perfectly matches the 
	description. Evaluators were encouraged to search the Internet for 
	necessary information if they encountered unfamiliar concepts.
	
	\subsubsection{Results}
	We obtained 750 scores from our human evaluation in which each generated code had three scores from three evaluators; average scores were used.
	
\begin{figure}[htbp]
		\centering
		\includegraphics[width=0.75\linewidth]{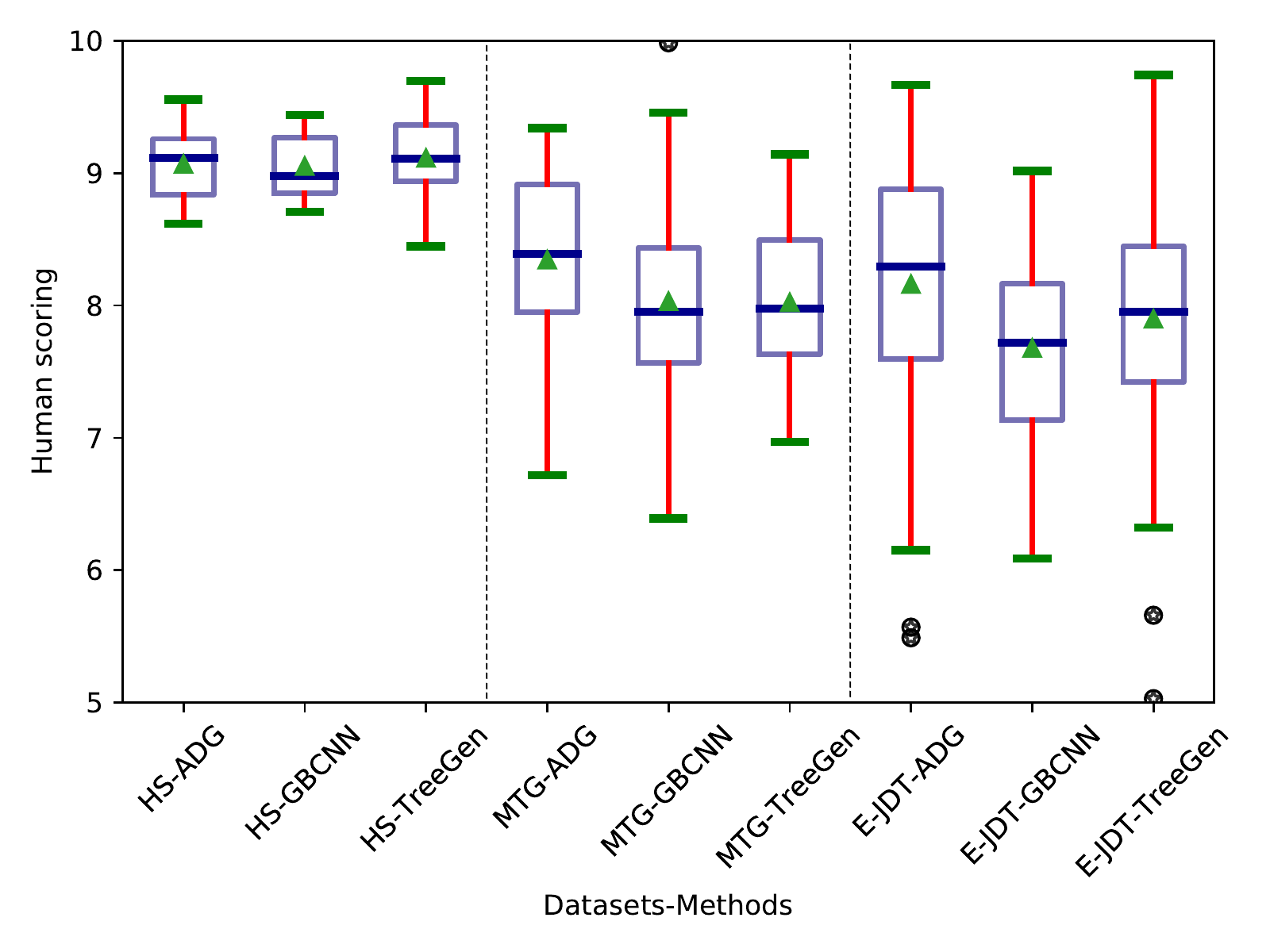}
		\caption{\textbf{Results of the human evaluation.}}
		\label{fig:box}
	\end{figure}

	Fig.\ref{fig:box} provides the distribution of scores. The boxplots of HS are relatively narrow, and their medians are higher than those of MTG and E-JDT. This result suggests that the quality of code generated on HS is generally high, while code generation tasks run on Java datasets appear to be more challenging, which is unsurprising since their complexity 
	(as shown in Table~\ref{tab:table2}) and quantity (in terms of the length of descriptions and code shown in Table~\ref{tab:table1}) far exceed those of HS. GB-CNN, TreeGen, and ADG-Seq2Seq all perform well on the HS dataset, but the quality of the generated code varies on the MTG and E-JDT datasets. Overall, ADG-Seq2Seq generates better code 
	than GB-CNN or TreeGen.
	
	More specifically, the medians of MTG and E-JDT are the same; however, the boxplots in these examples show different distributions of views, especially a noticeable height difference between boxplots. To uncover the reasons behind this discrepancy, we analyse factors such as the description and characteristics of the code. On the one hand, the description of E-JDT is unstructured and is expressed as a natural language statement. Therefore, 
	 due to the polysemy and fuzziness of natural-language words,
	it is difficult to perform well in every case. On the other hand, E-JDT has a broad range of code lengths in terms of both tokens and methods (see Table \ref{tab:table1}), which increases the uncertainty in predictions.

	To assess the inter-rater agreement of scores, kappa coefficients~\cite{SimJ} were calculated by aggregating the scores of each group. The final kappa coefficients for the three groups were calculated to be 0.73, 0.69, and 0.67, respectively.  According to the classification of the kappa coefficient, scores rated by evaluators agree to substantial degree and can be used for experimental analysis.

	In summary, the human evaluation results confirm that the proposed model outperforms the compared methods.

	\section{Discussion}
		\label{sec:threats}%

	\subsection{Specialty of the ADG Embedding Algorithm}

	The ADG is proposed to model the semantic structure of a program. We define special reachability (see Definition 1) in an ADG that aims to describe two types of dependencies in API sequences, i.e., API invocation constraints and API invocation order. If there are APIs in the API sequence with more than one input parameter, a sequence containing these APIs will be in the form of a DAG rather than a chain. Graph embedding for generic graphs does not consider the special reachability expressed by sequences in the DAG form or provide a targeted strategy for preserving API dependencies during the embedding process. In contrast to other graph embedding algorithms, the ADG embedding algorithm addresses the above issues as follows.
	
	\begin{itemize}
		\item Encoding API invocation constraints: During embedding, we first classify (forward and backward nodes) and group (providers of different types of parameters) the neighbour nodes of the current API according to their parameter types. The virtualization operation is then used to aggregate the nodes of the same group. Finally, the algorithm sorts the aggregated nodes topologically. Through the above classification, grouping, and virtualization operations, the number of parameters, parameter types, and input and output that the current API depends on can be determined. Therefore, our algorithm can differentiate this information and encode the implicit API invocation constraints into each API node.
		
		\item Encoding the API invocation order: To encode the order of API calls, we first input the neighbour nodes of each API into LSTM after grouping and topological ordering. The topological order is then used as the temporal order for aggregation. Existing research~\cite{SutskeverI} has shown that LSTM-based sequential aggregation is an effective way to encode sequence order information.
	\end{itemize}
	
    In summary, through an in-depth analysis of the semantic structure of ADG-expressed programs, we introduce targeted improvement strategies such as classification, grouping, topological ordering, and sequential aggregation in the neighbour aggregation-based graph embedding algorithm. The API invocation constraints and API invocation order are then effectively encoded to guide the subsequent learning and code generation tasks. The existing graph embedding  techniques (such as GCN or GraphSAGE) do not consider such strategies for fusing and preserving the above two types of API dependencies. 
	
	\subsection{The Option of Applying AST in Conjunction with ADG}
	\label{sec:AST+ADG}
	AST-based approaches have performed well. In this paper, we attempt a different approach to explore other possibilities for code generation. A straightforward way is to represent the AST and ADG of the code as two 1 $\times$ n-dimensional embedding vectors and then aggregate them into a single vector. The aggregated vector is used to bootstrap the decoder to generate the code. This approach is similar to the research of Wan et al~\cite{wan2019multi}. The researchers treat the token sequence, AST and CFG as different modalities of code, and then represent each modality with embedding features via LSTM, Tree-LSTM, and GGNN. Specifically, they represent each modality as three 1 $\times$ 512 vectors and fuse them into a single vector for code retrieval. In contrast to the multimodal fusion strategy described above, the AST- or ADG-based code generation methods decompose ASTs or ADGs into paths or nodes and obtain fine-grained path embedding or node embedding representations. Existing research has shown that fine-grained embedding representations of code can aid in learning a more sophisticated syntactical or semantic structure.  
	
	Although incorporating AST into the proposed model can compensate for the loss of missing syntactical structure information, it is challenging to integrate such multimodal fine-grained embedded representations effectively. We plan to investigate this topic in our future studies.

	\subsection{Limitations of the Proposed Approach}
	\label{sec:Discussion}
	Our model still has some limitations. For example, the model must create an ADG, which could be time consuming if the size of dataset size is large.
	For the HS dataset, an epoch takes 32 seconds, while for MTG and EJDT, the training times are 2100+ seconds and 61,000+ seconds, respectively. In addition, when an ADG is being built, different API extraction methods should be designed for various programming languages. Furthermore, some nodes are more generic and have more neighbouring nodes, which require more time for preprocessing and post-training. In the future, we plan to follow the practice of GraphSAGE \cite{HamiltonW} to build graph models by neighbourhood sampling.
	
	In general, the accuracy of our model is still not satisfactory for practical use. In Section 8.2, we discuss several types of errors that already exist in the model. This suggests that our model is weak at learning vector representations of synthetic identifiers that are out of vocabulary (OOV).  
	In addition, the ability to reason about implicit semantics (e.g., descriptions that do not contain code keywords) is unsatisfactory. 
	In our future research, we plan to study the composition and splitting patterns of identifiers and to improve the encoding of various identifiers.

	\subsection{Threats to Validity}
	We have identified the following threats to validity.
	
	\noindent{\bf Internal Validity.} \quad A threat to internal validity stems from biases and errors in replications of other models. Although we used the source code provided by the authors for each model, the specific parameters reported in the respective papers may be unsuitable for our datasets. To mitigate this threat, we contacted some of the authors of related studies and used  parameter settings that could achieve optimal results. The parameter settings of the models were also optimized through repeated experiments. Therefore, we believe that there is only a minor threat to internal validity.
		
	\noindent{\bf External Validity.} \quad Threats to external validity include the quality and representativeness of datasets. The datasets used in this paper inevitably contain noise that may affect the prediction results. Although we attempted to eliminate noise through pre-processing, we cannot guarantee that all data noise has been eliminated. In addition, models in this paper are tested on only Python and Java datasets, which are not representative of a wide range of programming languages. Moreover, some programming languages (e.g., Java) involve third-party class libraries that contain packages used to extend functionality and cater to various types of applications. Such libraries are quite useful for developers working on different projects. Incorporating the API dependencies of such libraries into our model could further improve the quality of generated code.  Currently, we did not specifically collect third-party class libraries. 
	This may affect the performance of the proposed method to some extent. In the future, we plan to extend our approach to other programming languages and incorporate more third-party libraries. 
	Since our model is trained on specific datasets, its generalization ability could be affected. In contrast to transductive algorithms such as GCN, our ADG algorithm is an extension of GraphSAGE, which is a framework for inductive learning that efficiently generates unknown node embeddings using the attribute information of nodes. Our approach has demonstrated the ability to generate code pieces that do not appear in the codebase in our experiments. We present a concrete example to illustrate this in detail in our case study of the HS dataset (see Section 8.1). Therefore, our model is generalizable to a certain extent. However, the generalizability of our model can be enhanced if more training samples (a larger codebase) are provided.

	\section{Conclusions}
	\label{sec:Conclusion}
	
	In this paper, we have presented a novel ADG-based Seq2Seq approach to generate target code automatically based on textural program descriptions. This approach transforms all API signatures mined from the specific codebase into a large-scale ADG, where nodes represent API methods and edges with tags represent API dependencies among nodes. We design an embedder that embeds an ADG to exploit the global structural information it captures. By integrating the embedder into an attention-based Seq2Seq model, the proposed approach can automatically learn the ADG structure and generate the target code. We have provided quantitative and qualitative comparisons of the proposed method with state-of-the-art techniques, and the proposed method demonstrated significant improvements in terms of all eight metrics on the two Java datasets and in terms of five metrics on the Python dataset.
		
	Our tool, the datasets used in our experiments, and the detailed experimental results are all available at \textbf{\url{https://github.com/RuYunW/ADG-Seq2Seq}}.

    Our study can be regarded as one of the first steps towards neural code generation.
	Although our results are significantly better than baseline methods, further improvement is needed prior to practical application.
		In addition to the research directions identified in Section \ref{sec:threats},
		we will further improve the accuracy of the proposed model as follows:
    \begin{itemize}
        \item 
    	First, we plan to replace the LSTM with the Transformer to reshape the network architecture of the proposed model. Although the embedder is currently integrated within the model, it is feasible to make it standalone and incorporate it with the Transformer. The most crucial feature of the Transformer over the LSTM is that it enables the model to batch-process input data and more effectively alleviate the long-range dependency problem. However, due to the strong influence of the API invocation order on the code, the proposed model based on the ADG embedding is more dependent on temporal order information than are other approaches. The Transformer, based on position encoding, is unable to capture these timing relationships well~\cite{shiv2019novel}. Therefore, embedding an ADG into the Transformer straightforwardly suffers the limitation of the loss of temporal order information. To effectively integrate the ADG embedder and the Transformer, we will design unique mechanisms to address this challenge.  
	\item Second,	we plan to combine the tasks of automatic code generation and automatic program summary generation as a dual model \cite{Dual} in which the proposed model can be extended and improved by jointly training the two related tasks concurrently. 
	\item Third, we plan to extend the proposed approach to support more programming languages and incorporate the characteristics of language features. 
	\item Finally, we plan to apply ADG to different tasks such as code classification \cite{PhanA,zhang2019novel}, retrieval \cite{NaiyanaS} and summarization \cite{hu2020deep,Dual,zhang2020retrieval}. 
	In this study, we have focused on the application of ADG in code generation. The proposed ADG could also be applied to other tasks such as code classification. Some of the initial results have been reported on our project website\footnote{https://github.com/RuYunW/ADG-Seq2Seq/tree/master/code\%20classification}. We will explore this topic further in our future research.

	\end{itemize}	
	
	\begin{acknowledgements}
		This work was supported in part by the National Natural Science Foundation of China under Grant Nos. 61602286 and 61976127, in part by the Shandong Key Research and Development Program under Grant 2018GGX101003, and in part by the Shandong Province Higher Educational Science and Technology Program under Grant J16LN09.
	\end{acknowledgements}

	\bibliographystyle{unsrt} 
	\bibliography{ref}   

	\newpage
	\begin{appendices}

	\section{More Details about the Experimental Data and Settings Used in Comparisons}

	In Section \ref{ARQ1}, we describe our experimental data. For models with existing experimental data, we use the experimental results presented in the original paper\footnote{Most of the replication results are consistent with those reported in the original paper, and a few (BLEU scores for ASN and SNM) are slightly lower (the difference is about 0.8\% in BLEU). Considering the stochastic nature of deep learning, the differences in these results are acceptable. We choose to present the original results for each method in the paper, which help to maintain consistency with the results reported by the related studies \cite{YinP, RabinovichM, SunZ, SunZ2}. }. For those without, we retrain the model with the default parameters provided in the original paper and calculate the experimental results under different metrics.  
	
	More specifically, we compared six models: the attention-based Seq2Seq model~\cite{NeubigG} and the Transformer model~\cite{VaswaniA} were used as baselines; and SNM~\cite{YinP}, ASN~\cite{RabinovichM}, GB-CNN~\cite{SunZ}, and TreeGen~\cite{SunZ2} were used as the state-of-the-art (SOTA) methods.

All evaluation results of the baselines (the attention-based Seq2Seq model and the Transformer model) were calculated by training the models  on HS, MTG and E-JDT datasets (see Table 4 and Fig. 9), and on the split datasets (see Fig.s 12 and 13).

    In the original publications of the four SOTA methods (SNM, ASN, GB-CNN, and TreeGen) \cite{YinP, RabinovichM, SunZ, SunZ2}, HS was chosen as one of the experimental datasets, which is also used in our experiments.
The scores of BLEU and Acc metrics on HS in Table 4 and Fig. 9 were obtained directly from the respective papers (see Table 4 and Fig. 9 (a)-(b)). Since the papers describing SNM, ASN, GB-CNN, and TreeGen did not provide scores apart from Acc and BLEU, we have to retrain these models to obtain the remaining metric scores (see Table 4 HS: F1-RIBES). We use the same hyperparameter values as those described in the respective paper. As the authors of SNM provided the trained model, we calculated the scores on HS by using their trained model.

For datasets MTG, E-JDT, and the split datasets, 
since the papers describing the SOTA methods did not provide any evaluation scores, we retrained their models based on the hyperparameters provided in the original papers, and calculated scores for all metrics on these datasets 
(see Table 4, Fig 9 (i)-(x), Figs. 12 and 13).

	In detail, the experimental data and settings for each model are as follows:

	\begin{enumerate}
	    \item All evaluation results of the attention-based Seq2Seq model and the Transformer model are calculated by us. The attention-based Seq2Seq model is our basic model. The Transformer is a well-known NMT model in the field of NLP, and the relevant code is available at \url{https://github.com/jadore801120/attention-is-all-you-need-pytorch}. 
	    
	    \item The SNM model was proposed by Yin et al. \cite{YinP} and its code is available at \url{https://github.com/pcyin/NL2code}. We use the ACC and BLEU values from the paper and calculate other metrics by testing the results on HS from the trained model they provided. For MTG and E-JDT, all metrics' values are determined by retraining the model. The size of all embeddings is 128, except for node type embeddings, the size of which is 64. Dimensions of RNN states and hidden layers are 256 and 50, respectively.    
	    
	    \item The ASN model was proposed by Rabinovich et al. \cite{RabinovichM}. This paper does not provide the source code for the implementation, but since this study has had a large impact, we have been able to ascertain the ASN code implemented by other scholars via PyTorch at \url{https://github.com/xiye17/torchASN}. Hence, the Acc and BLEU scores were copied from the cited paper. Other scores on HS and scores on MTG and E-JDT were computed by us by retraining the model. For each experiment, all feedforward and LSTM hidden dimensions were set to the same value. We selected the dimension from \{50, 75, 100, 125, 150\} for all datasets. The dimensionality used for the inputs to the encoder was set to 100 in all cases. We applied dropout to non-recurrent connections of vertical and horizontal LSTMs, selecting the noise ratio from \{0.2, 0.3, 0.4, 0.5\}. All parameters were randomly initialized using Glorot initialization.
	    
	    \item GB-CNN is a CNN-based model proposed by Sun et al. \cite{SunZ}. We copied the StrAcc and BLEU figures for HS. GB-CNN is also an open-source model available at \url{https://github.com/zysszy/GrammarCNN}. We retrained the model following default settings to evaluate other metrics on all three datasets. For the input descriptions, we replaced all punctuations with a space; all letters were lowercase. For the neural network, we set the number of CNN layers L to 21, where the bottom layer does not have skipping connections. The layers of difference CNN modules were set to the same dimension and chosen by validation from \{128, 192, 256\} for each predictor network. We applied dropout (drop rate= 0.5) and $l_2$ penalty to regularize the fully connected layers. The network was trained by the Adam optimizer with default hyperparameters. 
	    
	    \item TreeGen was proposed by Sun et al. \cite{SunZ2} and is a Transformer-based model. Its implementation is available at \url{https://github.com/zysszy/TreeGen}. The Acc and BLEU numbers were copied from the cited paper, and other evaluation results were obtained by training the model on the three datasets. For neural networks, we set the number of NL reader layers $N_d$ = 6, and $N_1$ = $N_2$ = 5 for the AST reader as well as the decoder. The size of all embeddings was 256. The hidden sizes were all set to 256 except for each fully-connected layer and the first layer, which had 1024 dimensions. We applied dropout after each layer (including attention layers, the gating mechanism's layers, convolutional layers, and fully-connected layers, where the drop rate was 0.15). The model was optimized by Adafactor with default parameters. 
	\end{enumerate}

\end{appendices}

\newpage
\vskip 2 cm\begin{bibliographystyle} [\parpic[l]{\includegraphics[width=1in,height=1.2in,clip,keepaspectratio]{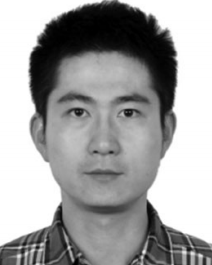}}
	{\textbf{Chen Lyu} received his Ph.D. from the Institute of Computing Technology, Chinese Academy of Sciences, Beijing, China, in 2015. He is currently an associate professor with the School of Information Science and Engineering, Shandong Normal University, Jinan, China. His research interests include program comprehension, software maintenance and evolution, and service computing.}
	
\end{bibliographystyle}

\vskip 2 cm\begin{bibliographystyle} [\parpic[l]{\includegraphics[width=1in,height=1.2in,clip,keepaspectratio]{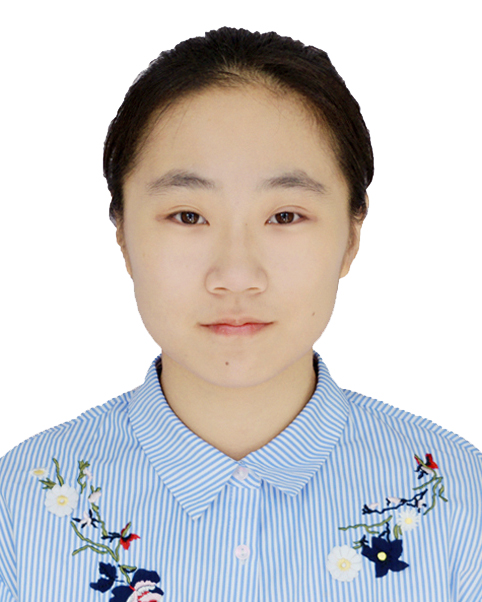}}
	{\textbf{Ruyun Wang} is pursuing a bachelor's degree in the School of Information Science and Engineering, Shandong Normal University, Jinan, China. Her research interests include code generation and component-based software development.}
	
\end{bibliographystyle}

\vskip 2 cm\begin{bibliographystyle} [\parpic[l]{\includegraphics[width=1in,height=1.2in,clip,keepaspectratio]{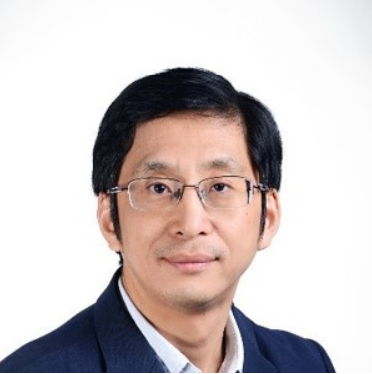}}
	{\textbf{Hongyu Zhang} is currently an associate professor with the University of Newcastle, Australia. Previously, he was a lead researcher at Microsoft Research Asia and an associate professor at Tsinghua University, China. He received his Ph.D. from the National University of Singapore in 2003. His research is in the area of software engineering, in particular, software analytics, data-driven software engineering, maintenance, and reuse. He has published more than 160 research papers in reputable international journals and conferences, including TSE, TOSEM, ICSE, FSE, ASE, ISSTA, POPL, AAAI, IJCAI, KDD, ICSME, ICDM, and USENIX ATC. He received four ACM Distinguished Paper awards. He has also served as a program committee member/track chair for many software engineering conferences.  He is a Distinguished Member of ACM. More information about him can be found at: \url{https://www.newcastle.edu.au/profile/hongyu-zhang}.}
	
\end{bibliographystyle}

\vskip 1 cm\begin{bibliographystyle} [\parpic[l]{\includegraphics[width=1in,height=1.2in,clip,keepaspectratio]{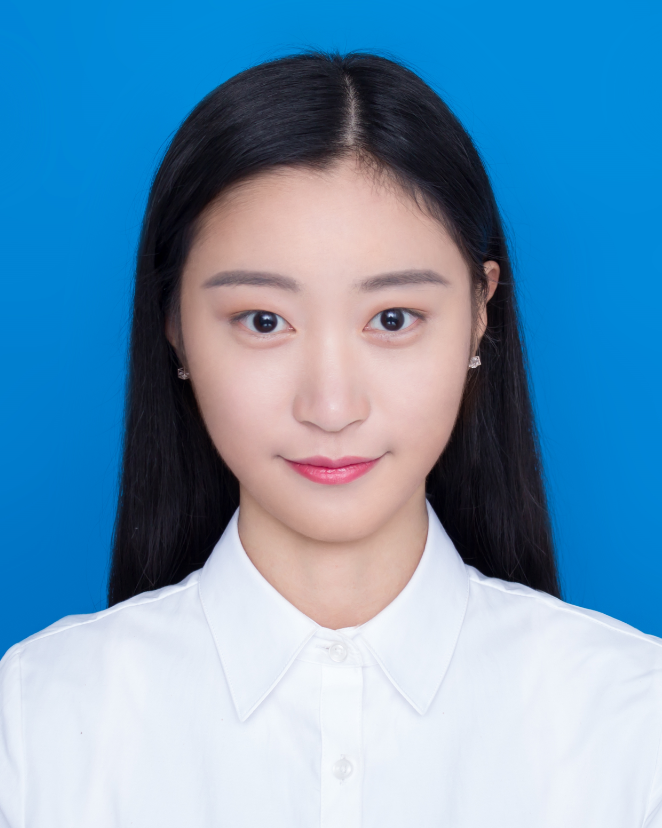}}
	{\textbf{Hanwen Zhang} 
received a master's degree in software engineering from Shandong Normal University, Jinan, China, supervised by Chen Lyu in 2020. She is currently working in the Big Data Center of Shandong province.  Her research interests include software reuse, natural language processing, and automatic code comment generation.}
	
\end{bibliographystyle}

\vskip 2 cm\begin{bibliographystyle} [\parpic[l]{\includegraphics[width=1in,height=1.2in,clip,keepaspectratio]{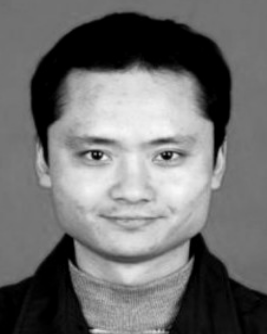}}
	{\textbf{Songlin Hu} received his Ph.D. from Beihang University, Beijing, China, in 2001. He was promoted to associate professor in 2002 in the Institute of Computing Technology, Chinese Academy of Sciences, Beijing, China, and is now a professor in the Institute of Information Engineering, Chinese Academy of Sciences, and School of Cyber Security, University of Chinese Academy of Sciences, Beijing, China. His research interests mainly include natural language processing, API mining, service computing, etc.}
	
\end{bibliographystyle}

\clearpage

\end{document}